%% file: Bianco-Boente-Chebi-v2.tex
\documentclass[12pt]{article}
\usepackage[english]{babel}
\usepackage{enumerate}
\usepackage{enumitem}
\usepackage{fancyhdr}
\usepackage{amssymb}
\usepackage{amsthm}
\usepackage{geometry}
\usepackage{amscd}
\usepackage{amsmath}
\usepackage{amsfonts}
\usepackage{graphicx}
\usepackage{amsmath, amsfonts, amssymb}
\usepackage{url}

\usepackage{algorithm}
\usepackage{algorithmic}
\usepackage[font=footnotesize]{caption}
\usepackage{tabularx}

\usepackage{pgf,pgfarrows,pgfnodes,pgfautomata,pgfheaps,pgfshade}
\usepackage{lscape}

\usepackage{hyperref}
\usepackage{geometry}

\usepackage{color}
\newtheorem{theorem}{Theorem}[section]
\newtheorem{lemma}[theorem]{Lemma}
\newtheorem{corollary}[theorem]{Corollary}

\newtheoremstyle{mythm}%
{3pt}
{3pt}
{}
{}
{\bfseries}
{}
{.5em}
{}%

\theoremstyle{mythm}
 \newtheorem{remarkex}{Remark}[section]
\newenvironment{remark}
  {\pushQED{\qed}\remarkex}
  {\popQED\endremarkex}

\newlength\tindent
\newcolumntype{L}[1]{>{\raggedright\arraybackslash}p{#1}}
\newcolumntype{C}[1]{>{\centering\arraybackslash}p{#1}}
\newcolumntype{R}[1]{>{\raggedleft\arraybackslash}p{#1}}

\include{definiciones}

\begin{document}

\title{Penalized robust estimators in logistic regression with  applications to  sparse models}
\author{Ana M. Bianco$^a$, Graciela Boente$^a$  and  Gonzalo Chebi$^a$ \\
\small $^a$ Facultad de Ciencias Exactas y Naturales, Universidad de Buenos Aires and CONICET, Argentina
}
\date{}
\maketitle

\small 
\begin{abstract}
Sparse covariates are frequent in  classification and regression problems and in these settings the task of variable selection is usually of interest.  As it is well known,  sparse statistical models correspond  to situations where there are only a small number of non--zero parameters and for that reason, they are  much easier to  interpret than  dense ones. In this paper, we focus on the logistic regression model and our aim is to address robust and penalized estimation for the regression parameter. We introduce a family of penalized weighted $M-$type estimators for the logistic regression parameter that are stable against atypical data. We explore different penalizations functions and we introduce the so--called Sign penalization. This new penalty has the advantage that  it depends only on one penalty parameter, avoiding arbitrary tuning constants.
We  discuss the variable selection capability of the given proposals as well as their asymptotic behaviour. Through a numerical study, we compare the finite sample performance of the proposal corresponding to different penalized estimators either robust or classical, under different scenarios. A robust cross--validation criterion is also presented. The analysis of two  real data sets enables to investigate the stability of the penalized estimators to the presence of outliers.

\end{abstract}

\section{Introduction}
   Sparse regression models  assume that the number of actually relevant predictors, $k$, is lower than the number of measured covariates.   Hastie \textsl{et al.} (2015) describe that \textsl{a sparse
statistical model is one in which only a relatively small number of parameters (or predictors) play an important role}, leading to models that are much easier to  interpret than  dense ones. This type of models has raised a paradigm shift in Statistics, since the traditional approach to classical issues such as regression or classification assumes that no restrictions are imposed when estimating the parameters.  In particular, for linear regression models, least squares estimators have all their coordinates non--null even when dealing with sparse models. In such a situation, selecting  relevant variables becomes an essential issue and for this task, different selection criteria, such as the traditional step--wise ones, have been developed.   As discussed in Fan and Li (2002) best subset selection suffers from several drawbacks, including their lack of stability, which are avoided using penalized least squares.   Penalized regression estimators are a useful tool when the practitioner is interested in automatic variable selection, see, for instance,  Efron and Hastie (2016) for an overview of adapted inference methods. 

More research has been developed in these directions in the area of linear regression models, where the least squares estimator will typically over-fit the data  becoming unreliable in presence of multicollinearity or when a sparse model is assumed.  
Different strategies have been proposed to overcome these difficulties and the experience show that even in these cases good predictions may be possible. For multi-collinear predictors, Hoerl and Kennard (1970) propose Ridge estimators in order to reduce the variance by introducing an $\ell_2$ penalty as in Tikhonov (1963). Ridge estimators do not produce sparse models, but the idea beneath regularization is useful and it is a well-suited tool to deal with sparse regression by choosing an adequate penalization. For instance, LASSO penalty  bets on the sparsity principle, which assumes that only a few $k$ covariates are enough to explain the response variable. As it is well known, the $\ell_1$ regularization, related to LASSO estimators, is effective for variable selection, under suitable conditions,  but tends to choose too many features. Zou and Hastie (2005)  introduced an alternative regularization, namely the Elastic Net penalty, which combines both $\ell_1$ and $\ell_2$ norms. Elastic Net preserves the sparsity of LASSO and  maintains  some of the desirable predictive properties of Ridge regression.  Tibshirani (1996), Fan and Li (2001) and Zhang (2010) proposed different penalties leading to sparse estimators following these strategies. See, for example, Hastie \textsl{et al.} (2015) for more details.

Logistic regression is a widely studied problem in statistics and has been useful to classify data. It is well known that in the non--sparse scenario the maximum likelihood estimator (MLE) of the regression coefficients is very sensitive to outliers, meaning that we cannot accurately classify a new observation based on these estimators, neither identify those covariates with important information for assignation. Robust methods for logistic regression bounding the deviance have been introduced and discussed   in Pregibon (1982) and Bianco and Yohai (1996). In particular, Croux and Haesbroeck (2003)   introduced a loss function that warranties the existence of the resulting robust estimator  when the maximum likelihood estimators does exist.   The proposal due to Basu \textsl{et al.} (2017) on the basis of minimum divergence can also be seen as a particular case of the Bianco and Yohai (1996) estimator with a properly defined loss function. Other approaches were given in   Cantoni and Ronchetti (2001) and Bondell (2005, 2008). However, these methods are not reliable under collinearity and   they do not allow for automatic variable selection when only a few number of covariates are relevant.  
 The previous  ideas on regularization can be directly extended to logistic regression.

Recently, some robust estimators for logistic regression in the sparse regressors framework have   been   proposed in the literature. 
Among others, we can mention Chi and Scott (2014) who considered a least squares estimator with a Ridge and Elastic Net penalty and Kurnaz \textsl{et al.} (2018) who proposed estimators based on a trimmed sum of the deviances with an Elastic Net penalty. It is worth noticing that the least squares estimator  in logistic regression corresponds to a particular choice of the loss function considered in Bianco and Yohai (1996).  Finally, Tibshirani and Manning (2013) introduced a real--valued shift factor to  protect against the possibility of mislabelling, while Park and Konishi (2016) considered a weighted deviance approach with weights based on the Mahalanobis distance computed over a lower--dimensional principal component space and includes an Elastic Net penalty.   In these circumstances, the statistical challenge of obtaining sparse and robust estimators for logistic regression that are computationally feasible and provide variable selection should be complemented with the study of their asymptotic properties.  Most of the asymptotic results for robust sparse estimators have been given under the linear regression model (see, for example, Smucler and Yohai, 2017) or when considering a convex loss function (see, for instance, van de Geer and M\"uller, 2012). Recently, Avella-Medina and  Ronchetti (2018)   treats the situation of general penalized $M-$estimators in shrinking neighbourhoods, when the parameter dimension $p$ is fixed, i.e., does not increases with the sample size. In this setting,  the penalty function considered by these authors is a deterministic sum of univariate functions.

In this paper, we introduce a general family of robust estimators in the sparse scenario that involves both a loss and a weight function to control influential points and also a general penalty term to produce sparse estimators. At this point, the choice of the penalty does matter. 
It is worth noticing that  in our objective function the loss function keeps  bounded the terms related to the deviance. For this reason, on a second side,  it  seems wise to consider a bounded penalty, otherwise, the regularization term tends to dominate in the minimization problem. In this sense, SCAD  or MCP, due by Fan and Li (2001) and Zhang (2010), respectively,  are appealing choices which must be tuned by the user with an additional parameter. Keeping these ideas in mind, we also introduce the new regularization Sign, that is bounded and, unlike  SCAD  and  MCP, does not depend on an extra parameter. This new penalty acts like the LASSO penalty applied to the direction of the regression vector, that is why,  it does not shrink the estimated coefficients to 0 as   LASSO does. 

 A primary focus of this paper is to provide a rigorous theoretical foundation for our approach to robust sparse logistic regression when the dimension of the covariates is fixed. In a first step, under very general conditions, we establish consistency results for a wide family of penalty functions, which may be random to include   the adaptive LASSO (ADALASSO) penalty.   Besides, to study variable selection and oracle properties, we distinguish  the case  of  Lipschitz functions, such as the Sign, from that of penalties that can be written as a sum of twice differentiable univariate functions, eventually random, such as SCAD and MCP and ADALASSO. 
  These two points make a difference with respect to Section 2 in Avella-Medina and  Ronchetti (2018).  It should be highlighted that a similar strategy to the one proposed herein could be followed in the high dimensional scenario. However, in the case where the dimension $p$ increases with the sample size $n$, particular considerations and developments should be done in order to obtain theoretical properties. This interesting topic will be part of future research.

The rest of this paper is organized as follows. In Section \ref{sec:prelim}, we recall  basic ideas 
of robust estimation in logistic regression in a non--sparse scenario. 
In Section {\ref{sec:proposal}},  the robust penalized logistic regression estimators are introduced, while   Sections {\ref{sec:consist}} and {\ref{sec:asdist}} summarize the asymptotic properties of the proposal.  Section {\ref{sec:monte}}   reports the results of a Monte Carlo study and describes an algorithm to effectively compute the estimators. In Section {\ref{sec:realdata}}, we present the analysis of two real datasets, while   Section \ref{sec:conclusiones} contains some concluding remarks. Proofs are relegated to the Appendix.

\section{Preliminaries: Robust estimators in the non--sparse setting}{\label{sec:prelim}}

Throughout this paper, we consider a logistic regression model, that is, we have a sample of i.i.d. observations   $\left(y_i, \bx_i \right)$, $1\le i \le n$      such that  $\bx_i \in \real^p$, $y_i\in \{0,1\}$ is a binary variable such that $y_i|\bx_i \sim Bi(1, \pi_{0,i})$, where  
\begin{equation*}
\pi_{0,i}= F(\bx_i \trasp \bbe_0)=\frac{\exp\left(\bx_i \trasp \bbe_0\right)}{1+\exp\left(\bx_i \trasp \bbe_0\right)}\,,  
\end{equation*}
with $\bbe_0 \in \real^p$  the true logistic regression vector.

Recall that the maximum likelihood estimator of $\bbe_0$ is defined as
$
\wbbe_{\mle} = \argmin_{\bbech}  \sum_{i = 1}^n d(y_i, \bx_i\trasp\bbe)$, 
where $d(y,t) =  - \log(F(t))  y - \log(1 - F(t)) (1-y)$ is the deviance function. 
  A corrected version of  Pregibon's (1982) proposal is given in  Bianco and Yohai (1996). More precisely, let $\rho:\real_{\ge 0}\to \real$ be  a bounded, differentiable and  nondecreasing function with derivative $\psi = \rho^{\prime}$. The  $M-$estimators defined in  Bianco and Yohai (1996) are given by
\begin{equation}
\wbbe= \argmin_{\bbech} \frac{1}{n} \sum_{i = 1}^n \phi(y_i, \bx_i\trasp \bbe) \,,
\label{eq:wbbeBY}
\end{equation} 
with 
\begin{eqnarray}
\phi(y, t)&=& \rho(d(y, t)) + G(F(t)) + G(1 - F(t))\nonumber\\
&=& y\rho\left(\,-\,\log\left[F(t)\right]\right)+(1-y)\rho\left(\,-\,\log\left[1-F(t)\right]\right) + G(F(t)) + G(1 - F(t))\,,
\label{phiBY}
\end{eqnarray}
where $G(t) = \int_0^t \psi(-\log u) \, du$ is the correction factor needed to guarantee Fisher--consistency.

To ensure the  existence of the estimators under the same conditions that guarantee existence for the maximum likelihood estimators, Croux and Haesbroeck (2003)   suggest to use the loss function $\rho=\rho_c$ 
 \begin{eqnarray}
\rho_c\left(  t\right)  =\left\{
\begin{array}{ll}
te^{-\sqrt{c}} & \hbox{if }  \,\, t\leq c\\
-2e^{-\sqrt{t}}\left(  1+\sqrt{t}\right)  +e^{-\sqrt{c}}\left(  2\left(
1+\sqrt{c}\right)  + c \right)  & \hbox{if }  \,\, t>c \, ,
\end{array}
\right. \label{rocroux}
\end{eqnarray}
where $c$ is a positive tuning constant. 
 Moreover, Croux and Haesbroeck (2003) show that the influence function of the functional related to the estimator $\wbbe$ defined in \eqref{eq:wbbeBY} is not bounded. To obtain bounded influence estimators, these last authors propose a weighted version, namely
\begin{equation}
\label{ESTGEN}
\wbbe =  \argmin_{\bbech\in \real^p} L_n(\bbe)\,,
\end{equation} 
 where 
\begin{equation}
\label{P_n}
 L_n(\bbe) = \frac{1}{n} \sum_{i = 1}^n \phi(y_i, \bx_i\trasp \bbe) w(\bx_i) \,. 
 \end{equation}
The weights  $ w(\bx_i)$ are usually  based  on a robust Mahalanobis distance of the explanatory variables, that is, they depend on the distance between $\bx_i^{\star}$    and a robust center of the data, where $\bx=(1,\bx^{\star \traspi})\trasp$   when an intercept is included in the model and $\bx=\bx^{\star}$  when no intercept is considered.  

It is worth noticing that the estimators introduced in  \eqref{ESTGEN}  represent a wide family which includes the estimators given in \eqref{eq:wbbeBY}, by taking   $w(\bx)=1$. In particular, by choosing  $\rho(t)= 1-\exp(-t)$ this family contains the least squares estimator that minimizes $\sum_{i = 1}^n (y_i - F(\bx_i \trasp \bbe ))^2$, while   the maximum likelihood estimators  correspond to   $\rho(t)=t$ which is not bounded and  the minimum divergence estimators defined in  Basu \textsl{et al.} (2017)  to $\rho (t)= \rho_{\basu}(t)= (1+1/c)\{1-\exp(-ct)\}$.  In other words, the general framework given in  our paper will allow  to include, among others,  penalized minimum divergence estimators as well.

Theorem \ref{theo:BY_fisher_consistency} in the appendix   shows that the estimators defined in \eqref{ESTGEN}  are indeed Fisher--consistent, which is a condition ensuring that the procedure is asymptotically unbiased and estimates the target quantities. It will also play a central role in the results presented in this paper. 

Furthermore, as the function $\phi$ is continuously differentiable with respect to its second argument, the weighted $M-$estimator $\wbbe$ is the solution of the estimating equations $\sum_{i = 1}^n \Psi(y_i, \bx_i\trasp \bbe)\,  w(\bx_i)\,\bx_i=\bcero$,
where $\Psi(y,t)={\partial}\phi(y,t)/{\partial t} $ can be written as
 \begin{equation}
\Psi(y,t) =\,-\,\left[y-F(t)\right] \nu(t)\, 
\label{funcionPsi} 
\end{equation}  
with
 \begin{equation}
\nu(t)= \psi\left(-\log F(t)\right)\left[1- F(t)\right] + \psi\left(-\log\left[1- F(t)\right]\right) F(t) \,.
\label{funcion-nu}
\end{equation}
Note that the function $\phi$ satisfies   $\phi(0,s) = \phi(1,-s)$, while  $\Psi(0,s) = -\Psi(1,-s)$. Furthermore, using \eqref{funcionPsi}  we get that
\begin{equation}
\esp \left[ \Psi(y_1, \bx_1\trasp \bbe_0)\,\Big|\bx_1\right]=\bcero\,,
\label{eq:FC}
\end{equation}
  which is usually known as the conditional  Fisher--consistency condition.

\vskip0.1in

In the non-sparse scenario, the asymptotic behaviour of the estimators $\wbbe$ defined in \eqref{ESTGEN} has been studied  in Bianco and Mart\'{\i}nez (2009), while Basu \textsl{et al.} (2017) consider the particular case of the minimum divergence estimators and $w(\bx) \equiv 1$. More precisely, the above mentioned authors have shown that
  $\sqrt{n}(\wbbe-\bbe)\convdist N_p(\bcero, \bSi)$ with $\bSi=\bA^{-1}\bB\bA^{-1}$, where the matrices $\bA$ and $\bB$ are given by
\begin{eqnarray}
\bA &=& \esp \left(F(\bx \trasp\bbe_0)\left[1-F(\bx \trasp\bbe_0)\right]  \nu(\bx \trasp\bbe_0)   \, w(\bx)  \,\bx  \bx \trasp\right)
\label{eq:A}\\
\bB &=&  \esp \left(F(\bx \trasp\bbe_0)\left[1-F(\bx \trasp\bbe_0)\right] \nu^2(\bx \trasp\bbe_0)   \, w^2(\bx)\,  \bx  \bx \trasp\right)\,.
\label{eq:B}
\end{eqnarray}

\section{Robust penalized estimators}{\label{sec:proposal}}
The robust estimators that we have reviewed in the previous section do not lead to sparse estimators. This entails that they do not allow to make variable selection and may have a bad performance regarding robustness and efficiency. In this setting a usual way to improve the behaviour of existing estimators is to include a regularization term that penalizes candidates without few non--zero components. For that reason, a penalty term is needed to obtain sparse  estimators. The penalized estimators  are defined as 
\begin{equation}
\label{BYPEN}
\wbbe_n = \argmin_{\bbech\in \real^p} \frac{1}{n} \sum_{i = 1}^n \phi(y_i, \bx_i\trasp \bbe)\, w(\bx_i)  + I_{\lambda_n}(\bbe) = \argmin_{\bbech\in \real^p} L_n(\bbe) + I_{\lambda_n}(\bbe) ,
\end{equation}
 where  $ L_n(\bbe)$ is given in \eqref{P_n}, $\phi$ is defined in \eqref{phiBY}  and $I_{\lambda_n}(\bbe)$ is a penalty function, chosen by the user, depending on a tuning parameter $\lambda_n$ which measures the estimated logistic regression model complexity.  The intercept  is not usually penalized, when the model contains one. For that reason and for the sake of simplicity,  when deriving the asymptotic properties of the estimators, we will assume that the model has no intercept.  If the penalty function is properly chosen, the penalized $M-$estimator defined in \eqref{BYPEN}   will lead to sparse models.

The estimators defined in Chi and Scott (2014)  belong to the family \eqref{BYPEN} just by taking $\rho(t)= 1-\exp(-t)$ and 
choosing the Elastic Net penalty  $I_{\lambda}(\bbe)=\lambda \left(\theta \|\bbe\|_1+[({1-\theta})/{2}]\|\bbe\|_2^2\right) $, 
with $\theta\in [0,1]$. Note that Elastic Net reduces  to the LASSO penalty for $\theta=1$ and to the Ridge penalty for $\theta=0$. The main drawbacks of this penalization is that it introduces an extra parameter that must be chosen besides the penalty factor $\lambda$ and that it produces estimators of the non--null components with a large bias.

Some other penalties considered in the linear regression model are the Bridge penalty introduced in Frank and Friedman (1993) and defined as $I_{\lambda}(\bbe)=\lambda \|\bbe\|_q^q$. For linear models the Bridge penalty leads to sparse estimations when   $q\le 1$.

A distinguishing feature in logistic regression is    that the response variable   is bounded. This implies that when considering the penalized least squares estimators the first term in  \eqref{BYPEN} is always smaller than 1 and hence the penalty term may dominate the behaviour of the objective function, unless the  regularization function is bounded.

For that reason, we will consider bounded penalties such as  the SCAD penalty defined in Fan and Li (2001) as 
\begin{eqnarray*}
I_{\lambda}(\bbe) & = &\sum_{j = 1}^p \lambda  |\beta_j|\; \indica_{\{|\beta_j| \leq \lambda\}}+\sum_{j = 1}^p \frac{a \lambda  |\beta_j| - 0.5(\beta_j^2 + \lambda^2)}{a- 1}\; \indica_{\{\lambda < |\beta_j| \leq a\lambda\}}\,   + \sum_{j = 1}^p \frac{\lambda^2(a^2 - 1)}{2(a- 1)}\; \indica_{\{|\beta_j| >a \lambda\}}\,,
\end{eqnarray*}
 for $a>2$, where $\indica_A$ is the indicator function of the set $A$,  and the MCP penalty proposed by Zhang (2010) in the linear regression model which is given by
 $$
I_{\lambda}(\bbe) = \sum_{j = 1}^p \left (\lambda |\beta_j| - \frac{\beta_j^2}{2 \, a}\right )\, \indica_{\{|\beta_j| \leq 
a \, \lambda\}} + \frac{1}{2} \, a \, \lambda^2\, \indica_{\{|\beta_j| > a \, \lambda\}}.
$$ 
 Furthermore, a main objective under a sparse setting is variable selection, that is, to identify variables related to non--null coefficients. Hence, it is more relevant to determine that   $\beta_j\ne 0$ than its size. For that purpose, we introduce a penalty that shrinks the coefficients by pulling the vector $\bbe$ to the unit euclidean ball before applying a LASSO penalty.  This results in the so--called  Sign penalty  defined as
$$I_{\lambda}(\bbe)=\lambda\left\Vert\frac{\bbe}{\|\bbe\|_2}\right\Vert_1\indica_{\bbech\ne \bcero}=\lambda \frac{\|\bbe\|_1}{\|\bbe\|_2}\indica_{\bbech\ne \bcero}\,,$$
that gives a new proposal. Note that the Sign penalty works like LASSO over all unit vectors and in this sense, it enables the selection of a direction, more than raw variable selection. The Sign penalty produces a  thresholding rule, that is, it estimates some coefficients as non--zero. It reaches the minimum when only one of its components is not zero and its maximum when all its components are equal and different from zero. Two important features of this new penalty are  that  it is scale invariant, so it does not shrink the estimated coefficients as the Elastic Net penalty, and it does not require to select an extra parameter as SCAD and MCP. 

\subsection{Selection of the penalty parameter}{\label{KCV}}
As it is well known, the selection of the penalty parameter is an important practical issue when fitting sparse models, since in some sense it tunes the complexity of the model. This problems has been discussed, among others, in Efron \textsl{et al.} (2004), Meinshausen (2007) and  Chi and Scott (2014). In this paper, a robust $K-$fold criterion is used to select the penalty parameter. 

As usual, first randomly split the data set into $K$ 
disjoint subsets of approximately
equal sizes, with indices $\itC_j$, $1 \le j \le K$, the $j-$th subset having size $n_j\ge 2$, so that $\bigcup_{j=1}^K \itC_j
= \{ 1, \ldots, n \}$ and $\sum_{j=1}^K n_j=n$. 
Let $\widetilde{\Lambda}\subset \real$ be the set of possible values for $\lambda$ to be considered, and let 
$\wbbe_{\lambda}^{(j)}$  be an estimator of $\bbe_0$, computed 
with penalty parameter $\lambda\in \widetilde{\Lambda}$ and 
without using the observations with indices in $\itC_j$.
For each $i=1,\dots, n$, the prediction residuals $\widehat{d}_i$ are 
$$
\widehat{d}_{i,\lambda} \, = \, d(y_i, \bx_i\trasp \wbbe_{\lambda}^{(j)} )\, , \quad i \in\itC_j \, , \  j = 1, \, \ldots, K \, . 
$$
 The classical cross--validation criterion constructs   adaptive data--driven estimators by minimizing
\begin{equation}
CV(\lambda)=\frac 1n \sum_{i=1}^n\widehat{d}_{i,\lambda}\,, \label{CVclas}
\end{equation}
an objective function  that is usually employed for the classical estimators that involve the minimization of the deviance. However, this criterion is very sensitive to the presence of outliers.  In fact,  even when $\bbe_0$ is estimated by means of a robust method, the traditional cross--validation criterion may lead to poor variable selection results since atypical data may have large prediction residuals that could be very influential on $CV(\lambda)$. To overcome this problem, when using robust estimators, it seems natural to use the same loss function $\phi$ as in \eqref{BYPEN}. Hence,  the robust cross-validation criterion selects the penalty parameter by minimizing over $\widetilde{\Lambda}$ 
\begin{equation}
RCV(\lambda)=\frac 1n \sum_{1\le j\le K} \sum_{i \in\itC_j} \phi(y_i, \bx_i\trasp \wbbe_{\lambda}^{(j)} )\,w(\bx_i)\,. \label{CVrob}
\end{equation}
The particular case $K=n$ leads to leave-one-out cross-validation  which is a popular choice with a more expensive computational cost. In Section \ref{sec:monte}, we illustrate through a numerical example, the importance of considering a bounded loss in the cross validation criterion when performing the selection of the penalty parameter.

\section{Consistency and order of convergence}{\label{sec:consist}}
In this section, we study the asymptotic behaviour of the estimators defined in \eqref{BYPEN} when $p$ is fixed. Even though we are mainly concerned with bounded penalties, our results are general and include among others the Bridge and Elastic Net penalties.

\subsection{Assumptions}
When considering the function $\phi$ given in  \eqref{phiBY}, the following set of assumptions on the function   $\rho$ are needed.
\begin{enumerate}[label=\textbf{R\arabic*}]
\item \label{ass:rho_bounded_derivable} $\rho: \real_{\geq 0} \to \real$ is a bounded, continuously differentiable function with   bounded derivative $\psi$ and $\rho(0) = 0$.
 
\item \label{ass:rho_derivative_positive}$\psi(t) \geq 0$ and there exists some $c \geq \log 2$ such that $\psi(t) > 0$ for all $0 < t < c$.
 
\item \label{ass:rho_two_times_derivable_bounded} $\rho$ is twice continuously differentiable with bounded derivatives, i.e., $\psi$ and  $\psi^{\prime} = \rho^{\prime\,\prime}$   are bounded.

\end{enumerate}

\vskip0.1in
\begin{remark} 
Recall that for the estimators defined in \eqref{BYPEN}, $\phi(y,t) = \rho(d(y, t)) + G(F(t)) + G(1 - F(t))$
and $\Psi(y,t) = {\partial} \phi(y,t)/{\partial t}= -[y-F(t)] \nu(t)$ with $\nu(t)$ given in \eqref{funcion-nu}. Note also that, under \ref{ass:rho_bounded_derivable} and \ref{ass:rho_derivative_positive}, the function $\Psi(y, \cdot)$ is continuous and strictly positive.

 Denote as  $\chi(y,t)=   \partial  \Psi(y,t)/\partial t= F(t)(1-F(t))\nu(t)  -(y-F(t))\nu^{\prime}(t)$ and note that   $\chi(0,s) = \chi(1,-s)$. The function $\chi(y,t)$ always exists for the minimum divergence estimators and is well defined for any function $\rho$ satisfying  \ref{ass:rho_two_times_derivable_bounded}.  

It is worth noticing that when $\psi(t) > 0$ the constant $c$ in \ref{ass:rho_derivative_positive} may be taken as $\infty$. For instance, this happens when choosing the function $\rho_c$ given in  \eqref{rocroux} or the loss function $\rho=\rho_{\basu}$ related to the divergence estimators, since their derivatives are always positive.  
Moreover, when considering the penalized minimum divergence estimators,   $\rho $ automatically satisfies  conditions  \ref{ass:rho_bounded_derivable}, \ref{ass:rho_derivative_positive} and \ref{ass:rho_two_times_derivable_bounded}.
\end{remark}

On the other hand, for the results in this section,   the following assumptions regarding the distribution of $\bx$ are needed.

\begin{enumerate}[label = \textbf{H\arabic*}]
\item \label{ass:X_not_singular} For all $\balfa \in \real^p$, $\balfa \neq \bcero$, we have  $\prob(\bx \trasp \balfa = 0) =0$.
\item \label{ass:funcionw} $w$ is a non--negative bounded function with support $\itC_w$ such that $\prob(\bx\in \itC_w)>0$.  Without loss of generality, we assume that $\|w\|_{\infty}=1$.
\item \label{ass:X_second_moments} $\esp[w(\bx)\|\bx\|^2] < \infty$.  
\item \label{ass:X_w_positive_definite} The matrix $\bA$ given in \eqref{eq:A} is non--singular.
 
\end{enumerate} 

\vskip0.1in
\begin{remark} 
Assumptions \ref{ass:X_not_singular} and  \ref{ass:funcionw} entail that the estimators defined in \eqref{ESTGEN}  are Fisher--consistent and  will allow to derive consistency results for the estimators defined in \eqref{BYPEN}.  \ref{ass:X_not_singular} holds for instance, when $\bx$ has a density with support $\itS$ such that $\itS\cap \itC_w\ne \emptyset$. In fact, the weaker assumption \linebreak $\prob(\bx \trasp \balfa = 0 \cup w(\bx)=0) < 1$ for any $\balfa\ne \bcero$ is enough for obtaining Fisher--consistency. However, in order to ensure consistency a stronger requirement is needed to warranty that the infimum is not attained at infinity.  It is worth noticing that \ref{ass:X_not_singular} entails that  $\esp[w(\bx)\,\bx\bx \trasp]$ is a positive definite matrix. Furthermore, when considering the minimum divergence estimators   the matrix $\bA $  is non--singular, since $\prob(\nu(\bx \trasp\bbe_0)  >0)=1 $, so \ref{ass:X_w_positive_definite} holds. Similarly, when $\prob(\bx \trasp \balfa = 0) < 1$ for any $\balfa\ne \bcero$, and $\phi$   is given by \eqref{phiBY} and $\psi(t)>0$ for all $t$, as is the case with the loss function  introduced in Croux and Haesbroeck (2003), $\bA$ is non--singular. On the other hand, when \ref{ass:rho_derivative_positive} holds for some finite positive constant $c \geq \log 2$,   $\bA$ is positive definite when \ref{ass:X_not_singular} holds.  Moreover, define $\Upsilon(t)= F(t) (1-F(t))\nu(t)$, straightforward arguments allow to see that $\bA$ is also non--singular when $\prob(\bx \trasp \balfa = 0) < 1$ holds, for any $\balfa\ne \bcero$,  and at least one of the following conditions is fulfilled:
a) the function $\esp[w(\bx)\bx\bx\trasp \indica_{\Upsilon(\bx\trasp \bbech_0) \geq \eta}]$ is continuous in $\eta$ or 
b) there exists some $c > 0$ such that $\prob(\Upsilon(\bx\trasp \bbe_0) > c) = 1$. 
\end{remark}

\subsection{Consistency and rate of convergence}{\label{sec:consist}}

\vskip0.1in

It is worth noticing that, in Theorem    \ref{teo:consistency}, the parameter $\lambda_n$ may be deterministic or random and in the latter situation, the only requirement  is that $I_{\lambda_n}(\bbe_0) \convpp 0 $. In particular,  for the penalties  LASSO, Sign, Ridge, Bridge, SCAD and MCP mentioned in Section \ref{sec:proposal}  this condition holds when $\lambda_n\convcs 0$.

The next Theorem  states the strong consistency of the estimators defined in \eqref{BYPEN}, when considering as function  $\phi$ the function controlling large values of the  \textsl{deviance} residuals given in \eqref{phiBY}.

\vskip0.1in
\begin{theorem}\label{teo:consistency}
Let $\phi:\real^2\to \real$ be the function given in \eqref{phiBY}, where the function  $\rho:\real_{\ge 0}\to \real$ satisfies  \ref{ass:rho_bounded_derivable} and \ref{ass:rho_derivative_positive}.   Then, if $I_{\lambda_n}(\bbe_0) \convpp 0$ when $n \to \infty$ and   \ref{ass:X_not_singular} and \ref{ass:funcionw} hold, we have that  the estimator $\wbbe_n$ defined in \eqref{BYPEN} is strongly consistent for $\bbe_0$.
 \end{theorem}

In order to prove the $\sqrt{n}-$consistency of the proposed estimators, we need the following assumption on the penalty function. From now on, $\itB(\bbe,\epsilon)$ stands for the closed ball, with respect to the usual $\|\cdot\|_2$ norm, centred at $\bbe$ with radius $\epsilon$, i.e., $\itB(\bbe,\epsilon)=\{\bb\in \real^p: \|\bb-\bbe\|_2\le \epsilon\}$.

\begin{enumerate}[label = \textbf{P\arabic*}]
\item \label{ass:penalty_locally_lipschitz} $I_{\lambda}(\bbe)/\lambda$ is Lipschitz in a neighbourhood of $\bbe_0$, that is,  there exists $\epsilon > 0$ a constant $K$, which does not depend on $\lambda$, such that if $\bbe_1, \bbe_2 \in \itB(\bbe_0,\epsilon)$ then $|I_{\lambda}(\bbe_1) - I_{\lambda}(\bbe_2)| \leq \lambda K\|\bbe_1 - \bbe_2\|_1 $.
\end{enumerate}

\vskip0.1in
\begin{remark} Note that the Ridge, Elastic Net, SCAD and MCP  penalties satisfy  \ref{ass:penalty_locally_lipschitz}, since $\|\bbe\|_2\le\|\bbe\|_1\le  \sqrt{p}\, \|\bbe\|_2$. Furthermore, the Sign penalty also satisfies \ref{ass:penalty_locally_lipschitz} if $\|\bbe_0\|_2\ne 0$.  Moreover, if $I_{\lambda}(\bbe) = \lambda\, \sum_{\ell=1}^p J_{\ell}(|\beta_{\ell}|)  $, where $ J_{\ell}(\cdot)$ is a continuously differentiable function, then $I_{\lambda}$ satisfies \ref{ass:penalty_locally_lipschitz}, which implies that the Bridge penalty satisfies \ref{ass:penalty_locally_lipschitz} for $q\ge 1$.  
\end{remark}

\vskip0.1in
\begin{theorem}\label{teo:rate}
Let $\wbbe_n$ be the estimator defined in \eqref{BYPEN} with $\phi(y,t)$ given in  \eqref{phiBY}, where the function  $\rho:\real_{\ge 0}\to \real$ satisfies    \ref{ass:rho_two_times_derivable_bounded}. Furthermore, assume that $\wbbe_n \convprob \bbe_0$ and that assumptions \ref{ass:funcionw} to \ref{ass:X_w_positive_definite} hold. 
\begin{enumerate}
\item[(a)] If assumption \ref{ass:penalty_locally_lipschitz} holds,  $\|\wbbe_n - \bbe_0\|_2=O_{\prob}(\lambda_n\,+\,1/\sqrt{n})$. Hence, if  $\lambda_n = O_\prob(1/\sqrt{n})$, we have that $\|\wbbe_n - \bbe_0\|_2=O_{\prob}(1/\sqrt{n})$, while if  $\lambda_n \sqrt{n}\to \infty$,  $\|\wbbe_n - \bbe_0\|_2=O_{\prob}(\lambda_n)$. 
\item[(b)]Suppose $I_{\lambda_n}(\bbe) = \sum_{\ell = 1}^p J_{\lambda_n}(|\beta_{\ell}|)$ where $J_{\lambda_n}(\cdot)$ is two times differentiable in $(0, \infty)$, takes nonnegative values, $J^{\prime}_{\lambda_n}(|\beta_{0,\ell}|)\ge 0 $ and $J_{\lambda_n}(0) = 0$. Let
\begin{equation*}
a_n = \max\, \left \{J^{\prime}_{\lambda_n}(|\beta_{0,\ell}|) : 1 \leq \ell \leq p \;\; \text{and} \;\; \beta_{0, \ell} \neq 0 \right \} \quad \text{and} \quad \alpha_n = \frac{1}{\sqrt{n}} + a_n.
\end{equation*} 
In addition, assume that there exists some $\delta > 0$ such that 
$$\sup\{|J_{\lambda_n}^{\prime\,\prime}(|\beta_{0,\ell}| + \tau \delta)| : \tau \in [-1,1] \;, \;  1 \leq \ell \leq p \;\; \text{and} \;\; \beta_{0,\ell} \neq 0 \} \convprob 0.$$ 
Then, $\|\wbbe_n - \bbe_0\|_2 = O_{\prob}(\alpha_n)$. 
\end{enumerate}
\end{theorem}

\vskip0.1in
\begin{remark} \label{remrate}
 Theorem \ref{teo:rate}(a) shows that, when the penalty satisfies assumption \ref{ass:penalty_locally_lipschitz}, the estimator rate of convergence  depend on the convergence rate of $\lambda_n$ to 0. In particular, if $\lambda_n \sqrt{n}$ is bounded in probability, then the robust penalized  consistent estimator has rate $\sqrt{n}$, while if $\lambda_n \sqrt{n}\to \infty$, the convergence rate of $\wbbe_n$ is slower  than  $\sqrt{n}$.  This result is analogous to the one obtained, under a linear regression model, in Zou (2006)  for the penalized least squares estimator  when a LASSO penalty is considered. Note that, for the LASSO penalty, the convergence rates obtained in (a) and ( b) are equal since  $J_{\lambda_n}(v)=\lambda_n\, v$.  which entails that $a_n=\lambda_n$ and for any   $\beta_{0,\ell} \neq 0$, $\tau \in [-1,1]$, $J_{\lambda_n}''(|\beta_{0,\ell}| + \tau \delta)=0$ for  a smaller enough $\delta>0$. 	

Instead, SCAD and MCP penalties are not only Lipschitz, but based on univariate twice continuously differentiable functions $J_{\lambda_n}(t)$ satisfying the requirements asked in Theorem \ref{teo:rate}(b) when  $\lambda_n \to 0$. 
Indeed, for these penalties $J'_{\lambda_n}(t)$ and  $J_{\lambda_n}''(t)$ are 0 if $t>\,a\,\lambda_n$ where $a$ is their second tuning constant which is assumed to be fixed. Hence, if  $\lambda_n \convprob 0$ for any $\delta>0$ there exists $n_0$ such that, for any $n\ge n_0$, we have that  $\prob(a \lambda_n <  m_{0})> 1-\delta$ with $m_0=\min\{|\beta_{0,\ell}|) : 1 \leq \ell \leq p \;\; \text{and} \;\; \beta_{0, \ell} \neq 0 \}$. Thus, for $n\ge n_0$, $\prob(   a_n = 0 \mbox{ and  } b_n = 0) > 1-\delta$ and therefore, $\alpha_n=O_{\prob}(1/{\sqrt{n}})$, implying  that the root--$n$ rate may be achieved only assuming only that $\lambda_n \convprob 0$. It is worth noticing that, even when, the Ridge penalty is Lipschitz and it is also based on univariate twice continuously differentiable functions,   $J^{\prime}_{\lambda_n}(|\beta_{0,\ell}|)=\lambda_n |\beta_{0,\ell}|$, so that $a_n= O(1/\sqrt{n} +\lambda_n)$, leading to root-$n$ consistency rate with the additional requirement $\lambda_n = O_\prob(1/\sqrt{n})$.  The different behaviour of the estimators related to   Lipschitz penalties or  penalties related to  twice continuously differentiable functions with null first derivative for $n$ large enough,  plays an important role regarding the variable selection properties of the procedure.   
\end{remark}

\section{Asymptotic distribution results}{\label{sec:asdist}}

The first result in this Section concerns the variable selection properties for our estimator. As shown below the result depends on the behaviour of the penalty function. 

Without loss of generality, assume that $\bbe_0 = (\bbe_{0,A}\trasp, \bcero_{p-k}\trasp)\trasp$ and $\bbe_{0,A} \in \real^k$, $k\ge 1$, is the subvector with \textbf{active} coordinates of $\bbe_0$ (i.e. the subvector of non--zero elements of $\bbe_0$). We will make use of the notation $\bbe = (\bbe_A\trasp,\bbe_B\trasp)\trasp$, where $\bbe_A \in \real^k$ with $k\ge 1$ and $\bbe_B \in \real^{p-k}$.

When the estimator automatically selects variables, we will be able to show an oracle property, that is, that the penalized $M-$estimator of the non--null components of $\bbe_0$, $\wbbe_{n,A}$ has the same asymptotic distribution as that of the estimator obtained assuming  that  the last components of $\bbe_0$ are equal to $0$ and using this restriction in the logistic regression model.

\subsection{Variable selection property}{\label{sec:selectvar}}

\begin{theorem}
\label{teo:orac_gen}
Let $\wbbe_n = (\wbbe_{n,A}\trasp, \wbbe_{n,B}\trasp)\trasp$ be the estimator defined in \eqref{BYPEN}, where $\phi(y,t)$ is  given in \eqref{phiBY}  and the function  $\rho:\real_{\ge 0}\to \real$ satisfies       \ref{ass:rho_two_times_derivable_bounded}. Furthermore, assume that \ref{ass:funcionw} and \ref{ass:X_second_moments} hold and that $\sqrt{n}\|\wbbe_n - \bbe_0\|_2=O_{\prob}(1)$. Furthermore, assume that for every $C > 0$ and $\ell \in \{k+1, \dots, p\}$,  there exist a constant $K_{C, \ell}$ and $N_{C, \ell} \in \natu$ such that if $\|\bu\|_2 \leq C$ and $n \geq N_{C,\ell}$, then
\begin{equation} 
\label{penalty_varsel_condition}
I_{\lambda_n}\left (\bbe_0 + \frac{\bu}{\sqrt{n}}\right ) - I_{\lambda_n}\left (\bbe_0 + \frac{\bu^{(-\ell)}}{\sqrt{n}}\right ) \geq K_{C,\ell}\, \frac{\lambda_n}{\sqrt{n}}\,  |u_\ell |,
\end{equation} where $\bu^{(-\ell)}$ is obtained by replacing the $\ell-$th coordinate of $\bu$ with zero and $u_\ell $ is the $\ell-$th coordinate of $\bu$.
\begin{enumerate}
\item[(a)] For every $\tau > 0$, there exists $b > 0$ and $n_0\in \natu$ such that if $\lambda_n = b/\sqrt{n}$, we have that, for any $n \geq n_0$,
\begin{equation*}
  \prob(\wbbe_{n,B} = \bcero_{p-k}) \geq 1-\tau \,.
\end{equation*}
\item[(b)] If $\lambda_n \, \sqrt{n} \to \infty$, then
 \begin{equation*}
  \prob(\wbbe_{n,B} = \bcero_{p-k}) \to 1 \,.
\end{equation*}
\end{enumerate}
\end{theorem}

\vskip0.1in
To prove variable selection properties for our estimators, it only remains to show that condition \eqref{penalty_varsel_condition}  holds for the different penalties mentioned above. First note that \eqref{penalty_varsel_condition} is clearly satisfied for the LASSO penalty.  In the proof of Corollary \ref{coro:orac} we show that SCAD, MCP and the Sign penalty   also verify  \eqref{penalty_varsel_condition}.

\vskip0.1in
\begin{corollary}{\label{coro:orac}}
Let $\wbbe_n = (\wbbe_{n,A}\trasp, \wbbe_{n,B}\trasp)\trasp$ be the estimator defined in \eqref{BYPEN}  with 
 $\phi(y,t)$   given by   \eqref{phiBY}  where the function  $\rho:\real_{\ge 0}\to \real$ satisfies       \ref{ass:rho_two_times_derivable_bounded}. Assume that \ref{ass:funcionw} and \ref{ass:X_second_moments} hold  and $\sqrt{n} \|\wbbe_n - \bbe_0\|_2=O_{\prob}(1)$.
\begin{enumerate}
\item[(a)] If $I_{\lambda_n}(\bbe)$ is the Sign penalty, then for every $\tau > 0$ there exist   $b>0$ and $n_0\in \natu$ such that if $\lambda_n = b/\sqrt{n}$, we have that, for any $n\ge n_0$,   
\begin{equation*}
  \prob(\wbbe_{n,B} = \bcero_{p-k}) \geq 1-\tau \,.
\end{equation*}
\item[(b)] If $I_{\lambda_n}(\bbe)$ is taken as the SCAD or MCP penalties and $\sqrt{n}\lambda_n \to \infty$, then
 \begin{equation*}
  \prob(\wbbe_{n,B} = \bcero_{p-k}) \to 1.
\end{equation*}
\end{enumerate}
\end{corollary}

\vskip0.1in
\begin{remark}
A consequence of Corollary \ref{coro:orac} is that the penalties SCAD and MCP  have the property of automatically selecting variables when$\sqrt{n}\lambda_n \to \infty$.  In contrast, when using the LASSO and Sign penalties, we cannot ensure the variable selection property when the estimator is root-$n$ consistent. Recall that, for these two penalties,  Theorem \ref{teo:rate} entails that the estimator converges at a rate slower than   $\sqrt{n}$ when $\lambda_n \sqrt{n}\to \infty$.  For that reason, we can only guarantee that for a given $0<\tau<1$, we can choose a sequence of penalty parameters  $\lambda_n=b/\sqrt{n}$ (in order to ensure that the estimator has a root-$n$ rate) and such that the penalized  $M-$estimator selects variables with probability larger than  $1-\tau$. 

The results in the asymptotic distribution given below will allow to conclude that,  for the LASSO and Sign penalties, when the estimator has convergence rate  $\sqrt{n}$, then $\limsup_n\prob(\itA_n=\itA)<1$, where $\itA=\{j: \beta_{0,j}\ne 0\}=\{1,\dots, k\}$ and $\itA_n=\{j: \wbeta_{n,j}\ne 0\}$  are the set of index related to the active components of  $\bbe_0$ and  to the non--null coordinates of  $\wbbe_n$, respectively. This result is analogous to Proposition 1 in Zou (2006), which shows that the   LASSO estimator leads to inconsistent variable selection in the linear regression model, when $\lambda_n=O(1/\sqrt{n})$. 

It is worth noticing that  $ \wbbe_{n,B} = \bcero_{p-k} $ if and only if $\itA_n\subset \itA$, hence, if  $ \prob(\wbbe_{n,B} = \bcero_{p-k}) \to 1$ we have that $ \prob(\itA_n\subset \itA)\to 1$. Note that when $\itA_n \subsetneq \itA$, the penalized   $M-$estimator may select a submodel with less predictors than the original one, shrinking the estimation of some of the active to $0$; however, the oracle property of the estimators based on   SCAD or MCP given in  Theorem \ref{theo:ASDIST_scad_mcp} will allow to conclude that $ \prob(\itA_n=\itA)\to 1$.  
\end{remark}

\subsection{Asymptotic distribution}{\label{sec:asdist}}
In this section, we derive separately the asymptotic distribution of our estimator depending on the choice of the penalty. As the rate of convergence to $0$ of   $\lambda_n$ required to obtain root$-n$ estimators for the Sign  is different from that of   SCAD or MCP  penalties, we will study these two situations separately. Even though most results on penalized estimators assume that the sequence of penalty parameters is deterministic, in this section, as in   Theorem \ref{teo:rate}, we will allow   random penalty  parameters  $\lambda_n$, having in this sense a more realistic point of view. 

It is worth noticing that, under \ref{ass:X_w_positive_definite},    the  matrix   $\bA$ defined in (\ref{eq:A}) is positive definite, so the submatrix corresponding to the active coordinates of   $\bbe_0$ is also positive definite.

From now on, $\be_\ell$ stands for the $\ell-$th canonical vector and $\signo(z)$  is the univariate sign function, that is,  $\signo(z)=z/|z|$ when $z\ne 0$ and $\signo(0)=0$.

\vskip0.1in
\begin{theorem}{\label{theo:ASDIST}}
Let $\wbbe_n$ be the estimator defined in \eqref{BYPEN} with $\phi(y,t)$   given in  \eqref{phiBY},  where the function  $\rho:\real_{\ge 0}\to \real$ satisfies   \ref{ass:rho_two_times_derivable_bounded}.  Assume that \ref{ass:funcionw} to \ref{ass:X_w_positive_definite}  hold, $\sqrt{n}(\wbbe_n-\bbe_0)=O_{\prob}(1)$  and $\sqrt{n} \, \lambda_n \convprob b$. Consider the Sign penalty given by 
$$I_{\lambda }(\bbe) = \lambda \,\frac{\|\bbe\|_1}{\|\bbe\|_2}=\lambda \,\sum_{\ell=1}^p \frac{|\beta_\ell|}{\|\bbe\|_2}\,.$$
Then,  if $\|\bbe_0\|\ne 0$,  $\sqrt{n}(\wbbe_n - \bbe_0) \convdist \argmin_{\bz} R(\bz)$,  where the process $R:\real^p\to \real$ is defined as
\begin{equation*}
R(\bz) = \bz \trasp \bw + \bz \trasp \bA \bz + b \; \bz \trasp \bq(\bz) \,,
\end{equation*}
with     $\bw  \sim N_p(\bcero, \bB)$, $\bA$  and $\bB$ are given  in \eqref{eq:A} and \eqref{eq:B}, respectively and  
\begin{eqnarray*}
\bq(\bz) &= & \sum_{\ell = 1}^p \nabla J_\ell(\bbe_0) \indica_{\{\beta_{0,\ell} \neq 0\}} + \frac{\signo(z_\ell)}{\|\bbe_0\|_2} \indica_{\{\beta_{0,\ell} = 0\}}\, \be_\ell\,,\\
\nabla J_\ell(\bbe) &= &\left (- \frac{|\beta_\ell|\beta_1}{\|\bbe\|_2^3}, - \frac{|\beta_\ell|\beta_2}{\|\bbe\|_2^3}, \dots, \signo(\beta_\ell) \frac{\|\bbe\|_2^2 - \beta_\ell^2}{\|\bbe\|_2^3} ,\dots, - \frac{|\beta_\ell|\beta_p}{\|\bbe\|_2^3}\right )\\
&=& \,-\,\frac{|\beta_\ell|}{\|\bbe\|_2^3}\, \bbe \,+\, \frac{\signo(\beta_\ell)}{\|\bbe\|_2}\,\be_\ell
\end{eqnarray*} 
where $J_\ell(\bbe)=|\beta_\ell|/\|\bbe\|_2  $.
\end{theorem}
\vskip0.1in

The following result generalizes Theorem \ref{theo:ASDIST} to differentiable penalties and includes, among others, the LASSO and Ridge penalties, and any convex combination of them, in particular the   Elastic Net. 

\vskip0.1in
\begin{theorem}{\label{theo:ASDISTJdif}}
Let $\wbbe_n$ be the estimator defined in  \eqref{BYPEN} with $\phi(y,t)$   given by  \eqref{phiBY},  where the function  $\rho:\real_{\ge 0}\to \real$ satisfies   \ref{ass:rho_two_times_derivable_bounded} and let  $\bA$ and $\bB$ be the matrices  defined in \eqref{eq:A} and \eqref{eq:B}, respectively. Let us consider the penalty given by 
\begin{equation}
I_{\lambda }(\bbe) = \lambda\,\left\{(1-\alpha)\sum_{\ell=1}^p J_{\ell}(|\beta_{\ell}|)+\alpha \sum_{\ell=1}^p |\beta_{\ell}|\right\}\,,
\label{eq:Iconvexa}
\end{equation}
where $ J_{\ell}(\cdot)$ is a continuously differentiable function such that $J_\ell^{\prime}(0)=0$.  Assume that \ref{ass:funcionw} to \ref{ass:X_w_positive_definite} hold, $\sqrt{n}(\wbbe_n-\bbe_0)=O_{\prob}(1)$ and that $\sqrt{n} \, \lambda_n \convprob b$.
Then,  if $\|\bbe_0\|\ne 0$,  $\sqrt{n}(\wbbe_n - \bbe_0) \convdist \argmin_{\bz} R(\bz)$ where the process $R:\real^p\to \real$ is defined as
\begin{equation*}
R(\bz) = \bz \trasp \bw + \frac{1}{2}\, \bz \trasp \bA \bz + b \; \bz \trasp \bq(\bz) \,,
\end{equation*}
   with $\bw  \sim N_p(\bcero, \bB)$   and  $\bq(\bz) = (q_1(\bz),\dots, q_p(\bz))\trasp$ being
\begin{eqnarray*}
q_{\ell}(\bz)&=& (1-\alpha)   J_\ell^{\prime}(|\beta_{0, \ell}|)\;    \signo(\beta_{0,\ell}) + \alpha   \left\{  \signo(\beta_{0,\ell}) \indica_{\{\beta_{0,\ell} \ne  0\}} + \signo(z_{\ell}) \indica_{\{\beta_{0,\ell} = 0\}}\right\} \,.
\end{eqnarray*} 
\end{theorem}

\vskip0.2in
\begin{remark}
Note that when  $\sqrt{n} \lambda_n \convprob 0$ ($b=0$), the penalized estimators based on the Sign penalty or on a penalty of the form  \eqref{eq:Iconvexa} have the same asymptotic distribution as the $M-$estimators defined through \eqref{ESTGEN}. If $b>0$ and $\alpha>0$ in  \eqref{eq:Iconvexa}, analogous arguments to those considered in linear regression by Knight and Fu (2000), allow to show that the asymptotic distribution  of the coordinates of $\wbbe_n$ corresponding to null coefficients of $\bbe_0$, that is, the asymptotic distribution of $\wbbe_{n,B}$ puts positive probability at zero. On the other hand, if $\alpha=0$ and $b>0$, the amount of shrinkage of the estimated regression coefficients increases with the magnitude of the true regression coefficients. Hence, for ``large" parameters, the bias introduced by the differentiable  penalty $J_{\ell}(\cdot)$ may be large.   

It is worth noticing that Theorem {\ref{theo:ASDISTJdif}} implies that, when $I_{\lambda }(\bbe) = \lambda\,\sum_{\ell=1}^p J_{\ell}(|\beta_{\ell}|)$ and $\sqrt{n} \lambda_n \convprob b$, $\sqrt{n}(\wbbe_n - \bbe_0) \convdist \bA^{-1} \left(\bw+ b \ba \right)$, where $\ba=(a_1,\dots,a_p)\trasp$ is such that $a_\ell=J_\ell^{\prime}(|\beta_{0, \ell}|)\;    \signo(\beta_{0,\ell})$, which shows the existing asymptotic bias introduced in the limiting distribution, unless $b=0$. In particular, the robust Ridge $M-$estimator, that provides a robust alternative under collinearity,  is asymptotically normally distributed as $N_p(2\,b\,\bA^{-1}\bbe_0,\bA^{-1} \bB \bA^{-1})$.
\end{remark}

\vskip0.1in

When considering   the Sign and LASSO penalties, analogous arguments to those considered in the proof of   Proposition 1 in Zou (2006), together with  Theorems \ref{theo:ASDIST} and \ref{theo:ASDISTJdif} allow to see that, if the penalized $M-$estimator has a root$-n$ rate of convergence, then it is inconsistent for variable selection (see Corollary \ref{coro:inconsist}). Furthermore,  from the proof we may conclude that if $\sqrt{n} \lambda_n \convprob 0$, then $\prob(\itA_n=\itA)\to 0$, that is, we need regularization parameters that converge to $0$, but not too fast in order to select variables with non--null probability.

\begin{corollary}{\label{coro:inconsist}}
Let $\wbbe_n = (\wbbe_{n,A}\trasp, \wbbe_{n,B}\trasp)\trasp$ be the estimator defined in  (\ref{BYPEN}), where $\phi(y,t)$ is given through (\ref{phiBY})  with the function  $\rho:\real_{\ge 0}\to \real$ satisfying \ref{ass:rho_two_times_derivable_bounded}. Assume that $\|\bbe_0\|\ne 0$,  $\sqrt{n} \lambda_n \convprob b$, $\sqrt{n}\|\wbbe_n - \bbe_0\|_2=O_{\prob}(1)$ and that \ref{ass:funcionw} to \ref{ass:X_w_positive_definite} hold. Then, for the Sign or LASSO penalties, there exists $c<1$ such that  $\limsup_n\prob(\itA_n=\itA)\le c<1$, where $\itA=\{j: \beta_{0,j}\ne 0\}$ is the set of indexes corresponding to the active coordinates of $\bbe_0$ and $\itA_n=\{j: \wbeta_{n,j}\ne 0\}$.
\end{corollary}

Similar arguments to those considered in the proof of Theorem\ref{theo:ASDIST},  allow to obtain the asymptotic distribution of the penalized   $M-$estimator with Sign penalty when $\sqrt{n} \lambda_n \to \infty$. A similar result holds for penalizations satisfying  \eqref{eq:Iconvexa}, as the LASSO one.

 \begin{theorem}{\label{theo:ASDISTlamdainfty}}
Let $\wbbe_n $ be the estimator defined in  (\ref{BYPEN}), where $\phi(y,t)$ is given through (\ref{phiBY})  with the function  $\rho:\real_{\ge 0}\to \real$ satisfying \ref{ass:rho_two_times_derivable_bounded}.  Assume that $\|\bbe_0\|\ne 0$,  $\sqrt{n} \lambda_n \convprob \infty$, $\wbbe_n-\bbe_0 =O_{\prob}(\lambda_n)$ and that \ref{ass:funcionw} to \ref{ass:X_w_positive_definite} hold.   Let  $\bA$   be the matrix defined in (\ref{eq:A}) and consider the Sign penalty
$$I_{\lambda }(\bbe) = \lambda\,\frac{\|\bbe\|_1}{\|\bbe\|_2}\,.$$ 
 Then,   $(1/\lambda_n)\;(\wbbe_n - \bbe_0) \convprob \argmin_{\bz} R(\bz)$, where the function $R:\real^p \to \real$ is defined through
\begin{equation*}
R(\bz) =  \frac 12\bz \trasp \bA \bz +   \bz \trasp \bq(\bz) \,,
\end{equation*}
with    $\bq(\bz)  $ the function defined in Theorem \ref{theo:ASDIST}.
\end{theorem}

 \begin{remark}
Lemma 3 in Zou (2006) provides a result analogous to Theorem  \ref{theo:ASDISTlamdainfty}  for the LASSO least squares estimator, under a linear regression model. As in that result, the rate of convergence of  $\wbbe_n$ is slower than  $\sqrt{n}$ and the limit is a non--random quantity. As noted in  Zou (2006), the optimal rate for   $\wbbe_n$ is obtained when $\lambda_n=O_{\prob}(1/\sqrt{n})$, but at expenses of not selecting variables.
\end{remark}

\vskip0.1in
Finally, the following theorem gives the asymptotic distribution of $\wbbe_{n,A}$ when the penalty is consistent for variable selection, that is, when  $\prob(\wbbe_{n,B} = \bcero_{p-k}) \to 1$. For that purpose, recall that $\bbe_0 = (\bbe_{0,A}\trasp, \bcero_{p-k}\trasp)\trasp$ where $\bbe_{0,A} \in \real^k$, $k\ge 1$, is the vector of active coordinates of  $\bbe_0$ and for  $\bb\in \real^k$, define
$$\nabla I_{\lambda }(\bb)=\frac{\partial I_{\lambda }\left((\bb\trasp, \bcero_{p-k}\trasp)\trasp\right)}{\partial \bb}\,.$$

\vskip0.1in
\begin{theorem}\label{theo:ASDIST_scad_mcp}
Let $\wbbe_n$ be the estimator defined in \eqref{BYPEN} with $\phi(y,t)$ given in \eqref{phiBY},  where the function  $\rho:\real_{\ge 0}\to \real$ satisfies   \ref{ass:rho_two_times_derivable_bounded} and assume that   \ref{ass:funcionw} and \ref{ass:X_second_moments} hold. Suppose that there exists some $\delta > 0$ such that
\begin{equation}
\label{cond:ASDIST_scad_mcp}
\sup_{\|\bbech_A - \bbech_{0,A}\|_2 \leq \delta}\|\nabla I_{\lambda_n}(\bbe_{A})\|_2 = o\left(\frac{1}{\sqrt{n}} \right ),
\end{equation} $\prob(\wbbe_{n,B} = \bcero_{p-k}) \to 1$ and $\wbbe_n \convprob \bbe_0$. Let $\wtbA$ and $\wtbB$ be the $k \times k$ submatrices  of $\bA$ and $\bB$, respectively, corresponding to the first $k$ coordinates of $\bbe_0$, where $\bA$ and $\bB$ were defined in \eqref{eq:A}  and \eqref{eq:B}. Then, if $\wtbA$ is invertible,
\begin{equation*}
\sqrt{n} (\wbbe_{n,A} - \bbe_{0,A}) \convdist N_k(\bcero, \wtbA ^{-1} \wtbB \wtbA^{-1}).
\end{equation*}

\end{theorem}

\begin{remark}
Penalties  SCAD and MCP fulfil  (\ref{cond:ASDIST_scad_mcp}) when  $\lambda_n \to 0$.  Effectively, recall that any of them may be written as
$I_{\lambda}(\bbe) = \sum_{j = 1}^p J_{\lambda}(|\beta_j|)$, where $J_{\lambda}(t)$ is constant  in $[a \lambda, \infty)$, with $a > 0$  the second tuning constant of these penalties. Using that $J_{\lambda}(0)=0$, we obtain that, for any $\bb\in \real^k$, $I_{\lambda}\left((\bb\trasp, \bcero_{p-k}\trasp)\trasp\right) =\sum_{j = 1}^k J_{\lambda}(|b_j|)$ and 
$\nabla I_{\lambda}(\bb)=\sum_{j = 1}^k J'_{\lambda}(|b_j|)$.
Since $ \|\wbbe - \bbe_0\|_2=O_{\prob}\left( 1/\sqrt{n}\right)$, given $\delta > 0$ there exists $C_1> 0$ such that $\prob(\itD_n)>1 - \delta $ for $n \ge n_0$, with $\itD_n = \{\|\wbbe - \bbe_0\|_2 \leq C_1/\sqrt{n}\}$. 

Let $n_1$ be such that $C_1/\sqrt{n}\le m_0/2$. Then,  for any  $\omega\in \itD_n$, $n \ge n_1$ and   $1 \leq j \leq k$,  we have that
\begin{equation*}
|\wbeta_j| \geq |\beta_{0,j}| - |\wbeta_j - \beta_{0,j}| \geq m_{0} -  \frac{C_1}{\sqrt{n}} \geq \frac{m_{0}}{2} \,.
\end{equation*} 
Using that $\lambda_n \to 0$ we get that for $n\ge \max\{n_0,n_1\}$, we have that $j=1,\ldots, k$, 
$|\wbeta_j| > a \lambda_n$, implying that $\itD_n\subset \{\|\nabla I_{\lambda_n} (\wbbe_A)\|_2 = 0\}$ as desired.

Hence,  using Corollary \ref{coro:orac}, we get that the penalized $M-$estimators defined through (\ref{BYPEN}) have the oracle property when using SCAD or MCP  and $\lambda_n \to 0$ with $\sqrt{n}\;\lambda_n \to \infty$.
\end{remark}

\small
\section{Monte Carlo study}{\label{sec:monte}}

In this section, we present the results of a Monte Carlo study designed to compare the small sample performance  of classical and robust penalized estimators. 
Section \ref{sec:algo} describes the algorithm used to compute the robust penalized estimators, while the simulation settings and the obtained results are summarized in Section \ref{sec:numerical}.

\subsection{Algorithm}{\label{sec:algo}}
The algorithm used to compute the  estimators proposed in Section \ref{sec:proposal} is an implementation of the cyclical descent algorithm. Taking into account that the estimators defined in (\ref{BYPEN}) depend on a penalty parameter which is usually chosen by cross--validation , we set a grid of candidates $ \widetilde{\Lambda} = \{\lambda_1, \lambda_2, \dots,  \lambda_K\}$. Given  $\lambda_k \in  \widetilde{\Lambda}$ and a subset of indexes $\itI_k \subset \{1, \dots, p\}$, the algorithm consists on the following steps:
\begin{enumerate}[label=(\alph*)]
	\item Obtain an initial estimator $\wbbe_{\ini}$ and define $M^{(0)}=M_n(\wbbe_{\ini})$, where $M_n(\bbe)=L_n(\bbe)+I_{\lambda}(\bbe)$ is given in \eqref{BYPEN}. Fix initially  $\ell = 0$.
	\item $\ell  \leftarrow \ell + 1$
	\begin{enumerate}[label = \textbf{\small Step \arabic*}]
		\item Choose a random permutation of the elements of $\itI_k$, say $i_1, \dots, i_s$, where $s$ is the number of elements of $\itI_k$. Following the order given by the indexes of the permutation, minimize the function  $M_n(\bbe) $  over one coordinate of $\bbe$ each at a time, while leaving fixed the remaining ones. This procedure  involves $s$ minimization steps. Denote as $\wtbbe$ the obtained value after passing through all the coordinates in $\itI_k$.
		\item Choose the value $c > 0$   that  minimizes the function $M_n(c\, \wtbbe)$. Denote $ \widetilde{c} $ the resulting value, $\bbe^{(\ell)}=\widetilde{c}\,\wtbbe$ and $M^{(\ell)}=M_n(\bbe^{(\ell)})$
		
		\item Compute $
		R^{(\ell)}=  |M^{(\ell-1)}- M^{(\ell)}|/M^{(\ell)}$.
		
	\end{enumerate}
	\item If the ratio $R^{(\ell)}$ is smaller than a fixed tolerance parameter, define  $\wbbe=\bbe^{(\ell)}$, otherwise  go back to   (b).  
\end{enumerate}

When the model includes an intercept, an intermediate step  between Steps 1 and 2 is  introduced in order to get an intercept estimate by minimizing the objective function  only over the intercept parameter, while keeping fixed the value of $\wbbe=\bbe^{(\ell)}$ obtained in Step 2.

It is worth noting that, in our numerical studies,  the univariate optimization   in Steps 1  and 2 are carried out using the function \texttt{optim} in \texttt{R}, which implements the Nelder--Mead optimization method. A key point is the choice of the initial estimator $\wbbe_{\ini}$. Recall that Chi and Scott (2014) based their initial estimator choice on the Karush--Kuhn--Tucker (KKT)  conditions for the problem of  minimizing $\argmin_{\bbech \in \real^p} (1/n) \sum_{i = 1}^n (y_i - F(\bx_i \trasp \bbe))^2 + \lambda \|\bbe\|_1\,$. 
For $j = 1, \dots, p$, those authors compute the scores $z_j = \left |\overline{y}(1 - \overline{y})\bx_{(j)}\trasp (\by - \overline{y} \buno_n)\right |$, where $\overline{y} = \sum_{i = 1}^n y_i/n$, $\by = (y_1, \dots, y_n)\trasp$ and $\bx_{(j)}$ is the $j-$th column of $\bx$. Their initial estimator is based on the response variables with the highest scores (in absolute value) which are set to 1, the remaining ones being set to 0. 

When the sample has no contamination, this initial estimator seems to be a good choice (see Chi and Scott, 2014). However, this method may be influenced by outliers as those added  in our numerical study. To overcome this drawback, we first compute the quantities {$\kappa_{ij} = x_{ij} (y_i - \overline{y})$}, $1\le j\le p$, $1\le i\le n$. Then, for every $j = 1, \dots, p$, the score $\wtz_j$ is evaluated as the absolute value of the $\alpha-$trimmed mean of $\{\kappa_{1j}, \kappa_{2j}, \dots, \kappa_{nj}\}$. In our simulation study, we fixed $\alpha = 0.15$. Finally,  we choose the proportion  $0<\tau<1$ of variables with highest absolute \textsl{score} and we apply the weighted estimator introduced in \eqref{P_n} with the $\rho-$function of Croux and Haesbroeck (2003) given in \eqref{rocroux} on these selected variables. In our experimental results we have chosen the proportion $\tau = 0.1$.

As mentioned above, the penalty parameter selection is an important step to effectively implement the computation of the procedure in sparse models. This fact and the necessity of a robust cross--validation criterion are discussed in Section {\ref{KCV}}.
Therefore, in our study we determine the penalty parameter by  minimizing of the traditional criterion  $CV(\lambda)$ given in \eqref{CVclas} when using the classical estimators, while for the robust ones we consider $RCV(\lambda)$ given in \eqref{CVrob}.

Several parts of the code were implemented in \texttt{C++} and integrated with the \texttt{Rcpp} package in \texttt{R}. 

\subsection{Numerical Experiments}{\label{sec:numerical}}
In this section,  we report the results of a Monte Carlo study for clean and contaminated samples with different choices of the loss function and penalties so as to cover a wide variety of possibilities. 
For that purpose, we have generated a training sample $\itM$ of i.i.d. observations $(y_i, \bx_i)$, $1\le i\le n$,  $\bx_i\in \real^p$ and $y_i|\bx_i \sim Bi(1, F(\gamma_0+\bx_i \trasp \bbe_0))$, where  the intercept $\gamma_0=0$ and varying the values  of $n$, $p$ and the true parameter $\bbe_0$. For clean samples the covariates distribution is $N_p(\bcero,\bI)$. Henceforth, the uncontaminated setting is denoted \textbf{C0}. 

To confront our estimators with  some challenging situations, we considered cases where the ratio $p/n$ is large. More precisely, we choose the pairs  $(n, p)$, with $n \in \{150,300\}$ and  $p \in \{40,80,120\}$.
 In order to generate a sparse scenario we chose the true regression parameter as $\bbe_0 = (1,1,1,1,1,0,0,\dots, 0)\trasp\in \real^p$, that is, only  the first five components are non--null and equal to one, yielding to values of $\esp(y_i)$ equal to 0.50. In all cases, the number of Monte Carlo replications was $NR=500$.

To study the impact of contamination, we have explored two settings by adding a proportion $\varepsilon=0.05$ or $0.10$ of atypical points.  
In the first contamination scheme, namely outliers of class \textbf{A}, we  generated misclassified points  $(\wty, \wtbx)$, where $\wtbx \sim N_p(0, 20 \, \bI)$ and
\begin{equation}\label{outlier_generation}
\wty = \begin{cases} 
1 &\quad \text{if } \gamma_0+\wtbx \trasp \bbe_0 < 0\\
0 &\quad \text{if } \gamma_0+\wtbx \trasp \bbe_0 \geq 0 \;.\\
\end{cases}
\end{equation}
Besides, outliers of class \textbf{B}, were obtained as in Croux y Haesbroeck (2003). This means that given $m >0$,  we fixed  $\wtbw = {m} \sqrt{p} \, \bbe_0 /{5}  $ and set $\wtbx = \wtbw + \wtbu$, where $\wtbu \sim N_p(\bcero, \bI/{100})$ is introduced so as to get distinct covariate values. The response  $\wty$, related to  $\wtbx$, is always taken equal to $0$. It is worth noticing that $\wtbw \trasp \bbe_0 \approx m$, thus the leverage of the added points increases with  $m$. The selected values of $m$ are 0.5, 1, 1.5, 2, 3, 4 and 5.

Summarizing, we consider the scenarios
\textbf{CA1} and \textbf{CA2} which correspond to adding, respectively, a proportion $\varepsilon = 0.05$ and $0.10$  of outliers of class \textbf{A} and 
\textbf{CB1} and \textbf{CB2} where we add outliers of class \textbf{B} in a proportion $\varepsilon = 0.05$ and $0.10$, respectively. 

We compare the performance of the estimators based on the deviance, that is, when $\rho(t)=t$, labelled \textsc{ml} in all Tables and Figures, with those obtained bounding the deviance and also with their robust weighted versions. The three bounded loss functions considered are  $\rho(t)=1-\exp(-t)$ that leads to the least squares estimators, the loss functions $\rho_c$ introduced by Croux and Haesbroeck (2003),  given in \eqref{rocroux}, and  $\rho(t)=(c + 1)(1 + \exp(-ct))$  related to the divergence estimators. For the last two loss functions, the tuning constant equals $c=0.5$. These estimators are indicated with the subscript \textsc{ls}, \textsc{m} and \textsc{div}, respectively. We have also considered weighted version of them that bound the leverage. For this purpose, define  $D^2(\bx,\bmu,\bSi^{-1})=(\bx-\bmu)\trasp \bSi^{-1}(\bx-\bmu)$, the square of the  Mahalanobis distance. We take weights $w(\bx)=W(D^2(\bx,\wbmu, \wbSi^{-1}))$,  where to adjust for robustness $\wbmu$ is the $L_1-$median, $\wbSi^{-1}$ is an estimator of $\bSi^{-1}$ computed using graphical LASSO and  $W$ is the hard rejection weight function $W(t)=\indica_{[0,c_w]}(t)$. The tuning constant $c_w$ is adaptive and based on the quantiles of  $d^2_i=D^2(\bx_i,\wbmu,\wbSi^{-1})$. These estimators are labelled with the subscript \textsc{wls}, \textsc{wm} or \textsc{wdiv}, according to the loss function considered.
For each loss function, different penalties are considered:   LASSO, Sign and , MCP, labelled with the superscript \textsc{l}, \textsc{s} and \textsc{mcp}, respectively. The non--sparse estimators without any penalization term  are indicated with no superscript. All estimators were computed using the algorithm defined on Section \ref{sec:algo}.

Under \textbf{C0} and scenarios \textbf{CA1} and \textbf{CA2}, we compare all the described estimators. However, in view of the results obtained for these three situations and for the sake of brevity, under \textbf{CB1} and \textbf{CB2} we only report  the results for $\wbbe_{\mle}$,  $\wbbe_{\mch}$ and $\wbbe_{\wmch}$ with penalties \textsc{s} and \textsc{mcp}.

To evaluate the performance of a given estimator $\wbbe$, we consider four summary measures. In the following, let  $\itT =\{ (y_{i,\itT}, \bx_{i,\itT}), i = 1, \dots, n_{\itT}\}$, $n_{\itT}=100$, be a new sample  generated independently from the training sample  and distributed as  \textbf{C0}.  Given the estimates $\wbbe$ of the slope and $\wgamma$ of the intercept computed  from $\itM$, denote $\wy_{i,\itT} = \indica_{\{\bx_{i,\itT} \trasp \wbbech +\wgamma> 0\}}$ and $\Pi = \prob(y_{i,\itT} = \indica_{\{\bx_{i,\itT} \trasp \bbech_0+\gamma_0\,>\,0\}})$.
We define the following quantities
\begin{itemize}
	\item \textbf{Probabilities Mean Squared Error} 
\begin{equation*}
\text{PMSE}  = \frac{1}{n_{\itT}} \sum_{i = 1}^{n_{\itT}} (F(\bx_{i,\itT} \trasp \bbe_0+\gamma_0) - F(\bx_{i,\itT} \trasp \wbbe+\wgamma))^2\, ,
\end{equation*} 
 	\item \textbf{Mean Squared Error}
	\begin{equation*}
	\text{MSE} = \|\wbbe - \bbe_0\|_2^2 \, ,
	\end{equation*}
	\item \textbf{True Positive Proportion}
	\begin{equation*}
	\text{TPP} = \frac{\#\{j : 1 \leq j \leq p,\; \beta_{0,j} \neq 0,\; \wbeta_j \neq 0 \}}{\#\{j : 1 \leq j \leq p,\;  \beta_{0,j} \neq 0 \}} \, ,
	\end{equation*}
	\item \textbf{True Null Proportion}
	\begin{equation*}
	\text{TNP} = \frac{\#\{j : 1 \leq j \leq p,\; \beta_{0,j} = 0,\; \wbeta_j = 0 \}}{\#\{j : 1 \leq j \leq p,\;  \beta_{0,j} = 0 \}} \,.
	\end{equation*}
\end{itemize} 

\subsection{Finite--sample performance of the cross-validation criteria}{\label{sec:monteCVlam}}

In this section, we report the results of a small simulation study regarding the selection of the regularization parameter $\lambda$. The analysis is twofold, on one hand we illustrate the importance of choosing $\lambda$ by means of a robust criterion when we deal with a robust estimator so as to ensure the stability of the resulting estimates. On the other one, we concern about the rate of convergence to $0$ of the penalization parameter $\lambda$ under \textbf{C0}.

For the first purpose, we generate clean samples and contaminated ones following scheme  \textbf{CB1} with slope $m = 4$. We compute the estimators
$\wbbe_{\weme}^{\mcppen}$ 
and $\wbbe_{\weme}^{\signpen}$ with the regularization parameter that minimizes the cross-validation criteria  $CV(\lambda)$ or its robust version $RCV(\lambda)$, given in \eqref{CVclas} and \eqref{CVrob}, respectively. 

The sample size $n$ and the dimension of the covariates $p$ are taken as $(n,p)=(150,40),(150,80),(300,80)$ and $(300,120)$. Table \ref{tab:cvrob_cvnorob_c0} 
shows the $10\%-$trimmed means of the measures PMSE, MSE, TPP and TNP  under \textbf{C0} for each pair $(n,p)$, while Table  \ref{tab:cvrob_cvnorob_cb1} presents the results obtained under \textbf{CB1}.  

\begin{table}[ht]
	\setlength\arrayrulewidth{1pt}        
	\centering
	\def\arraystretch{1.55}
	
	\begin{tabular}{|c|c|cc|cc|c|cc|cc|}
		\hline
		&  & \multicolumn{4}{c|}{ $n=150$} & &  \multicolumn{4}{c|}{ $n=300$} \\ \hline
		& & \multicolumn{2}{c|}{ $CV$} & \multicolumn{2}{c|}{ $RCV$} & &  \multicolumn{2}{c|}{ $CV$} & \multicolumn{2}{c|}{ $RCV$} \\
		\hline
		&$p$ &  $40$ & $80$ & $40$ & $80$ & & $80$ & $120$ & $80$ & $120$\\ 
		\hline
		PMSE &  $\wbbe_{\weme}^{\signpen}$ & 0.033 & 0.046 & 0.029 & 0.038 & $\wbbe_{\weme}^{\signpen}$ & 0.010 & 0.011 & 0.010 & 0.011 \\ 
		& $\wbbe_{\weme}^{\mcppen}$ & 0.022 & 0.028 & 0.022 & 0.027 & $\wbbe_{\weme}^{\mcppen}$ & 0.006 & 0.008 & 0.007 & 0.008 \\ \hline
		MSE & $\wbbe_{\weme}^{\signpen}$ & 1.708 & 2.239 & 1.642 & 2.014 & $\wbbe_{\weme}^{\signpen}$ & 0.501 & 0.546 & 0.507 & 0.539 \\ 
		& $\wbbe_{\weme}^{\mcppen}$ & 1.144 & 1.377 & 1.150 & 1.414 & $\wbbe_{\weme}^{\mcppen}$ & 0.318 & 0.348 & 0.327 & 0.359 \\ \hline
		TPP & $\wbbe_{\weme}^{\signpen}$ & 0.926 & 0.906 & 0.966 & 0.944 & $\wbbe_{\weme}^{\signpen}$ & 1.000 & 1.000 & 1.000 & 1.000 \\ 
		& $\wbbe_{\weme}^{\mcppen}$ & 0.932 & 0.924 & 0.942 & 0.936 & $\wbbe_{\weme}^{\mcppen}$ & 1.000 & 1.000 & 1.000 & 1.000 \\ \hline
		TNP & $\wbbe_{\weme}^{\signpen}$ & 0.968 & 0.965 & 0.949 & 0.963 & $\wbbe_{\weme}^{\signpen}$ & 0.961 & 0.955 & 0.960 & 0.953 \\ 
		& $\wbbe_{\weme}^{\mcppen}$ & 0.988 & 0.971 & 0.986 & 0.970 & $\wbbe_{\weme}^{\mcppen}$ & 0.977 & 0.972 & 0.977 & 0.971 \\ 
		\hline
	\end{tabular}
	\caption{\small \label{tab:cvrob_cvnorob_c0} $10\%-$trimmed means of measures PMSE, MSE, TPP and TNP  under  \textbf{C0} } 
\end{table}

\begin{table}[ht!]
	\setlength\arrayrulewidth{1pt}        
	\centering
	\def\arraystretch{1.55}
	\begin{tabular}{|c|c|cc|cc|c|cc|cc|}
		\hline
	&  & \multicolumn{4}{c|}{ $n=150$} & &  \multicolumn{4}{c|}{ $n=300$} \\ \hline
		& & \multicolumn{2}{c|}{ $CV$} & \multicolumn{2}{c|}{ $RCV$} & &  \multicolumn{2}{c|}{ $CV$} & \multicolumn{2}{c|}{ $RCV$} \\
		\hline
		&$p$ &  $40$ & $80$ & $40$ & $80$ & & $80$ & $120$ & $80$ & $120$\\ 
		\hline
		PMSE &  $\wbbe_{\weme}^{\signpen}$ & 0.097 & 0.095 & 0.035 & 0.051 & $\wbbe_{\weme}^{\signpen}$ & 0.090 & 0.090 & 0.010 & 0.012 \\ 
		& $\wbbe_{\weme}^{\mcppen}$ & 0.109 & 0.107 & 0.023 & 0.039 & $\wbbe_{\weme}^{\mcppen}$ & 0.105 & 0.109 & 0.009 & 0.011 \\ \hline 
		MSE & $\wbbe_{\weme}^{\signpen}$ & 4.191 & 4.160 & 2.023 & 2.542 & $\wbbe_{\weme}^{\signpen}$ & 4.089 & 4.097 & 0.519 & 0.558 \\ 
		& $\wbbe_{\weme}^{\mcppen}$ & 4.969 & 4.922 & 1.300 & 2.157 & $\wbbe_{\weme}^{\mcppen}$ & 4.804 & 5.000 & 0.435 & 0.509 \\ \hline
		TPP & $\wbbe_{\weme}^{\signpen}$ & 0.260 & 0.272 & 0.948 & 0.831 & $\wbbe_{\weme}^{\signpen}$ & 0.264 & 0.252 & 1.000 & 1.000 \\ 
		& $\wbbe_{\weme}^{\mcppen}$ & 0.084 & 0.140 & 0.956 & 0.876 & $\wbbe_{\weme}^{\mcppen}$ & 0.107 & 0.606 & 1.000 & 1.000 \\ \hline
		TNP & $\wbbe_{\weme}^{\signpen}$ & 0.972 & 0.969 & 0.926 & 0.957 & $\wbbe_{\weme}^{\signpen}$ & 0.972 & 0.972 & 0.954 & 0.950 \\ 
		& $\wbbe_{\weme}^{\mcppen}$ & 0.954 & 0.960 & 0.971 & 0.960 & $\wbbe_{\weme}^{\mcppen}$ & 0.967 & 0.947 & 0.972 & 0.968 \\ 
		\hline
	\end{tabular}
	\caption{\small \label{tab:cvrob_cvnorob_cb1} $10\%-$trimmed means of measures PMSE, MSE, TPP and TNP  under scheme \textbf{CB1} with slope $m= 4$.} 
\end{table}

As illustrated in Table \ref{tab:cvrob_cvnorob_c0}, for clean samples, the estimators obtained with $CV$ or $RCV$  are very similar. In some cases the latter ones show a slight advantage over those based on $CV$. For instance,   
$\wbbe_{\weme}^{\signpen}$ shows better results in measures PMSE and TPP when $n = 150$ when the criterion $RCV$ is used.

The substantial advantage of using the robust cross--validation procedure over the classical one is clearly  seen in Table \ref{tab:cvrob_cvnorob_cb1}.
The proportion of true positives (TPP) is strongly affected when the classical cross--validation criterion is used, even when robust estimators are computed, showing the important role played by the selection method for $\lambda$, so as to ensure the final resistance of the estimator to the presence of atypical data.
An example of the effect of the artificially introduced misclassified observations can be illustrated through the estimator $\wbbe_{\weme}^{\mcppen}$: when the $CV$ method is used, the proportion of true positives is less than 0.2 for several pairs $(n, p)$, while using $RCV$ criterion, the corresponding TPP values are always greater than 0.5.

The same conclusions can be applied to other weighted $M-$estimators. In general, the  weighted $M-$estimators obtained using the classic cross--validation procedure are excessively sparse. As expected, this fact severely affects the measures PMSE and MSE. In all cases, these quantities decrease when $RCV$ is used. For these estimators, when $n= 150$, the PMSE and MSE measures double their value when $CV$ is used instead of the robust alternative. On the other hand, when $n = 300$, this relationship is even greater: these measures are approximately 10 times larger when using $CV$ instead of $RCV$, also being almost 10 times larger than those obtained under \textbf{C0}.

\begin{figure}[ht!]
	\centering
	\begin{tabular}{ccc}
		& $CV$ & $RCV$ \\
		$\wbbe_{\weme}^{\signpen}$ &
		\raisebox{-.5\height}{\includegraphics[scale=0.4]{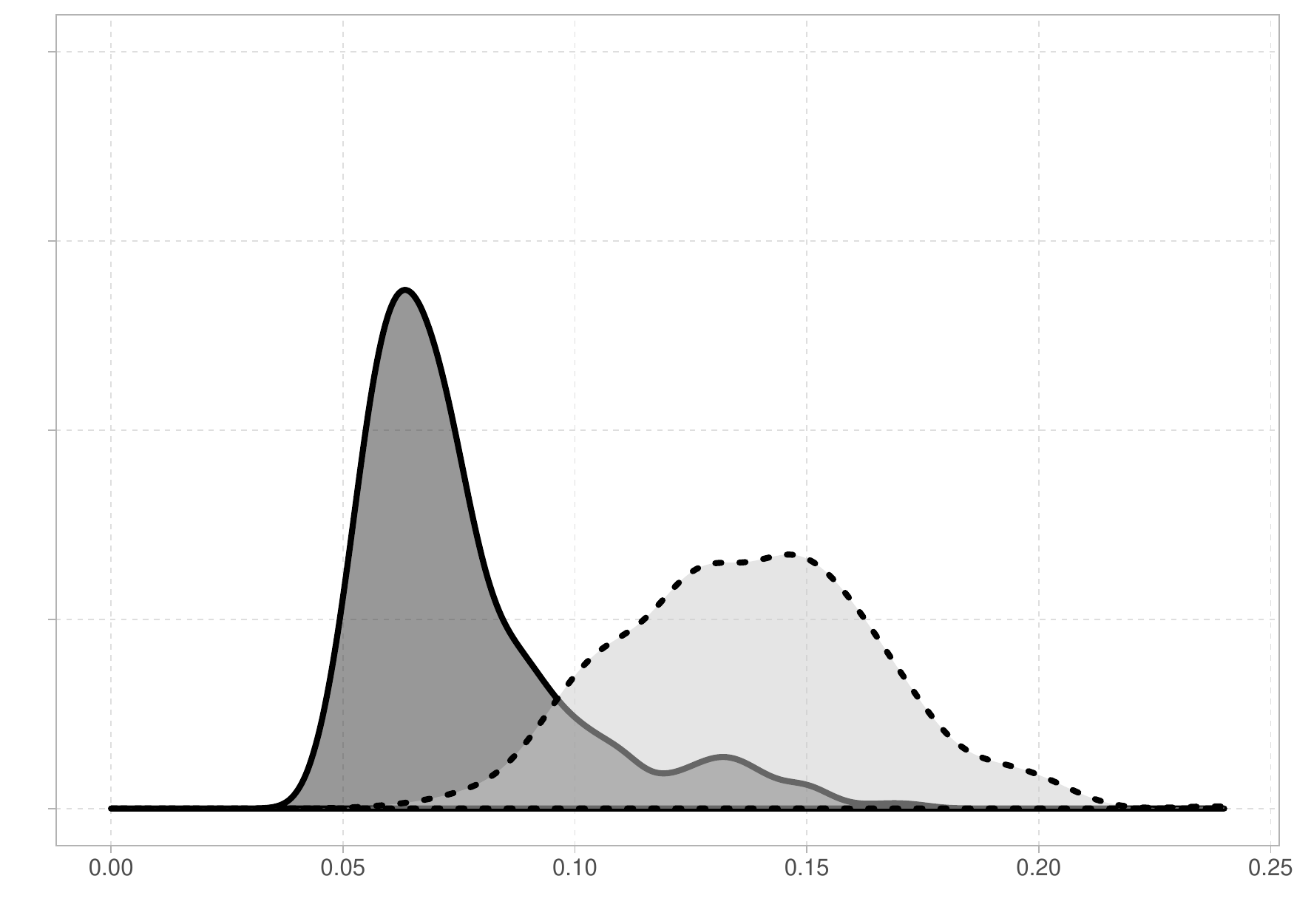}} &  
		\raisebox{-.5\height}{\includegraphics[scale=0.4]{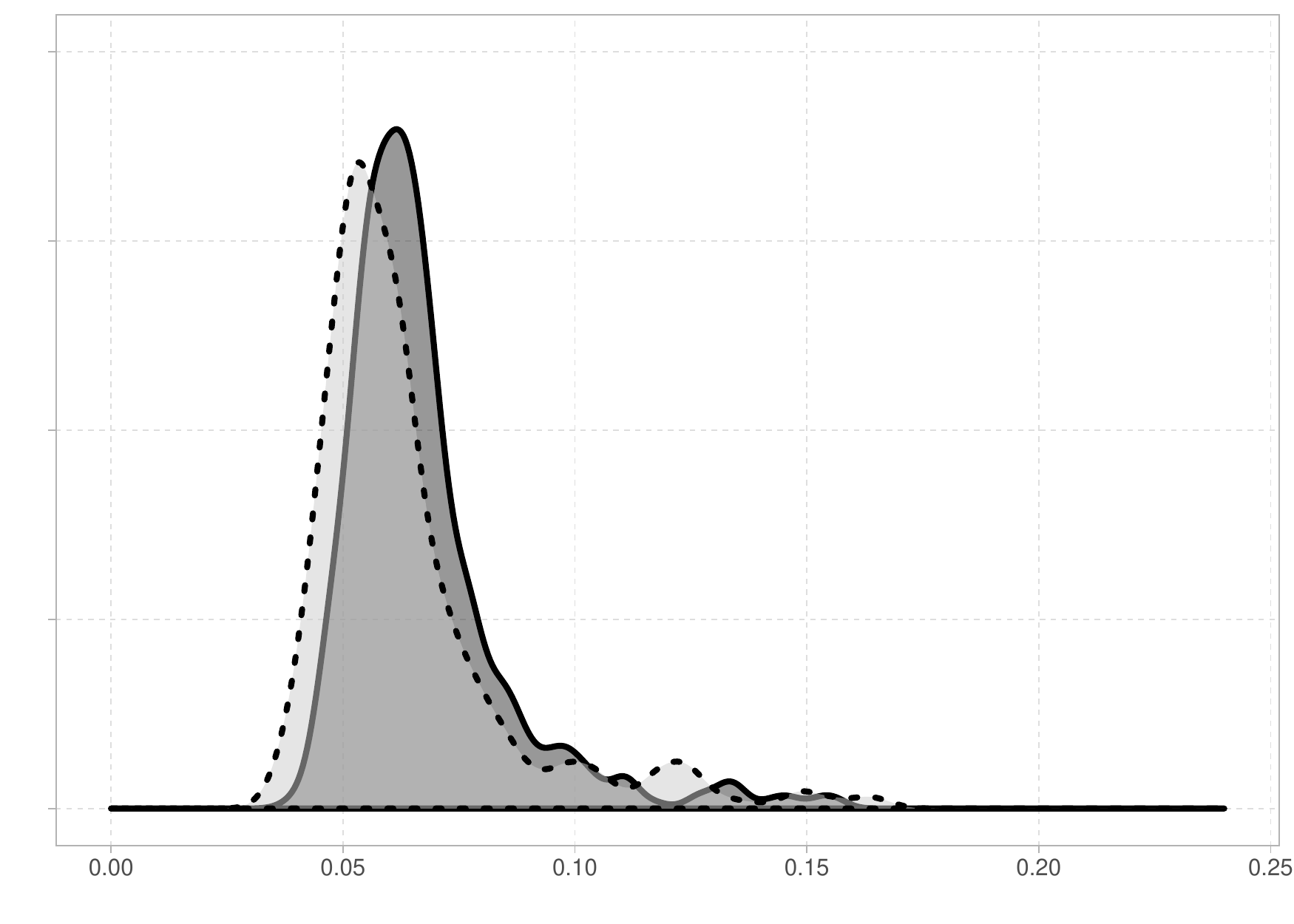}} \\ 
		$\wbbe_{\weme}^{\mcppen}$ &
		\raisebox{-.5\height}{\includegraphics[scale=0.4]{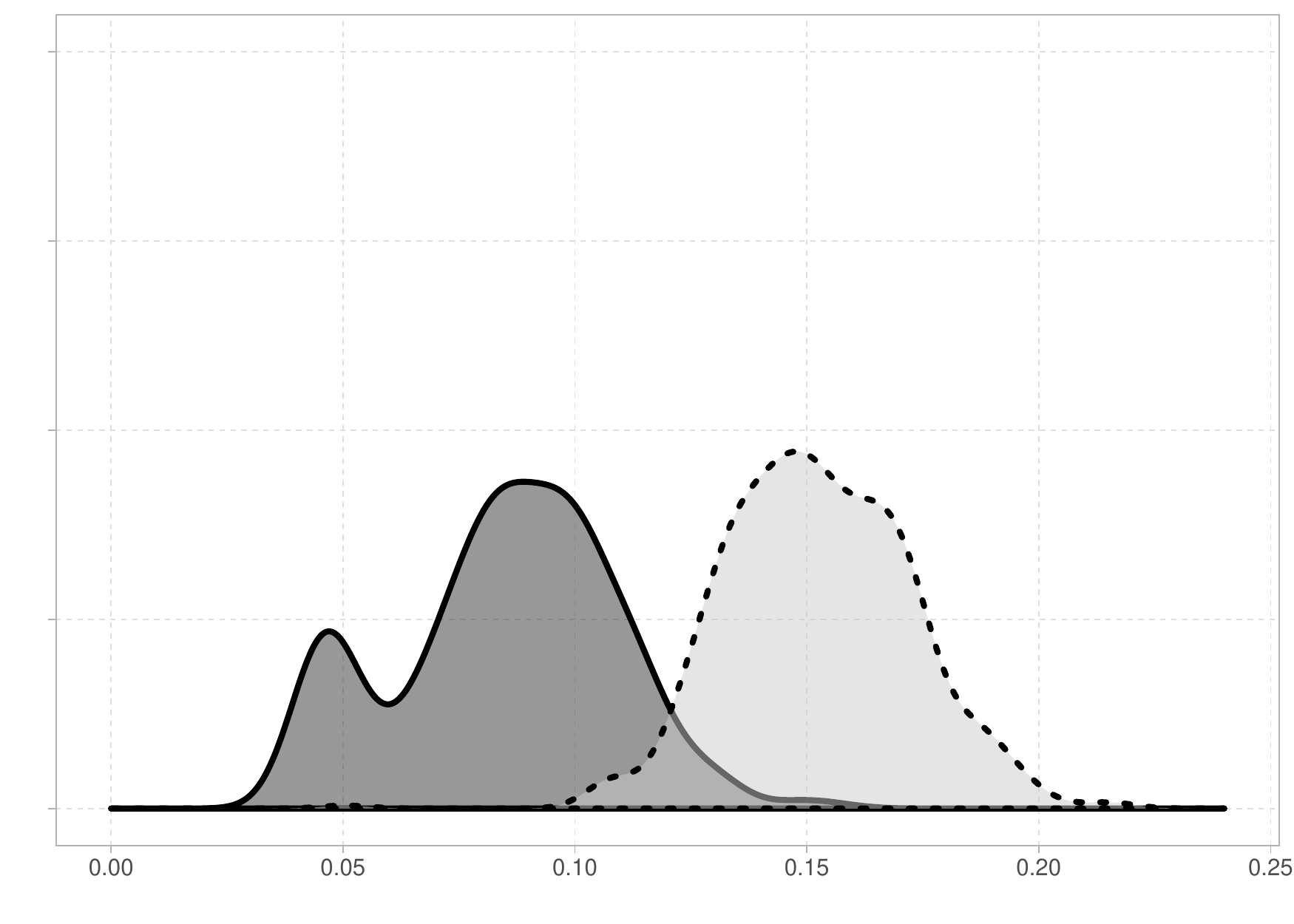}} &  
		\raisebox{-.5\height}{\includegraphics[scale=0.4]{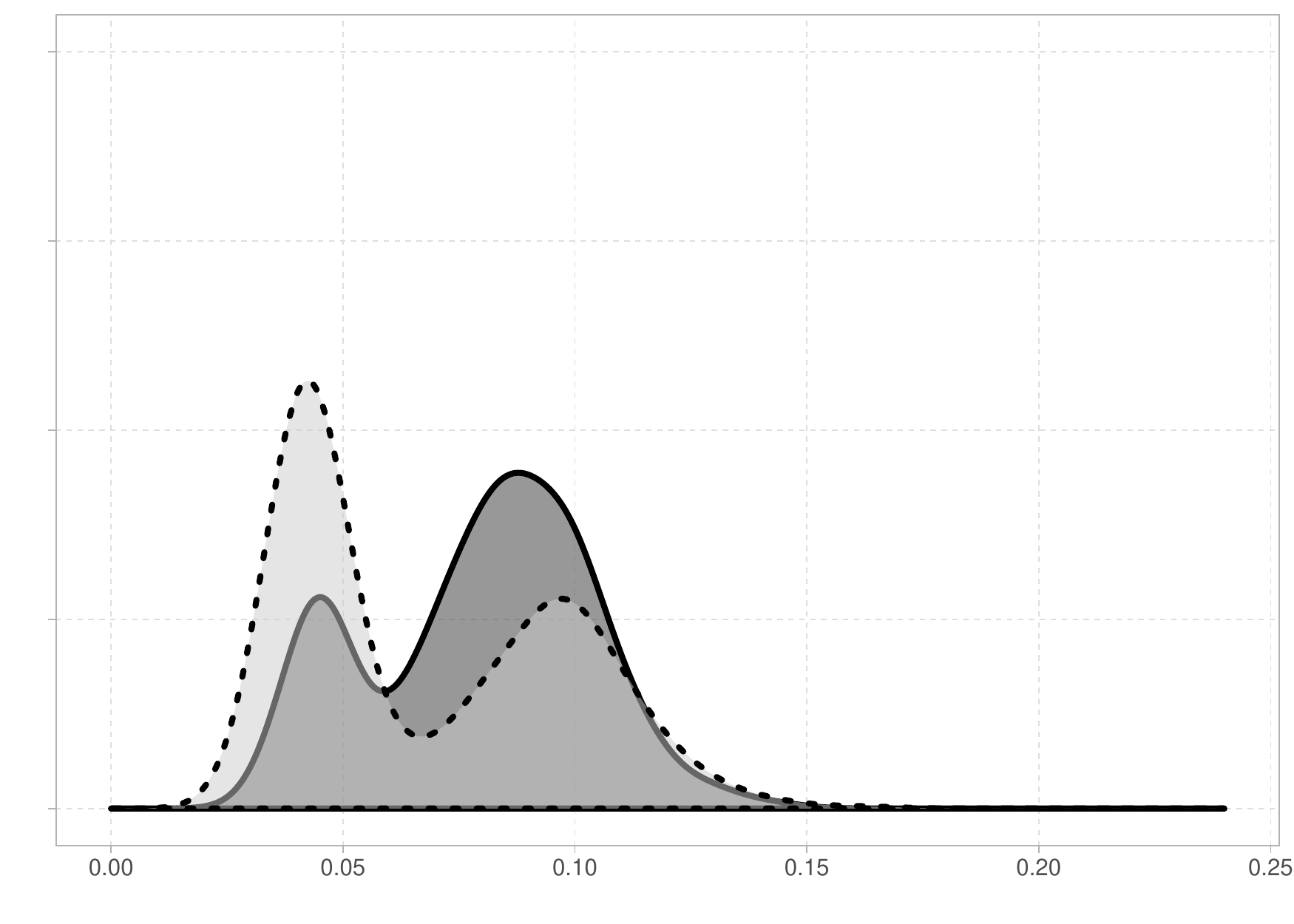}}
	\end{tabular}
	\caption{\small\label{fig:lambdas_cv_cvr_est_rob} Density estimators of the values of the penalizing parameter $\lambda$ obtained when $(n,p) = (150,80)$. The plots on the left correspond to the classical criterion $CV$, while those on the right to the robust version $RCV$. Solid lines filled in deep grey are related to clean samples,while broken lines filled in light grey to contamination scheme \textbf{CB1} with $m=4$.} 
\end{figure}

This numerical experiment shows that the presence of outliers affects the choice of the parameter $\lambda$ when the classical criterion $CV$ is used. For this reason, to study the effect produced in this parameter, Figure \ref{fig:lambdas_cv_cvr_est_rob} shows superimposed the density estimators of the values $\lambda$ chosen for each estimator and for each cross--validation method. 

As expected, by combining a robust estimation procedure with the associated robust cross--validation method, the  obtained values of $\lambda$ remain stable, giving rise to similar densities for both, clean and contaminated samples (see Figure \ref{fig:lambdas_cv_cvr_est_rob}). On the other hand, by choosing the regularization parameter $\lambda $ with the classical $CV$ procedure, the outliers severely affect the final selection. The effect of the contamination becomes evident from the estimated density which is shifted to the right, giving rise to higher values of the regularization parameter. This phenomenon is consistent with the results of Tables \ref{tab:cvrob_cvnorob_c0} and \ref{tab:cvrob_cvnorob_cb1} where the estimators result to be more sparse, since they lead to lower TPP values.

\begin{figure}[ht!]
	\centering
	\begin{tabular}{c}
		{\small(a)}   \\
		\includegraphics[scale=0.6]{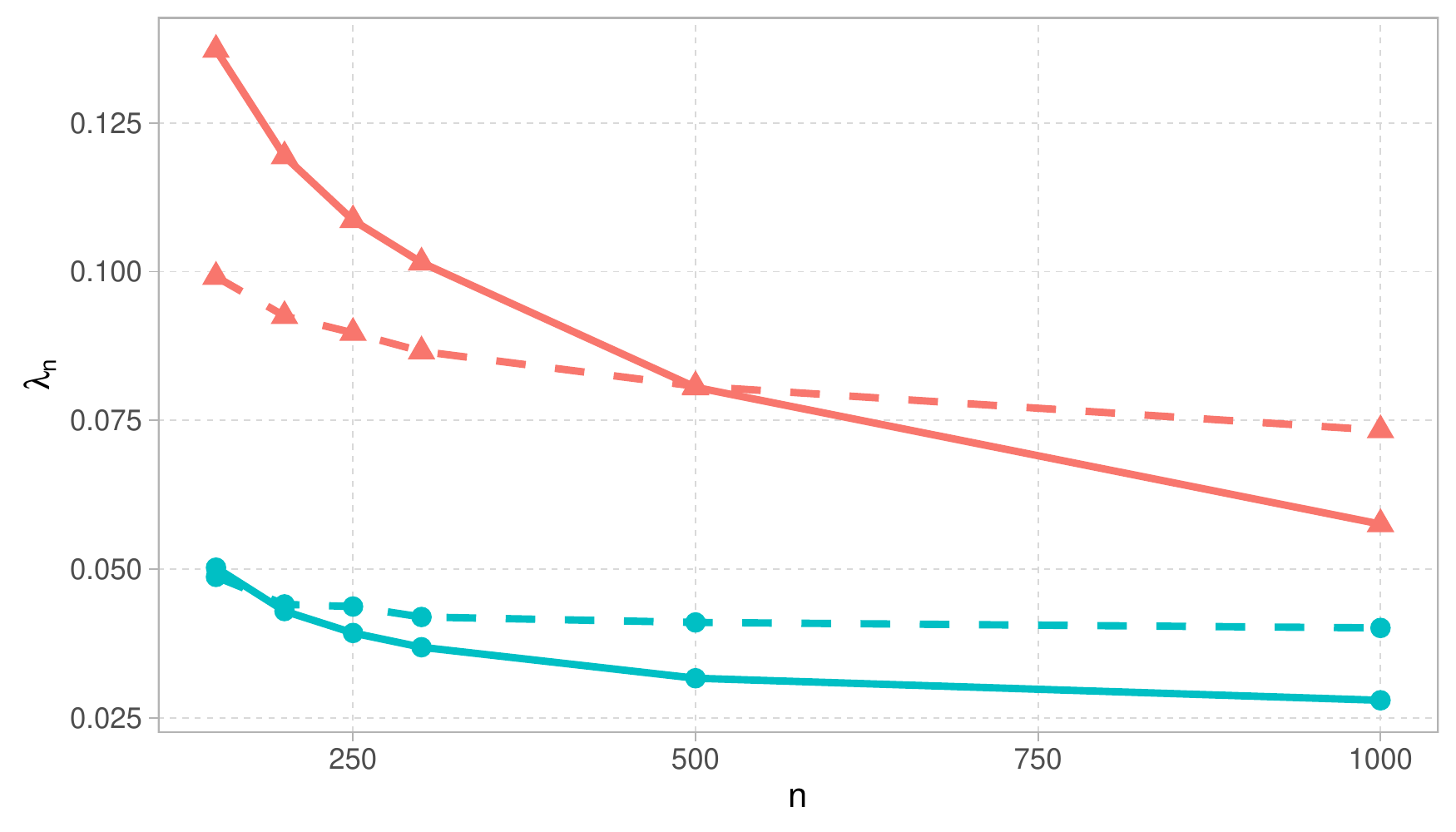}  \\
		{\small(b)}\\ 
		 \includegraphics[scale=0.6]{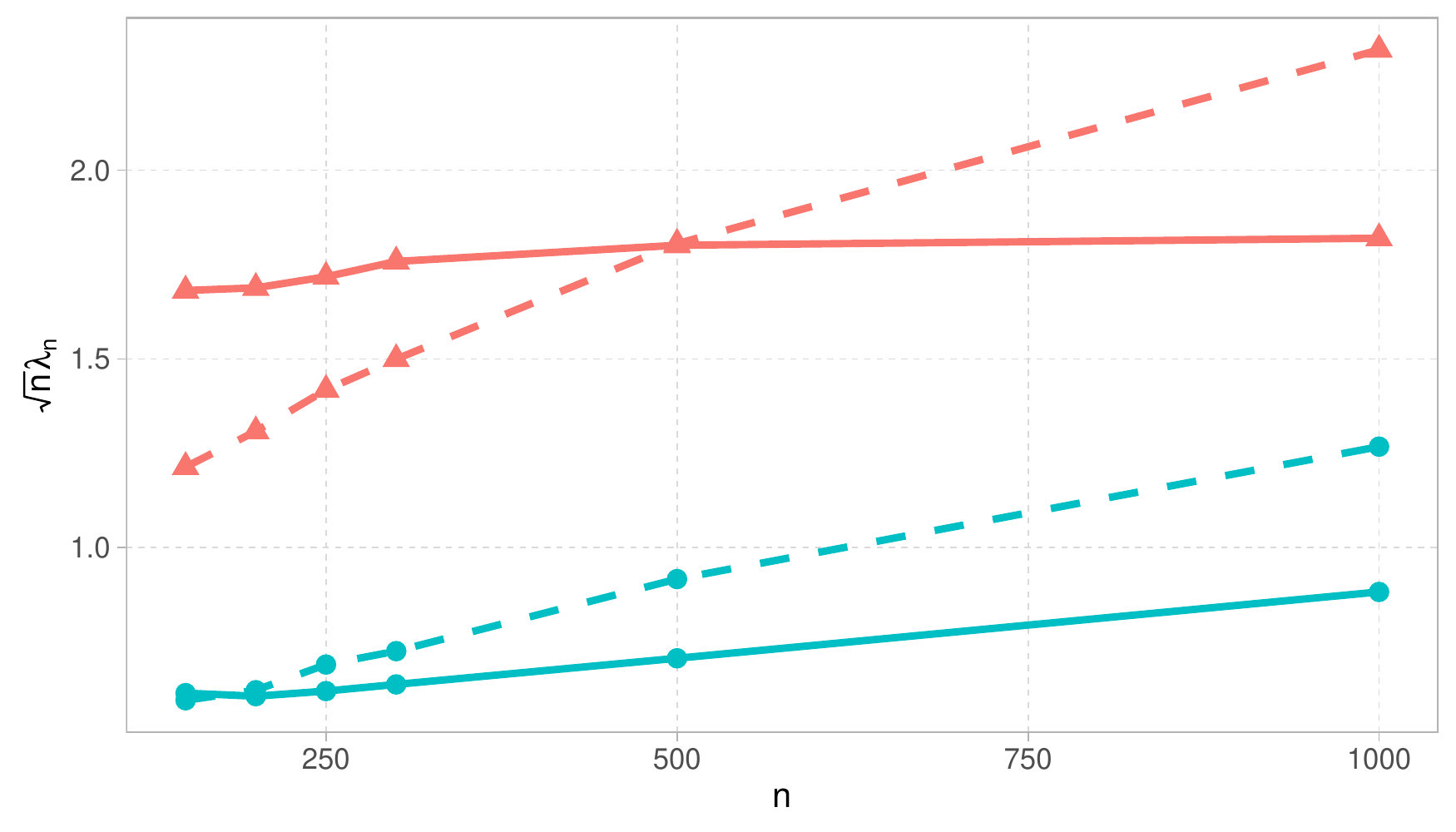}
	\end{tabular}
	\caption{\small\label{fig:srqtla} Means over 500 replications of the cross-validation penalty parameter $\lambda_n$ versus the sample size $n$ are represented on panel (a), while the means of $\sqrt{n} \, \lambda_n$ versus $n$ are given  on panel (b). Red triangles correspond to the estimator based on $\rho(t)=t$ and blue circles to that based on  $\rho=\rho_c$. The solid and dashed lines represent  the values obtained using the Sign and MCP penalties, respectively.}  
\end{figure}

In order to evaluate the convergence rate to $0$ of $\lambda$, a numerical study was conducted for different sample sizes $n = 150, 200,$ $250, 300,$ $500$ and $1000$ under \textbf{C0}. For simplicity, only two loss functions are considered, i.e., the loss $\rho(t)=t$ that gives rise to the maximum likelihood estimator and that introduced by Croux and Haesbroeck (2003). Each of them is combined with the Sign and MCP  penalties. For the classical estimator, the classical cross--validation procedure $CV(\lambda)$ was used, while the robust version $RCV(\lambda)$  is employed for the robust estimator. On the upper panel of Figure \ref{fig:srqtla},  the means over 500 replications of $\lambda_n$ versus the sample size $n$ are represented, while those of $\sqrt{n} \, \lambda_n$ are plotted on panel (b).
Figure \ref{fig:srqtla} (b) shows that for the classical procedure with the Sign penalty, the values of $\sqrt{n} \, \lambda_n$ quickly stabilize around $ 1.80$, which suggests that $\sqrt{n} \, \lambda_n$ is bounded  and therefore, the method leads to estimators with $\sqrt{n}-$rate. In contrast, when using the MCP penalty, both for the classical and the robust estimators, it is observed that $\sqrt {n} \, \lambda_n $ grows with the sample size, while $\lambda_n$ decreases to $0$ (more slowly than with the Sign penalty, as expected). This last fact suggests that in this case we get an estimator with a $\sqrt{n}-$rate (see Remark \ref{remrate}) that also consistently selects variables, according to Corollary \ref{coro:orac}.

\subsection{Results of the numerical study}{\label{sec:monteresultados}}

As usual when considering robust methods, we report pruned averages by considering 
10\%-trimmed means over 500 replications. All tables are presented in Appendix \textbf{B}.  Tables \ref{tab:pmse-c0} to \ref{tab:tpp-tnp-c0} sum up the results corresponding to PMSE, MSE, TPP and TNP quantities under \textbf{C0}, Tables \ref{tab:pmse-a1-a2} to \ref{tab:tpp-tnp-a2} summarize contaminations \textbf{CA1} and \textbf{CA2}, while  Tables \ref{tab:bmse-b1-b2-n150} to \ref{tab:tnp-b1-b2-n300} present the results obtained under scenarios \textbf{CB1} and \textbf{CB2}. 

Figures \ref{fig:ECMP_n_150} to \ref{fig:PVN_n_300} graphically show the 10\%-trimmed means of measures PMSE, TPP and TNP, under \textbf{C0}, \textbf{CA1} and \textbf{CA2}. In these figures, the solid line corresponds to  \textbf{C0}, while the dotted ones with triangles and the dashed lines with squares to \textbf{CA1} and \textbf{CA2}, respectively. In addition, the blue, violet, red and green lines are related  to the following penalized estimators: those  minimizing the \textsl{deviance} ($\rho (t) = t $), the  least squares estimators ($\rho(t) = 1- \exp(-t) $), the $M-$estimators obtained using the loss function $\rho = \rho_c$ introduced in Croux and Haesbroeck (2003) given in \eqref{rocroux} and those based on $\rho =\rho_ {\basu} $ that gives rise to the divergence estimators, respectively. The upper graphs show the results when $w \equiv 1$ and the lower ones when $w(\bx) = W(D^2(\bx,\wbmu,\wbSi ^{- 1})) $. Finally, the plots on the left side of each figure correspond to the estimators with LASSO penalty, those of the center are associated with the Sign penalty and those on the right to MCP.

Besides, Figures \ref{fig:out_B_ECMP_Signo} to \ref{fig:out_B_PVN_MCP} display the values of PMSE, TPP and TNP, when considering clean samples and contaminated ones according to schemes \textbf{CB1} and \textbf {CB2}. In all cases, the solid line corresponds to \textbf{C0}, while the dotted line with triangles and the dashed line with squares are related to \textbf{CB1} and \textbf{CB2}, respectively. Moreover, the blue, red and green lines correspond to $ \wbbe_{\mle}^{\signpen}$, $\wbbe_{\eme}^{\signpen}$ and $\wbbe_ {\weme}^{\signpen}$ in the case of Figures \ref{fig:out_B_ECMP_Signo}, \ref{fig:out_B_PVP_Signo},  \ref{fig:out_B_PVN_Signo} and to $\wbbe_{\mle}^{\mcppen}$, $\wbbe_{\eme}^{\mcppen}$, $\wbbe_{\weme}^{\mcppen}$ in the remaining ones.

Table \ref{tab:pmse-c0} and Figure \ref{fig:ECMP_n_150} show that, for samples without contamination, the estimators penalized with MCP achieve lower PMSE values than with the other penalties. In particular, for samples of size $n = 300$, the maximum likelihood estimators using the MCP penalty come to have PMSE values that are a third of those obtained with the LASSO penalty. That difference is even greater for the least squares estimator and for the $M-$estimators calculated with the function  $\rho = 
\rho_c $ given in \eqref{rocroux}. Under \textbf{C0}, the  robust weighted estimators give similar results to the unweighted ones, not only with respect to the mean squared error of the probabilities, but also with respect to all the other measures (see Tables \ref{tab:pmse-c0} to \ref {tab:tpp-tnp-c0}).

As Tables \ref{tab:pmse-c0} and \ref{tab:bmse-c0} reveal, the $M-$estimator penalized with LASSO loses more efficiency than with the other penalties, reaching PMSE and MSE values that at least double those obtained with $\wbbe_{\mle}^{\lpen}$. Indeed, Figure  \ref{fig:ECMP_n_300}   shows this phenomenon, when $n =  300$ and  the sample is clean. In  this case, the Sign and MCP penalties give lower PMSE values than the LASSO penalty. This fact can be explained by the non--negligible bias, already discussed in this paper, introduced by the LASSO penalty even when the ratio $n/p$ is large. For both bounded penalties, all loss functions give very similar results.  

As expected, the non--penalized estimators give worse results than those obtained by regularizing the estimation procedure. In addition, the PMSE and MSE errors grow when the dimension increases. In particular, this growth is greater when using the Sign penalty for $ n = 150$ and $p =  120$, where PMSE values almost double those obtained with $ n = 150$ and $p = 40$ for most estimators. As mentioned above, the case $(n, p) = (150, 120) $ poses a great challenge to the estimation of the regression parameter and to the selection of variables, as well.

It should be mentioned that the behaviour of measures PMSE and MSE do not always coincide. For some cases, a very high estimation error of the estimates of $\bbe_0$ is obtained together with a low prediction error of the probabilities. This happens, for example, with $\rho = \rho_{\basu}$ and the loss function that gives rise to the least squares estimators. Indeed, for some dimensions the MSE values of these estimators take such large values that they are reported with a $\bigstar$. A possible explanation of this fact could be that these losses, unlike what happens with the proposal given in Croux and Haesbroeck (2003), do not meet the conditions that guarantee the existence of the non--penalized maximum likelihood estimator. The obtained results suggest that introducing a penalty does not solve this existence problem, so these estimators may explode in some samples. The procedures based on the loss introduced by Croux and Haesbroeck (2003), both weighted and unweighted versions, produce better average squared errors, MSE$(\wbbe)$, than other bounded losses, in particular, when using the penalty Sign.

Regarding the proportion of correct classifications and the proportions of true positives and null coefficients, all penalized estimators give similar results. It should be mentioned that, when the LASSO penalty is used, lower TNP values are obtained than with other penalties, giving rise to less sparse estimators. This procedure seems to be less skilled than MCP to identify as $0$ those coefficients associated with explanatory variables that are not involved in the model. This drawback is also observed, although to a lesser extent, when considering the unweighted divergence estimator or the maximum likelihood one, both combined with the Sign penalty (see Table \ref{tab:tpp-tnp-c0}).

The sensitivity to atypical data of estimators based on $\rho(t) = t $ and $ w \equiv 1$,   combined with any of the considered penalties, becomes evident all along the tables. On one hand, Tables \ref{tab:pmse-a1-a2} and \ref{tab:bmse-a1-a2} show that, when outliers following schemes \textbf{CA1} or \textbf{CA2} are introduced, PMSE and MSE are at least three times those obtained for uncontaminated samples. On the other hand,  in some situations under contaminations \textbf{CB1} and \textbf{CB2}, the MSE become five times larger than the corresponding value under \textbf{C0} (see Tables \ref{tab:bmse-b1-b2-n150} and \ref{tab:bmse-b1-b2-n300}).

Figures \ref{fig:ECMP_n_150} and \ref{fig:ECMP_n_300} reveal that, under contamination patterns \textbf{CA1} and \textbf{CA2}, the best behaviour is attained by the penalized weighted $M-$estimators. In fact, their probability mean squared errors (PMSE) are close to those obtained for clean samples with the bounded penalties Sign and MCP. The benefits of using  weighted estimators is also reflected in the proportions of true positives and zeros, as illustrated in Figures \ref{fig:PVP_n_150} to \ref{fig:PVN_n_300}. In the case of these latter measures, the LASSO penalty gives higher values of the probability of true positives in detriment of the TNP values since, as we mentioned, this penalty has more difficulties in the identification of non-active explanatory variables.

Worth noticing that, under \textbf{CA1} and \textbf{CA2}, unweighted estimators have  higher PMSE values than their weighted versions, especially when $n = 150$. Under \textbf{CA2}, these values can  double those obtained with the estimators that control the leverage of the covariates. Among the estimators with $ w \equiv 1$, those that give lower PMSE values are the procedures  corresponding to $\rho = \rho_{\basu}$ and those based on the least squares method when combined with the Sign and MCP penalties, in particular when $ n = 300$.

In scenario \textbf{CA1}, the most stable estimators are those based on bounded loss functions. For example, Figure \ref{fig:PVP_n_300} shows that the procedure based on  $\rho (t) = t$ is   the only one  having problems with this level of contamination. On the other hand, the loss function introduced by Croux and Haesbroeck (2003) leads to more sparse estimators than those obtained with $\rho = \rho_{\basu}$ and $\rho (t) = 1- \exp(-t)$.

Table \ref{tab:tpp-tnp-a2} shows that  as the level of contamination increases (scheme \textbf {CA2}), all estimators seem to become too sparse. This effect directly  impacts on measure TPP that decreases by almost half in unweighted estimators. As expected, this behaviour is more pronounced when using the Sign and MCP penalties combined with $\rho (t) = t$. Although to a lesser extent, the $M-$estimators with $\rho = \rho_c$ given in \eqref{rocroux} are affected by this contamination scheme.  With respect to the ability to detect active variables, weighted estimators achieve similar results to those obtained under \textbf{C0}.

When considering contamination schemes \textbf{CB1} and \textbf{CB2}, Tables \ref{tab:tpp-b1-b2-n150} and \ref{tab:tpp-b1-b2-n300} show a decrease in the probability of true positives, in particular, for small values of the slope $m$ ($m = 0.5, 1.2$) corresponding to mild outliers that are the most difficult ones to be detected. The TPP values of the weighted $M-$estimators, based on the function introduced by Croux and Haesbroeck (2003), recover their good performance when the slope  $m$ increases. This effect is also observed in the results obtained for the PMSE measure that grows for low values of $m$ and decreases as the leverage of the outliers increases (see Figures \ref{fig:out_B_ECMP_Signo} and \ref{fig:out_B_ECMP_MCP}). It is worth mentioning that the TNP values  obtained under \textbf{CB1} and \textbf{CB2} are similar to those obtained for uncontaminated samples, except for estimator $\wbbe_{\mle}^{\mcppen}$  that  seems to be the most affected by this type of contamination. In summary, these results show that the Sign and MCP penalties manage to identify the non--active variables (see Tables \ref{tab:tnp-b1-b2-n150} and \ref{tab:tnp-b1-b2-n300})

From Figures \ref{fig:out_B_ECMP_Signo} and \ref{fig:out_B_ECMP_MCP} it follows that,  under \textbf{CB1},  the estimators based on $\rho = \rho_c$ combined with for the Sign or MCP penalties have much lower PMSE values than those obtained with $\rho(t) = t$, particularly when considering weighted $M-$estimators. This effect is clearer when $m$ is greater than 3 since the estimators $\wbbe_{\mle}^{\signpen}$ and $\wbbe_{\mle}^{\mcppen}$ yield  PMSE values larger than $0.10$, that is, 5 times larger than those obtained under \textbf{C0}. The best behaviour of the weighted estimators is due to the fact that these methods detect most of the atypical data when the slope $m$ is large. In some cases,  when the MCP penalty is used, the advantages of the robust weighted $M-$estimators  are strengthened. For example, when $(n,p) = (300,80)$, the PMSE of $\wbbe_{\weme}^{\signpen} $ and $\wbbe_{\weme}^{\mcppen}$ is very similar to that obtained for clean data.

Summarizing, for the studied contaminations, the weighted $M-$estimators based on the function $\rho = \rho_c$ given in \eqref{rocroux} combined with the MCP and Sign penalties, turn out to be the most stable and reliable among the considered procedures.

\section{Real Data Analysis}{\label{sec:realdata}}
In this section, we consider two real data sets: the Diagnostic Wisconsin Breast Cancer and the Single Positron Emission Computed Tomography (SPECT) data. Based on the results obtained in the numerical experiments reported in Section  \ref{sec:numerical},   we only illustrate the performance of the  $M-$estimators  computed with the  Croux and Haesbroeck (2003) loss function and  of the classical ones by using different penalties. For the robust estimators,   the tuning constants are equal to those considered in Section \ref{sec:numerical}.

\subsection{Breast cancer diagnosis}{\label{sec:cancer}}
We study a dataset   corresponding to the Diagnostic Wisconsin Breast Cancer Database  available at  \url{https://archive.ics.uci.edu/ml/datasets/Breast+Cancer+Wisconsin+\%28Diagnostic\%29}. 

Ten real-valued features are computed from a digitized image of a fine needle aspirate (FNA) of a breast mass and they describe characteristics of the cell nuclei present in the image. 
Measured attributes are related to:  radius (mean of distances from centre to points on the perimeter),  texture (standard deviation of grey-scale values),
perimeter,  area, smoothness (local variation in radius lengths),  compactness ($perimeter^2 / area - 1.0$), concavity (severity of concave portions of the contour), concave points (number of concave portions of the contour),  symmetry and  fractal dimension.
For each of these features  the mean, the standard deviation and  the maximum among all the nuclei of the image were computed, generating a total of $p = 30$ covariates for each image.
From the $n = 569$ tumours, 357 were benign and  212 malignant and the goal is to predict the type of tumour from the  $p = 30$ covariates.

\begin{figure} 
	\renewcommand{\arraystretch}{0.1}
	\newcolumntype{M}{>{\centering\arraybackslash}m{\dimexpr.97\linewidth-0.\tabcolsep}}
	\newcolumntype{G}{>{\centering\arraybackslash}m{\dimexpr.26\linewidth-0.1\tabcolsep}}
	\begin{center}
	\begin{tabular}{G p{0.01in} M}
		$\wbbe_{\mle}^{\lpen}$ && $\wbbe_{\mle}^{\mcppen}$\\
\includegraphics[scale=0.35]{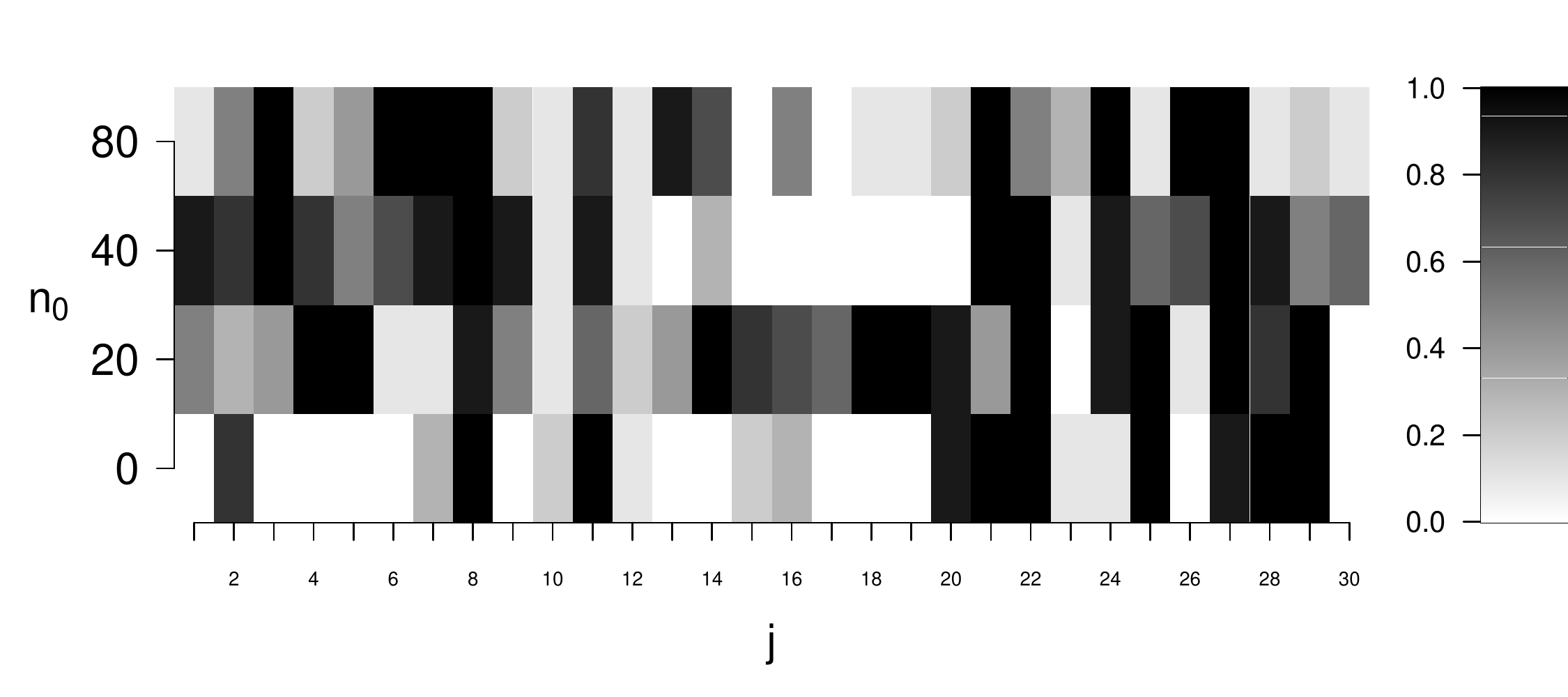} && 
 \includegraphics[scale=0.35]{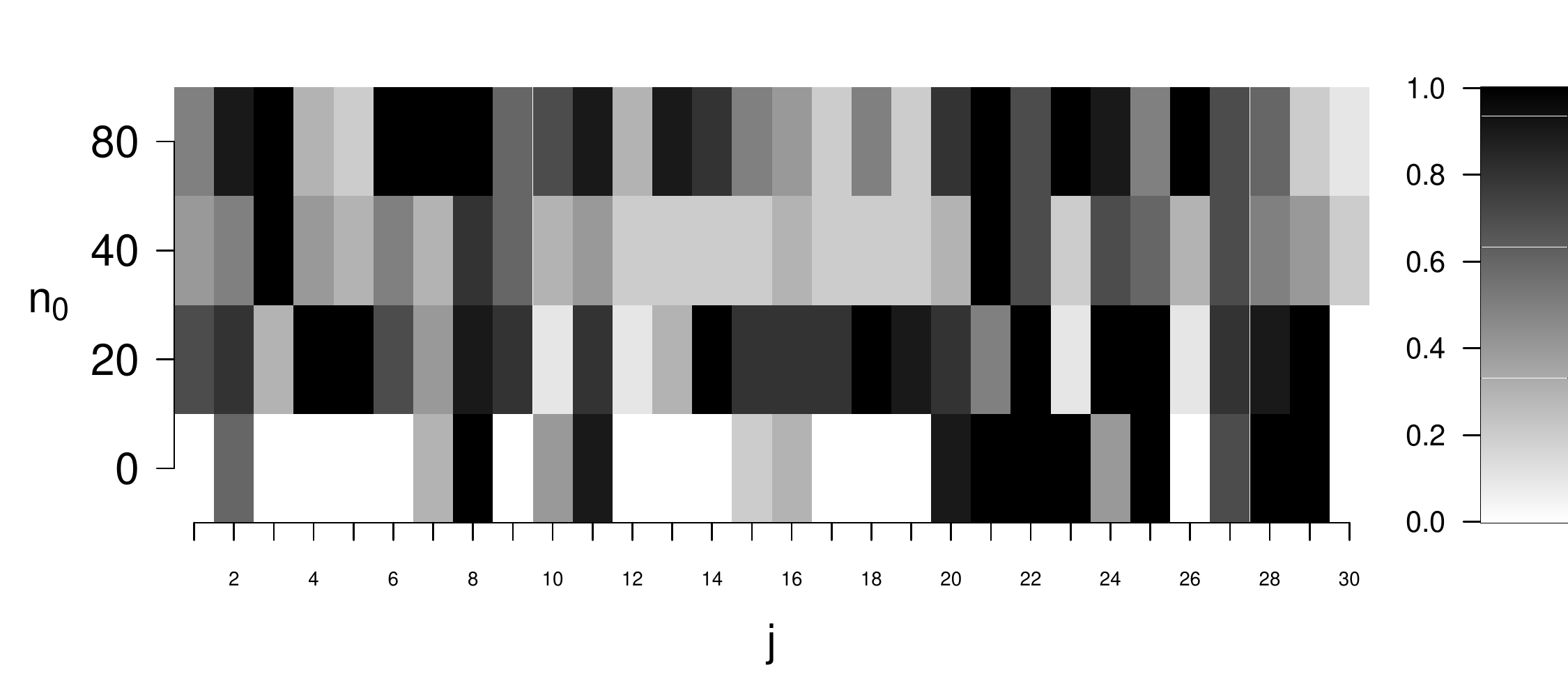}\\
  $\wbbe_{\eme}^{\signpen}$ && $\wbbe_{\eme}^{\mcppen}$\\
   \includegraphics[scale=0.35]{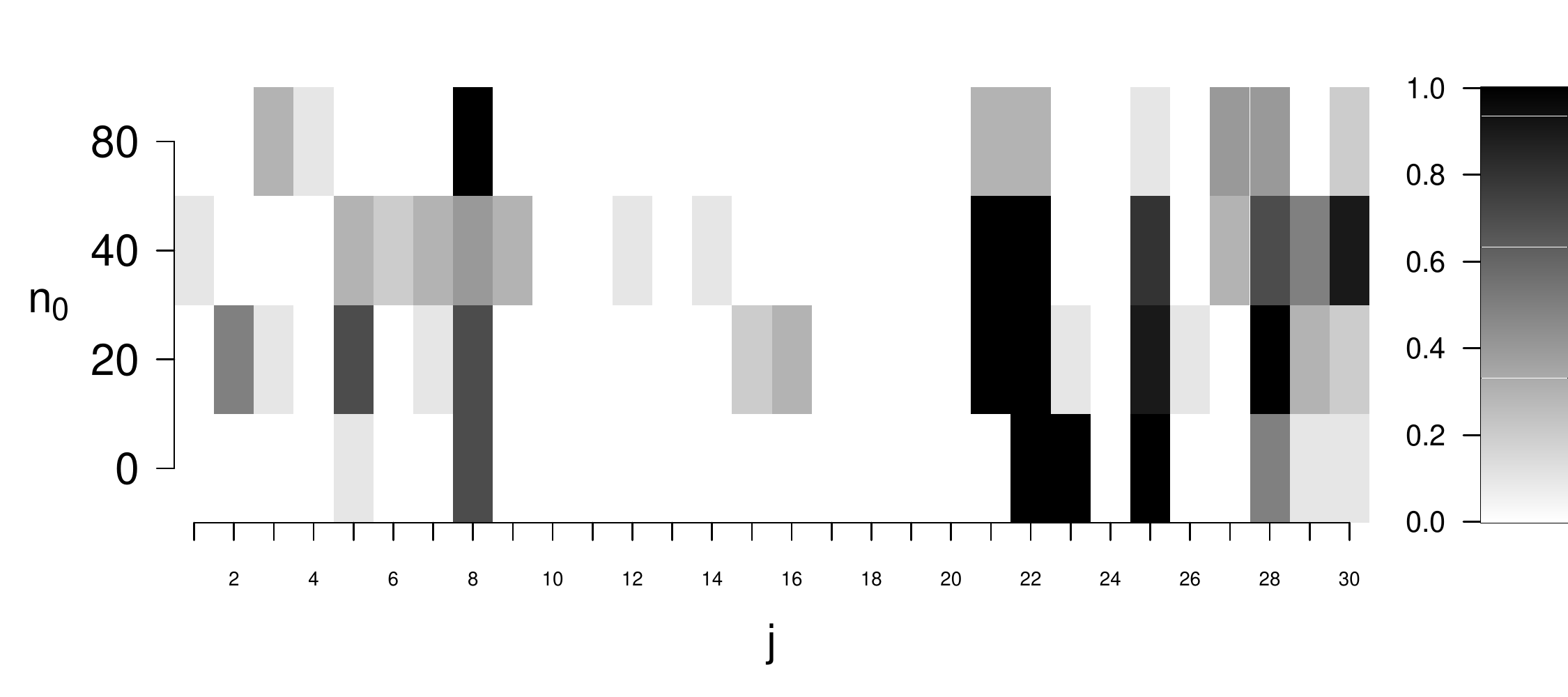} &&  
   \includegraphics[scale=0.35]{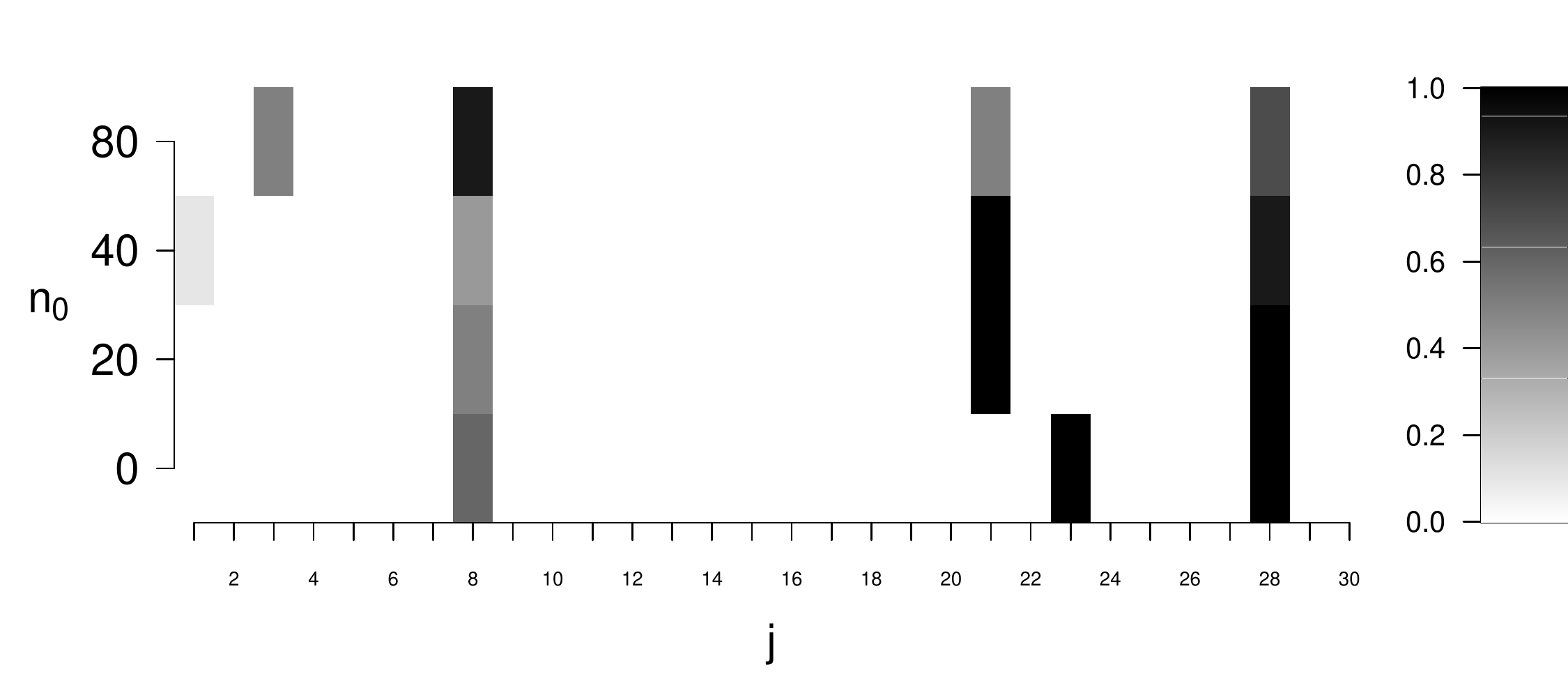} \\
    $\wbbe_{\weme}^{\signpen}$ && $\wbbe_{\weme}^{\mcppen}$\\
      \includegraphics[scale=0.35]{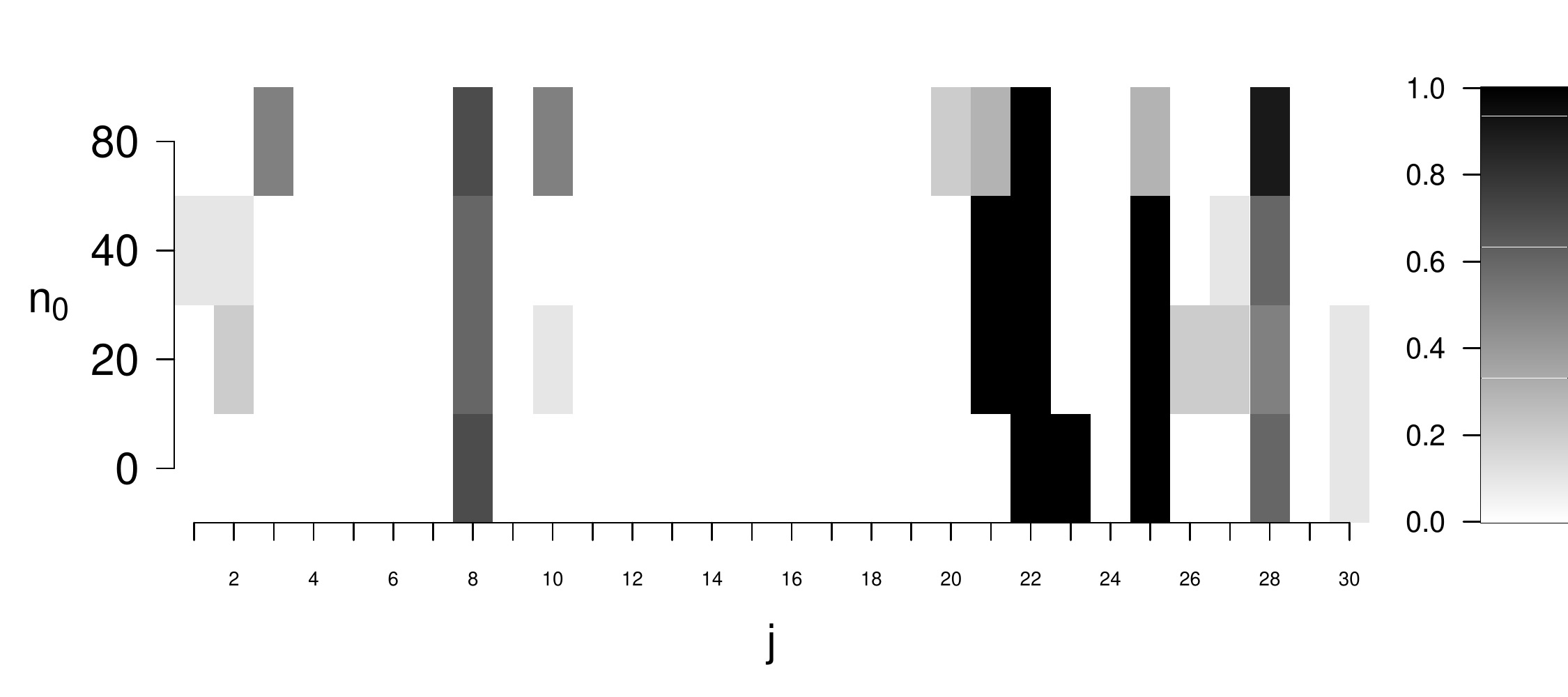} &&  
      \includegraphics[scale=0.35]{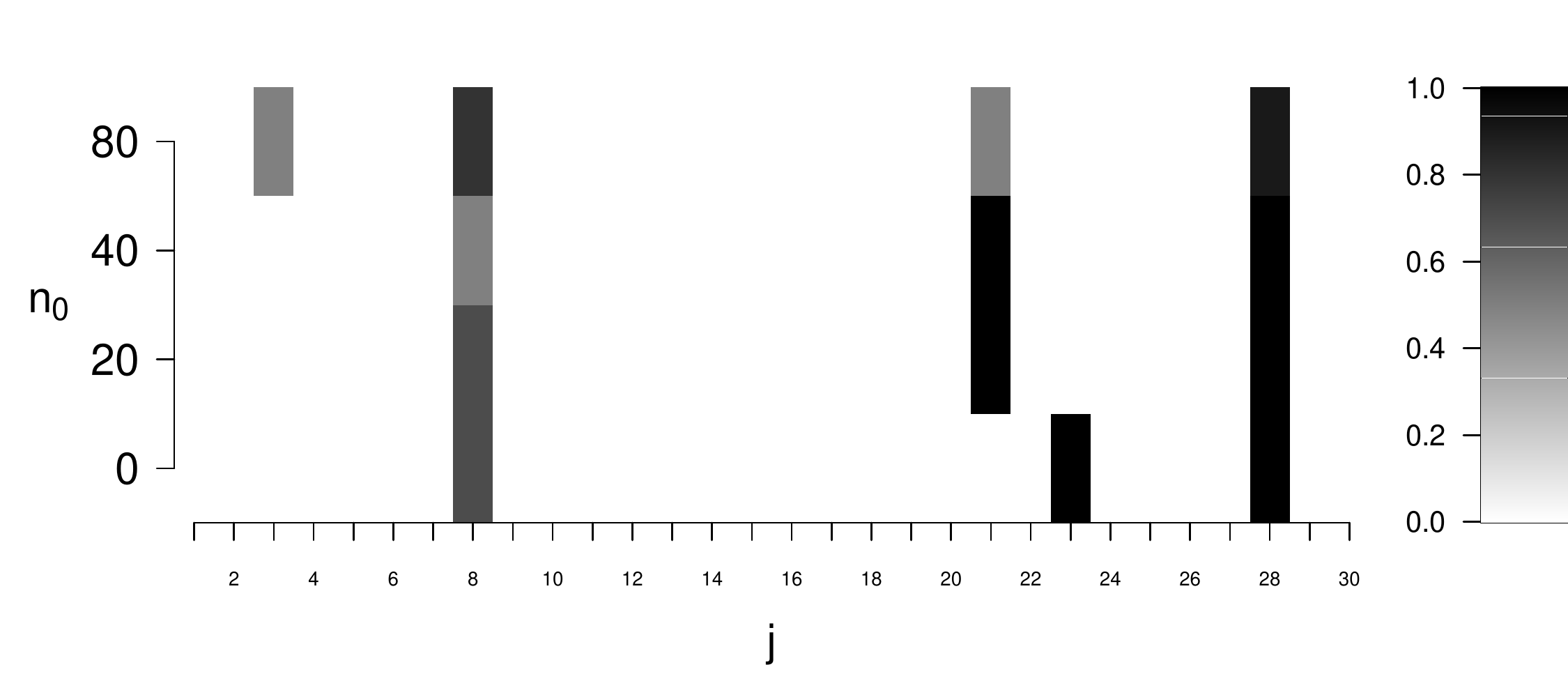}   
	\end{tabular}
\vskip-0.15in
	\caption{\small	\label{fig:wdbc}Grey--scale representation of measures  $\Pi_{a,j}, 1\le j\le 30$ for each method and number of atypical points introduced artificially.}
\end{center}
\end{figure}

From this dataset, we want to assess the impact of artificial outliers on the variable selection capability of different methods. For this purpose, we add $n_0$ atypical observations artificially. Each outlier  $(\wty, \wtbx)$ was generated as follows. 
In a first step we compute the weighted $M-$estimator with MCP penalty, $(\wbbe_{\wmch}^{\mcppen}, \wgamma_{\wmch}^{\mcppen})$, with the original points and then, we generate $\wtbx \sim N_p(\bcero, 100\, \bI)$ and define a bad classified observations as
$	\wty = 1$ when $ \wtbx \trasp \, \wbbe_{\wmch}^{\mcppen} +\wgamma_{\wmch}^{\mcppen}< 0 $ and $0$, otherwise.
We add $n_0= 0,20,40$ and $80$ outliers. Given each contaminated set, we split the data in 10 folds of approximately the same size. For each estimation method and each  subset $i$ ($1 \leq i \leq 10$),   we obtain $\wbbe^{(-i)} $ and $ \wgamma^{(-i)}$, the slope and intercept estimates computed without the observations that lie in the $i-$th subset. 
  Then, for each variable, we evaluate the fraction of times that it is detected as active among the $10$ folds as
$\Pi_{a,j} =  {\#\{i: \wbbe^{(-i)}_j \neq 0\}}/{10}$ for $ 1 \leq j \leq 30 $.
Note that this quantity depends on the estimator that is used and on $n_0$ and, regarding variable selection, it attempts to capture the stability of each method against outliers. 
In each row of the plots of Figure \ref{fig:wdbc}, for each estimator and each value of $n_0$,  we show a grey--scale  representation of the measures $\Pi_{a,1}, \dots, \Pi_{a,30}$.

As illustrated in Figure \ref{fig:wdbc}, for the considered contamination, the non--robust estimators $\wbbe_{\mle}^{\lpen}$ and $\wbbe_{\mle}^{\mcppen}$ show a very unstable and erratic variable selection, making evident their sensitivity to outliers. The results regarding $\wbbe_{\mle}^{\signpen}$ are not included just for brevity since they lead to similar conclusions. In contrast, the robust procedures based on the Croux and Haesbroeck (2003) loss function select approximately the same subset of covariates, regardless of the amount  $n_{0}$ of added outliers, showing a stable identification of active variables. In particular, the hard rejection weighted estimators are more stable than their unweighted counterparts, when using the Sign penalty. The robust estimators with MCP penalty are more sparse than when using the Sign penalty, which can be explained by means of the  theoretical properties studied in Section \ref{sec:selectvar}.

\subsection{SPECT dataset}{\label{sec:spect}}
Single Positron Emission Computed Tomography (SPECT) imaging is used as a diagnostic tool for myocardial perfusion. This technique is very popular due to its high signal--to--noise rate and relative low cost. However, subjective interpretations of these images are often inaccurate so a computational procedure is needed as a complement in order to obtain an semi--automatic classification. The data are available at the UCI repository (\url{https://archive.ics.uci.edu/ml/datasets/SPECT+Heart}).

In order to semi--automate the SPECT diagnostic process, $p = 44$ features were generated from each image, as described in Kurgan \textsl{et al.} (2001).  We aim to classify each image into one of the two categories referring to the patient's cardiac situation: \textit{Normal} and \textit{Abnormal}. The dataset consists on $n = 267$ observations, where $212$ were classified as \textit{Normal} and the remaining 55 as \textit{Abnormal}.  A feature of  this  data set is that it is highly unbalanced. Hence, one may suspect that some difficulties may be encountered when classifying the observations.

The dataset was split in 10 folds of approximately the same size as in Section \ref{sec:cancer}. For each subset $i$ ($1 \leq i \leq 10$) and for each estimator, we obtain $\wbbe^{(-i)} $ and $ \wgamma^{(-i)}$, the slope and intercept computed without the $i-$th subsample. We consider the classical maximum likelihood estimator with the LASSO penalty and the weighted $M-$estimators  with the three penalties: LASSO, Sign and MCP.  After this step, each observation $s$ in the $i-$th subsample  is classified according to the sign of  $\bx_s \trasp \wbbe^{(-i)}+\wgamma^{(-i)}$. Within the $i-$th subset, we define the following quantities:  $t_i$   the proportion of observations correctly classified, $m_i$ the proportion of observations correctly assigned that belong to the class \textit{Abnormal} and $b_i$ those correctly 	assigned to the category \textit{Normal}. Besides, $a_i$ denotes  the active coordinates of  $\wbbe^{(-i)}$, i.e., those different from 0. 

\begin{figure}  
\begin{center}
	\begin{tabular}{ccc}
		$\wbbe_{\mle}^{\lpen}$ &   $\wbbe_{\weme}^{\lpen}$\\
\includegraphics[scale=0.3]{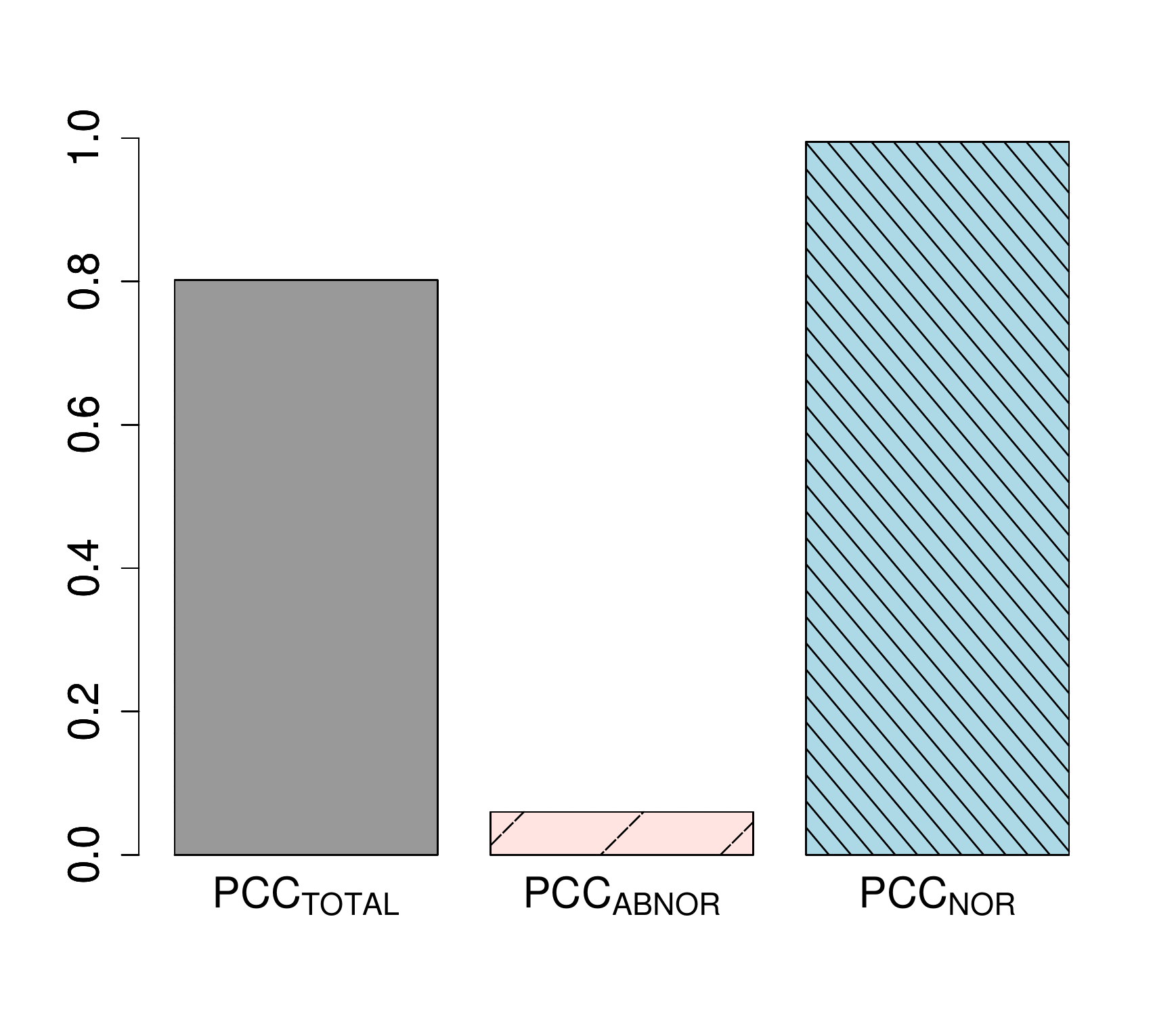}   & 
\includegraphics[scale=0.3]{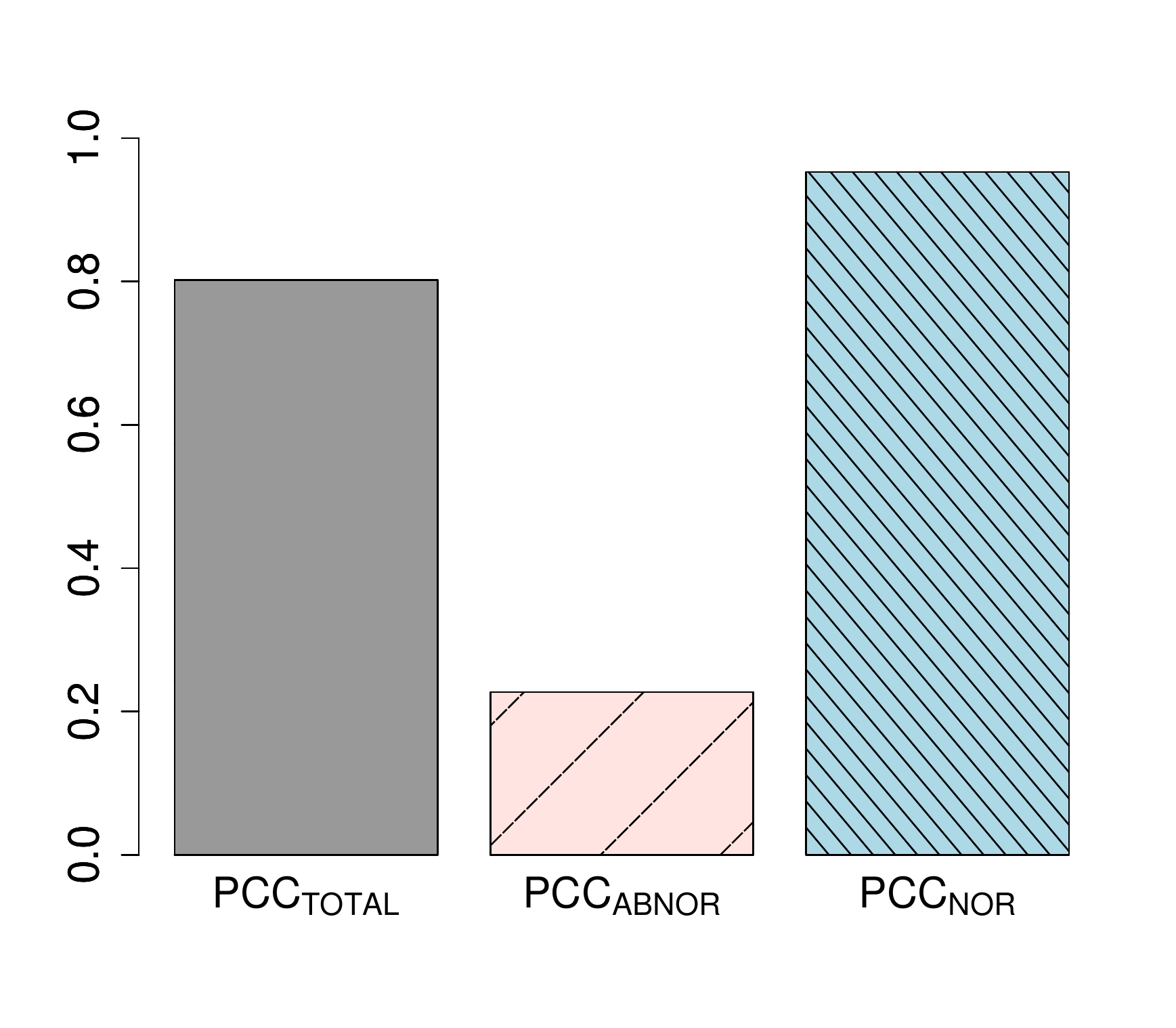}
\\
$\wbbe_{\weme}^{\signpen}$ & $\wbbe_{\weme}^{\mcppen}$   \\
\includegraphics[scale=0.3]{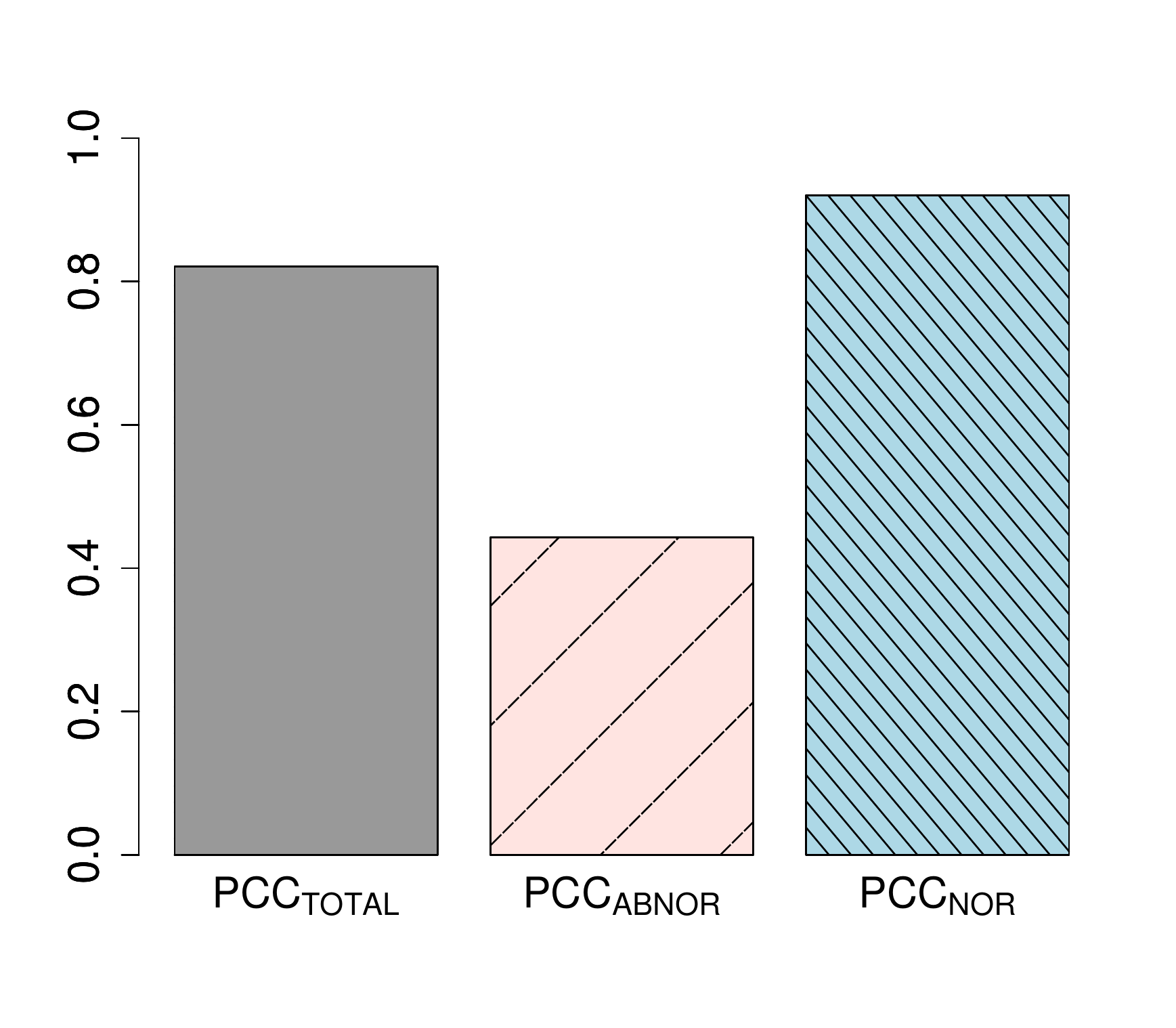} & 
\includegraphics[scale=0.3]{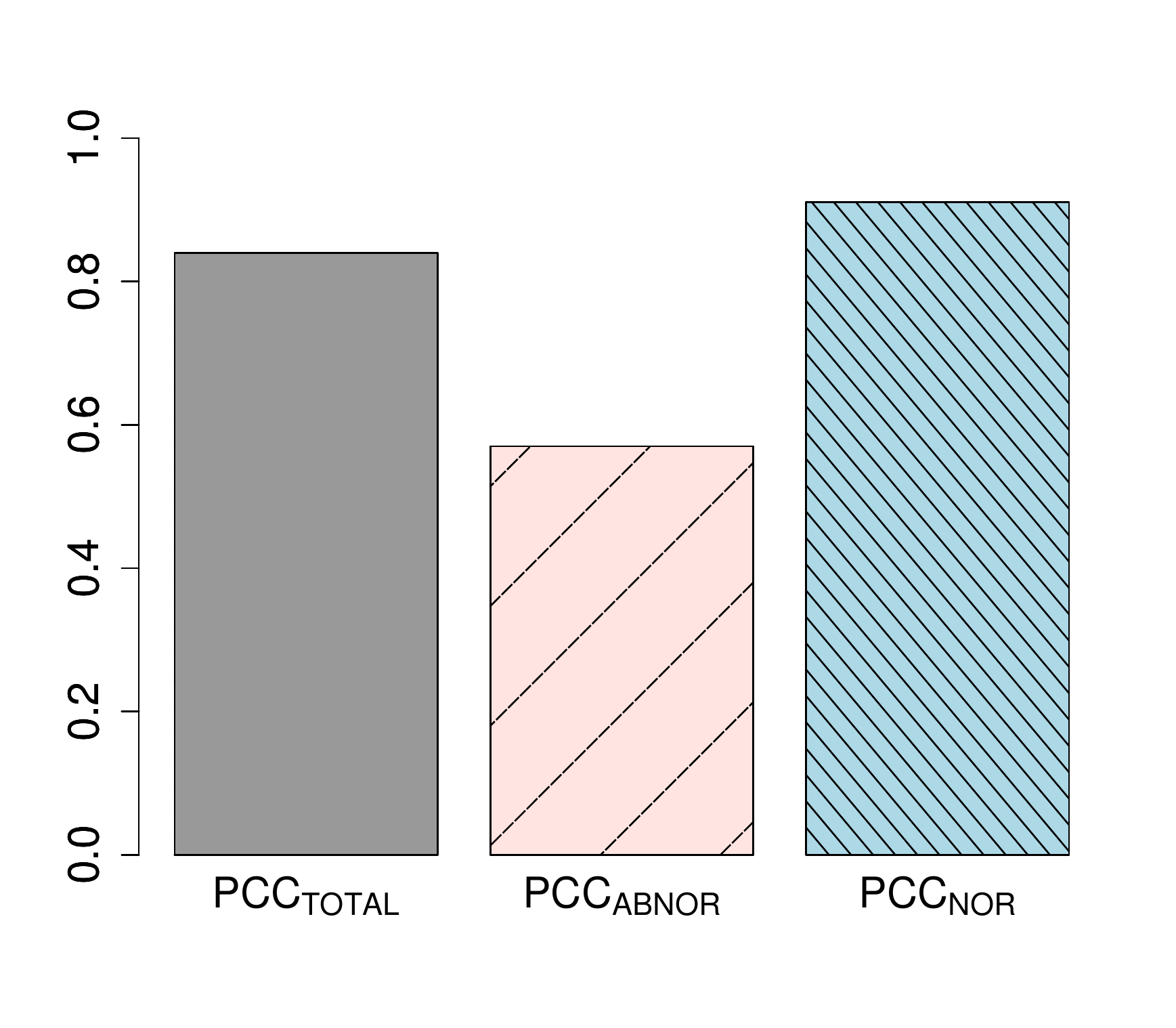}    
\end{tabular}
\end{center}
\vskip-0.2in
\caption{Barplot corresponding to $\mbox{PCC}_{\mbox{\tiny \sc total}} $ (in grey), $\mbox{PCC}_ {\mbox{\tiny \sc abnor}}$ (in pink with lines with positive slope) and $\mbox{PCC}_ {\mbox{\tiny \sc nor}}$ (in blue with lines with negative slope) for the SPECT dataset.}\label{fig:barplot_spect}
\end{figure}

\begin{figure}  
\begin{center}
\includegraphics[scale=0.3]{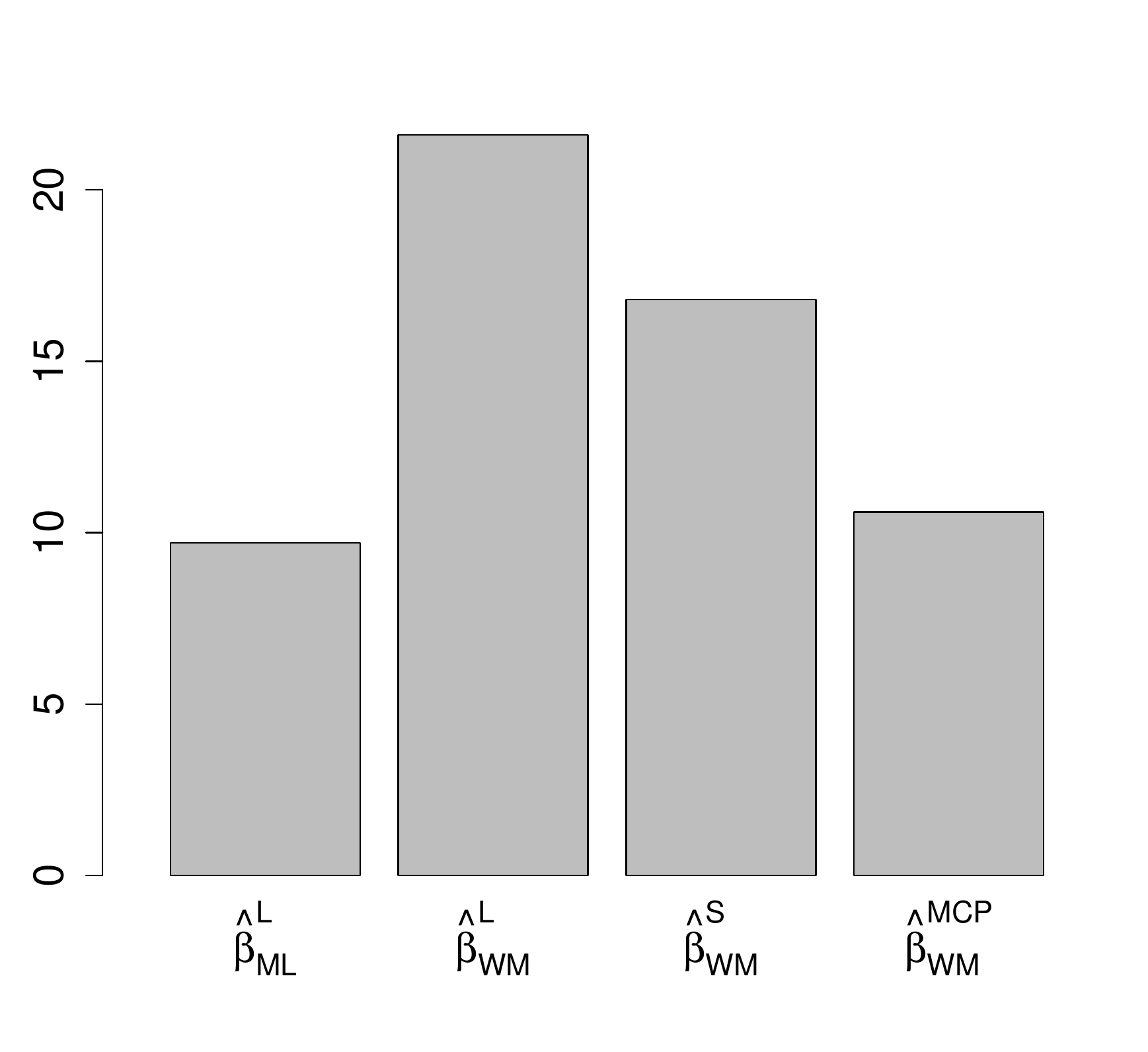} 
\end{center}
\vskip-0.2in
\caption{Barplot corresponding to $\mbox{N}_{\mbox{\tiny \sc act}}$ for the SPECT dataset.}\label{fig:Nact_spect}
\end{figure}

We label $\mbox{PCC}_{\mbox{\tiny \sc total}}$, $\mbox{PCC}_{\mbox{\tiny \sc abnor}}$, $\mbox{PCC}_{\mbox{\tiny \sc nor}}$ and $\mbox{N}_{\mbox{\tiny \sc act}}$, the mean over the $10$ folds of  the quantities  $\{t_i\}$, $\{m_i\}$, $\{b_i\}$ and $\{a_i\}$, respectively. Figures \ref{fig:barplot_spect} and \ref{fig:Nact_spect} summarize  these averages for the considered estimators  as a barplot.

The obtained results   show that the overall classification proportion,  $\mbox{PCC}_{\mbox{\tiny \sc total}} $,  is quite similar in all cases. Note that, when using the LASSO penalty, this total correct classification proportion is close to 1 at the expense of obtaining a very low proportion of correct classifications in the \textit{Abnormal} category, in particular for the classical procedure the value of  $\mbox{PCC}_{\mbox{\tiny \sc abnor}}$ is 0.060. In other words, when using LASSO the   procedures tend to classify almost all the observations as \textit{Normal}. Better classification proportions for the class  \textit{Abnormal} are obtained when considering $\wbbe_{\weme}^{\lpen}$ than for $\wbbe_{\mle}^{\lpen}$. This phenomenon may be explained by the fact that these estimators give rise to  less sparse estimators of $\bbe$ (see Figure \ref{fig:Nact_spect}) and are less sensitive to outliers into a data set than the classical ones (see Section \ref{sec:monteresultados}).

The weighted $M-$estimators with the  Croux and Haesbroeck (2003) loss function combined either with the Sign or MCP penalties lead to a better classification in the \textit{Abnormal} category, with only a slight decrease in $\mbox{PCC}_{\mbox{\tiny \sc nor}}$ and at the same they obtain  higher values of  the total proportion of correct classification. For example, the estimator $\wbbe_{\weme}^{\mcppen} $ correctly classifies more than 50 \% of the observations in the category \textit{Abnormal}  and  90 \% of those in the \textit{Normal} class,   assessing the best results in this example, besides the MCP penalty is the one that gives also the most sparse robust estimator.

\section{Concluding remarks}{\label{sec:conclusiones}}

The logistic regression model may be used for  classification purposes when  covariates with predictive capability are observed for each of the classes. When the regression coefficients are assumed to be sparse, i.e., when only a few coefficients are nonzero,  the problem of joint estimation and automatic variable selection needs to be considered. For this reason and with the goal of  obtaining more reliable estimates in the presence of atypical data, under a logistic regression model,  we addressed the problem of estimating and selecting variables using weighted penalized $M-$procedures.
The obtained results are derived for   a broad family of penalty functions, which include the LASSO, ADALASSO, Ridge, SCAD and MCP penalties. In addition to these known penalties, we define a new one called Sign, which has an intuitive motivation and a simple expression,  depending only on a single adjustment parameter.

An in-depth study of the theoretical properties of the proposed methods is presented. In particular,   we showed that the penalized weighted $ M-$estimators   are consistent and, for some families of penalties, they select variables consistently. In addition, we obtain expressions for its asymptotic distribution. In particular, it is shown that the choice of the penalty function plays a fundamental role in this case. Specifically, we obtain that by using  the random penalty ADALASSO or   penalties which are constant from one point onwards (such as SCAD or MCP), the estimators have the desired oracle  property. The assumptions required to derived these results are very undemanding, which shows that these methods can be applied in very diverse contexts. 
 
We also  proposed a robust  cross-validation procedure and numerically showed its advantage over the classical one. Through an extensive simulation study, we compared the behaviour of classical and robust estimators for different choices for the loss function and penalty. The obtained results illustrate that  robust methods   have a  performance similar to the classical ones   for clean samples and behave much better  in contaminated scenarios, showing greater reliability. On the other hand, we showed that the results obtained when using penalties bounded as the Sign or MCP were remarkably better than those obtained when using convex penalties such as LASSO. Finally, the penalized weighted $ M-$estimators based on the function  $ \rho = \rho_c $ given in 
\eqref{rocroux} combined with the MCP and Sign penalties, were the most stable and reliable among the considered procedures.
Finally, the proposed methods are applied to two data sets, where the robust estimators combined with  bounded penalties showed their  advantages over the classical ones.

{\setcounter{equation}{0}
\renewcommand{\theequation}{A.\arabic{equation}}
{\setcounter{section}{0}
\renewcommand{\thesection}{\Alph{section}}

\section{Appendix: Proofs}{\label{sec:app}}

\subsection{Fisher--consistency}{\label{sec:FC}}

Theorem \ref{theo:BY_fisher_consistency} states the Fisher--consistency of the estimators defined  through  \eqref{ESTGEN}. When considering the estimators with $w\equiv 1$, Theorem \ref{theo:BY_fisher_consistency} follows from Theorem 2.2 in Bianco and Yohai (1996), while for the weighted estimators the proof is relegated to the appendix. Henceforth, for simplicity, let $(y,\bx)$ be a random vector with the same distribution as $(y_i,\bx_i)$, that is, $y|\bx\sim Bi(1, F(\bx\trasp \bbe_0))$.

 \vskip0.1in
\begin{theorem}\label{theo:BY_fisher_consistency}
Let $\phi:\real^2\to \real$ be the function given by \eqref{phiBY}, satisfying  \ref{ass:rho_bounded_derivable}  and \ref{ass:rho_derivative_positive} and let $w$ be a non--negative bounded function. 
 Furthermore, assume that 
 \begin{equation}
       \prob\left(\bx\trasp\balfa=0 \cup w(\bx)=0\right)<1, \qquad \forall  \balfa\in \real^p, \quad \balfa\ne   \bcero .
       \label{eq:identifiafuerte}
     \end{equation}
  Then, for all $\bbe \in \real^p$, $\bbe \neq \bbe_0$, we have $\esp[\phi(y,\bx\trasp\bbe_0)w(\bx)] < \esp[\phi(y,\bx\trasp\bbe)w(\bx)]$. 
\end{theorem}

\noi \textsc{Proof.} 
As in Theorem 2.2 in Bianco and Yohai (1996), taking conditional expectation, we have that
$$\esp \phi(y,\bx\trasp\bbe)w(\bx)= \esp \left[\esp \phi(y,\bx\trasp\bbe)w(\bx)|\bx\right]=\esp  \phi(F(\bx\trasp\bbe_0),\bx\trasp\bbe) w(\bx)\,.$$
For a fixed value $\bx$, denote $t=\bx\trasp\bbe $ and $t_0=\bx\trasp\bbe_0$, we will show that the function $\phi( F(t_0),t)$ reaches its unique minimum when $t=t_0$. For simplicity, denote $ \Phi(t)=  \phi( F(t_0),t)$, then, straightforward calculations allow to show that 
\begin{eqnarray*}
\Phi^{\prime}(t)=  \,-\, (F(t_0)-F(t)) \nu(t) \qquad \Phi^{\prime\,\prime}(t)= F(t)(1-F(t)) \nu(t)\,-\, (F(t_0)-F(t))   \nu^{\prime}(t)\,.
\end{eqnarray*}
Hence, $ \Phi^{\prime}(t_0)=0$ and $ \Phi^{\prime\,\prime}(t_0)= F(t_0)(1-F(t_0)) \nu(t_0)>0$. Furthermore, $\Phi^{\prime}(t)>0$, for $t>t_0$, and $\Phi^{\prime}(t)<0$ for $t<t_0$ which entails that $\Phi$ has a unique minimum at $t_0$, so $ \phi(F(\bx\trasp\bbe_0),\bx\trasp\bbe)> \phi(F(\bx\trasp\bbe_0),\bx\trasp\bbe_0)$ for any $\bx\trasp(\bbe-\bbe_0)\ne \bcero$,
which concludes the proof, since \eqref{eq:identifiafuerte} holds. $\square$

\subsection{Proof of Theorem \ref{teo:consistency}.}

The following result will be needed to derive Theorem \ref{teo:consistency}. It provides a consistency result for the estimators defined in \eqref{BYPEN} for a general function $\phi:\real^2 \to \real$.

\begin{theorem}{\label{teo:CONSIS}}
Let $\wbbe_n$ be the estimator defined in  \eqref{BYPEN}. Assume that $L(\bbe)=\esp \phi(y,\bx,\bbe)\, w(\bx)$ reaches its unique minimum at $\bbe = \bbe_0$ and that $I_{\lambda_n}(\bbe_0) \convpp 0 $ when $n \to \infty$. Furthermore, assume that, for any $\epsilon>0$,
\begin{equation}
 \inf_{\|\bbech - \bbech_0\| > \epsilon} L(\bbe)> L(\bbe_0)
 \label{aux-mas}
\end{equation}
and that the following uniform Law of Large Numbers holds  
\begin{equation}
\prob\left(\lim_{n \to \infty} \sup_{\bbech\in \real^p} \left |\frac{1}{n} \sum_{i = 1}^n \phi(y_i, \bx_i\trasp \bbe) w(\bx_i)- \esp \phi(y,\bx,\bbe)\,w(\bx) \right | = 0\right)=1\,.
\label{eq:UNIF}
\end{equation} 
Then, $\wbbe_n$ is strongly consistent for $\bbe_0$.
\end{theorem}
\noi \textsc{Proof.}
The fact that $\wbbe_n$ minimizes $L_n(\bbe) + I_{\lambda_n}(\bbe)$ entails that
\begin{equation*}
L_n( \wbbe_n) \le L_n( \wbbe_n) + I_{\lambda_n}(\wbbe_n) \leq L_n( \bbe_0) + I_{\lambda_n}(\bbe_0)\,. 
 \end{equation*} 
Therefore, 
\begin{equation*}
\limsup_{n \to \infty} L_n( \wbbe_n) \leq \limsup_{n \to \infty} L_n( \bbe_0) + I_{\lambda_n}(\bbe_0).
 \end{equation*} 
Using the law of large numbers and the fact that $I_{\lambda_n}(\bbe_0) \to 0 $ when $n \to \infty$, we have that, with probability one,
\begin{equation} 
\label{aux_3}
\limsup_{n \to \infty} L_n( \wbbe_n) \leq \esp \phi(y,\bx\trasp\bbe_0)\, w(\bx)\,.
\end{equation} 
Recall that $L(\bbe)=\esp \phi(y,\bx\trasp\bbe) w(\bx)$.  Using \eqref{eq:UNIF}, we get that with probability $1$, for any $\epsilon>0$, 
\begin{equation} 
\label{aux_1}
\lim_{n \to \infty}  \sup_{\|\bbech - \bbech_0\| > \epsilon} \left |L_n(\bbe) - L(\bbe)\right | =0\,.
\end{equation}
Note that $ L_n(\bbe)= L_n(\bbe)-L(\bbe)+L(\bbe)\ge -|L_n(\bbe)-L(\bbe)|+L(\bbe)$, hence  for any fixed $\epsilon > 0$, we have that
$$
\inf_{\|\bbech - \bbech_0\| > \epsilon} L_n( \bbe) \geq \,-\,\sup_{\|\bbech - \bbech_0\| > \epsilon} \left |L_n(\bbe) - L(\bbe)]\right | + \inf_{\|\bbech - \bbech_0\| > \epsilon} L(\bbe)\,.
$$
Hence, using \eqref{aux_1} we get that with probability one
\begin{equation}
\label{aux_2}
\liminf_{n \to \infty} \inf_{\|\bbech - \bbech_0\| > \epsilon} L_n(\bbe) \ge   \inf_{\|\bbech - \bbech_0\| > \epsilon} L(\bbe)> L(\bbe_0)\,,
\end{equation}
where  the first inequality follows from   \eqref{aux_1}  and the second one from \eqref{aux-mas}. Therefore,  from  \eqref{aux_2} and  \eqref{aux_3}, we obtain that with probability one there exists $n_0 \in \natu$ such that $\|\wbbe_n - \bbe_0\| \leq \epsilon$ for all $n \geq n_0$, concluding the proof. 
$\square$

\vskip0.1in
The next lemma provides a bound for the Vapnik-Chervonenkis (VC) dimension for the set defined by the functions $\phi(y,\bx\trasp \bbe)$      when the vector $\bbe$ varies in $\real^p$ and $\phi$ is given in \eqref{phiBY}. This result will be used to obtain a uniform Law of Large Numbers that guarantees consistency of our proposal.

\vskip0.1in
\begin{lemma} \label{VC}
Let $\phi:\real^2 \to \real$ be the function given in   \eqref{phiBY} where the function  $\rho:\real_{\ge 0}\to \real$ satisfies   \ref{ass:rho_bounded_derivable} and \ref{ass:rho_derivative_positive}. Then, the   class of functions
\begin{equation*}
\itF = \{f_{\bbech}(y,\bx)=\phi(y,\bx\trasp\bbe)w(\bx) : \bbe \in \real^p\}
\end{equation*}
is VC-subgraph with index $V(\itF) \leq 2p + 4$.
\end{lemma}
 \vskip0.1in

\noi \textsc{Proof.}
Taking into account that multiplying by a fixed function preserves the index of a class, it is enough to derive the result when $w(\bx)\equiv 1$. Suppose this class is not VC-subgraph, that is, the sub--graphs of the functions in $\itF$ are not a VC family, or that its VC--index is greater than $2p + 4$. Then, there exists a set $\itC_0 = \{(y_i,\bx_i,r_i)\,, \, i=1, \dots, 2p+5\}$, where $r_i \in \real$ and $\phi$ that can be shuttered by the sub--graphs of the functions in $\itF$. Since there are only two possible values for $y$, we can take a subset $\itC \subset \itC_0$ such that $|\itC| = \ell = p+3$ and the corresponding values of  $y$ is the same for all the elements in  $\itC$. 

Without loss of generality, assume that this common value is 0 and that the corresponding indexes in $\itC$ are $i=1,\dots, \ell$. Let $ \phi_0(\bx \trasp \bbe)=\phi(0,\bx\trasp\bbe)$ and  $ \phi_1(\bx \trasp \bbe)=\phi(1,\bx\trasp\bbe)$. Then,    $\phi_0(s)$ is a strictly increasing function, while $\phi_1(s)$  is   strictly decreasing.

Suppose $\{(\bx_1,r_1) ,\dots,(\bx_{\ell}, r_{\ell})\}$ are the second and third arguments corresponding to the  $\ell$ elements in $\itC$. Then, for each subset $\itI \subset \{1,\dots,\ell\}$, there exists $\bbe_\itI$ such that $\phi_0(\bx_i \trasp \bbe_\itI) \geq r_i$ if and only if $
i \in \itI$ which implies that $\bx_i \trasp \bbe_\itI \geq \phi_0^{-1}(r_i)$ if and only if $i \in \itI$. 

Let $\wtbx_i = (\bx_i,\phi_0^{-1}(r_i))$, $1\le i\le \ell$, and $\wtbbe_\itI = (\bbe_\itI,-1)$. Then, $\wtbx_i\trasp \wtbbe_\itI \geq 0$ if and only if $i \in \itI$. However, this would imply that, either for $y = 0$ or $y = 1$, the family of half--spaces of dimension $p+1$ can shutter an $\ell = (p+3)-$element set, which is known to be false, see, for instance, Van de Geer (2000), Example 3.7.4c. Thus, we conclude that $\itF$ is a VC-subgraph family and its index satisfies $V(\itF) \leq 2p+4$.
$\square$

 \vskip0.1in
 The following Lemma corresponds to Lemma 6.3 in Bianco and Yohai (1996) when $w(\bx)\equiv 1$, its proof for a general weight function follows using similar arguments and assumption \ref{ass:funcionw}.

\begin{lemma}\label{lema:lema63}
Let $\phi:\real^2\to \real$ be the function given by   \eqref{phiBY}, where the function  $\rho:\real_{\ge 0}\to \real$ satisfies  \ref{ass:rho_bounded_derivable} and \ref{ass:rho_derivative_positive}.  Then, if \ref{ass:X_not_singular} and \ref{ass:funcionw}  hold, for any $\|\bu\|_2=1$ there exists $\epsilon_\bu$ such that 
$$
\esp \left(\liminf_{a\to \infty} \inf_{ \bv\in \itV(\bu, \epsilon_\bu)} \phi(y, a  \, \bx\trasp \bv)\, w(\bx)\right) > L(\bbe_0) \,,
$$
where  $\itV(\bu, \epsilon)=\{\bv: \|\bv-\bu\|_2 < \epsilon \}$.
 \end{lemma}
  
  \vskip0.1in
\noi \textsc{Proof of Theorem \ref{teo:consistency}.}
It is enough to show that the conditions of Theorem \ref{teo:CONSIS} are satisfied. Theorem \ref{theo:BY_fisher_consistency} implies that $L(\bbe)$ has a unique minimum at $\bbe=\bbe_0$. On the other hand, Corollary 3.12 in Van de Geer (2000), Lemma \ref{VC} and the fact that $|\phi(y,\bx\trasp\bbe)|$ is uniformly bounded and $w(\bx)$ is a bounded function implies that \eqref{eq:UNIF}  holds.

It remains to show that \eqref{aux-mas} holds. 
Assume that it does not hold, that is, assume that, for some $\epsilon>0$,
\begin{equation}
 \inf_{\|\bbech - \bbech_0\| > \epsilon} L(\bbe)\le  L(\bbe_0)\,.
 \label{aux-mas2}
\end{equation}
Let  $(\bbe_n)_{n\ge 1}$ be a sequence such that 
$$\lim_{n\to \infty} L(\bbe_n) = \inf_{\|\bbech - \bbech_0\| > \epsilon} L(\bbe)\,.$$
Assume first that the sequence $(\bbe_n)_{n\ge 1}$ is bounded. Then,  there exists a subsequence $(\bbe_{n_j})_{j\ge 1}$ of $(\bbe_n)_{n\ge 1}$ converging to a value $\bbe^{\star}$.  The  continuity of $L(\bbe)$ entails that 
$$\lim_{j\to \infty} L(\bbe_{n_j})=L(\bbe^{\star})>L(\bbe_0)\,,$$
where the last inequality follows from Theorem \ref{theo:BY_fisher_consistency} leading to a contradiction with  \eqref{aux-mas2}. 

Hence, $\limsup_{n\to \infty} \|\bbe_n\|_2=\infty$ and $\lim_{n\to \infty} L(\bbe_n) = \inf_{\|\bbech - \bbech_0\| > \epsilon} L(\bbe)\le  L(\bbe_0)$. 

Define $ \bbe_n^{\star}= \bbe_n/\|\bbe_n\|_2$. Assume, eventually taking a subsequence,  that 
$\lim_{n\to \infty} \bbe_n^{\star}=\bbe^{\star}$ with $\|\bbe^{\star}\|=1$. 

Recall that we have denoted $\itV(\bu, \epsilon)=\{\bv: \|\bv-\bu\|_2 < \epsilon \}$. Lemma \ref{lema:lema63} entails that   there exists $\epsilon^{\star}=\epsilon_{\bbech^{\star}}$   such that 
\begin{equation}
\esp \liminf_{a\to \infty} \inf_{ \bv\in \itV(\bbech^{\star}, \epsilon^{\star})} \phi(y, a  \bx\trasp \bv)\, w(\bx)> L(\bbe_0) \,.
\label{eq:lema63}
\end{equation}
Choose  $n_0\in \natu$ such that  $\bbe_n^{\star}\in \itV(\bbe^{\star}, \epsilon^{\star})$ and $\|\bbe_n\|_2>M$, for $n\ge n_0$.  Then,
$$\phi(y, \|\bbe_n\|_2 \bx\trasp \bbe_n^{\star}) \ge \inf_{a>M} \inf_{ \bv\in \itV(\bbech^{\star}, \epsilon^{\star})} \phi(y, a \,\bx\trasp \bv)$$
which implies, using Fatou's Lemma, that 
\begin{eqnarray*}
 \lim_{n\to \infty} L(\bbe_n) &=&  \lim_{n\to \infty} L(\|\bbe_n\|_2 \, \bbe_n^{\star}) \ge   \esp \liminf_{a\to +\infty} \inf_{ \bv\in \itV(\bbech^{\star}, \epsilon^{\star})} \phi(y, a\, \bx\trasp \bv)\, w(\bx)
\end{eqnarray*}  
 Hence, using \eqref{eq:lema63}, we get that  $ \lim_{n\to \infty} L(\bbe_n)> L(\bbe_0)$ which again contradicts our assumption implying that \eqref{aux-mas} holds. $\square$

\subsection{Proof of Theorem \ref{teo:rate}}
To prove Theorem \ref{teo:rate}, we will need the following Lemma which is a refinement of Lemma 4.1 in Bianco and Mart\'{\i}nez (2009).

\vskip0.1in
\begin{lemma}\label{lema:convAn}
Let $\wtbbe_n$ be such that $\wtbbe_n\convpp \bbe_0$ and    $\phi(y,t)$ the function given by   \eqref{phiBY}  with  $\rho:\real_{\ge 0}\to \real$ satisfying  \ref{ass:rho_two_times_derivable_bounded}.  Then,  if \ref{ass:X_second_moments} holds, we have that $\bA_n(\wtbbe_n)\convpp \bA$ where $\bA$ is given in \eqref{eq:A} and 
\begin{equation}
\bA_n(\bbe)=   \frac{1}{n} \sum_{i = 1}^n \chi(y_i, \bx_i \trasp \bbe) w(\bx_i)\,\bx_i \bx_i \trasp\,.
\label{eq:Anbeta}
\end{equation}
\end{lemma}

\vskip0.1in
\noi \textsc{Proof of Lemma \ref{lema:convAn}.}
It is enough to show that $ \|\bA_n(\wtbbe_n) - \bA_n(\bbe_0)\| \convpp 0$, since $\bA_n(\bbe_0)\convpp \bA$. It suffices to show that for every $1 \leq k,j \leq p$,
 \begin{equation}\label{aux_4}
 A_{n,kj}=\frac{1}{n}\sum_{i = 1}^n |\chi(y_i, \bx_i \trasp \wtbbe_n) - \chi(y_i, \bx_i \trasp \bbe_0)| \, w(\bx_i) \,|x_{ik}|\, |x_{ij}| \convpp 0.
 \end{equation}
 Let $\epsilon > 0$. Taking into account that $\phi$ is given in  \eqref{phiBY}, assumption \ref{ass:rho_two_times_derivable_bounded} implies that the function $\chi$ is bounded. Choose $M > 0$ such that
 \begin{equation}
 \esp[\indica_{\{\|\bx\| > M\}} \, w(\bx)\,\|\bx\|^2] < \frac{\epsilon}{4\|\chi\|_{\infty}}. 
 \label{eq:cota}
 \end{equation} 
 Then, we have that 
 \begin{eqnarray*}
 A_{n,kj} &\le &
  \frac{1}{n}  \sum_{i = 1}^n |\chi(y_i, \bx_i \trasp \wtbbe_n) - \chi(y_i, \bx_i \trasp \bbe_0)|\, w(\bx_i) \,\|\bx_i\|^2 \left (\indica_{\{\|\bx_i\| > M\}} + \indica_{\{\|\bx_i\| \le  M\}}\right )\\
 &\leq & 2\,\|\chi\|_{\infty} \, \frac{1}{n}  \sum_{i = 1}^n   w(\bx_i)\, \|\bx_i\|^2 \indica_{\{\|\bx_i\| > M\}} + \frac{1}{n} \sum_{i=1}^n |\chi(y_i, \bx_i \trasp \wtbbe_n) - \chi(y_i, \bx_i \trasp \bbe_0)|\,   w(\bx_i)\, \|\bx_i\|^2 \indica_{\{\|\bx_i\| \le M\}}.
 \end{eqnarray*} 
 Note that if $\itC_w$ is compact, for $M$ large enough $\esp[\indica_{\{\|\bx\| > M\}} \, w(\bx)\,\|\bx\|^2]=0$ and the first term on the right hand side of the above equation also equals $0$, simplifying the arguments below.
 
 The function $\chi(y, t)$ is uniformly continuous in $t$ when restricting $t$ to be in a compact set. Since there are only two possible values for the first coordinate $y$, one can choose a value $\delta > 0$ such that if $|s| \leq M (\|\bbe_0\| + 1)$ and $|s'| \leq M (\|\bbe_0\| + 1)$ are such that $|s - s'| < \delta$ then $|\chi(y,s) - \chi(y,s')| < \epsilon / (2M^2)$.

Using that  $\wtbbe_n\convpp \bbe_0$ and  the strong law of large numbers,  we get that there exists a null probability set $\itN$, such that, for any $\omega \notin \itN$, $\wtbbe_n\to \bbe_0$ and 
$(1/n)\sum_{i = 1}^n  w(\bx_i)\,\|\bx_i\|^2 \indica_{\{\|\bx_i\| > M\}} \to \esp[\indica_{\{\|\bx\| > M\}}  w(\bx)\,\|\bx\|^2] $.
Let $n_1 = n_1(\omega)$ be such that, for $n\ge n_1$,  $\|\wtbbe_n - \bbe_0\| \leq \min\{1, \delta/M\}$  and
\begin{equation*}
\frac{1}{n} \sum_{i = 1}^n  w(\bx_i)\,\|\bx_i\|^2 \indica_{\{\|\bx_i\| > M\}} < \frac{\epsilon}{4 \|\chi\|_{\infty}}.
 \end{equation*} 
 Suppose $\|\bx_i\|\le M$ and $n \geq n_1$. Then, using that $\|\wtbbe_n - \bbe_0\| \le  1$ it is easy to see that both $\bx_i \trasp \wtbbe_n$ and $\bx_i \trasp \bbe_0$ have absolute value not greater than $M (\|\bbe_0\| + 1)$. Moreover, since $\|\wtbbe_n - \bbe_0\| \le \delta /M$, we also have that
 $|\bx_i \trasp \wtbbe_n - \bx_i \trasp \bbe_0| \leq \delta$, so
 \begin{equation*}
  |\chi(y_i, \bx_i \trasp \wtbbe_n)  - \chi(y_i,\bx_i \trasp \bbe_0)| \leq \frac{\epsilon}{2M^2}.
  \end{equation*}
Hence, using that $\|w\|_{\infty}=1$, we have 
$$
\frac{1}{n}  \sum_{i = 1}^n |\chi(y_i, \bx_i \trasp \wtbbe_n) - \chi(y_i, \bx_i \trasp \bbe_0)| \, w(\bx_i)\, |x_{ik}\|x_{ij}|  
\leq  2 \|\chi\|_{\infty} \frac{\epsilon}{4\|\chi\|_{\infty}} + \frac{1}{n} \sum_{i=1}^n \frac{\epsilon}{2M^2}M^2 = \epsilon, 
$$
concluding the proof of \eqref{aux_4}. $\square$

\vskip0.1in
The following result is an extension of Lemma \ref{lema:convAn} and can be derived using similar arguments to those considered in Lemma 1 in Bianco and Boente (2002). Note that a direct consequence of Lemma \ref{lema:convAn2} is that $\bA_n(\wtbbe_n)\convprob \bA$, whenever $\wtbbe_n\convprob \bbe_0$.

\vskip0.1in
\begin{lemma}\label{lema:convAn2}
Let   $\phi(y,t)$ the function given by   \eqref{phiBY}  with  $\rho:\real_{\ge 0}\to \real$ satisfying    \ref{ass:rho_two_times_derivable_bounded}. Then, if \ref{ass:X_second_moments} holds, for any $\delta>0$
\begin{itemize}
\item[a)] $\lim_{\bbech\to \bbech_0} \esp  \chi(y, \bx \trasp \bbe)\,  w(\bx)\, \bx \bx \trasp=\bA$,
\item[b)]  $\sup_{\|\bbech-\bbech_0\|<\delta}\left|\bA_n(\bbe)- \esp \chi(y, \bx \trasp \bbe) \,  w(\bx)\,\bx \bx \trasp\right|\convprob 0$ 
\end{itemize}
where $\bA$ and $\bA_n(\bbe)$ are given in \eqref{eq:A}  and  \eqref{eq:Anbeta}, respectively.
\end{lemma}

\vskip0.1in 

\noi \textsc{Proof of Theorem \ref{teo:rate}.}
Let $W_n(\bbe) = L_n(\bbe) + I_{\lambda_n}(\bbe) $
where $L_n$ is defined in \eqref{P_n}. 
Using a Taylor's expansion of order 2 of  $L_n(\wbbe_n)$ around $\bbe_0$, we get
\begin{equation*}
W_n(\wbbe_n) = L_n(\bbe_0) + (\wbbe_n - \bbe_0)\trasp \nabla L_n(\bbe_0) + \frac{1}{2} (\wbbe_n - \bbe_0)\trasp \bA_n(\wtbbe_n)(\wbbe_n - \bbe_0) + I_{\lambda_n}(\wbbe_n),
\end{equation*}
 where $\wtbbe_n=\bbe_0 + \tau_n(\wbbe_n - \bbe_0)$ is an intermediate point between $\bbe_0$ and $\wbbe_n$,  $\tau_n \in [0,1]$, $\nabla L_n(\bbe)$ is the gradient $L_n(\bbe)$  given by 
$$ \nabla L_n(\bbe)= \frac 1n \sum_{i=1}^n \Psi(y_i, \bx_i\trasp \bbe)\,  w(\bx_i)\,\bx_i $$
 and   $\bA_n(\bbe)$ is defined in \eqref{eq:Anbeta} and corresponds to  the  Hessian of  $L_n(\bbe)$, that is, 
 $$\bA_n(\bbe) = \frac{\partial^2}{(\partial \bbe)^2}L_n(\bbe)  = \frac{1}{n} \sum_{i = 1}^n \chi(y_i, \bx_i \trasp \bbe) \,  w(\bx_i)\,\bx_i \bx_i \trasp\,.$$
 
Let $\varepsilon$ be a positive constant and $\zeta_1$ be the smallest eigenvalue of the matrix $\bA$ which is strictly positive from \ref{ass:X_w_positive_definite}. Since $\wbbe_n \convprob \bbe_0$, from Lemma \ref{lema:convAn2} we have that $\bA_n(\wtbbe_n) \convprob \bA$, so there exists $n_1$ such that for every $n \geq n_1$, $
\prob (\itB_n) > 1 - \varepsilon/4$, where $\itB_n = \left \{\|\bA_n(\wtbbe_n) - \bA\| < \zeta_1 / 2 \right \}$.  

On the other hand, the Central Limit Theorem together with \eqref{eq:FC} leads to 
\begin{equation*}
\sqrt{n} \nabla L_n(\bbe_0) = O_\prob(1),
\end{equation*} 
so there exists a constant $M_1$ such that, for all $n$, 
$\prob(\itC_n) > 1- \varepsilon/4$, where $\itC_n = \{\|\sqrt{n} \nabla L_n(\bbe_0)\|_2 < M_1\}$. 
By the definition of $\wbbe_n$, we have that in $\itB_n \cap \itC_n$,
\begin{align}
0 &\geq W_n(\wbbe_n) - W_n(\bbe_0) \nonumber \\
&= (\wbbe_n - \bbe_0)\trasp \nabla L_n(\bbe_0) + \frac{1}{2} (\wbbe_n - \bbe_0)\trasp \bA_n(\wtbbe_n)(\wbbe_n - \bbe_0) + I_{\lambda_n}(\wbbe_n) - I_{\lambda_n}(\bbe_0) \nonumber \\
& \geq - \|\wbbe_n - \bbe_0\|_2 \frac{1}{\sqrt{n}} \|\sqrt{n} \nabla L_n(\bbe_0)\|_2 - \|\bA_n(\wtbbe_n) - \bA\| \|\wbbe_n - \bbe_0\|_2^2 + \zeta_1 \|\wbbe_n - \bbe_0\|_2^2 + I_{\lambda_n}(\wbbe_n) - I_{\lambda_n}(\bbe_0) \nonumber \\
& \geq - \|\wbbe_n - \bbe_0\|_2 \frac{1}{\sqrt{n}} M_1 + \frac{\zeta_1}{2} \|\wbbe_n - \bbe_0\|_2^2 + I_{\lambda_n}(\wbbe_n) - I_{\lambda_n}(\bbe_0) \label{rate_ineq}
\end{align}

To prove (a), define the event $\itD_n = \{I_{\lambda_n}(\wbbe_n) - I_{\lambda_n}(\bbe_0) \leq K \, \lambda_n \|\wbbe_n - \bbe_0\|_2\}$. Observe that \ref{ass:penalty_locally_lipschitz} together with the fact that $\wbbe_n \convprob \bbe_0$ implies that there exists a constant $K$ and $n_2 \in \natu$ such that for $n \geq n_2$, $\prob(\itD_n) \geq 1 - \varepsilon/2$. Hence, for $n \geq \max\{n_1, n_2\}$ we have that $\prob(\itB_n \cap \itC_n \cap \itD_n) > 1 - \epsilon$. Besides, in  $\itB_n \cap \itC_n \cap \itD_n$ we have that
\begin{equation*}
0 \geq - \|\wbbe_n - \bbe_0\|_2 \frac{1}{\sqrt{n}} M_1 + \frac{\zeta_1}{2} \|\wbbe_n - \bbe_0\|_2^2 - K \, \lambda_n \|\wbbe_n - \bbe_0\|_2,
\end{equation*} 
which implies
$$
  \|\wbbe_n - \bbe_0\|_2  \le 2\, \left(\lambda_n+ \frac{1}{\sqrt{n}}\right) \frac{ M_1 +  K }{\zeta_1}\,.
$$
Hence, $\|\wbbe_n - \bbe_0\|_2=O_{\prob}( \lambda_n + 1/\sqrt{n})$, which completes the proof.

We now turn to prove (b). Suppose without loss of generality that $\bbe_0 = (\bbe_{0,A}\trasp, \bcero_{p-k}\trasp)\trasp$ and $\bbe_{0,A} \in \real^k$ is the subvector with active coordinates of $\bbe_0$ (this means the first $k$ coordinates of $\bbe_0$ are different from 0). Since $J_{\lambda_n}(0)=0$ and $J_{\lambda_n}(s)\ge 0$, we get that 
$$I_{\lambda_n}(\bbe) - I_{\lambda_n}(\bbe_0)= \sum_{\ell = 1}^k J_{\lambda_n}(|\beta_{\ell}|) - J_{\lambda_n}(|\beta_{0,\ell}|) + \sum_{\ell = l+1}^p J_{\lambda_n}(|\beta_{\ell}|) \ge \sum_{\ell = 1}^k J_{\lambda_n}(|\beta_{\ell}|) - J_{\lambda_n}(|\beta_{0,\ell}|)\,,$$
which together with \eqref{rate_ineq} leads to
\begin{equation*}
0 \geq - \|\wbbe_n - \bbe_0\|_2 \frac{1}{\sqrt{n}} M_1 + \frac{\zeta_1}{2} \|\wbbe_n - \bbe_0\|_2^2 + \sum_{\ell = 1}^k J_{\lambda_n}(|\wbeta_{n,\ell}|) - J_{\lambda_n}(|\beta_{0,\ell}|).
\end{equation*} 
Choose $\delta$ and $n_3$ such that for every $n \geq n_3$, $\prob(\itE_{n,1})>1-\varepsilon/4$ and $\prob(\itE_{n,2}) \geq 1 - \varepsilon/4$, where $\itE_{n,2} =  \{\|\wbbe_n - \bbe_0\|_2 \leq \delta \}$ 
$$\itE_{n,1}=\left\{\sup\{|J_{\lambda_n}^{\prime\,\prime}(|\beta_{0,\ell}| + \tau \delta)| : \tau \in [-1,1] \;, \;  1 \leq \ell \leq p \;\; \text{and} \;\; \beta_{0,\ell} \neq 0 \} \leq \frac{\zeta_1}{2}\right\}.$$
 Let $\itE_n=\itE_{n,1} \cap\itE_{n,2}$. Using a first order Taylor's  expansion, we have
\begin{equation*}
J_{\lambda_n}(|\wbeta_{n,\ell}|) - J_{\lambda_n}(|\beta_{0,\ell}|) = J'_{\lambda_n}(|\beta_{0,\ell}|)  ( |\wbeta_{n,\ell}|  - | \beta_{0,\ell}|) + \frac{1}{2} J''_{\lambda_n}(\xi_{n,\ell}) ( |\wbeta_{n,\ell} | -  |\beta_{0,\ell}| )^2\,,
\end{equation*} 
where $\xi_{n,\ell}$ lies between $|\wbeta_{n,\ell}| $ and $|\beta_{0,\ell}|$.
 Using that $|\, |a|-|b|\,|\le |a-b|$, $J'_{\lambda_n}(|\beta_{0,\ell}|) \ge 0$ and that in the event $\itB_n \cap \itC_n \cap \itE_n$, $|J''_{\lambda_n}(\xi_{n,\ell})|<\zeta_1/2$, since $\max(0, |\beta_{0,\ell}|-\delta) <\xi_{n,\ell} \le |\beta_{0,\ell}|+\delta$, we get that 
\begin{eqnarray*}
I_{\lambda_n}(\wbeta_{n}) - I_{\lambda_n}(\beta_{0})&\ge & \sum_{\ell=1}^k J_{\lambda_n}(|\wbeta_{n,\ell}|) - J_{\lambda_n}(|\beta_{0,\ell}|)  \\
 & \ge &  - \sum_{\ell=1}^k J'_{\lambda_n}(|\beta_{0,\ell}|)  |\wbeta_{n,\ell}   -   \beta_{0,\ell}| -
 \frac{1}{2} \sum_{\ell=1}^k |J''_{\lambda_n}(\xi_{n,\ell})| (  \wbeta_{n,\ell}  -   \beta_{0,\ell}  )^2 
  \\
 & \ge & - a_n \sum_{\ell=1}^k |\wbeta_{n,\ell}   -   \beta_{0,\ell}|- \frac{\zeta_1}{4}  \sum_{\ell=1}^k (  \wbeta_{n,\ell}  -   \beta_{0,\ell}  )^2
  \\
  & \ge & - a_n \sqrt{k}\; \|\wbbe_n - \bbe_0\|_2 - \frac{\zeta_1}{4} \|\wbbe_n - \bbe_0\|_2^2\,. 
\end{eqnarray*}
Hence,  
\begin{equation*}
0 \geq - \|\wbbe_n - \bbe_0\|_2 \frac{1}{\sqrt{n}} M_1 + \frac{\zeta_1}{2} \|\wbbe_n - \bbe_0\|_2^2 - a_n \sqrt{k}\; \|\wbbe_n - \bbe_0\|_2 - \frac{\zeta_1}{4} \|\wbbe_n - \bbe_0\|_2^2\;, 
\end{equation*} 
which implies that $4\alpha_n (M_1 + \sqrt{k})/\zeta_1 \geq \|\wbbe_n - \bbe_0\|_2$.  Now, for $n \geq \max_{1\le i\le 3} n_i$, 
$\prob(\itB_n \cap \itC_n \cap \itE_n) \geq 1 - \varepsilon$, 
so the desired result follows. 
$\square$

\subsection{Proofs of the results in Section \ref{sec:selectvar}}

\noi \textsc{Proof of Theorem \ref{teo:orac_gen}.}
Consider the decomposition $\bbe_0 = (\wtbbe_0\trasp, 0)\trasp$ where $\wtbbe_0 \in \real^{p-1}$ and define
\begin{equation*}
V_n(\bu_1, u_2) = L_n\left (\wtbbe_0 + \frac{\bu_1}{\sqrt{n}}  ,  \frac{u_2}{\sqrt{n}}\right ) + I_{\lambda_n}\left (\wtbbe_0 + \frac{\bu_1}{\sqrt{n}}, \frac{u_2}{\sqrt{n}}\right ),
\end{equation*}
where $L_n(\bbe)$ was defined in \eqref{P_n}. Fix $\tau>0$ and define $\tau^* = \tau / (2(p-k))$ . Let $C > 0$ be such that the event $\itB_n = \{\sqrt{n}\|\wbbe_n - \bbe_0\|_2 \leq C\}$ satisfies $\prob(\itB_n) \geq 1 - \tau^*$, for all $n$. Then, for any $\omega \in \itB_n$,  
$$\wbbe_n = \left(\wtbbe_0\trasp + \frac{\bup_{1,n}\trasp}{\sqrt{n}},  \frac{u_{2,n}}{\sqrt{n}}\right)\trasp$$ 
where $\bu_{1,n} \in \real^{p-1}$, $u_{2,n} \in \real$, $\|\bu_n\|_2\le C$ with $\bu_n = (\bu_{1,n}\trasp, u_{2,n})\trasp$ such that 
\begin{equation*}
\bu_n = (\bu_{1,n}\trasp, u_{2,n})\trasp=\argmin_{\|(\bu_1,u_2)\|_2\le C}V_n(\bu_1, u_2) \,.
\label{eq:Un}
\end{equation*} 
Our aim is to prove that for $\|\bu_1\|^2+u_2^2\le C^2$ and $u_2\ne 0$, $V_n(\bu_1, u_2) - V_n(\bu_1,0) > 0$ with high probability. Take $\bu_1 \in \real^{p-1}$ and $u_2 \neq 0$ such that $\|\bu_1\|^2+u_2^2\le C^2$. Observe that $V_n(\bu_1, u_2) - V_n(\bu_1,0) = S_{1,n}(\bu)+ S_{2,n}(\bu)$ where  $\bu=(\bu_1,u_2)$ and
\begin{align*}
S_{1,n}(\bu) &= L_n\left (\wtbbe_0 + \frac{\bu_1}{\sqrt{n}}, \frac{u_2}{\sqrt{n}}\right ) - L_n\left (\wtbbe_0 + \frac{\bu_1}{\sqrt{n}}, 0\right ),\nonumber\\
S_{2,n}(\bu) &=  I_{\lambda_n}\left (\wtbbe_0 + \frac{\bu_1}{\sqrt{n}}, \frac{u_2}{\sqrt{n}}\right ) - I_{\lambda_n}\left (\wtbbe_0 + \frac{\bu_1}{\sqrt{n}}, 0\right ).
\label{eq:S2n}
\end{align*}
First, we have to bound $S_{1,n}(\bu)$. Denote as $\bu_{n}^{(0)}=\left (\bcero_{p-1}\trasp,  u_2/\sqrt{n}\right )\trasp$. Then, the mean value theorem entails that
\begin{equation*}
S_{1,n}(\bu) = \frac{1}{n}\sum_{i = 1}^n \Psi(y_i, \bx_i\trasp \bbe^*_n) w(\bx_i)\, \bx_i \trasp \bu_{n}^{(0)}, 
\end{equation*} 
where 
\begin{equation*}
\bbe_{n}^{*} = \left (\begin{array}{c}
\wtbbe_0 +   \dst\frac{\bu_1}{\sqrt{n}} \\
\\   \,\alpha_{n,1} \dst\frac{u_2}{\sqrt{n}}
\end{array}
\right )\,,
\end{equation*}
with $\alpha_{n,1} \in [0,1]$. Moreover, 
using again   the mean value theorem, we get that
\begin{eqnarray*}
\bu_{n}^{(0) \traspi}\frac{1}{n}\sum_{i = 1}^n [\Psi(y_i, \bx_i\trasp \bbe^*_n) - \Psi(y_i, \bx_i\trasp \bbe_0)] \,  w(\bx_i)\,\bx_i   &=& \bu_{n}^{(0) \traspi}\frac{1}{n}\sum_{i = 1}^n \chi(y_i, \bx_i\trasp \bbe_{n}^{**})\,  w(\bx_i)\, \bx_i  \bx_i \trasp (\bbe^*_n - \bbe_0) \\
&=& \bu_{n}^{(0) \traspi}\bA_n(\bbe_{n}^{**}) (\bbe^*_n - \bbe_0)\, ,
\end{eqnarray*}
where $\bA_n(\bbe)$ is given in \eqref{eq:Anbeta} and  
\begin{equation*}
\bbe_{n}^{**} = \left (\begin{array}{c}
\wtbbe_0 + \alpha_{n,2} \dst\frac{\bu_1}{\sqrt{n}} \\
\\ \alpha_{n,2} \,\alpha_{n,1} \dst\frac{u_2}{\sqrt{n}}
\end{array}
\right )
\end{equation*}
with $\alpha_{n,2} \in [0,1]$. Thus, noticing that
\begin{equation*}
S_{1,n}(\bu) = \left \{\frac{1}{n}\sum_{i = 1}^n \Psi(y_i, \bx_i\trasp \bbe_0)\,  w(\bx_i)\, \bx_i \trasp + \frac{1}{n}\sum_{i = 1}^n [\Psi(y_i, \bx_i\trasp \bbe^*_n) - \Psi(y_i, \bx_i\trasp \bbe_0)] \,  w(\bx_i)\,\bx_i \trasp  \right \} \bu_{n}^{(0)}. 
\end{equation*} 
we get $S_{1,n}(\bu)=S_{11,n}+S_{12,n}$ with 
\begin{eqnarray*}
S_{11,n} &= &\frac{1}{n}\sum_{i = 1}^n \Psi(y_i, \bx_i\trasp \bbe_0)\,  w(\bx_i)\, \bx_i \trasp  \bu_{n}^{(0)} =\frac{1}{n} \frac{1}{\sqrt{n}} \sum_{i = 1}^n \Psi(y_i, \bx_i\trasp \bbe_0) \,  w(\bx_i)\,\bx_i \trasp  (\bcero_{p-1}\trasp, u_2)\trasp
\\
S_{12,n}& = &(\bbe^*_n - \bbe_0) \trasp\, \bA_n(\bbe_{n}^{**}) \bu_{n}^{(0)}= \frac{1}{n} (\bu_1\trasp, \alpha_{n,1}\,u_2)\,  \bA_n(\bbe_{n}^{**}) (\bcero_{p-1}\trasp, u_2)\trasp\,.
\end{eqnarray*}
Using \eqref{eq:FC} and the Multivariate Central Limit Theorem, we have that $n \, |S_{11,n}| = O_{\prob}(1) \, |u_2|$. 
On the other hand,
\begin{eqnarray*}
|S_{12,n}| & \leq & \frac{1}{n} \left |(\bu_1, \alpha_n \, u_2) \trasp\,\frac 1n \sum_{i = 1}^n \chi(y_i, \bx_i\trasp \bbe_{n}^{**}) \,  w(\bx_i)\, \bx_i \bx_i \trasp (\bcero_{p-1}, u_2)\right | \\
& \leq &\frac{1}{n} \|\chi\|_{\infty} \|(\bu_1, \alpha_n \, u_2)\|_2 \left (\frac{1}{n}\sum_{i = 1}^n \,  w(\bx_i)\,\|\bx_i\|^2\right )|u_2| 
\le \frac{1}{n} \|\chi\|_{\infty} \, C\, \left (\frac{1}{n}\sum_{i = 1}^n \,  w(\bx_i)\, \|\bx_i\|^2\right )|u_2|  \,.
\end{eqnarray*} 
Therefore, using that   \ref{ass:X_second_moments}    entails that $(1/n) \sum_{i = 1}^n \,  w(\bx_i)\, \|\bx_i\|^2 \convprob \esp   w(\bx)\, \|\bx\|^2$, we get  that
$$C\, \|\chi\|_{\infty}   \left (\frac{1}{n}\sum_{i = 1}^n \,  w(\bx_i)\,\|\bx_i\|^2\right )=O_\prob(1)$$
which implies that $n |S_{12,n}| =  O_{\prob}(1) \, |u_2|$ which together with the fact that  $n |S_{11,n}| =  O_{\prob}(1) \, |u_2|$ leads to  $n |S_{1,n}(\bu)| = A_n \, |u_2|$ with $A_n\ge 0$ and $A_n= O_{\prob}(1)$.

Let $M_p$ be such that $\prob(   0\le  A_n  <  M_p) \ge 1-\tau^{*}$ for all $n$, note that $M_p$ depends on $C$ and so on $\tau$.  Hence, if   $\itD_n=\{ n \, S_{1,n} (\bu)>  \,-\,  M_p\, |u_2|\}$
\begin{equation*}
\prob(\itD_n ) > \prob(   0\le  A_n  <  M_p) \ge 1- \tau^{*}\;.
\end{equation*}
Take $N_{p}$ and $K_{p}$ (both depending on $C$) such that for $n \geq N_{p}$ and $\|\bu\|_2 \leq C$,
\begin{equation*}
I_{\lambda_n}\left (\bbe_0 + \frac{\bu}{\sqrt{n}}\right ) - I_{\lambda_n}\left (\bbe_0 + \frac{\bu^{(-p)}}{\sqrt{n}}\right ) \geq K_{p}\, \frac{\lambda_n}{\sqrt{n}}\,  |u_p|,
\end{equation*} 
which implies $n \, S_{2,n}(\bu) \geq  K_{p}\,  \lambda_n\, \sqrt{n}\,  |u_p|$. Then, in $\itB_n\cap \itD_n$ and for $n \geq N_p$, we have
\begin{equation}\label{aux_varsel}
V(\bu_{1,n} , u_{2,n}) - V(\bu_{1,n}, 0) = S_{1,n}(\bu_{1,n},u_{2,n}) + S_{2,n}(\bu_{1,n},u_{2,n}) \geq \frac{1}{n} |u_{2,n}| (K_{p} \,\lambda_n \, \sqrt{n} - M_p).
\end{equation} 
Thus, if $b \geq M_p/K_p$,
$$\prob(\wbeta_{n,p} = 0) \geq 1-2\tau^* = 1- \tau / (p-k).$$
We can proceed sequentially with the same reasoning for every non active coordinate, obtaining values $N_s\in \natu$, $K_s \in \real$, $M_s\in \real$  for $s = p, p-1, \dots, k+1$. 

Item (a) is proved by taking  $n_0=\max(N_p, N_{p-1}, \dots, N_{k+1})$ and $b>  \max(A_p, A_{p-1}, \dots, A_{k+1})$ with $A_j=M_j/K_j$. 

On the other hand, if $\sqrt{n}\lambda_n \to \infty$, then there exists some $\widetilde{n}_0$ such that for $n \geq \widetilde{n}_0$, $\sqrt{n} \lambda_n > \max\{M_p/K_p, \dots, M_{k+1} / K_{k+1}\}$. Then, if $n_0 = \max\{\widetilde{n}_0, N_p, \dots, N_{k+1}\}$, for every $n \geq n_0$, we have
\begin{equation*}
\prob(\wbeta_{n,p} = 0 \cap \wbeta_{n,p-1} = 0 \cap \dots \cap  \wbeta_{n,k+1} = 0) \geq 1 - \tau,
\end{equation*} 
so item (b) follows. $\square$

\vskip0.1in

\noi \textsc{Proof of Corollary \ref{coro:orac}.}
In order to use Theorem \ref{teo:orac_gen}, it only remains to show that condition \eqref{penalty_varsel_condition} holds. Without loss of generality, we will prove condition \eqref{penalty_varsel_condition} holds only for the last coordinate, this is, for $\ell = p$. Fix $C > 0$ and take $\bu = (\bu_1, u_2)$ with $\bu_1 \in \real^{p-1}$, $u_2 \in \real$ and $\|\bu_1\|_2^2 + u_2^2 \leq C^2$. 

We first prove part (a) of the Corollary. For a fixed value $\wtbu \in \real^{p-1} - \{\bcero_{p-1}\}$ consider the function  
$$h(u)=h_{\wtbu}(u) = J(\wtbu,u)\,,$$ 
where $J(\bbe) = \|\bbe\|_1/\|\bbe\|_2$. 

To have a better insight of the behaviour of this even function, let us consider the case when $\wtbu = \buno_{15}$. Figure \ref{fig:hplot} gives the we plot of  the function $h_{\wtbu}(u)$.

\begin{figure}[ht!]
\begin{center}
\includegraphics[scale=0.3]{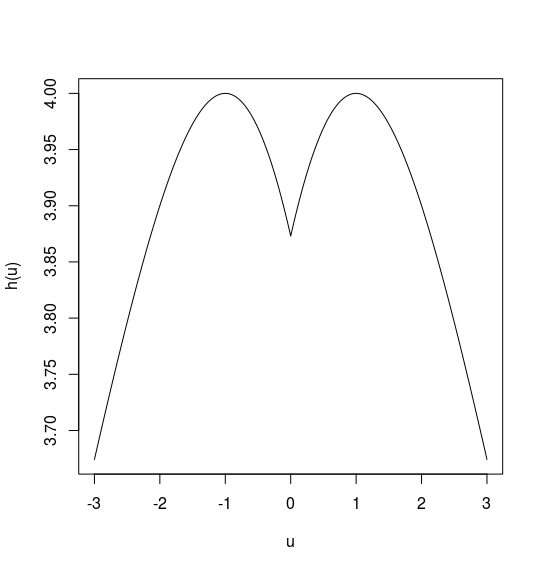}
   \caption{\small\label{fig:hplot} Plot of the function $h_{\wtbu}(u) = J(\wtbu,u)$ for $\wtbu = \buno_{15}$. }
 \end{center}
 \end{figure}

As shown in the plot,  function $h(u)$ is clearly not derivable on $u = 0$. However, for $u\ne 0$ the  derivative of $h$ can be computed and is given by
$$h^{\prime}(u)=\frac{\signo(u) \left(\sum_{j=1}^{p-1} \wtu_{\ell}^2 + u^2\right)- u \left(\sum_{j=1}^{p-1} |\wtu_{\ell}| + |u|\right)}{\left(\sum_{j=1}^{p-1} \wtu_{\ell}^2 + u^2\right)^{\frac 32}}\,,$$
so that the critical points of $h$ are $\pm \|\wtbu\|_2^2 / \|\wtbu\|_1$, both of them being local maxima. Hence, $h_{\wtbu}$ is an increasing function of $|u|$ when $|u| \le \|\wtbu\|_2^2 / \|\wtbu\|_1$. Moreover, 
\begin{equation*}
\lim_{u \to 0^+} h^{\prime}(u) =\frac{1}{\|\wtbu\|_2} \qquad \text{and} \qquad \lim_{u \to 0^-} h^{\prime}(u) = -\frac{1}{\|\wtbu\|_2} .
 \end{equation*} 
For a given $\bu_1$, denote as 
$$h_{n, \bu_1}(u) = h_{\wtbbech_0+\frac{\bu_1}{\sqrt{n}}}(u)= J\left( \wtbbe_0 + \frac{\bu_1}{\sqrt{n}}, u\right)\,.$$
Using that the critical points of $h_{\wtbu}$ are $\pm \|\wtbu\|_2^2 / \|\wtbu\|_1$, we get that those of $h_{n, \bu_1}$ are 
$c_n^{+}(\bu_1)=  \|\wtbbe_0+ {\bu_1}/{\sqrt{n}}\|_2^2/ \|\wtbbe_0+ {\bu_1}/{\sqrt{n}}\|_1$ and $c_n^{-}(\bu_1)= - c_n^{+}(\bu_1)$ 
which, respectively,   converge to   $c^{+}$ and $c^{-}= -c^{+}$, , uniformly over compact sets, with
$$c^{+}= \frac{\|\wtbbe_0\|_2^2}{\|\wtbbe_0\|_1}=\frac{\|\bbe_{0,A}\|_2^2}{\|\bbe_{0,A}\|_1}$$  
  that is, $\lim_{n\to \infty} \sup_{\|\bu_1\|\le C}\left | c_n^{+}(\bu_1) -  c^{+}\right| =0$.
 Furthermore, 
 \begin{equation*}
 \lim_{n \to \infty} \lim_{u \to 0^+} h_{n, \bu_1}^{\prime}(u) = \frac{1}{\|\wtbbe_0\|_2}  \qquad \text{and} \qquad  \lim_{n \to \infty} \lim_{u \to 0^-} h_{n, \bu_1}^{\prime}(u) =  - \, \frac{1}{\|\wtbbe_0\|_2}\,,
 \end{equation*} 
 where the convergence is again uniform over any compact set since $\|\wtbbe_0\|_2=\|\bbe_{0,A}\|_2\ne 0$.
 
 Let  $n_1 \in \natu$ and $\delta>0$ be such that for $n\ge n_1$ and $0<|u|<\delta$, we have
 $$\sup_{\|\bu_1\|\le C}\left |h_{n, \bu_1}^{\prime}(u)- \frac{\signo(u)}{\|\wtbbe_0\|_2}\right| < \frac{1}{2\|\wtbbe_0\|_2} \,,$$ 
and choose $n_2$  such that $ C n^{-1/2}  \le  \min(  { c^{+}}/{2}, \delta) $ and   $\sup_{\|\bu_1\|\le C}\left |  c_n^{+}(\bu_1) -  c^{+}\right|  \le    c^{+}/2$ for $n\ge n_2$.
Then, for $n\ge N_p=\max(n_1,n_2)$ we have that  for any  $ \bu_1 \in \{\bv\in \real^{p-1}: \|\bv\|\le C\}$ 
\begin{equation}
   C n^{-1/2}  \le  \min\left(\frac{ c^{+}}{2}, \delta\right) \;, c_n^{+}(\bu_1)>  \frac{\;c^{+}}{2} 
   \quad  \mbox{and} \quad  
  \left|h_{n, \bu_1}^{\prime}(u)\right|\indica_{\{0<|u|<\delta\}}< \frac{1}{2\|\wtbbe_0\|_2} \, .
  \label{eq:desig}
  \end{equation}  
 In particular, we have that  the functions $h_{n, \bu_1}(u)$ are increasing functions  of $|u|$ when restricted to the interval $[-{ c^{+}}/{2}, { c^{+}}/{2}]$. 
Using that $\|\bu\|_2\le C$, which implies that $\|\bu_1\|\le C$ and $0<|u_2|\le C$, from \eqref{eq:desig} we get that
$$h_{n, \bu_1}\left(\frac{u_2}{\sqrt{n}}\right)> h_{n, \bu_1}(0)   \,.$$
Furthermore, if $\xi_n$ is an intermediate point between $0$ and $u_2/\sqrt{n}$, we have that $0<|\xi_n|\le C n^{-1/2} <\delta$ so
$$h_{n, \bu_1}\left(\frac{u_2}{\sqrt{n}}\right)- h_{n, \bu_1}(0)= |h_{n, \bu_1}^{\prime}(\xi_n)| \frac{|u_2|}{\sqrt{n}} > \frac{1}{2\|\wtbbe_0\|_2}\; \frac{|u_2|}{\sqrt{n}}\,.$$
Finally, we have 
\begin{equation*}
I_{\lambda_n}\left (\bbe_0 + \frac{\bu}{\sqrt{n}}\right ) - I_{\lambda_n}\left (\bbe_0 + \frac{\bu^{(-p)}}{\sqrt{n}}\right ) = \lambda_n \left \{h_{n, \bu_1}\left(\frac{u_2}{\sqrt{n}}\right)- h_{n, \bu_1}(0)\right \},
\end{equation*} so condition \eqref{penalty_varsel_condition}  holds by taking $K_{C,p} = 1/(2 \|\wtbbe_0\|_2) = 1/(2 \|\bbe_0\|_2)$ and $N_{C,p} = N_p$. The desired results follows from using item (a) from Theorem \ref{teo:orac_gen}.

We now turn to prove item (b). For the SCAD penalty, it is easy to see that 
\begin{equation*}
I_{\lambda_n}\left (\bbe_0 + \frac{\bu}{\sqrt{n}}\right ) - I_{\lambda_n}\left (\bbe_0 + \frac{\bu^{(-p)}}{\sqrt{n}}\right ) = \scadgrande_{\lambda_n, a}\left (\frac{u_2}{\sqrt{n}}\right ),
\end{equation*} 
where
$$
\scadgrande_{\lambda, a}(\beta) = \begin{cases}
\lambda |\beta| \quad &\text{if} \quad |\beta| \leq \lambda \\
a \lambda |\beta| - \frac{\beta^2 + \lambda^2}{2} \quad &\text{if} \quad \lambda  < |\beta| \leq a \lambda \\
\frac{\lambda ^2 (a^2 -1)}{2(a-1)} \quad &\text{if} \quad |\beta| > a \lambda \,.
\end{cases}
$$
Take $n_0$ such that for $n \geq n_0$, $\sqrt{n} \lambda_n \geq C$. If this is the case, $\scadgrande_{\lambda_n,a}(u_2 / \sqrt{n}) = \lambda_n |u_2| / \sqrt{n}$, so condition \eqref{penalty_varsel_condition}  holds for $K_{C,p} = 1$ and $N_{C,p} = n_0$.

For the MCP penalty, the proof is very similar. In this case, 
\begin{equation*}
I_{\lambda_n}\left (\bbe_0 + \frac{\bu}{\sqrt{n}}\right ) - I_{\lambda_n}\left (\bbe_0 + \frac{\bu^{(-p)}}{\sqrt{n}}\right ) = \mcpgrande_{\lambda_n, a}\left (\frac{u_2}{\sqrt{n}}\right ),
\end{equation*} 
where
$$
\mcpgrande_{\lambda , a}(\beta) = \begin{cases}
\lambda  |\beta| - \frac{\beta^2}{2a}  \quad &\text{if} \quad |\beta| \leq a\lambda  \\
\frac{a \lambda ^2}{2} \quad &\text{if} \quad |\beta| > a \lambda\,.
\end{cases}
$$
As before, take $n_0$ such that for $n \geq n_0$, $\sqrt{n}  \lambda_n \geq C/a$. If this is the case, 
$$\mcpgrande_{\lambda_n,a}\left (\frac{u_2}{\sqrt{n}}\right ) = \lambda_n \frac{|u_2|}{\sqrt{n}} - \frac{u_2^2}{2na} = \lambda_n \frac{|u_2|}{\sqrt{n}} \left (1 - \frac{u_2}{2 \sqrt{n} \lambda_n a}\right ) \geq \frac{1}{2} \lambda_n \frac{|u_2|}{\sqrt{n}},$$ 
so condition \eqref{penalty_varsel_condition} holds for $K_{C,p} = 1/2$ and $N_{C,p} = n_0$.

For both SCAD and MCP penalties, item (b) from Theorem \ref{teo:orac_gen} gives the desired result. $\square$
 
 \subsection{Proof of the results in Section \ref{sec:asdist}}
In order to prove Theorem \ref{theo:ASDIST} the following two Lemmas are useful.

\begin{lemma}{\label{lema:ASDISTRn1}}
Let   $\phi(y,t)$   given by   \eqref{phiBY}  where the function  $\rho:\real_{\ge 0}\to \real$ satisfies   \ref{ass:rho_two_times_derivable_bounded}.  Assume that the matrix $\bA$ defined in \eqref{eq:A} is non-singular and define the process $R_1:\real^p\to \real$ as
\begin{equation*}
R_1(\bz) = \bz \trasp \bw + \frac{1}{2}\bz \trasp \bA \bz  \,,
\end{equation*}
where   $\bw  \sim N_p(\bcero, \bB)$ with $\bB$ given in \eqref{eq:B}. Furthermore, let $R_{n,1}(\bz)$ be
\begin{eqnarray}
R_{n,1}(\bz) &= & \sum_{i =1 }^n\left\{ \phi\left (y_i, \bx_i\trasp \left[\bbe_0 + \frac{\bz}{\sqrt{n}}\right]\right ) \,  w(\bx_i)  - \phi(y_i, \bx_i\trasp \bbe_0)\,  w(\bx_i) \right\}\,, 
\label{eq:Rn1}
\end{eqnarray}
Then, the process $R_{n,1} $ converges in distribution to $R_1 $.
\end{lemma}

\vskip0.1in
\noi \textsc{Proof.} According to Theorem 2.3 in Kim and Pollard (1990) it is enough to show the finite--dimensional convergence and the stochastic equicontinuity, i.e., 
\begin{enumerate}
\item[(a)] For any $\bz_1,\cdots, \bz_s$  $(R_{n,1}(\bz_1), \cdots, R_{n,1}(\bz_s))\trasp \convdist (R_1(\bz_1), \cdots, R_1(\bz_s))\trasp $.
\item[(b)] Given $\epsilon > 0$, $\eta > 0$ and $M < \infty$ there exists $\delta > 0$ such that
\begin{equation*}
\limsup_{n \to \infty} \prob^*\left(\sup_{\stackrel{\|\bu\|\le M, \|\bv\|\le M}{\|\bu-\bv\|_2<\delta}} |R_{n,1}(\bu) - R_{n,1}(\bv)| > \epsilon\right) < \eta,
\end{equation*} 
where $\prob^*$ stands for outer probability.
\end{enumerate} 

Let us show  (a), we will only consider the situation $s=1$ since for any $s$ the proof follows similarly using the Cramer--Wald device, that is projecting over any $\ba\in \real^s$. Hence, we fix $\bz \in \real^p$. 

Using a first order Taylor's expansion, 
we get that
\begin{equation*}
 R_{n,1}(\bz) = \sqrt{n} \, \bz \trasp \nabla L_n(\bbe_0) + \frac{1}{2} \bz \trasp \bA_n (\wtbbe_{\bz} \bz  ),
 \end{equation*} 
 where $L_n(\bbe)$ and $\bA_n(\bbe)$ where defined in \eqref{P_n} and \eqref{eq:Anbeta}, respectively and 
 $\wtbbe_{\bz}= \bbe_0 +  {\tau_n \bz}/{\sqrt{n}}$, 
is an  the intermediate point with $\tau_n\in [0,1]$. As above, the conditional Fisher--consistency given in \eqref{eq:FC} and the Multivariate Central Limit Theorem entail that $ \sqrt{n} \nabla L_n(\bbe_0) \convdist N_p(\bcero, \bB)$, 
since $\var[\Psi(y, \bx\trasp \bbe_0)w(\bx)\, \bx]=\bB$. On the other hand,  Lemma  \ref{lema:convAn} implies that $\bA_n (\wtbbe_{\bz} ) \convprob \bA$,    so  using Slutsky's Theorem we obtain that $R_{n,1} \convdist \bz \trasp \bw + \frac{1}{2}\bz \trasp \bA \bz$, concluding the proof of (a). 

To derive (b),  we perform a first order Taylor's expansion of $R_{n,1}(\bu)$ and $R_{n,1}(\bv)$ around $\bbe_0$ obtaining
$$
R_{n,1}(\bu) - R_{n,1}(\bv) = \sqrt{n} \nabla L_n(\bbe_0) \trasp (\bu - \bv) + \frac{1}{2} \bu \trasp \bA_n ( \wtbbe_{\bu} ) \bu - \frac{1}{2}\bv \trasp \bA_n ( \wtbbe_{\bv}  ) \bv,
$$
where $ \wtbbe_{\bu}$ and $ \wtbbe_{\bv}$ are defined as above, that is,
$$\wtbbe_{\bv}= \bbe_0 + \frac{\tau_{\bv,n} \bv}{\sqrt{n}}\qquad\qquad  \wtbbe_{\bu}= \bbe_0 + \frac{\tau_{\bu,n} \bu}{\sqrt{n}}\,,$$ 
with $\tau_{\bv,n}$, $\tau_{\bu,n} \in [0,1]$. Noting that $\sqrt{n} \nabla L_n(\bbe_0) \trasp (\bu - \bv) \leq O_\prob(1) \|\bu - \bv\|_2$ and 
\begin{eqnarray*}
 \bu \trasp \bA_n  (\wtbbe_{\bu} ) \bu - \bv \trasp \bA_n (\wtbbe_{\bv} ) \bv &=&  \bu \trasp \bA_n (\wtbbe_{\bu} ) \bu - \bu \trasp \bA_n(\wtbbe_{\bv} ) \bu  + \bu \trasp \bA_n (\wtbbe_{\bv} ) \bu \\
 &- & \bu \trasp \bA_n (\wtbbe_{\bv} ) \bv + \bu \trasp\bA_n(\wtbbe_{\bv} ) \bv - \bv \trasp \bA_n (\wtbbe_{\bv} ) \bv\,,   
\end{eqnarray*} 
we obtain that, if $\|\bu\|, \|\bv\| \leq M$, 
\begin{eqnarray*}
|R_{n,1}(\bu) - R_{n,2}(\bv)| &\leq&  O_p(1) \|\bu - \bv\|_2 + M^2  \left \|\bA_n (\wtbbe_{\bu}  ) - \bA_n (\wtbbe_{\bv} )\right \| \\
 && + 2 \|\bu - \bv\|_2 M  \left \|\bA_n (\wtbbe_{\bv} )\right \|\,,
\end{eqnarray*}
where   $\|\bC\|$ stands for the Frobenius norm of the matrix  $\bC$. Lemma \ref{lema:convAn2} entails that $\bA_n (\wtbbe_{\bu} ) - \bA_n (\wtbbe_{\bv}  ) \convprob 0$ and $\bA_n  (\wtbbe_{\bv}  )\convprob \bA$,
uniformly over all $\{\bu, \bv \in \real^p : \max\{\|\bu\|, \|\bv\|\} \leq M\}$ and (b) follows easily, concluding the proof. $\square$

\vskip0.1in
It is worth noticing that Theorem 2.3 in Kim and Pollard (1990) states that if   conditions a) and b)  hold, then the limiting stochastic process exists and its finite dimensional projections are those of $R_1(\bz)$. However, since stochastic processes that concentrates its paths in $\itC_b(\real^p)$ are determined by its finite dimensional projections, we can conclude that $R_1$ must be this limiting stochastic process. 

In the next Lemma as in Theorem \ref{theo:ASDIST}, we allow the penalty constant $\lambda_n$ to be random.

\begin{lemma}{\label{lema:equicontpenalty}}
Let $I_{\lambda_n}(\bbe)$ be a penalty satisfying \ref{ass:penalty_locally_lipschitz} and such that $\sqrt{n} \, \lambda_n=O_{\prob}(1)$. Define 
\begin{equation}
R_{n,2}(\bz) = n\,  \left \{I_{\lambda_n}\left (\bbe_0 + \frac{\bz}{\sqrt{n}}\right ) - I_{\lambda_n}(\bbe_0)\right \}\,.
\label{eq:Rn2} 
\end{equation}
Then, the process $R_{n,2}(\bz)$ is equicontinuous, i.e.,  for any $\epsilon > 0$, $\eta > 0$ and $M < \infty$ there exists $\delta > 0$ such that
\begin{equation*}
\limsup_{n \to \infty} \prob^*\left(\sup_{\stackrel{\|\bu\|\le M, \|\bv\|\le M}{\|\bu-\bv\|_2<\delta}} |R_{n,2}(\bu) - R_{n,2}(\bv)| > \epsilon\right) < \eta,
\end{equation*} 
\end{lemma}

\vskip0.1in
\noi \textsc{Proof.} It is enough to note that \ref{ass:penalty_locally_lipschitz} implies that
\begin{eqnarray*}
|R_{n,2}(\bu) - R_{n,2}(\bv)| &=& n\,   \left |I_{\lambda_n}\left (\bbe_0 + \frac{\bu}{\sqrt{n}}\right ) - I_{\lambda_n}\left (\bbe_0 + \frac{\bv}{\sqrt{n}}\right )\right | \leq n\, \lambda_n K\,  \frac{\|\bu - \bv\|_1}{\sqrt{n}} \\
& \leq & \sqrt{n}\,\lambda_n \, K\, \sqrt{p} \,\|\bu - \bv\|_2 \, ,   
\end{eqnarray*}
which together with the fact that $\sqrt{n} \, \lambda_n=O_{\prob}(1)$ concludes the proof. $\square$

\vskip0.1in

\noi \textsc{Proof of Theorem \ref{theo:ASDIST}.} 
Let us consider the stochastic process indexed in $\bz$ defined by $R_n(\bz) = R_{n,1}(\bz) + R_{n,2}(\bz)$, where $R_{n,1}(\bz)$ and $R_{n,2}(\bz)$ are defined in \eqref{eq:Rn1} and \eqref{eq:Rn2}, respectively with  $I_{\lambda_n}(\bbe)=\lambda_n J(\bbe)$ and $J(\bbe) = \|\bbe\|_1/\|\bbe\|_2$. Observe that $\argmin_{\bz} R_n(\bz) = \sqrt{n}(\wbbe_n - \bbe_0)$. 

To show that $ \sqrt{n}(\wbbe_n - \bbe_0)=\argmin_{\bz} R_n(\bz)\convdist \argmin_{\bz} R(\bz)$, we will use Theorem 2.7 in Kim and Pollard (1990). Condition (iii) is trivially verified and condition (ii) is a direct consequence of Theorem \ref{teo:rate}. Thus, it is enough to show that  the process  $R_n(\bz)$ converges in distribution to the process $ R(\bz)$ which corresponds to condition (i) in Theorem 2.7 in Kim and Pollard (1990). For that purpose, it is enough to show   the finite--dimensional convergence and the stochastic equicontinuity, that is,   that the following two conditions hold 
\begin{enumerate}
\item[(a)] For any $\bz_1,\cdots, \bz_s$  $(R_n(\bz_1), \cdots, R_n(\bz_s))\trasp \convdist (R(\bz_1), \cdots, R(\bz_s))\trasp $.
\item[(b)] Given $\epsilon > 0$, $\eta > 0$ and $M < \infty$ there exists $\delta > 0$ such that
\begin{equation*}
\limsup_{n \to \infty} \prob^*\left(\sup_{\stackrel{\|\bu\|\le M, \|\bv\|\le M}{\|\bu-\bv\|_2<\delta}} |R_n(\bu) - R_n(\bv)| > \epsilon\right) < \eta,
\end{equation*} 
where $\prob^*$ stands for outer probability.
\end{enumerate} 

Using that $\bbe_0\ne \bcero$, we get that the Sign penalty satisfies \ref{ass:penalty_locally_lipschitz}, hence the equicontinuity stated in (b) follows easily from  Lemmas \ref{lema:ASDISTRn1} and  \ref{lema:equicontpenalty}.

It only remains to derive  (a). As noted in the proof of Lemma \ref{lema:ASDISTRn1} it is enough to  consider the situation $s=1$, for that reason, we fix $\bz \in \real^p$. 
 
From Lemma \ref{lema:ASDISTRn1}, we get that $R_{n,1}(\bz) \convdist \bz \trasp \bw + \frac{1}{2}\bz \trasp \bA \bz$, so we only have to study the convergence of $R_{n,2}(\bz)$. Observe that, for $J(\bbe) = \|\bbe\|_1/\|\bbe\|_2$, one has 
\begin{equation*}
J(\bbe) = \sum_{\ell = 1}^p \frac{|\beta_\ell|}{\|\bbe\|_2} = \sum_{\ell = 1}^p J_\ell(\bbe)
\end{equation*} 
and $J_\ell$ is differentiable everywhere except on the hyperplane $\{\bbe : \beta_\ell = 0\}$. Suppose that $\beta_{0,\ell} \neq 0$. Then, for $n$ large enough, $\beta_{0,\ell} + z_\ell/ \sqrt{n}$ stays away from zero and   has the same sign as $\beta_\ell$. Therefore,   the Mean Value Theorem for $J_\ell(\bbe_0 + \bz / \sqrt{n}) - J_\ell(\bbe_0)$  yields
\begin{equation*}
 J_\ell\left (\bbe_0 + \frac{\bz}{\sqrt{n}}\right ) - J_\ell(\bbe_0) = \left [\nabla J_\ell\left (\bbe_0 + \alpha_{n,\ell} \frac{\bz}{\sqrt{n}}\right )\right ]\trasp \frac{\bz}{\sqrt{n}},
 \end{equation*}
  with $\alpha_{n,\ell} \in [0,1]$. When $\beta_{0,\ell} = 0$, $J_\ell(\bbe_0) = 0$, so
 \begin{equation*}
  J_\ell\left (\bbe_0 + \frac{\bz}{\sqrt{n}}\right ) - J_\ell(\bbe_0) = \frac{|z_\ell|}{\sqrt{n}}\frac{1}{\left \|\bbe_0 + \dst\frac{\bz}{\sqrt{n}}\right \|_2}.
 \end{equation*} 
Hence, $ R_{n,2}(\bz) $ can be written as
 \begin{eqnarray*}
 R_{n,2}(\bz) 
 &=& \sqrt{n}\, \lambda_n \left\{\sum_{\ell=1}^p \left [\nabla J_\ell\left (\bbe_0 + \alpha_{n,\ell} \frac{\bz}{\sqrt{n}}\right )\right ]\trasp {\bz} \indica_{\{\beta_{0.\ell}\ne 0\}}+ \sum_{\ell=1}^p  |z_\ell| \frac{1}{\left\|\bbe_0 + \dst\frac{\bz}{\sqrt{n}}\right\|_2} \indica_{\{\beta_{0.\ell}= 0\}}\right\}
  \end{eqnarray*}
 which implies that $ R_{n,2}(\bz) \convprob b \; \bz \trasp \bq(\bz)$  so condition (a) holds.    $\square$

\vskip0.1in

\noi \textsc{Proof of Theorem \ref{theo:ASDISTJdif}.} 
The proof follows the same steps as that of   Theorem \ref{theo:ASDIST}. We only point out the differences.  As in  Theorem \ref{theo:ASDIST}, let us consider the stochastic process indexed in $\bz$ defined by $R_n(\bz) = R_{n,1}(\bz) + R_{n,2}(\bz)$, where $R_{n,1}(\bz)$ and $R_{n,2}(\bz)$ are defined in \eqref{eq:Rn1} and \eqref{eq:Rn2}, respectively with  
$$I_{\lambda_n}(\bbe) = \lambda_n\,\left\{(1-\alpha)\sum_{\ell=1}^p J_{\ell}(|\beta_{\ell}|)+\alpha \sum_{\ell=1}^p |\beta_{\ell}|\right\}\,.$$
Then, arguing as in Theorem \ref{theo:ASDIST}, we have to show that  the process  $R_n(\bz)$ converges in distribution to the process $ R(\bz)$, that is, it is enough to show   that the conditions (a) and (b) in the proof of  Theorem \ref{theo:ASDIST} hold.

Using that $ J_{\ell}(\cdot)$ is a continuously differentiable function, we get that $I_{\lambda_n}(\bbe)$ satisfies \ref{ass:penalty_locally_lipschitz}, hence the equicontinuity stated in (b) follows again from  Lemmas \ref{lema:ASDISTRn1} and  \ref{lema:equicontpenalty}.

It only remains to derive  (a). As noted in the proof of Lemma \ref{lema:ASDISTRn1} it is enough to  consider the situation $s=1$, for that reason, we fix $\bz \in \real^p$. 
 
From Lemma \ref{lema:ASDISTRn1}, we get that $R_{n,1}(\bz) \convdist \bz \trasp \bw + \frac{1}{2}\bz \trasp \bA \bz$, so we only have to study the convergence of $R_{n,2}(\bz)$. 
Note that $R_{n,2}(\bz)= R_{n,2,1}(\bz)+ R_{n,2,1}(\bz)$ where 
\begin{eqnarray*}
R_{n,2,1}(\bz) &=& n\, \lambda_n (1-\alpha)  \left \{\sum_{\ell=1}^p J_{\ell}\left(\left|\beta_{0,\ell}+\frac{z_\ell}{\sqrt{n}}\right|\right)   - J_{\ell}(|\beta_{0,\ell}|)\right \}\\
R_{n,2,2}(\bz) &=& n\, \lambda_n \alpha  \left \{\sum_{\ell=1}^p \left |\beta_{0,\ell}+\frac{z_\ell}{\sqrt{n}} \right|   - |\beta_{0,\ell}|\right \}\,.\\
\end{eqnarray*}
Standard arguments allow to show that
\begin{eqnarray*}
R_{n,2,1}(\bz) & \convprob & b \,(1-\alpha) \, \sum_{\ell=1}^p J_{\ell}^{\prime}\left(\left|\beta_{0,\ell} \right|\right)\, \signo(\beta_{0,\ell})  \, z_\ell  \\
R_{n,2,2}(\bz) & \convprob & b \,\alpha\, \sum_{\ell=1}^p \left\{z_\ell\, \signo(\beta_{0,\ell})  \indica_{\{\beta_{0,\ell} \ne  0\}}+ |z_{\ell}|\indica_{\{\beta_{0,\ell} = 0\}}\right\} \,,
\end{eqnarray*}
 uniformly over compact sets. Hence,  $ R_{n,2}(\bz) \convprob b \; \bz \trasp \bq(\bz)$  concluding the proof.  $\square$
 
\vskip0.1in

\noi \textsc{Proof of Corollary \ref{coro:inconsist}.} We will show the result for the Sign penalty, the case of the LASSO one being similar.  

 Let
$$\alpha_\ell= \frac{\signo(\beta_{0,\ell})}{\|\bbe_0\|_2} + \beta_{0,\ell}\;\sum_{j=1}^k \frac{|\beta_{0,j}|}{\|\bbe_0\|_2^3}\qquad \qquad \bA=\left(\begin{array}{cc}
\bA_{11} & \bA_{12}\\
\bA_{21} & \bA_{22}
\end{array}\right)\quad \bA_{11}\in \real^{k\times k}, \bA_{22}\in \real^{(p-k)\times (p-k)}\,.$$
Using that $\beta_{0,\ell}=0$ for $\ell=k+1, \dots, p$, from Theorem \ref{theo:ASDIST} we get that $\sqrt{n}(\wbbe_n - \bbe_0) \convdist \bz^{\star}=\argmin_{\bz} R(\bz)$ where
\begin{align*}
R(\bz) 
&= \bz \trasp \bw + \frac 12\bz \trasp \bA \bz + \frac{b}{\|\bbe_0\|_2}  \; \sum_{\ell=k+1}^p  |z_\ell| + b \; \sum_{\ell=1}^k z_{\ell} \;\alpha_{\ell}
\end{align*}
with  $\bw  \sim N_p(\bcero, \bB)$. 

 Note that if $\itA=\itA_n$, then $\wbbe_{n,j}=0$ for $j\notin \itA$ so
$$\prob(\itA=\itA_n)\le \prob(\sqrt{n}\; \wbbe_{n,j}=0; \forall j\notin \itA)\,.$$ 
Besides, since $\itA=\{1,\dots, k\}$, we get that   $\sqrt{n}\wbbe_{n, B} \convdist \bz^{\star}_{B}$, which implies that
$$\limsup_{n\to \infty} \prob(\itA=\itA_n)\le \limsup_{n\to \infty}\prob(\sqrt{n}\; \wbbe_{n,B}=\bcero_{p-k})\le  \prob(\bz^{\star}_{B}=\bcero_{p-k})\,.$$ Hence, it will be enough to show that $\prob(\bz^{\star}_{B}=\bcero_{p-k})<1$. We will consider separately the situation $b=0$ and $b>0$. 

When $b=0$, $R(\bz) = \bz \trasp \bw + \frac 12\bz \trasp \bA \bz $, thus $ \bz^{\star}= -\bA^{-1} \bw\sim N_p(\bcero, \bA^{-1} \,\bB \,\bA^{-1}) $, leading to  $\prob(\bz^{\star}_{B}=\bcero_{p-k})=0$. 

When $b>0$, $R(\bz)$ is not differentiable with respect to $z_\ell$ whenever $z_\ell=0$, $\ell=k+1,\dots, p$. The Karush–Kuhn–Tucker (KKT) optimality condition entail that
\begin{align*}
w_\ell+ (\bA \bz^{\star})_\ell + b  \; \alpha_\ell&= 0 \qquad \qquad  \qquad  \ell=1,\dots k\\
|w_\ell+ (\bA \bz^{\star})_\ell| & \le \frac{b}{\|\bbe_0\|_2}  \qquad \qquad \ell\notin \itA\,,
\end{align*}
es decir, si $\balfa=(\alpha_1, \dots, \alpha_p)\trasp$,
$$
\bw_{A}+ \bA_{11} \bz_{A}^{\star} + \bA_{12} \bz_{B}^{\star} + b  \; \balfa_A  = 0 \qquad \mbox{and}\qquad 
|\bw_{B}+ \bA_{21} \bz_{A}^{\star} + \bA_{22} \bz_{B}^{\star} |  \le \frac{b}{\|\bbe_0\|_2} \;, 
$$
where the inequality is understood component-wise. Hence, if $\bz^{\star}_{B}=\bcero_{p-k}$ we get that $
\bw_{A}+ \bA_{11} \bz_{A}^{\star}  + b  \; \balfa_A  = 0 $ and $
|\bw_{B}+ \bA_{21} \bz_{A}^{\star}   |  \le  {b}/{\|\bbe_0\|_2} $. Using that $\bA$ is positive definite, we conclude that  $\bz_{A}^{\star} = - \bA_{11}^{-1} (\bw_{A}+ b  \; \balfa_A)  $. Therefore, if we denote as $\bv$ the random vector   $\bv=\bw_{B}- \bA_{21} \bA_{11}^{-1} (\bw_{A}+ b  \; \balfa_A)=(v_{1},\dots, v_{p-k})\trasp$, we obtain that   $\bv$ is normally distributed and 
$$\prob(\bz^{\star}_{B}=\bcero_{p-k}) \le \prob \left( |v_{\ell}|  \le \frac{b}{\|\bbe_0\|_2} , \quad \forall \ell=1,\dots, p-k\right)<1\,,$$
conclusing the proof. \qed 
\vskip0.1in

 \noi \textsc{Proof of Theorem \ref{theo:ASDISTlamdainfty}.}  
As in the proof of Theorem  \ref{theo:ASDIST} define   $R_n(\bz) = R_{n,1}(\bz) + R_{n,2}(\bz)$, where now
\begin{align*}
R_{n,1}(\bz) &=   \frac{1}{n\; \lambda_n^2}\sum_{i =1 }^n\left\{ \phi\left (y_i, \bx_i\trasp \left[\bbe_0 + \lambda_n\; \bz\right]\right ) - \phi(y_i, \bx_i\trasp \bbe_0)\right\}\, w(\bx_i)\\
R_{n,2}(\bz) &= \frac{1}{\lambda_n^2}  \left\{I_{\lambda_n}\left (\bbe_0 + \lambda_n\; \bz\right ) - I_{\lambda_n}(\bbe_0)\right \}\,,
\end{align*}
with $I_{\lambda}(\bbe)=\lambda   \|\bbe\|_1/\|\bbe\|_2$. Note that  $\argmin_{\bz} R_n(\bz) = (1/\lambda_n)\;(\wbbe_n - \bbe_0)$. We will begin by showing the equicontinuity of  $R_{n,\ell}(\bz)$, $\ell=1,2$, i.e.,   given  $\epsilon > 0$, $\eta > 0$ and $M < \infty$ we will show that there  exists  $\delta > 0$ such that
\begin{equation}
\limsup_{n \to \infty} \prob^*\left(\sup_{\stackrel{\max(\|\bu\|_2, \|\bv\|_2)\le M}{\|\bu-\bv\|_2<\delta}} |R_{n,\ell}(\bu) - R_{n,\ell}(\bv)| > \epsilon\right) < \eta\;.
\label{eq:equinew}
\end{equation} 
As in Lemma \ref{lema:equicontpenalty}, we have that \ref{ass:penalty_locally_lipschitz} implies that
$$I_{\lambda_n}\left (\bbe_0 + \lambda_n\; \bu \right ) - I_{\lambda_n}\left (\bbe_0 + \lambda_n\; \bv \right )\le  \lambda_n K \; (\lambda_n\; \|\bu - \bv\|_1)$$
so $|R_{n,2}(\bu) - R_{n,2}(\bv)|\le    K  \|\bu - \bv\|_1   \le \, K\, \sqrt{p} \,\|\bu - \bv\|_2 $, concluding the proof of  \eqref{eq:equinew} for $\ell=2$.

Let us show that $R_{n,1}(\bz)$ is equicontinuous. As in the proof of Lemma \ref{lema:ASDISTRn1}, through a Taylor's expansion of order one of $R_{n,1}(\bu)$ and $R_{n,1}(\bv)$ around $\bbe_0$, we get
$$
R_{n,1}(\bu) - R_{n,1}(\bv) = \frac{1}{\lambda_n\; \sqrt{n}}\;\sqrt{n} \nabla L_n(\bbe_0) \trasp (\bu - \bv) + \frac{1}{2} \bu \trasp \bA_n ( \wtbbe_{\bu} ) \bu - \frac{1}{2}\bv \trasp \bA_n ( \wtbbe_{\bv}  ) \bv,
$$
where $ \wtbbe_{\bu}$ and $ \wtbbe_{\bv}$ are intermediate point defined as
$$\wtbbe_{\bv}= \bbe_0 + \lambda_n\; \tau_{\bv,n} \bv\qquad\qquad  \wtbbe_{\bu}= \bbe_0 + \lambda_n\; \tau_{\bu,n} \bu \,,$$ 
with $\tau_{\bv,n}$, $\tau_{\bu,n} \in [0,1]$. As in the proof of Lemma \ref{lema:ASDISTRn1}, using that $\sqrt{n} \nabla L_n(\bbe_0) \convdist N_p(\bcero, \bB)$, we conclude that  $\sqrt{n} \nabla L_n(\bbe_0) \trasp (\bu - \bv) \leq O_\prob(1) \|\bu - \bv\|_2$. On the other hand, 
\begin{eqnarray*}
 \bu \trasp \bA_n  (\wtbbe_{\bu} ) \bu - \bv \trasp \bA_n (\wtbbe_{\bv} ) \bv &=&  \bu \trasp \bA_n (\wtbbe_{\bu} ) \bu - \bu \trasp \bA_n(\wtbbe_{\bv} ) \bu  + \bu \trasp \bA_n (\wtbbe_{\bv} ) \bu \\
 &- & \bu \trasp \bA_n (\wtbbe_{\bv} ) \bv + \bu \trasp\bA_n(\wtbbe_{\bv} ) \bv - \bv \trasp \bA_n (\wtbbe_{\bv} ) \bv\,,   
\end{eqnarray*} 
so, if $\|\bu\|_2, \|\bv\|_2 \leq M$,  
\begin{eqnarray*}
|R_{n,1}(\bu) - R_{n,2}(\bv)| &\leq&  \frac{1}{\lambda_n\; \sqrt{n}}\; O_p(1) \|\bu - \bv\|_2 + M^2  \left \|\bA_n (\wtbbe_{\bu}  ) - \bA_n (\wtbbe_{\bv} )\right \| \\
 && + 2 \|\bu - \bv\|_2 M  \left \|\bA_n (\wtbbe_{\bv} )\right \|\,.
\end{eqnarray*}
Using that   $\lambda_n\; \sqrt{n}\to \infty$ and that Lemma \ref{lema:convAn2} entails that $\bA_n (\wtbbe_{\bu} ) - \bA_n (\wtbbe_{\bv}  ) \convprob 0$   and $\bA_n  (\wtbbe_{\bv}  )\convprob \bA$ uniformly over   ${\{\bu, \bv \in \real^p : \max\{\|\bu\|, \|\bv\|\} \leq M\}}$, we get that \eqref{eq:equinew} holds for $\ell=1$. 
 
It remains to see that given $\bz_1,\dots, \bz_s \in \real^p$,  $(R_{n}(\bz_1), \dots, R_{n}(\bz_s))\trasp \convdist (R(\bz_1), \dots, R(\bz_s))\trasp $, where $R(\bz)= (1/2) \bz\trasp \bA \bz +\bz\trasp \bq(\bz)$. As in the proof of Theorem \ref{theo:ASDIST}, it is enough to show the result when $s=1$. Fix $\bz \in \real^p$. Using again a Taylor's expansion of order one, we obtain that
\begin{equation*}
 R_{n,1}(\bz) =  \frac{1}{\lambda_n\; \sqrt{n}}\;\sqrt{n} \nabla L_n(\bbe_0) \trasp (\bu - \bv)  + \frac{1}{2} \bz \trasp \bA_n (\wtbbe_{\bz}) \bz  \,,
 \end{equation*} 
where $L_n(\bbe)$ and $\bA_n(\bbe)$ are defined in (\ref{P_n})and (\ref{eq:Anbeta}), respectively and $\wtbbe_{\bz}= \bbe_0 + \lambda_n\; \tau_n \bz$, 
with $\tau_n\in [0,1]$. The Fisher--consistence given in (\ref{eq:FC}) and the multivariate central limit theorem imply that 
 \begin{equation*}
 \sqrt{n} \nabla L_n(\bbe_0) \convdist N_p(\bcero, \bB)\,,
 \end{equation*} 
where $\bB$ is defined in (\ref{eq:B}). The fact that $\lambda_n\; \sqrt{n}\to \infty$, implies that
$$\frac{1}{\lambda_n\; \sqrt{n}}\;\sqrt{n} \nabla L_n(\bbe_0) \trasp (\bu - \bv)  \convprob 0\,.$$
Finally, from Lemma \ref{lema:convAn2}, we conclude that  $\bA_n (\wtbbe_{\bz} ) \convprob \bA$, so $R_{n,1} \convprob  (1/2)\bz \trasp \bA \bz$.  
 
On the other hand,
$R_{n,2}(\bz)\convprob   \bz\trasp \bq(\bz)$. As in the proof of Theorem \ref{theo:ASDIST} if $J(\bbe) = \|\bbe\|_1/\|\bbe\|_2$, we get that  $I_{\lambda}(\bbe)=\lambda \; J(\bbe)$ 
\begin{equation*}
J(\bbe) = \sum_{\ell = 1}^p \frac{|\beta_\ell|}{\|\bbe\|_2} = \sum_{\ell = 1}^p J_\ell(\bbe)
\end{equation*} 
where $J_\ell$ is differentiable everywhere except on the hyperplane  $\{\bbe : \beta_\ell = 0\}$.  Recall that
$$R_{n,2}(\bz)=\frac{1}{\lambda_n} \left(J\left(\bbe_0 + \lambda_n\; \bu \right )- J(\bbe_0)\right)\;.$$
Assume that $\beta_{0,\ell} \neq 0$. As in the proof of Theorem \ref{theo:ASDIST}, for   $n$ large enough, $\beta_{0,\ell} + \lambda_n\; z_\ell$ has the same sign than  $\beta_{0,\ell}$. Hence, using the mean value theorem, we get 
\begin{equation*}
 J_\ell\left (\bbe_0 + \lambda_n\; \bz \right ) - J_\ell(\bbe_0) = \left [\nabla J_\ell\left (\bbe_0 + \alpha_{n,\ell} \;\lambda_n\; \bz \right )\right ]\trasp \lambda_n\; \bz \;,
 \end{equation*}
 with $\alpha_{n,\ell} \in [0,1]$. When $\beta_{0,\ell} = 0$, $J_\ell(\bbe_0) = 0$, so
 \begin{equation*}
  J_\ell\left (\bbe_0 + \lambda_n\; \bz \right ) - J_\ell(\bbe_0) = \frac{\lambda_n\; |z_{\ell}|}{\left \|\bbe_0 + \lambda_n\; \bz \right \|_2}\;.
 \end{equation*} 
Therefore, $ R_{n,2}(\bz) $ can be written as
 \begin{eqnarray*}
 R_{n,2}(\bz) 
 &=& \left\{\sum_{\ell=1}^p \left [\nabla J_\ell\left (\bbe_0 + \alpha_{n,\ell} \lambda_n\; \bz \right )\right ] \trasp {\bz}\; \indica_{\{\beta_{0.\ell}\ne 0\}}+ \sum_{\ell=1}^p  |z_\ell| \frac{1}{\left \|\bbe_0 + \lambda_n\; \bz \right\|_2} \;\indica_{\{\beta_{0.\ell}= 0\}}\right\}\,,
  \end{eqnarray*}
 which entails that $ R_{n,2}(\bz) \convprob  \bz \trasp \bq(\bz)$. Therefore, the process $ R_{n}\convprob R$ and using Theorem 2.7 in Kim y Pollard (1990), we conclude that $ (1/\lambda_n)\;(\wbbe_n - \bbe_0)=\argmin_{\bz} R_n(\bz)\convprob \argmin_{\bz} R(\bz)$. 
 
\vskip0.1in

\noi \textsc{Proof of Theorem \ref{theo:ASDIST_scad_mcp}.}   
Let $\gamma = \min\{|\beta_{0,j}| : 1 \leq j \leq k\} / 2$. Consider the event
\begin{equation*}
   \itA_n = \{\wbbe_{n,B} = \bcero_{p-k} \; \wedge \|\wbbe_{n,A} - \bbe_{0,A}\|_2 \leq \gamma \}.
   \end{equation*} 
Since $\prob(\itA_n) \to 1$ and by definition of $\wbbe_n$, we have that
   \begin{equation}
   \bcero_k = \frac{1}{n} \sum_{i = 1}^n \Psi(y_i, \bx_{i,A}\trasp \wbbe_{n,A}) \,  w(\bx_i)\,\bx_{i,A} + \nabla I_{\lambda_n}(\wbbe_{n,A}) + \br_n
   \label{eq:cero}
   \end{equation} 
   where $\bx_{i,A} \in \real^k$ is the subvector of $\bx_i$ corresponding to its first $k$ coordinates and  $\prob(\br_n = \bcero_k) \to 1$. 
   
Fix $\ba \in \real^k$. Then, \eqref{eq:cero} and the mean value theorem implies that
\begin{equation*}
   \ba\trasp  \bcero_k = \ba\trasp \frac{1}{n} \sum_{i = 1}^n \Psi(y_i, \bx_{i,A}\trasp \bbe_{0,A})\, w(\bx_i)\bx_{i,A} + \ba\trasp \wtbA_{n} (\wbbe_{n,A} - \bbe_{0,A}) +\ba\trasp \nabla I_{\lambda_n}(\wbbe_{n,A}) + \ba\trasp\br_n\,,
   \end{equation*} 
   where $\wtbA_n  = ({1}/{n}) \sum_{i = 1}^n \chi(y_i, \bx_{i,A} \trasp \wtbbe_{n,A}) \, w(\bx_i)\bx_{i,A} \bx_{i,A} \trasp$  with
   $\wtbbe_{n,A} = \bbe_{0,A} + \tau_n (\wbbe_{n,A} - \bbe_{0,A})$ y $0 \leq \tau_n \leq 1$.
   Therefore, 
   \begin{align*}
   \sqrt{n}\ba\trasp \wtbA_n (\wbbe_{n,A} - \bbe_{0,A}) = &- \ba\trasp \frac{1}{\sqrt{n}} \sum_{i = 1}^n \Psi(y_i, \bx_{i,A}\trasp \bbe_{0,A})\, w(\bx_i)\bx_{i,A} - \ba\trasp \sqrt{n} \;\nabla I_{\lambda_n}(\wbbe_{n,A}) -\ba\trasp \sqrt{n} \br_n\,.
   \end{align*} 
The requirement (\ref{cond:ASDIST_scad_mcp}) implies that $\sqrt{n}\; \nabla I_{\lambda_n}(\wbbe_{n,A}) \convprob 0$. Using that $\prob(\br_n = \bcero_k) \to 1$ and the central limit theorem we conclude that
$ \sqrt{n}\;\ba\trasp \wtbA_n \; (\wbbe_{n,A} - \bbe_{0,A}) \convdist N(0, \ba\trasp \wtbB \ba)$, that leads to $\bc_n=\sqrt{n}  \wtbA_n (\wbbe_{n,A} - \bbe_{0,A}) \convdist N_k(\bcero,   \wtbB )$.

Lemma \ref{lema:convAn2} and the fact that  $\wbbe_{n,A} \convprob \bbe_{0,A}$, imply that $\wtbA_{n}^{-1} \convprob \wtbA^{-1}$. Hence, using that \linebreak $\sqrt{n}  (\wbbe_{n,A} - \bbe_{0,A})=  \wtbA_n^{-1} \bc_n$, the result follows from Slutsky's Lemma. $\square$

\newpage
\section{Appendix: Tables and Figures}{\label{sec:apptab}}

{\setcounter{table}{0}
\renewcommand{\thefigure}{B.\arabic{figure}}
\renewcommand{\thetable}{B.\arabic{table}}

\begin{table}[ht]
	\setlength\arrayrulewidth{1pt}        
	\centering
	\def\arraystretch{1.55}
	\small
\begin{tabular}{|c|ccc|ccc|c|ccc|ccc|}
		\hline
		$n$ & \multicolumn{3}{c|}{ $150$} & \multicolumn{3}{c|}{ $300$} & & \multicolumn{3}{c|}{ $150$} & \multicolumn{3}{c|}{ $300$} \\
		\hline
		$p$ &  $40$ & $80$ & $120$ & $40$ & $80$ & $120$ & & $40$ & $80$ & $120$ & $40$ & $80$ & $120$\\ 
		\hline
		$\wbthe_{\mle}$ & 0.095 & 0.162 & 0.170 & 0.032 & 0.091 & 0.160 & $\wbthe_{\wmle}$ & 0.096 & 0.162 & 0.170 & 0.033 & 0.092 & 0.160 \\ 
		$\wbthe_{\mle}^{\lpen}$ & 0.022 & 0.027 & 0.033 & 0.012 & 0.014 & 0.015 & $\wbthe_{\wmle}^{\lpen}$ & 0.022 & 0.027 & 0.034 & 0.012 & 0.014 & 0.015 \\ 
		$\wbthe_{\mle}^{\signpen}$ & 0.025 & 0.043 & 0.089 & 0.007 & 0.009 & 0.010 & $\wbthe_{\wmle}^{\signpen}$ & 0.026 & 0.044 & 0.092 & 0.007 & 0.009 & 0.010 \\ 
		$\wbthe_{\mle}^{\mcppen}$ & 0.020 & 0.024 & 0.031 & 0.004 & 0.005 & 0.005 & $\wbthe_{\wmle}^{\mcppen}$ & 0.020 & 0.024 & 0.032 & 0.004 & 0.005 & 0.005 \\ 
		\hline
		$\wbthe_{\lse}$ & 0.134 & 0.146 & 0.155 & 0.113 & 0.125 & 0.134 & $\wbthe_{\wmlse}$ & 0.134 & 0.146 & 0.156 & 0.114 & 0.126 & 0.134 \\ 
		$\wbthe_{\lse}^{\lpen}$ & 0.046 & 0.047 & 0.051 & 0.037 & 0.038 & 0.038 & $\wbthe_{\wmlse}^{\lpen}$ & 0.046 & 0.047 & 0.051 & 0.037 & 0.038 & 0.038 \\
\hline
	\end{tabular}
	\caption{\small\label{tab:pmse-c0} 10\%-trimmed means  of the mean squared error of probabilities (PMSE) under \textbf{C0}.} 
\end{table}

\begin{table}[ht]
	\setlength\arrayrulewidth{1pt}        
	\centering
	\def\arraystretch{1.5}
	\small
	\begin{tabular}{|c|ccc|ccc|c|ccc|ccc|}
		\hline
		$n$ & \multicolumn{3}{c|}{ $150$} & \multicolumn{3}{c|}{ $300$} & & \multicolumn{3}{c|}{ $150$} & \multicolumn{3}{c|}{ $300$} \\
		\hline
		$p$ &  $40$ & $80$ & $120$ & $40$ & $80$ & $120$ & & $40$ & $80$ & $120$ & $40$ & $80$ & $120$\\ 
		\hline
		$\wbbe_{\mle}$ & 49.54 & $\bigstar$ & $\bigstar$ & 2.28 & 26.32 & $\bigstar$ & $\wbbe_{\wmle}$ & 60.68 & $\bigstar$ & $\bigstar$ & 2.30 & 27.24 & $\bigstar$ \\ 
		$\wbbe_{\mle}^{\lpen}$ & 1.23 & 1.50 & 1.75 & 0.90 & 0.95 & 1.03 & $\wbbe_{\wmle}^{\lpen}$ & 1.23 & 1.50 & 1.76 & 0.90 & 0.96 & 1.03 \\ 
		$\wbbe_{\mle}^{\signpen}$ & 1.33 & $\bigstar$ & $\bigstar$ & 0.35 & 0.43 & 0.45 & $\wbbe_{\wmle}^{\signpen}$ & 1.38 & $\bigstar$ & $\bigstar$ & 0.35 & 0.44 & 0.45 \\ 
		$\wbbe_{\mle}^{\mcppen}$ & 1.02 & 1.18 & 1.52 & 0.20 & 0.25 & 0.24 & $\wbbe_{\wmle}^{\mcppen}$ & 1.03 & 1.16 & 1.54 & 0.21 & 0.27 & 0.24 \\ 
		\hline
		$\wbbe_{\lse}$ & $\bigstar$ & $\bigstar$ & $\bigstar$ & $\bigstar$ & $\bigstar$ & $\bigstar$ & $\wbbe_{\wmlse}$ & $\bigstar$ & $\bigstar$ & $\bigstar$ & $\bigstar$ & $\bigstar$ & $\bigstar$ \\ 
		$\wbbe_{\lse}^{\lpen}$ & 2.78 & 2.81 & 2.93 & 2.50 & 2.52 & 2.54 & $\wbbe_{\wmlse}^{\lpen}$ & 2.78 & 2.81 & 2.94 & 2.50 & 2.52 & 2.54 \\
\hline
	\end{tabular}
	\caption{\label{tab:bmse-c0} Mean Squared Error (MSE). No contamination model: scenario \textbf{C0}. 10\% trimmed average over 500 replications.} 
\end{table}

\begin{table}[ht]
	\setlength\arrayrulewidth{1pt}        
	\centering
	\footnotesize
	\def\arraystretch{1.4}
	\begin{tabular}{|l|ccc|ccc|}
		\hline
		& \multicolumn{3}{c|}{ $n = 150$} & \multicolumn{3}{c|}{ $n = 300$} \\
		\hline
		$p$ &  $ 40$ & $ 80$ & $120$ & $ 40$ & $80$ & $ 120$\\ 
		\hline
		$\wbbe_{\mle}$ & 1.00 / 0.00 & 1.00 / 0.00 & 1.00 / 0.00 & 1.00 / 0.00 & 1.00 / 0.00 & 1.00 / 0.00 \\ 
		$\wbbe_{\mle}^{\lpen}$ & 1.00 / 0.61 & 1.00 / 0.68 & 1.00 / 0.72 & 1.00 / 0.69 & 1.00 / 0.70 & 1.00 / 0.72 \\ 
		$\wbbe_{\mle}^{\signpen}$ & 0.98 / 0.87 & 0.98 / 0.82 & 0.97 / 0.76 & 1.00 / 0.93 & 1.00 / 0.92 & 1.00 / 0.90 \\ 
		$\wbbe_{\mle}^{\mcppen}$ & 0.96 / 0.95 & 0.95 / 0.93 & 0.90 / 0.92 & 1.00 / 0.98 & 1.00 / 0.92 & 1.00 / 0.92 \\ 
		\hline
		$\wbbe_{\lse}$ & 1.00 / 0.09 & 1.00 / 0.18 & 1.00 / 0.24 & 1.00 / 0.03 & 1.00 / 0.09 & 1.00 / 0.13 \\ 
		$\wbbe_{\lse}^{\lpen}$ & 1.00 / 0.84 & 1.00 / 0.85 & 0.98 / 0.85 & 1.00 / 0.92 & 1.00 / 0.92 & 1.00 / 0.91 \\
\hline
	\end{tabular}
	\caption{\label{tab:tpp-tnp-c0}True Positive Proportion / True Null Proportion. No contamination model: scenario \textbf{C0}. 10\% trimmed average over 500 replications.} 
\end{table}

\begin{table}[ht]
	\setlength\arrayrulewidth{1pt}        
	\centering
	\def\arraystretch{1.3}
	\footnotesize
	\begin{tabular}{|L{0.8cm}|C{0.8cm}C{0.8cm}C{0.8cm}|C{0.8cm}C{0.8cm}C{0.8cm}|C{0.8cm}C{0.8cm}C{0.8cm}|C{0.8cm}C{0.8cm}C{0.8cm}|}
		\hline
		& \multicolumn{6}{c|}{$\varepsilon = 0.05$} & \multicolumn{6}{c|}{$\varepsilon = 0.10$} \\ \hline
		& \multicolumn{3}{c|}{ $n = 150$} & \multicolumn{3}{c|}{ $n = 300$} & \multicolumn{3}{c|}{ $n = 150$} & \multicolumn{3}{c|}{ $n = 300$} \\
		\hline
		$p$ &  $ 40$ & $ 80$ & $120$ & $ 40$ & $80$ & $ 120$ &  $40$ & $ 80$ & $ 120$ & $ 40$ & $ 80$ & $ 120$\\ 
		\hline
		$\wbthe_{\mle}$ & 0.093 & 0.193 & 0.204 & 0.058 & 0.093 & 0.179 & 0.104 & 0.217 & 0.230 & 0.079 & 0.103 & 0.157 \\ 
		$\wbthe_{\mle}^{\lpen}$ & 0.077 & 0.079 & 0.082 & 0.069 & 0.071 & 0.071 & 0.105 & 0.106 & 0.107 & 0.104 & 0.104 & 0.105 \\ 
		$\wbthe_{\mle}^{\signpen}$ & 0.083 & 0.096 & 0.143 & 0.078 & 0.084 & 0.080 & 0.102 & 0.105 & 0.130 & 0.097 & 0.102 & 0.104 \\ 
		$\wbthe_{\mle}^{\mcppen}$ & 0.074 & 0.078 & 0.086 & 0.062 & 0.068 & 0.070 & 0.106 & 0.108 & 0.109 & 0.106 & 0.108 & 0.109 \\ 
		\hline
		$\wbthe_{\lse}$ & 0.146 & 0.165 & 0.177 & 0.116 & 0.138 & 0.150 & 0.173 & 0.194 & 0.206 & 0.100 & 0.162 & 0.175 \\
\hline
	\end{tabular}
	\caption{\label{tab:pmse-a1-a2} 10\%-trimmed means  of the mean squared error of probabilities (PMSE) under  \textbf{CA1} and \textbf{CA2}.} 
\end{table}

\begin{table}[ht]
	\setlength\arrayrulewidth{1pt}        
	\centering
	\def\arraystretch{1.3}
	\footnotesize
	\begin{tabular}{|L{0.8cm}|C{0.8cm}C{0.8cm}C{0.8cm}|C{0.8cm}C{0.8cm}C{0.8cm}|C{0.8cm}C{0.8cm}C{0.8cm}|C{0.8cm}C{0.8cm}C{0.8cm}|}
		\hline
		& \multicolumn{6}{c|}{$\varepsilon = 0.05$} & \multicolumn{6}{c|}{$\varepsilon = 0.10$} \\ \hline
		& \multicolumn{3}{c|}{ $n = 150$} & \multicolumn{3}{c|}{ $n = 300$} & \multicolumn{3}{c|}{ $n = 150$} & \multicolumn{3}{c|}{ $n = 300$} \\
		\hline
		$p$ &  $ 40$ & $ 80$ & $120$ & $ 40$ & $80$ & $ 120$ &  $40$ & $ 80$ & $ 120$ & $ 40$ & $ 80$ & $ 120$\\ 
		\hline
		$\wbbe_{\mle}$ & 7.43 & $\bigstar$ & $\bigstar$ & 2.37 & 6.50 & $\bigstar$ & 4.93 & $\bigstar$ & $\bigstar$ & 3.33 & 4.67 & 33.30 \\ 
		$\wbbe_{\mle}^{\lpen}$ & 3.78 & 3.88 & 3.93 & 3.61 & 3.67 & 3.66 & 4.86 & 4.86 & 4.92 & 4.83 & 4.85 & 4.89 \\ 
		$\wbbe_{\mle}^{\signpen}$ & 3.76 & $\bigstar$ & $\bigstar$ & 3.82 & 4.00 & 3.77 & 4.71 & 4.76 & $\bigstar$ & 4.62 & 4.78 & 4.83 \\ 
		$\wbbe_{\mle}^{\mcppen}$ & 3.22 & 3.30 & 3.66 & 3.03 & 3.20 & 3.21 & 4.84 & 4.86 & 4.92 & 4.89 & 4.98 & 4.99 \\ 
		\hline
		$\wbbe_{\lse}$ & $\bigstar$ & $\bigstar$ & $\bigstar$ & $\bigstar$ & $\bigstar$ & $\bigstar$ & $\bigstar$ & $\bigstar$ & $\bigstar$ & $\bigstar$ & $\bigstar$ & $\bigstar$ \\
\hline
	\end{tabular}
	\caption{\label{tab:bmse-a1-a2}  Mean Squared Error (MSE) for scenarios \textbf{CA1} and \textbf{CA2}. 10\% trimmed average over 500 replications.} 
\end{table}

\begin{table}[ht]
	\setlength\arrayrulewidth{0.8pt}        
	\centering
	\footnotesize
	\def\arraystretch{1.4}
	
	\begin{tabular}{|l|ccc|ccc|}
		\hline
		& \multicolumn{3}{c|}{ $n = 150$} & \multicolumn{3}{c|}{ $n = 300$} \\
		\hline
		$p$ &  $ 40$ & $ 80$ & $120$ & $ 40$ & $80$ & $ 120$ \\ 
		\hline
		$\wbbe_{\mle}$ & 1.00 / 0.00 & 1.00 / 0.00 & 1.00 / 0.00 & 1.00 / 0.00 & 1.00 / 0.00 & 1.00 / 0.00 \\ 
		$\wbbe_{\mle}^{\lpen}$ & 0.81 / 0.69 & 0.79 / 0.75 & 0.76 / 0.79 & 0.94 / 0.65 & 0.95 / 0.71 & 0.95 / 0.73 \\ 
		$\wbbe_{\mle}^{\signpen}$ & 0.54 / 0.85 & 0.75 / 0.73 & 0.91 / 0.67 & 0.60 / 0.94 & 0.61 / 0.88 & 0.61 / 0.81 \\ 
		$\wbbe_{\mle}^{\mcppen}$ & 0.70 / 0.79 & 0.71 / 0.82 & 0.68 / 0.84 & 0.81 / 0.78 & 0.78 / 0.81 & 0.77 / 0.82 \\ 
		\hline
		$\wbbe_{\lse}$ & 1.00 / 0.07 & 1.00 / 0.16 & 1.00 / 0.24 & 1.00 / 0.02 & 1.00 / 0.07 & 1.00 / 0.11 \\ 
		$\wbbe_{\lse}^{\lpen}$ & 0.93 / 0.70 & 0.93 / 0.72 & 0.92 / 0.75 & 0.99 / 0.75 & 1.00 / 0.75 & 1.00 / 0.75 \\
\hline
	\end{tabular}
	\caption{\label{tab:tpp-tnp-a1} True Positive Proportion / True Null Proportion. 5\% contamination model: scenario \textbf{CA1}. 10\% trimmed average over 500 replications.} 
\end{table}

\begin{table}[ht]
	\setlength\arrayrulewidth{1pt}        
	\centering
	\footnotesize
	\def\arraystretch{1.4}
	
	\begin{tabular}{|l|ccc|ccc|}
		\hline
		& \multicolumn{3}{c|}{ $n = 150$} & \multicolumn{3}{c|}{ $n = 300$} \\
		\hline
		$p$ &  $ 40$ & $ 80$ & $120$ & $ 40$ & $80$ & $ 120$ \\ 
		\hline
		$\wbbe_{\mle}$ & 1.00 / 0.00 & 1.00 / 0.00 & 1.00 / 0.00 & 1.00 / 0.00 & 1.00 / 0.00 & 1.00 / 0.00 \\ 
		$\wbbe_{\mle}^{\lpen}$ & 0.45 / 0.78 & 0.44 / 0.84 & 0.41 / 0.87 & 0.57 / 0.77 & 0.60 / 0.83 & 0.56 / 0.86 \\ 
		$\wbbe_{\mle}^{\signpen}$ & 0.33 / 0.94 & 0.38 / 0.88 & 0.60 / 0.78 & 0.46 / 0.98 & 0.42 / 0.94 & 0.33 / 0.92 \\ 
		$\wbbe_{\mle}^{\mcppen}$ & 0.34 / 0.81 & 0.33 / 0.87 & 0.33 / 0.89 & 0.40 / 0.82 & 0.31 / 0.90 & 0.34 / 0.91 \\ 
		\hline
		$\wbbe_{\lse}$ & 1.00 / 0.05 & 1.00 / 0.13 & 1.00 / 0.19 & 1.00 / 0.00 & 1.00 / 0.05 & 1.00 / 0.08 \\ 
		$\wbbe_{\lse}^{\lpen}$ & 0.60 / 0.77 & 0.59 / 0.82 & 0.58 / 0.84 & 0.81 / 0.69 & 0.84 / 0.73 & 0.83 / 0.75 \\
\hline
	\end{tabular}
	\caption{\label{tab:tpp-tnp-a2} True Positive Proportion / True Null Proportion. 10\% contamination model: scenario \textbf{CA2}. 10\% trimmed average over 500 replications.} 
\end{table}

\begin{table}[ht]
	\setlength\arrayrulewidth{1pt}        
	\footnotesize
	\def\arraystretch{1.5}
	
	\begin{tabular}{|l|cccccc|cccccc|}
		\hline
		$n = 150$ 	& \multicolumn{12}{c|}{ $p = 40$ }\\
		\hline
		& \multicolumn{6}{c|}{$\varepsilon = 0.05$} & \multicolumn{6}{c|}{$\varepsilon = 0.10$} \\
		\hline
		$m$ &  $0.5$ & $1$ & $2$ & $3$ & $4$ & $5$&  $0.5$ & $1$ & $2$ & $3$ & $4$ & $5$\\ 
		\hline
		$\wbbe_{\mle}^{\signpen}$ & 1.843 & 3.638 & 4.382 & 4.736 & 4.970 & 5.070 & 3.304 & 4.484 & 5.092 & 5.138 & 5.160 & 5.215 \\ 
		$\wbbe_{\mle}^{\mcppen}$ & 1.632 & 3.079 & 4.540 & 5.114 & 5.321 & 5.349 & 2.829 & 4.310 & 5.433 & 5.499 & 5.165 & 5.181 \\ 
		\hline
		$\wbbe_{\mch}^{\signpen}$ & 1.925 & 3.316 & 4.123 & 4.352 & 4.397 & 4.243 & 3.598 & 4.531 & 5.050 & 5.152 & 5.153 & 5.185 \\ 
		$\wbbe_{\mch}^{\mcppen}$ & 1.515 & 2.244 & 3.119 & 3.236 & 2.938 & 2.661 & 2.987 & 4.548 & 5.110 & 5.110 & 5.066 & 5.137 \\ 
		\hline
		$\wbbe_{\wmch}^{\signpen}$ & 1.999 & 3.348 & 4.167 & 1.871 & 2.023 & 2.016 & 3.597 & 4.563 & 5.061 & 3.597 & 3.880 & 3.967 \\
\hline
	\end{tabular}
	\caption{\label{tab:bmse-b1-b2-n150} Mean Squared Error (MSE)for scenarios \textbf{CB1} and \textbf{CB2} with $n=150$. 10\% trimmed average over 500 replications.}
\end{table}

\begin{table}[ht]
	\setlength\arrayrulewidth{1pt}        
	\footnotesize
	\def\arraystretch{1.5}
	
	\begin{tabular}{|l|cccccc|cccccc|}
		\hline
		$n = 300$		& \multicolumn{12}{c|}{  $p = 80$ }\\
		\hline
		& \multicolumn{6}{c|}{$\varepsilon = 0.05$} & \multicolumn{6}{c|}{$\varepsilon = 0.10$} \\
		\hline
		$m$ &  $0.5$ & $1$ & $2$ & $3$ & $4$ & $5$&  $0.5$ & $1$ & $2$ & $3$ & $4$ & $5$\\ 
		\hline
		$\wbbe_{\mle}^{\signpen}$ & 1.571 & 3.589 & 4.621 & 5.013 & 5.079 & 5.123 & 3.520 & 4.974 & 5.071 & 5.216 & 5.284 & 5.296 \\ 
		$\wbbe_{\mle}^{\mcppen}$ & 1.537 & 3.819 & 4.966 & 5.216 & 5.194 & 5.060 & 3.135 & 4.766 & 5.376 & 5.134 & 5.237 & 5.250 \\ 
		\hline
		$\wbbe_{\mch}^{\signpen}$ & 1.046 & 1.607 & 1.736 & 1.286 & 1.277 & 1.932 & 3.732 & 4.963 & 5.080 & 5.175 & 5.215 & 5.221 \\ 
		$\wbbe_{\mch}^{\mcppen}$ & 0.981 & 1.061 & 0.618 & 0.465 & 0.426 & 0.677 & 3.045 & 4.972 & 5.028 & 5.102 & 5.109 & 5.082 \\ 
		\hline
		$\wbbe_{\wmch}^{\signpen}$ & 1.061 & 1.612 & 1.749 & 0.509 & 0.519 & 0.507 & 3.768 & 4.976 & 5.074 & 2.681 & 2.354 & 1.719 \\
\hline
	\end{tabular}
	\caption{\label{tab:bmse-b1-b2-n300} Beta Squared Error (MSE) for scenarios \textbf{CB1} and \textbf{CB2} with $n = 300$. 10\% trimmed average over 500 replications.} 
\end{table}

\begin{table}[ht]
	\setlength\arrayrulewidth{1pt}        
	\footnotesize
	\def\arraystretch{1.5}
	
	\begin{tabular}{|l|cccccc|cccccc|}
		\hline
		$n = 150$ & \multicolumn{12}{c|}{  $p = 40$ }\\
		\hline
		& \multicolumn{6}{c|}{$\varepsilon = 0.05$} & \multicolumn{6}{c|}{$\varepsilon = 0.10$} \\
		\hline
		$m$ &  $0.5$ & $1$ & $2$ & $3$ & $4$ & $5$&  $0.5$ & $1$ & $2$ & $3$ & $4$ & $5$\\ 
		\hline
		$\wbbe_{\mle}^{\signpen}$ & 0.865 & 0.535 & 0.403 & 0.356 & 0.363 & 0.394 & 0.597 & 0.327 & 0.272 & 0.249 & 0.282 & 0.307 \\ 
		$\wbbe_{\mle}^{\mcppen}$ & 0.874 & 0.666 & 0.497 & 0.469 & 0.453 & 0.456 & 0.704 & 0.459 & 0.467 & 0.482 & 0.460 & 0.552 \\ 
		\hline
		$\wbbe_{\mch}^{\signpen}$ & 0.820 & 0.566 & 0.422 & 0.389 & 0.415 & 0.458 & 0.476 & 0.259 & 0.217 & 0.211 & 0.224 & 0.256 \\ 
		$\wbbe_{\mch}^{\mcppen}$ & 0.846 & 0.723 & 0.588 & 0.588 & 0.640 & 0.727 & 0.591 & 0.306 & 0.275 & 0.305 & 0.216 & 0.425 \\ 
		\hline
		$\wbbe_{\wmch}^{\signpen}$ & 0.805 & 0.551 & 0.418 & 0.965 & 0.948 & 0.933 & 0.469 & 0.254 & 0.224 & 0.619 & 0.546 & 0.535 \\
\hline
	\end{tabular}
	\caption{\label{tab:tpp-b1-b2-n150} True Positive Proportions for scenarios \textbf{CB1} and \textbf{CB2} with $n = 150$. 10\% trimmed average over 500 replications.  } 
\end{table}

\begin{table}[ht]
	\setlength\arrayrulewidth{1pt}        
	\footnotesize
	\def\arraystretch{1.5}
	
	\begin{tabular}{|l|cccccc|cccccc|}
		\hline
		$n = 300$ & \multicolumn{12}{c|}{  $p = 80$ }\\
		\hline
		& \multicolumn{6}{c|}{$\varepsilon = 0.05$} & \multicolumn{6}{c|}{$\varepsilon = 0.10$} \\
		\hline
		$m$ &  $0.5$ & $1$ & $2$ & $3$ & $4$ & $5$&  $0.5$ & $1$ & $2$ & $3$ & $4$ & $5$\\ 
		\hline
		$\wbbe_{\mle}^{\signpen}$ & 0.991 & 0.744 & 0.571 & 0.605 & 0.674 & 0.626 & 0.715 & 0.372 & 0.311 & 0.454 & 0.476 & 0.459 \\ 
		$\wbbe_{\mle}^{\mcppen}$ & 0.961 & 0.726 & 0.474 & 0.455 & 0.526 & 0.492 & 0.744 & 0.593 & 0.521 & 0.533 & 0.611 & 0.619 \\ 
		\hline	
		$\wbbe_{\mch}^{\signpen}$ & 1.000 & 0.952 & 0.887 & 0.912 & 0.908 & 0.837 & 0.673 & 0.338 & 0.289 & 0.391 & 0.356 & 0.359 \\ 
		$\wbbe_{\mch}^{\mcppen}$ & 0.987 & 0.970 & 0.981 & 0.996 & 1.000 & 0.978 & 0.736 & 0.312 & 0.401 & 0.491 & 0.530 & 0.532 \\ 
		\hline
		$\wbbe_{\wmch}^{\signpen}$ & 0.999 & 0.945 & 0.868 & 1.000 & 1.000 & 1.000 & 0.672 & 0.329 & 0.324 & 0.753 & 0.786 & 0.869 \\
\hline
	\end{tabular}
	\caption{\label{tab:tpp-b1-b2-n300} True Positive Proportions for scenarios \textbf{CB1} and \textbf{CB2} with $n = 300$. 10\% trimmed average over 500 replications.} 
\end{table}

\begin{table}[ht]
	\setlength\arrayrulewidth{1pt}        
	\footnotesize
	\def\arraystretch{1.5}
	
	\begin{tabular}{|l|cccccc|cccccc|}
		\hline
		$n = 150$ & \multicolumn{12}{c|}{  $p = 40$ }\\
		\hline
		& \multicolumn{6}{c|}{$\varepsilon = 0.05$} & \multicolumn{6}{c|}{$\varepsilon = 0.10$} \\
		\hline
		$m$ &  $0.5$ & $1$ & $2$ & $3$ & $4$ & $5$&  $0.5$ & $1$ & $2$ & $3$ & $4$ & $5$\\ 
		\hline
		$\wbbe_{\mle}^{\signpen}$ & 0.931 & 0.955 & 0.962 & 0.963 & 0.950 & 0.947 & 0.937 & 0.922 & 0.908 & 0.924 & 0.946 & 0.962 \\ 
		$\wbbe_{\mle}^{\mcppen}$ & 0.931 & 0.903 & 0.842 & 0.784 & 0.743 & 0.732 & 0.897 & 0.831 & 0.697 & 0.739 & 0.905 & 0.945 \\ 
		\hline
		$\wbbe_{\mch}^{\signpen}$ & 0.949 & 0.970 & 0.979 & 0.970 & 0.967 & 0.963 & 0.971 & 0.967 & 0.956 & 0.956 & 0.959 & 0.968 \\ 
		$\wbbe_{\mch}^{\mcppen}$ & 0.974 & 0.971 & 0.965 & 0.961 & 0.959 & 0.958 & 0.961 & 0.945 & 0.911 & 0.928 & 0.976 & 0.969 \\ 
		\hline
		$\wbbe_{\wmch}^{\signpen}$ & 0.951 & 0.973 & 0.979 & 0.918 & 0.926 & 0.931 & 0.971 & 0.966 & 0.956 & 0.938 & 0.935 & 0.935 \\
\hline
	\end{tabular}
	\caption{\label{tab:tnp-b1-b2-n150} True Null Proportions for scenarios \textbf{CB1} and \textbf{CB2} with $n = 150$. 10\% trimmed average over 500 replications.} 
\end{table}

\begin{table}[ht]
	\setlength\arrayrulewidth{1pt}        
	\footnotesize
	\def\arraystretch{1.5}
	
	\begin{tabular}{|l|cccccc|cccccc|}
		\hline
		$n = 300$ & \multicolumn{12}{c|}{  $p = 80$ }\\
		\hline
		& \multicolumn{6}{c|}{$\varepsilon = 0.05$} & \multicolumn{6}{c|}{$\varepsilon = 0.10$} \\
		\hline
		$m$ &  $0.5$ & $1$ & $2$ & $3$ & $4$ & $5$&  $0.5$ & $1$ & $2$ & $3$ & $4$ & $5$\\ 
		\hline
		$\wbbe_{\mle}^{\signpen}$ & 0.950 & 0.957 & 0.964 & 0.957 & 0.960 & 0.969 & 0.953 & 0.926 & 0.941 & 0.965 & 0.968 & 0.968 \\ 
		$\wbbe_{\mle}^{\mcppen}$ & 0.953 & 0.917 & 0.840 & 0.767 & 0.801 & 0.932 & 0.910 & 0.763 & 0.753 & 0.957 & 0.964 & 0.964 \\ 
		\hline
		$\wbbe_{\mch}^{\signpen}$ & 0.957 & 0.953 & 0.955 & 0.953 & 0.956 & 0.964 & 0.964 & 0.963 & 0.963 & 0.969 & 0.974 & 0.975 \\ 
		$\wbbe_{\mch}^{\mcppen}$ & 0.971 & 0.971 & 0.952 & 0.938 & 0.944 & 0.965 & 0.963 & 0.942 & 0.944 & 0.967 & 0.964 & 0.964 \\ 
		\hline
		$\wbbe_{\wmch}^{\signpen}$ & 0.957 & 0.953 & 0.956 & 0.951 & 0.954 & 0.953 & 0.963 & 0.964 & 0.960 & 0.944 & 0.943 & 0.940 \\
\hline
	\end{tabular}
	\caption{\label{tab:tnp-b1-b2-n300} True Null Proportions for scenarios \textbf{CB1} and \textbf{CB2} with $n = 300$. 10\% trimmed average over 500 replications.}
\end{table}
\restoregeometry

\newpage 
\newgeometry{left=1cm,right=1cm,vmargin=0.5cm,footnotesep=0.5cm}

\begin{landscape}
	\begin{figure}[ht!]
		\centering
	\begin{tabular}{ccc p{2.8cm}} 
 LASSO &  Sign & MCP & \\
			  \raisebox{-\totalheight}{\includegraphics[scale=0.58]{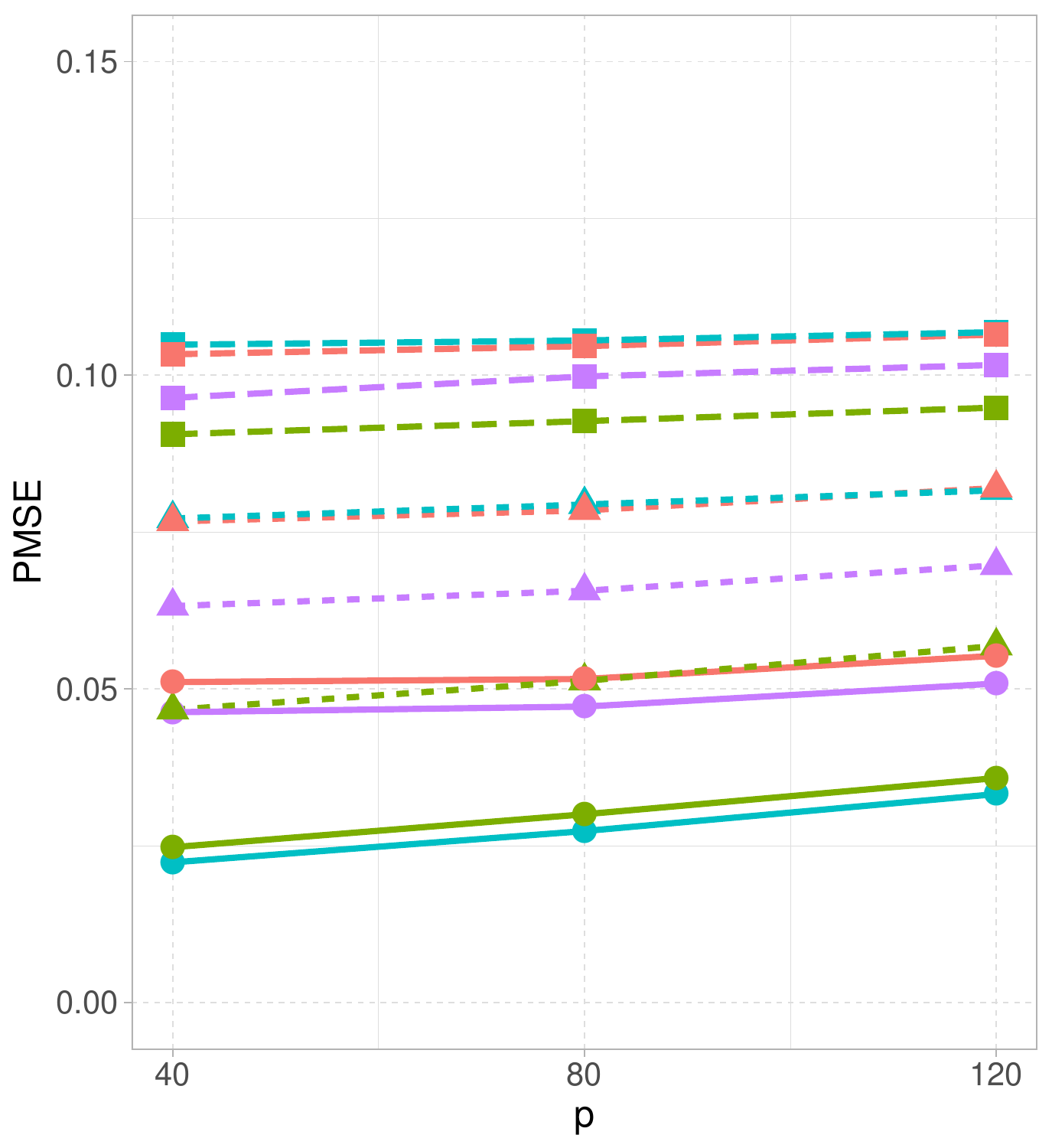} }& 
			  \raisebox{-\totalheight}{\includegraphics[scale=0.58]{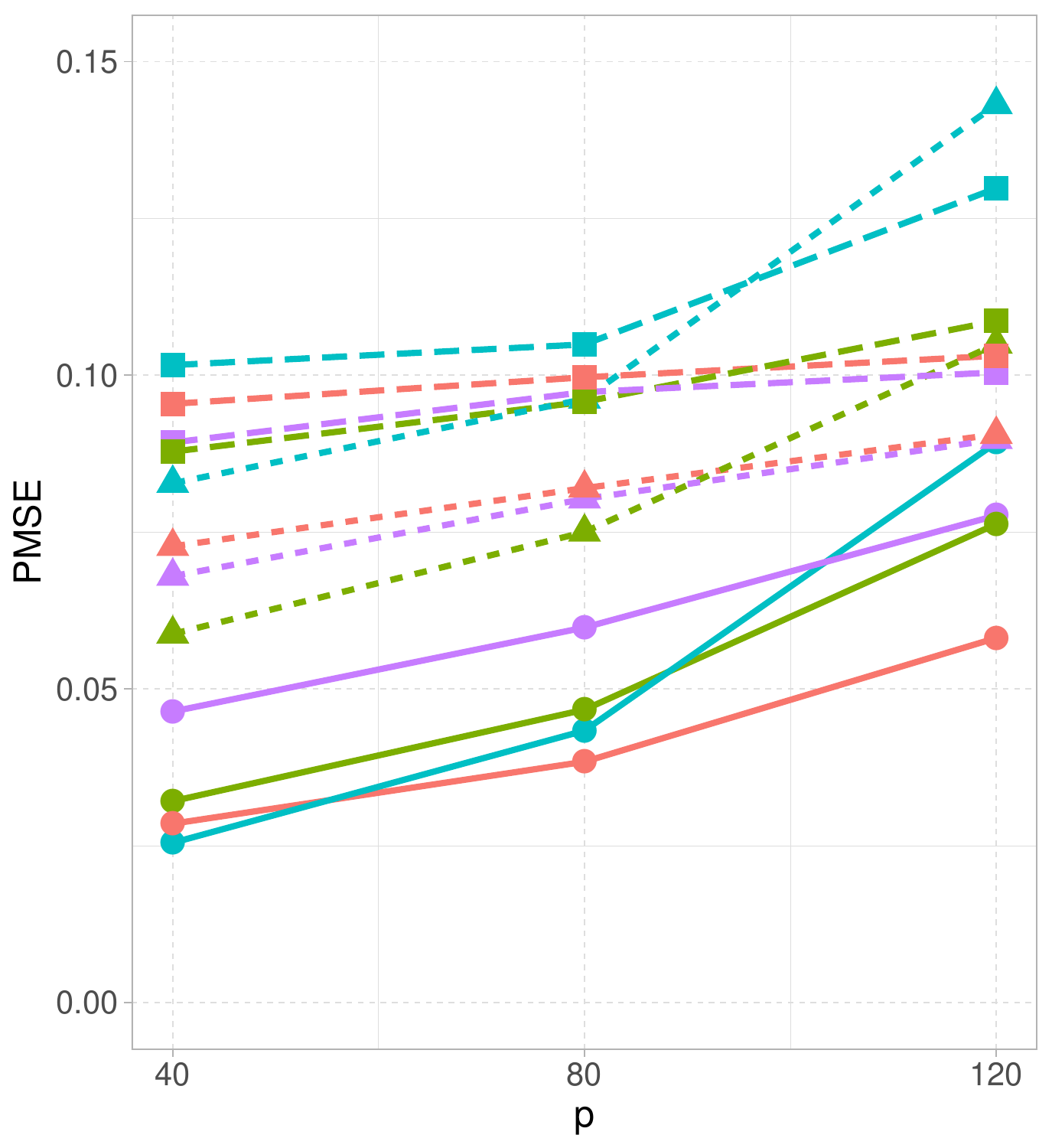}} & 
			  \raisebox{-\totalheight}{\includegraphics[scale=0.58]{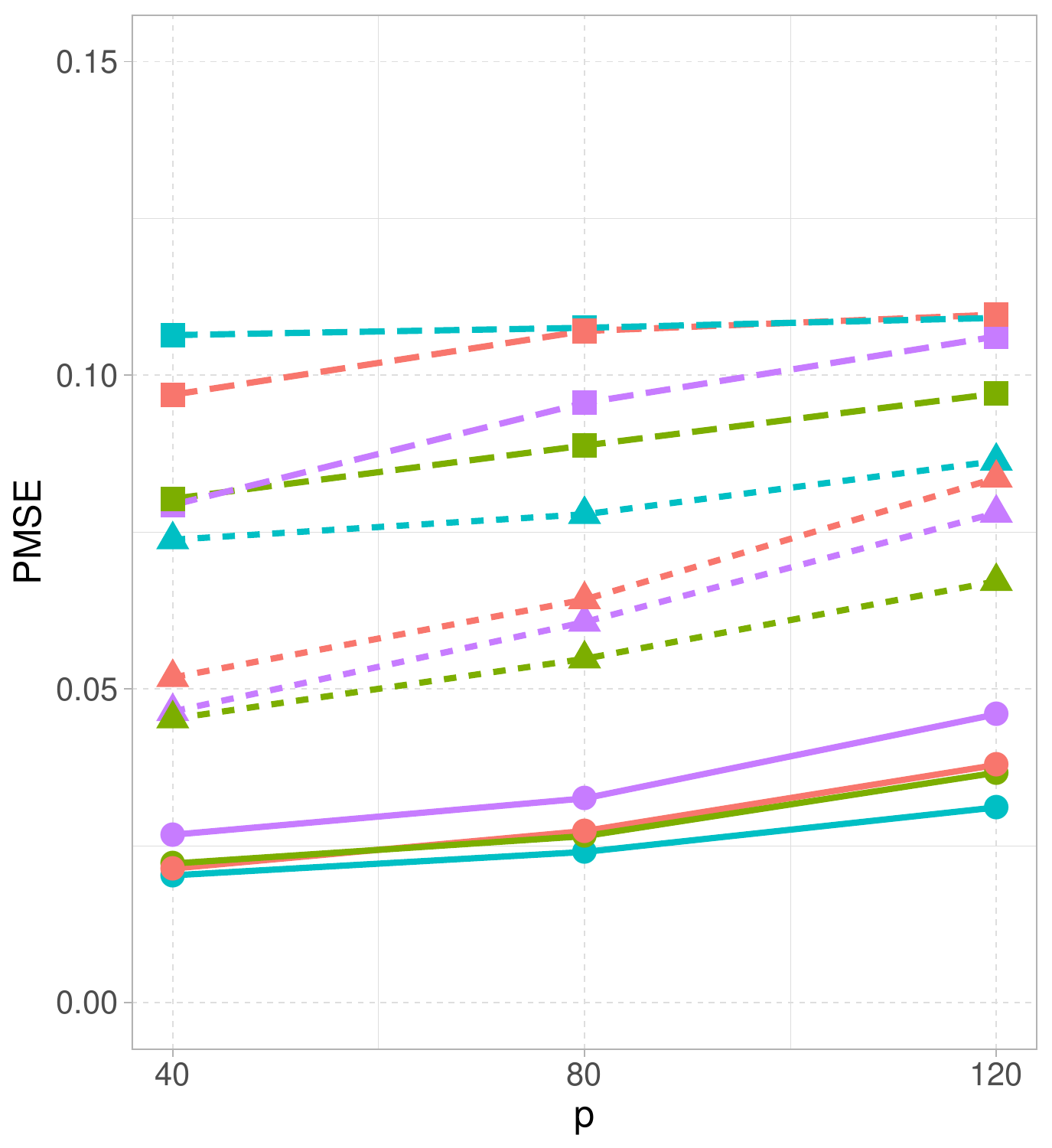}}&  
{\begin{itemize}\item[]\item[]\item[]
      \item[$w\equiv 1$]
      \end{itemize}
      }
      \\
			  \raisebox{-\totalheight}{\includegraphics[scale=0.58]{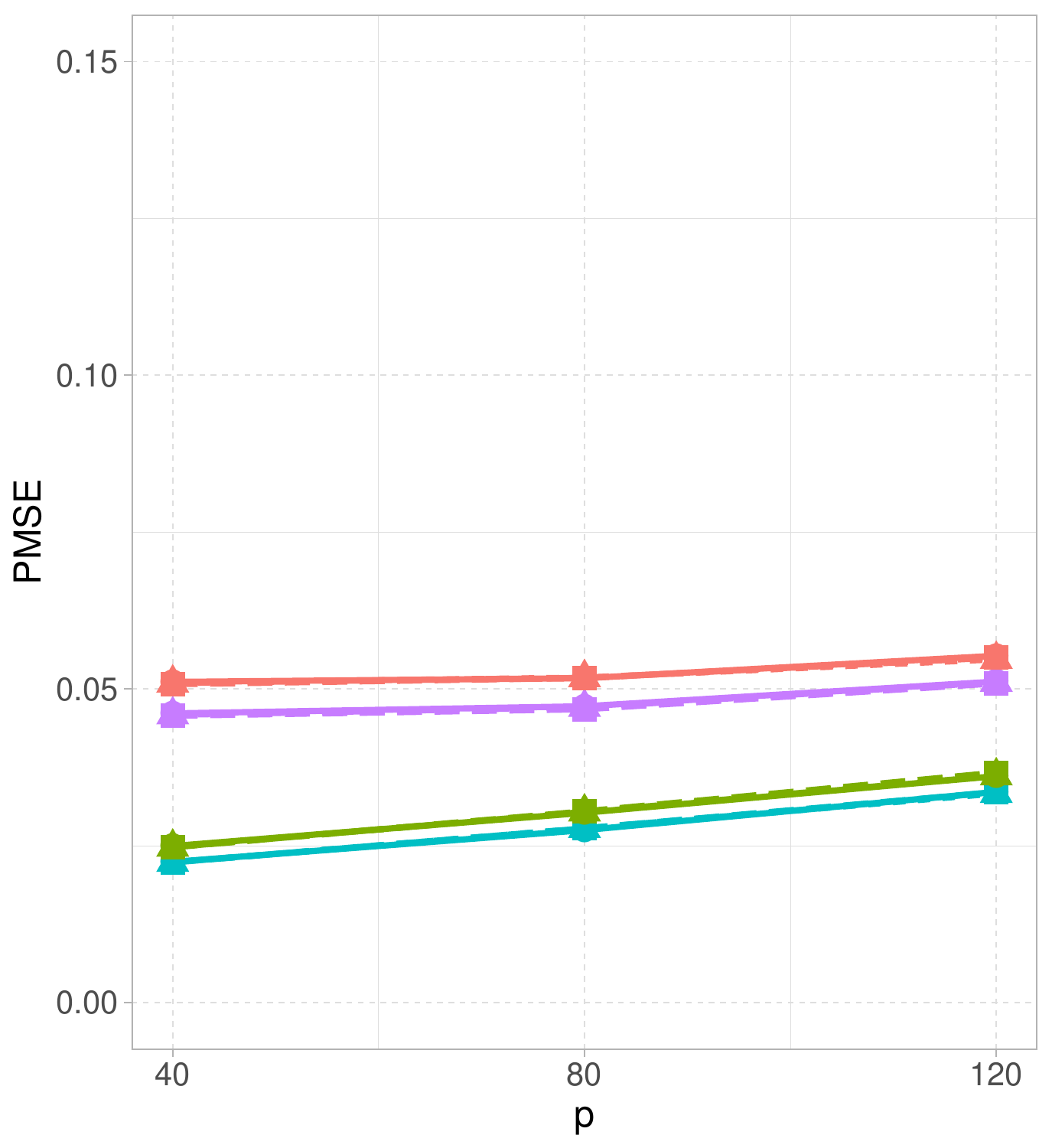}} & 
			  \raisebox{-\totalheight}{\includegraphics[scale=0.58]{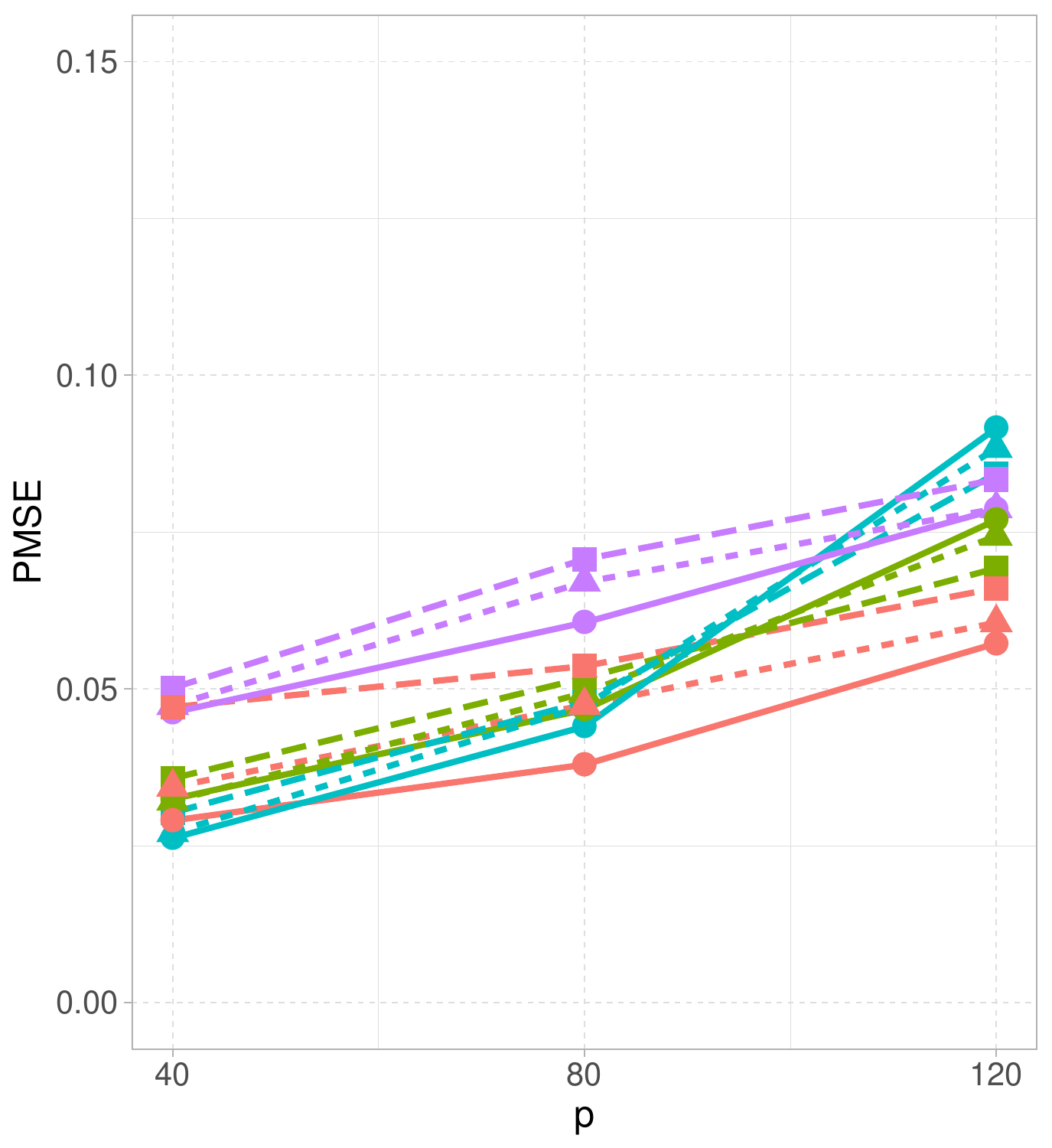}} & 
			  \raisebox{-\totalheight}{\includegraphics[scale=0.58]{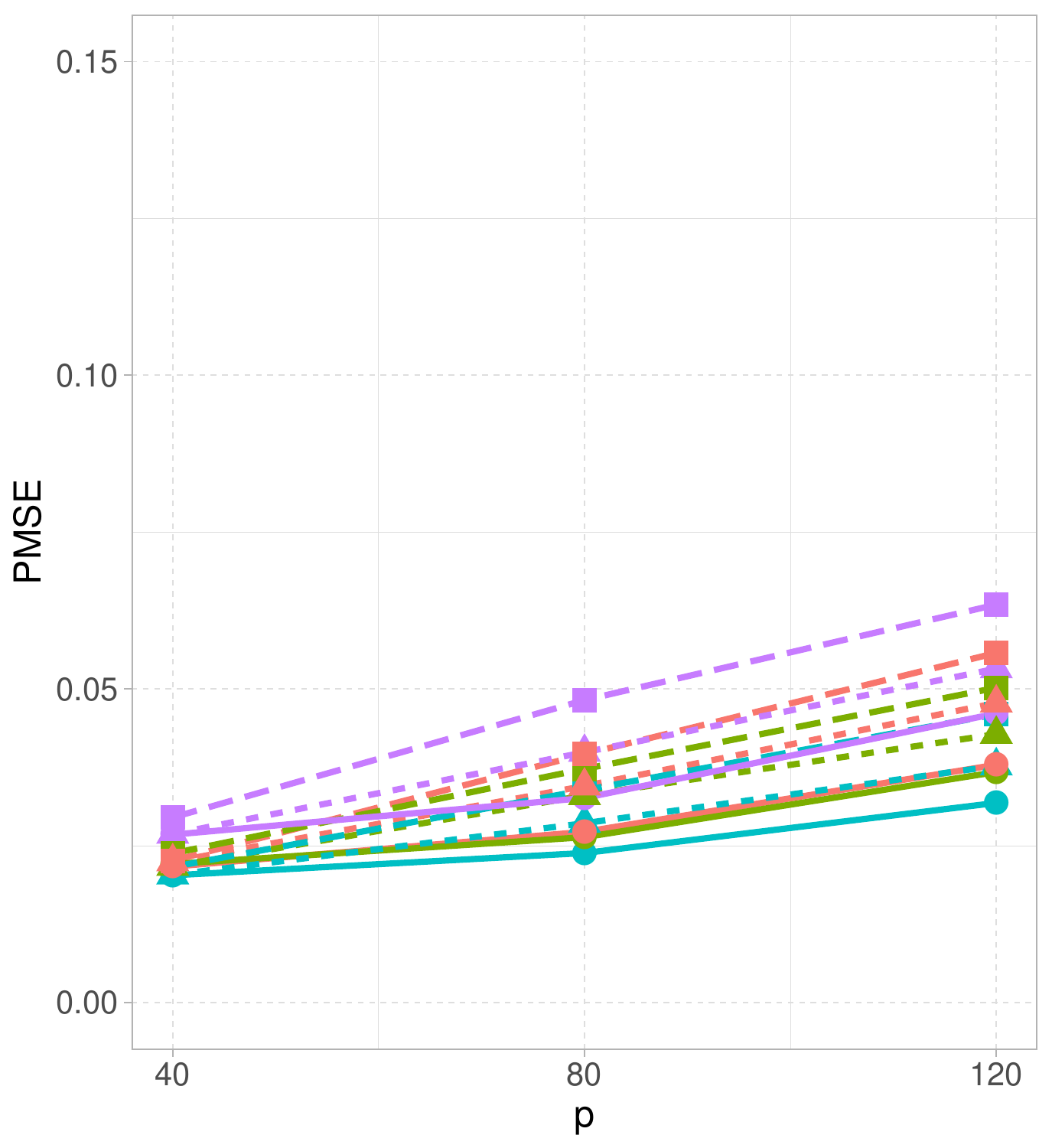}} & 
\begin{itemize}\item[]\item[]\item[]
      \item[$w(t)$]$=\indica_{t\le c_w}$ 
      \end{itemize}
      \\            
		\end{tabular}
		\caption{Probabilities Mean Squared Error for $n = 150$ for scenarios \textbf{C0}, \textbf{CA1} and \textbf{CA2}.}
		\label{fig:ECMP_n_150}
	\end{figure}
	
	\newpage
	\begin{figure}[ht!]
		\centering
\begin{tabular}{ccc p{2.8cm}} 
 LASSO &  Sign & MCP & \\
 			  \raisebox{-\totalheight}{\includegraphics[scale=0.58]{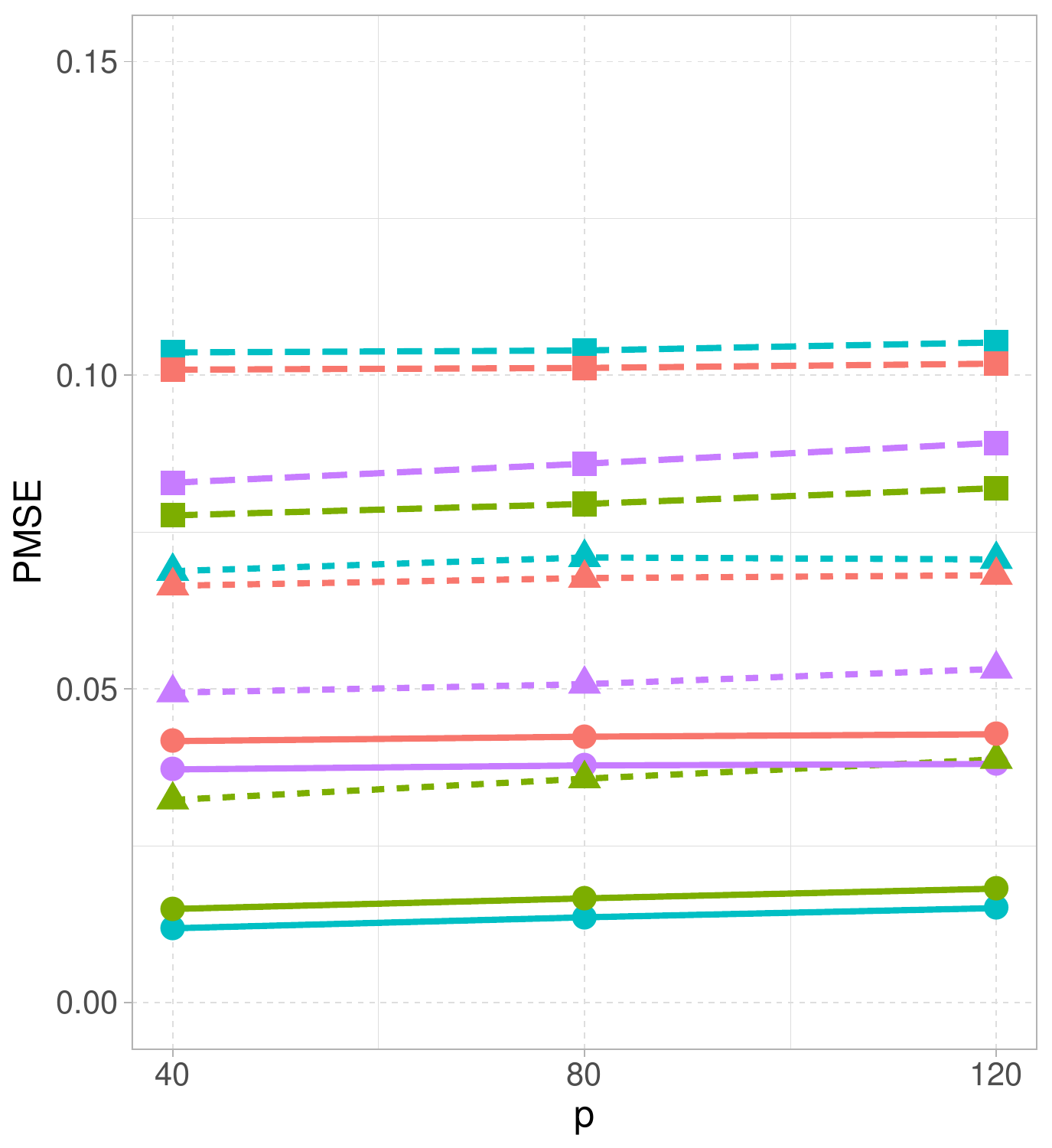}} & 
			  \raisebox{-\totalheight}{\includegraphics[scale=0.58]{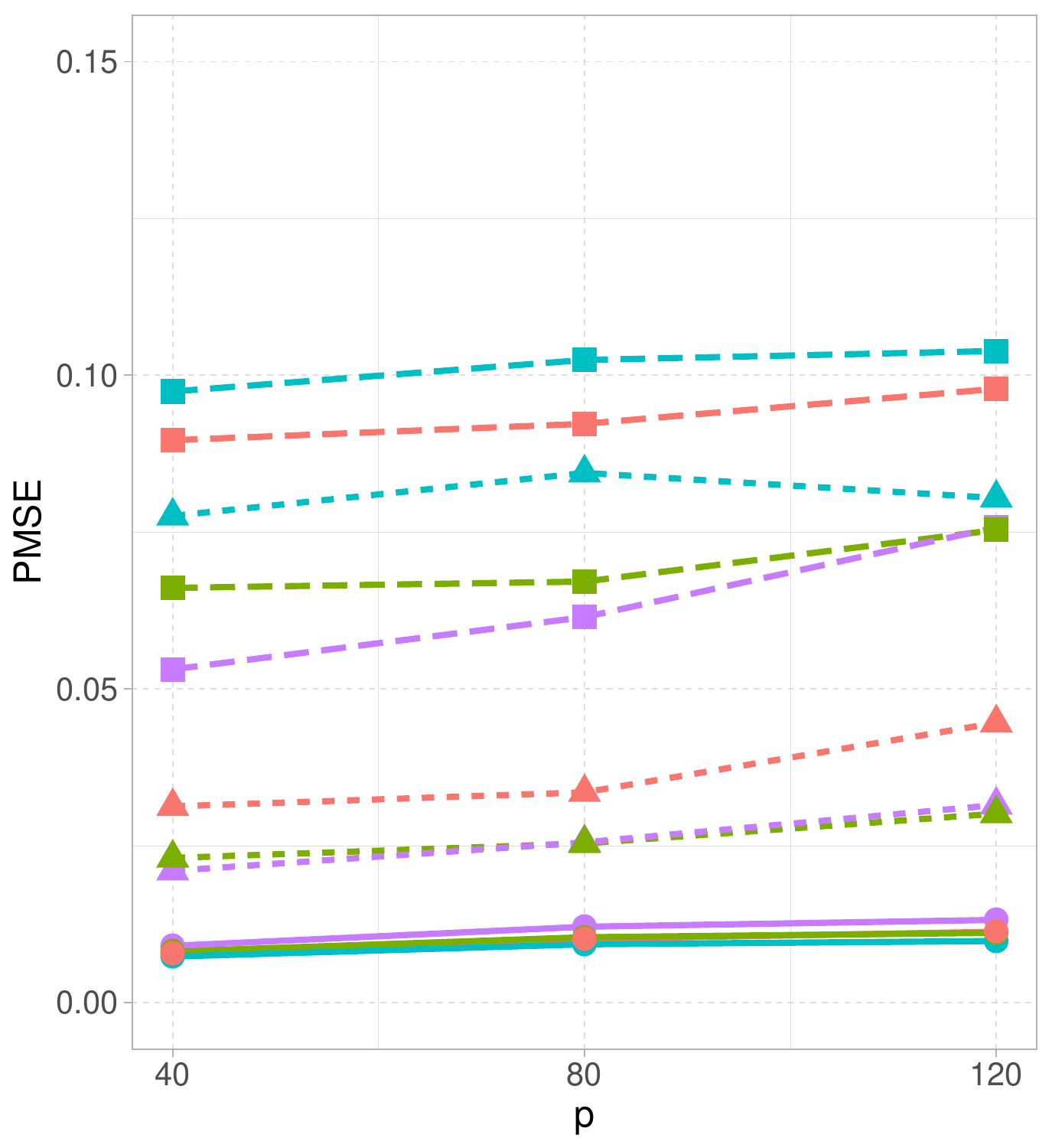}} & 
			  \raisebox{-\totalheight}{\includegraphics[scale=0.58]{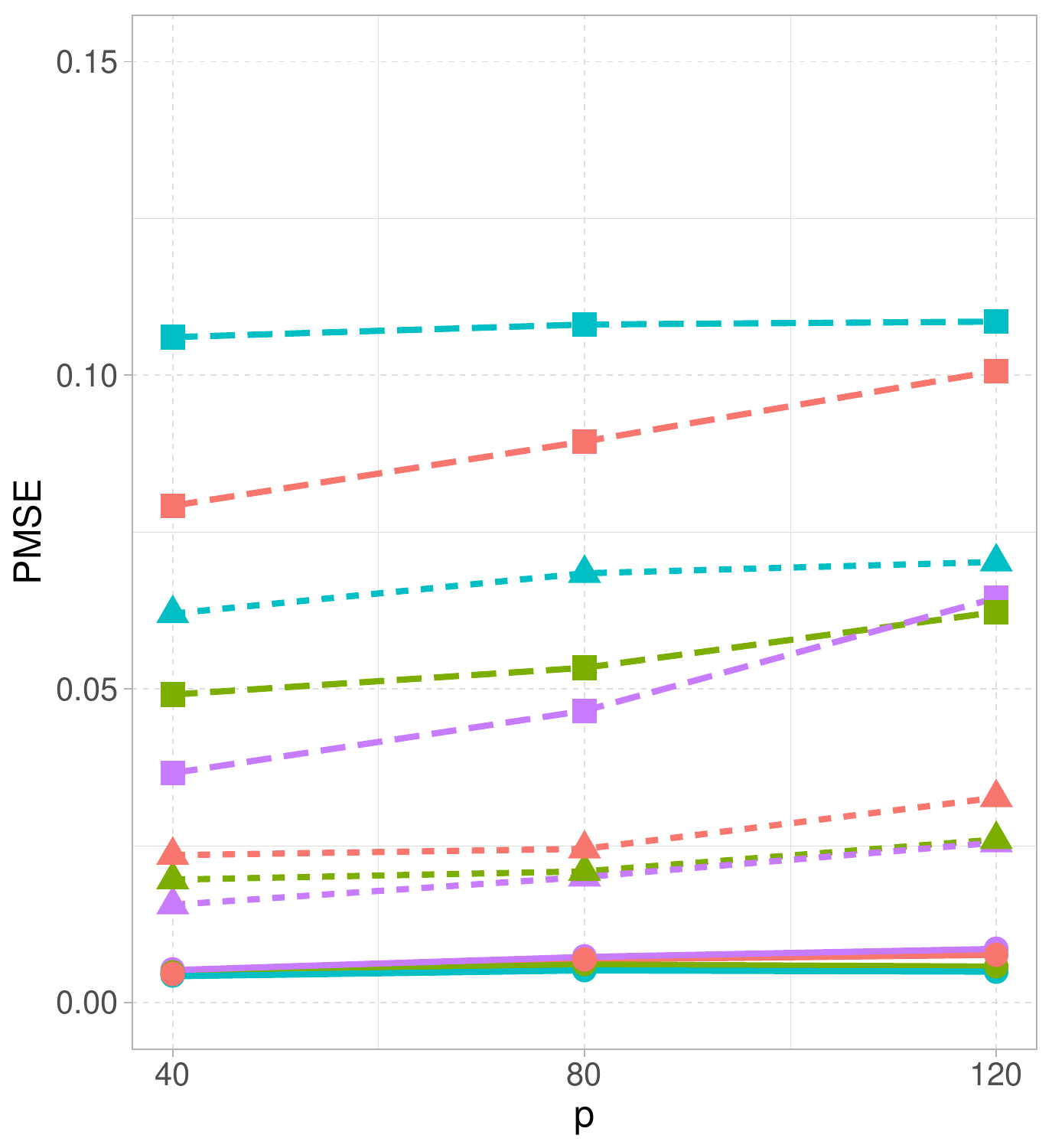}}&  
{\begin{itemize}\item[]\item[]\item[]
      \item[$w\equiv 1$]
      \end{itemize}
      }
      \\
			  \raisebox{-\totalheight}{\includegraphics[scale=0.58]{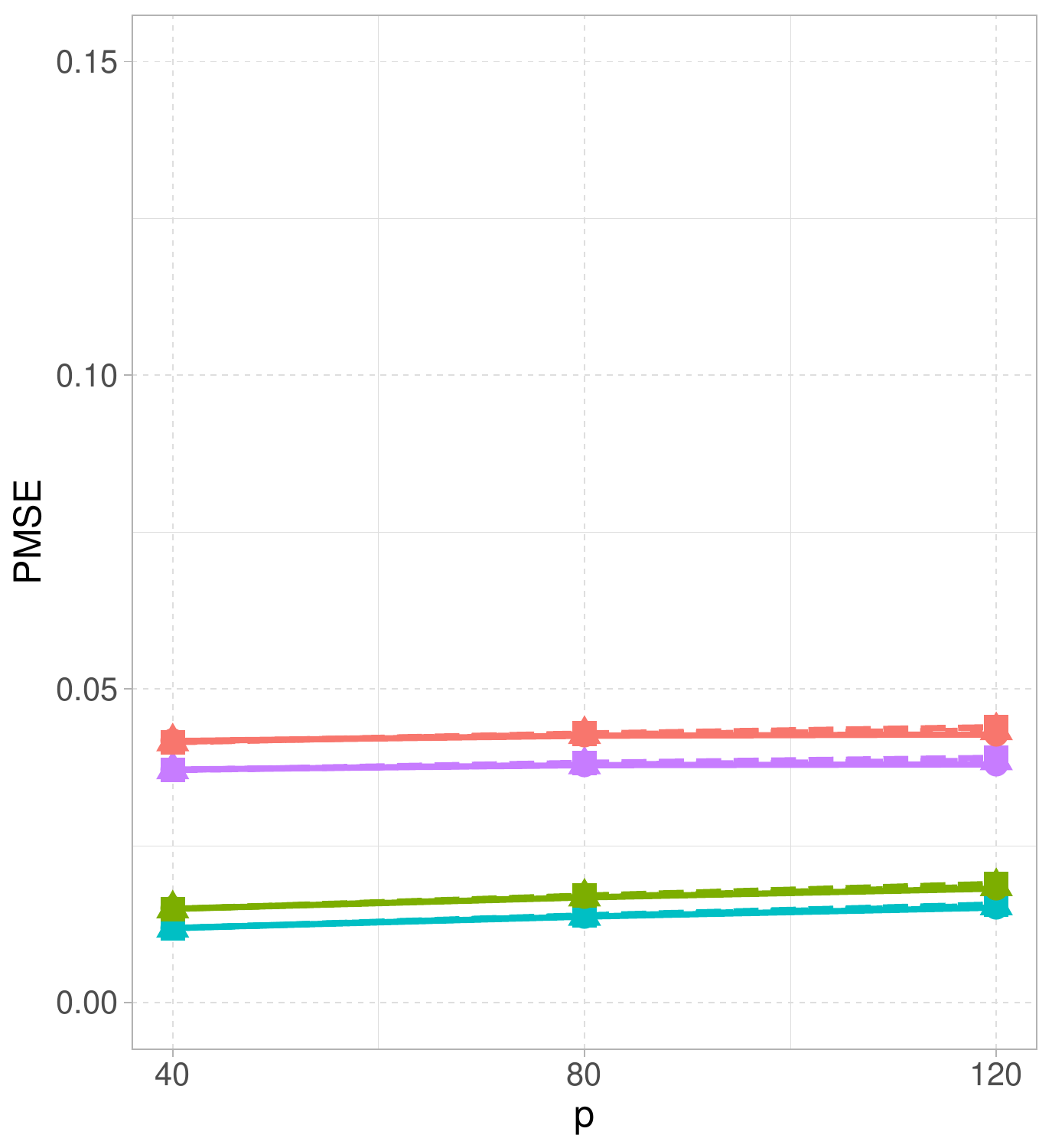}} & 
			  \raisebox{-\totalheight}{\includegraphics[scale=0.58]{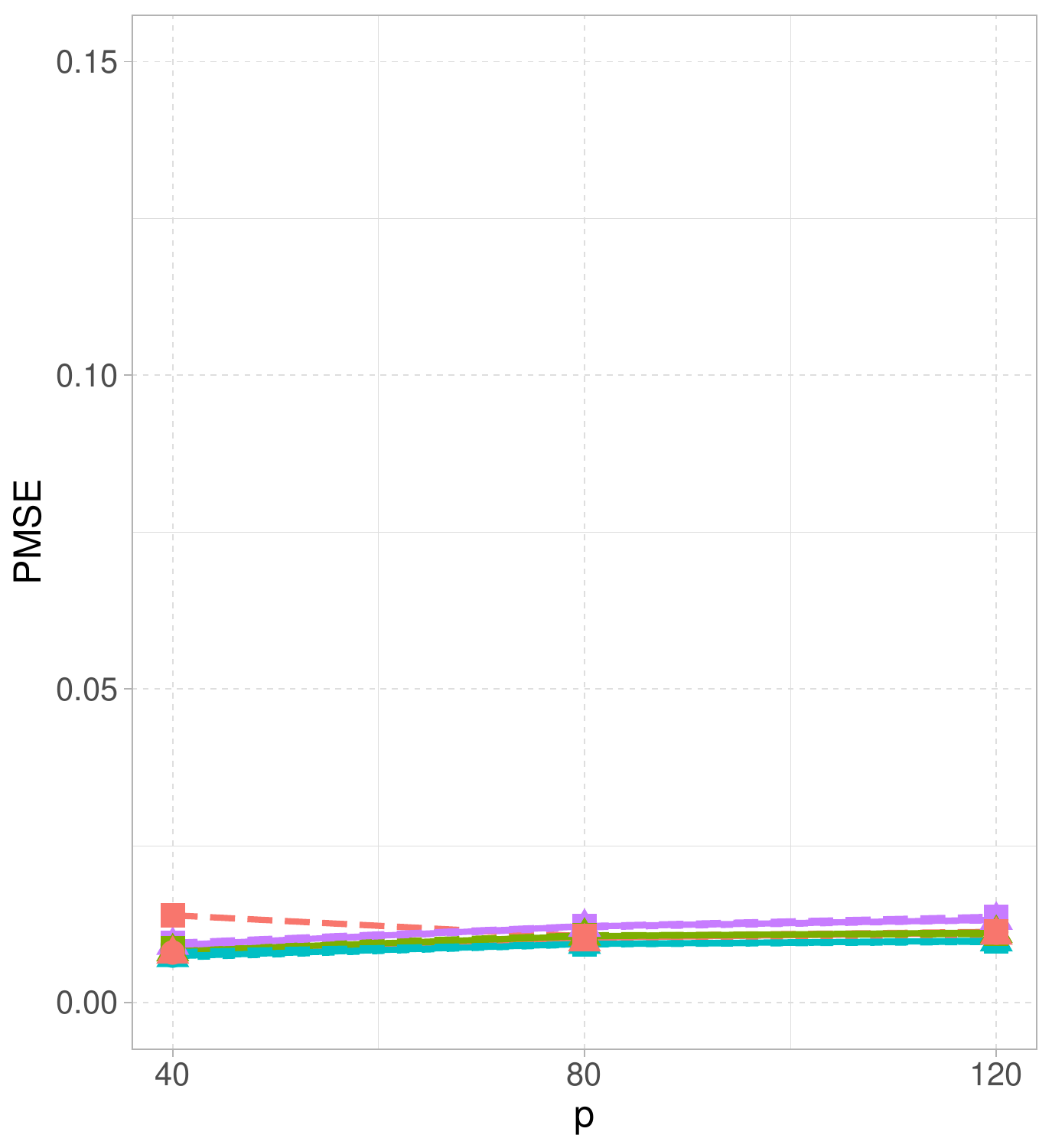}} & 
			  \raisebox{-\totalheight}{\includegraphics[scale=0.58]{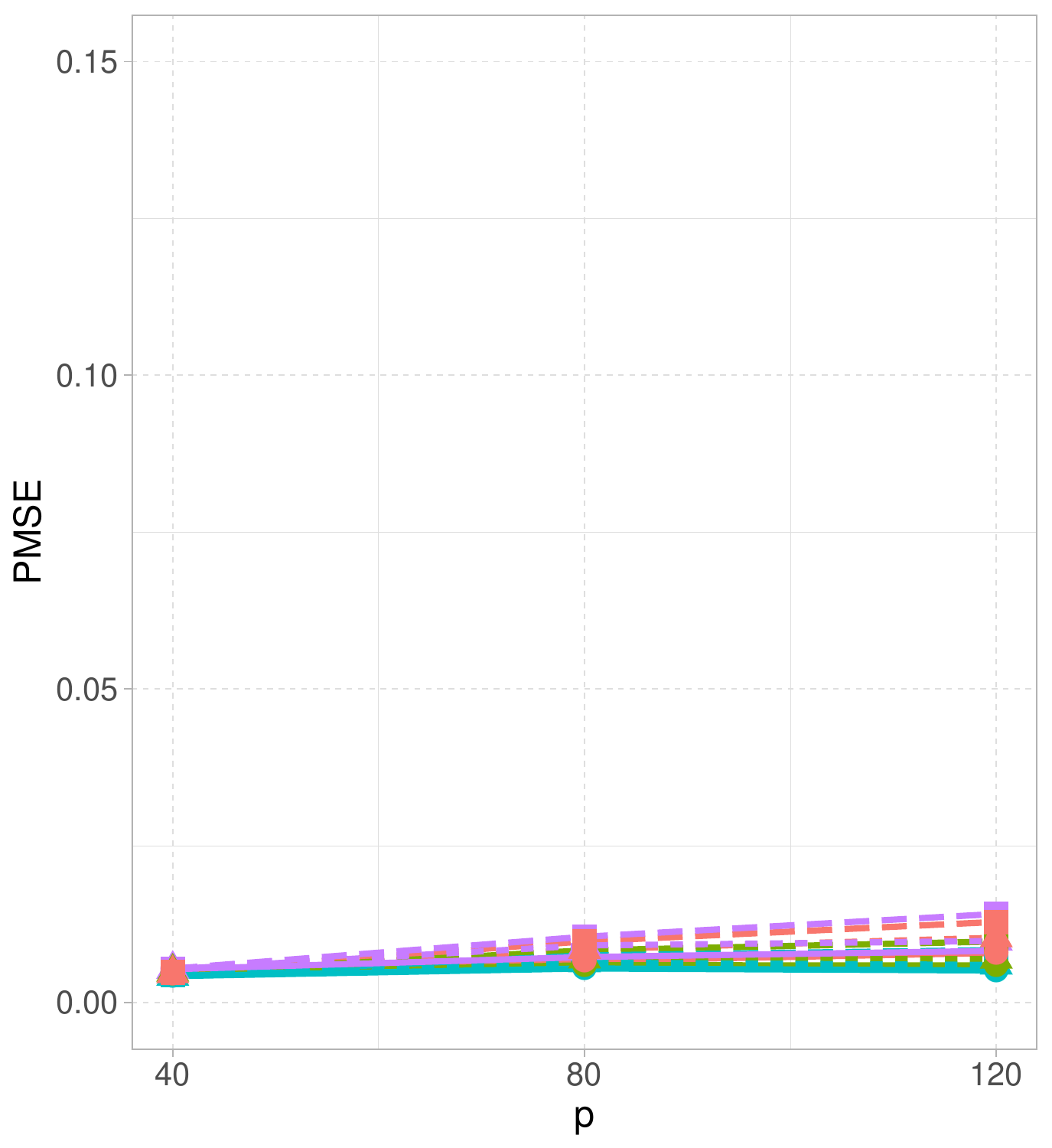}} & 
\begin{itemize}\item[]\item[]\item[]
      \item[$w(t)$]$=\indica_{t\le c_w}$ 
      \end{itemize}
      \\            
		\end{tabular}
		\caption{Probabilities Mean Squared Error for $n = 300$ for scenarios \textbf{C0}, \textbf{CA1} and \textbf{CA2}.}
		\label{fig:ECMP_n_300}
	\end{figure}
	
	\newpage
	
	\begin{figure}[ht!]
		\centering
	\begin{tabular}{ccc p{2.8cm}} 
 LASSO &  Sign & MCP & \\
			  \raisebox{-\totalheight}{\includegraphics[scale=0.58]{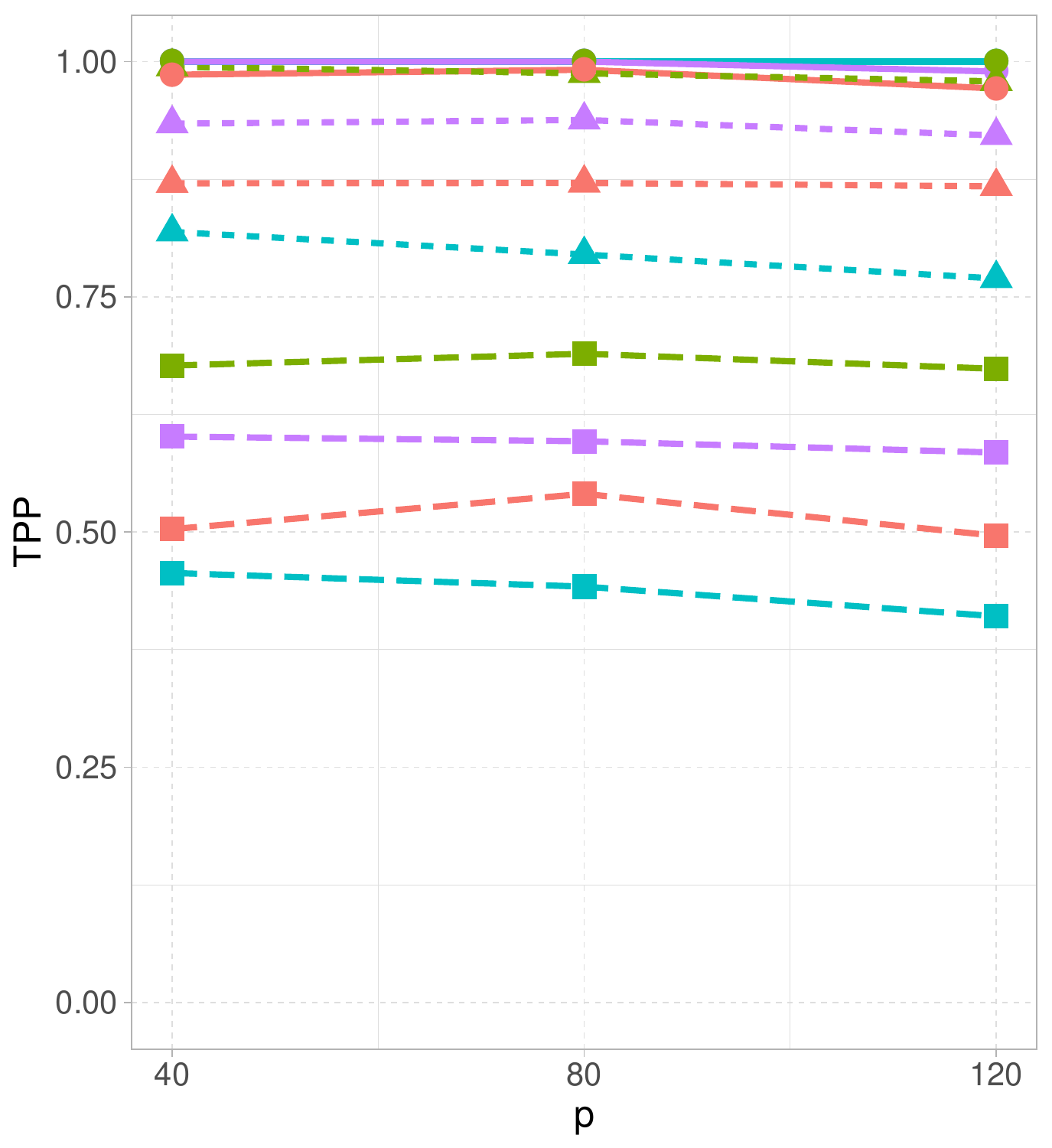}} & 
			  \raisebox{-\totalheight}{\includegraphics[scale=0.58]{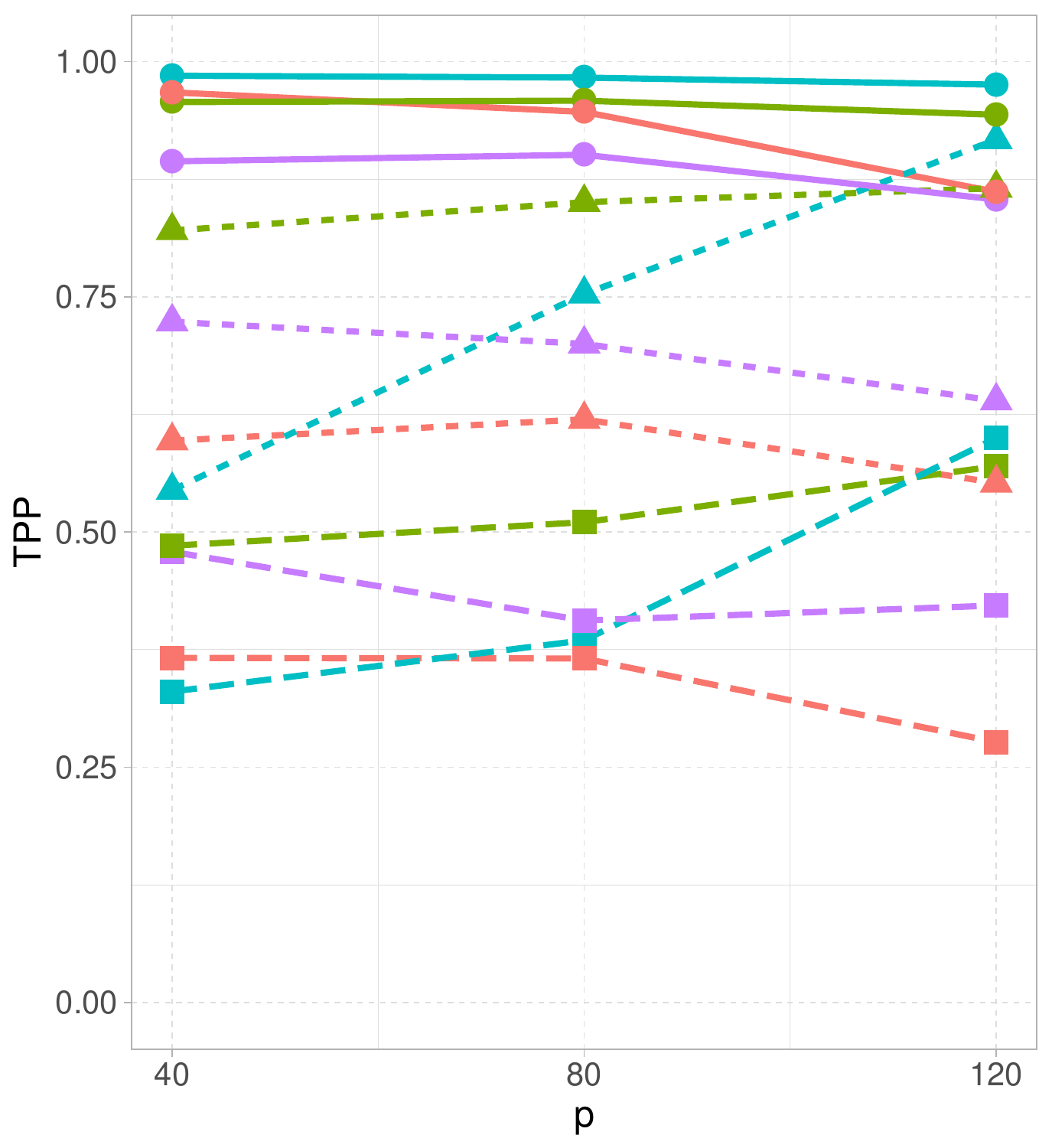}} & 
			  \raisebox{-\totalheight}{\includegraphics[scale=0.58]{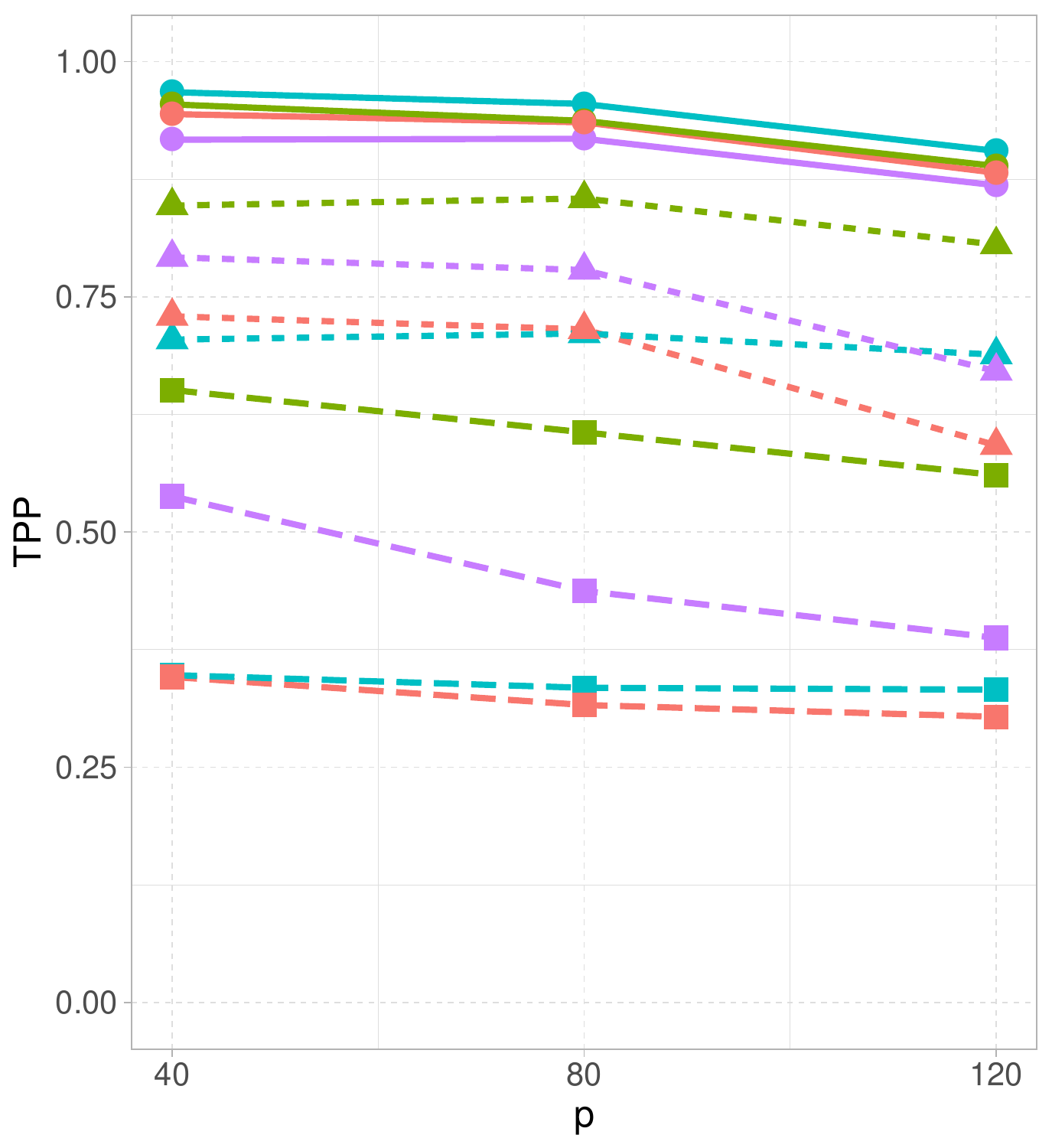}}&  
{\begin{itemize}\item[]\item[]\item[]
      \item[$w\equiv 1$]
      \end{itemize}
      }
      \\
			  \raisebox{-\totalheight}{\includegraphics[scale=0.58]{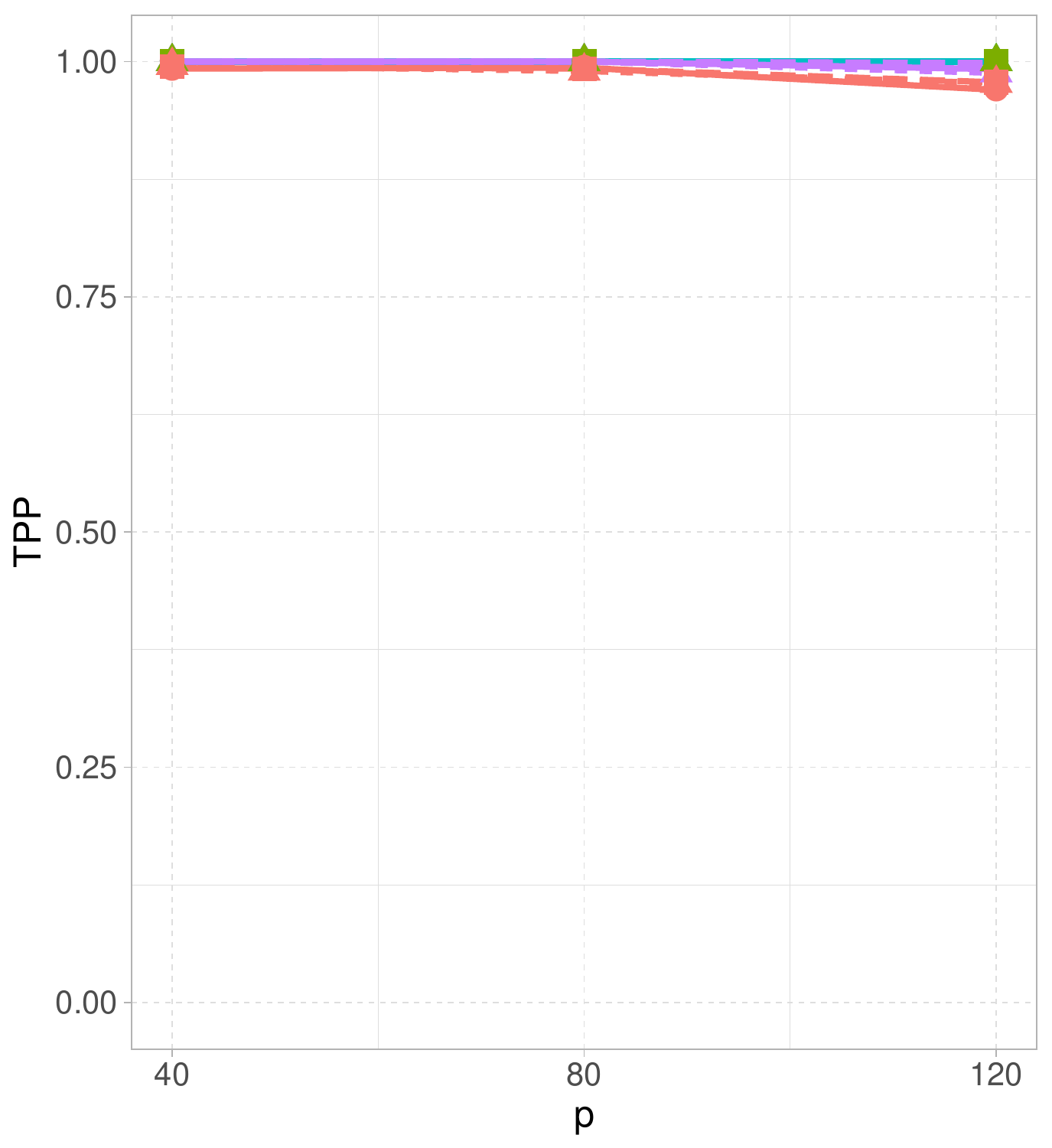}}& 
			  \raisebox{-\totalheight}{\includegraphics[scale=0.58]{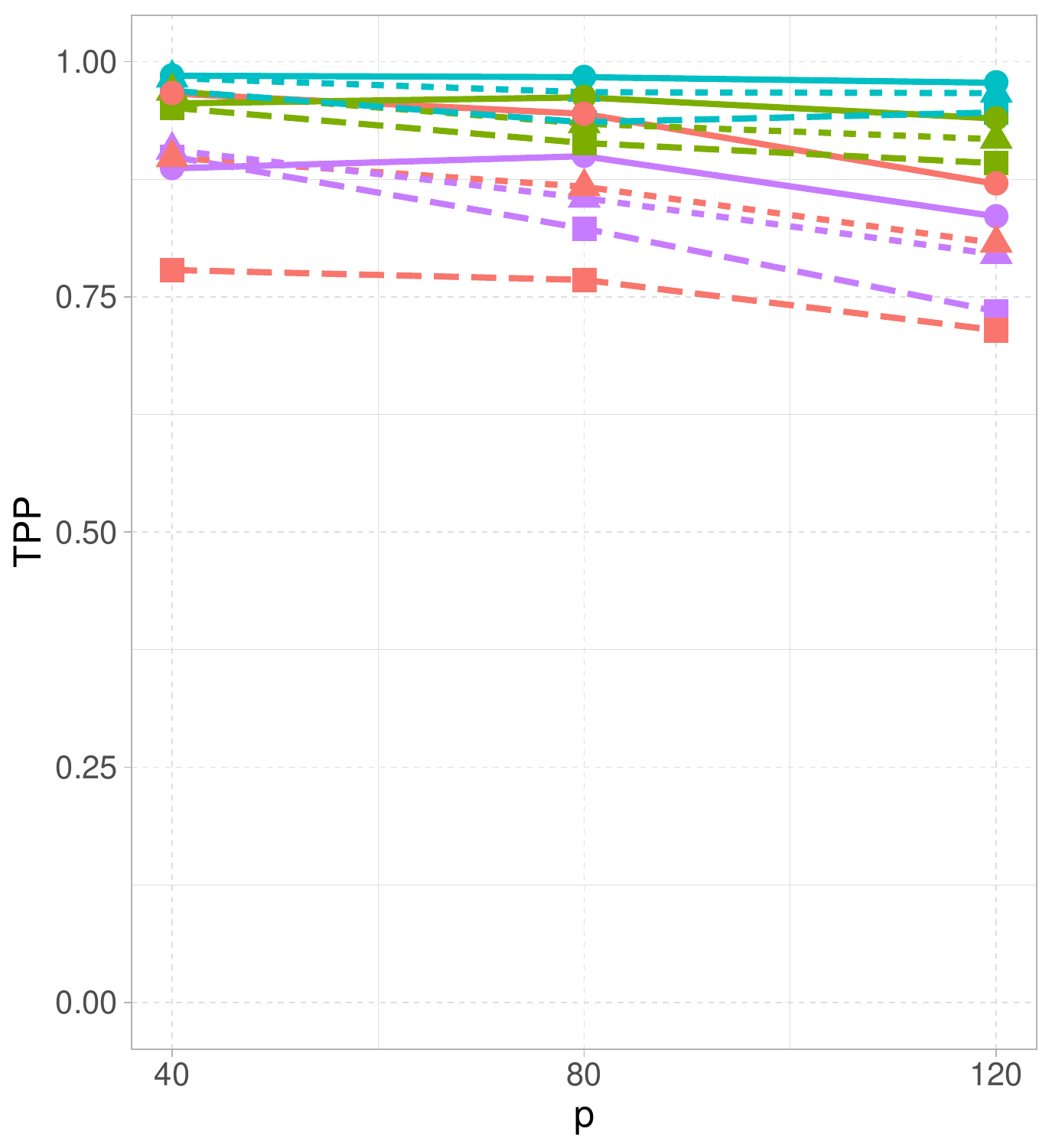}} & 
			  \raisebox{-\totalheight}{\includegraphics[scale=0.58]{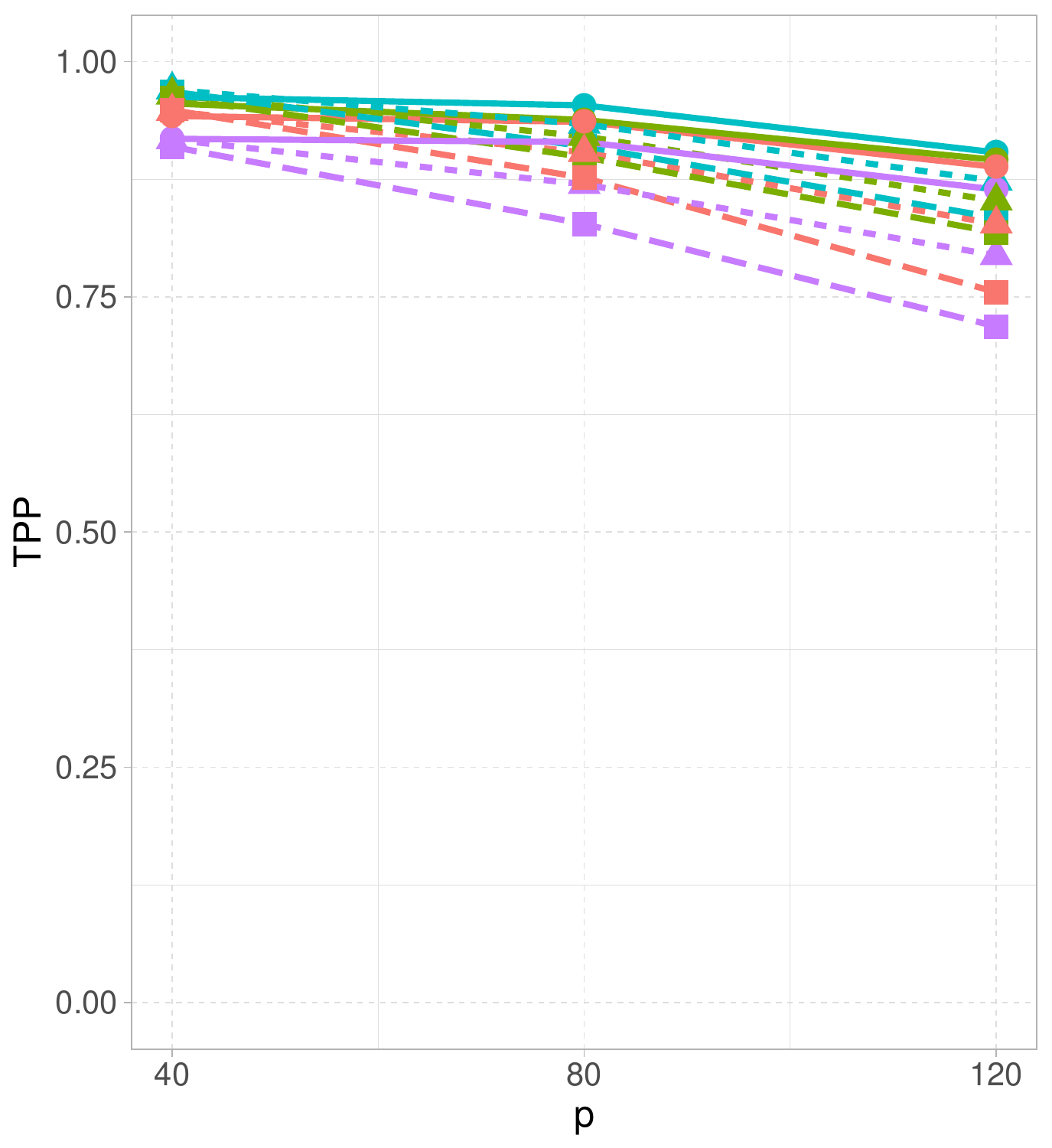}} & 
\begin{itemize}\item[]\item[]\item[]
      \item[$w(t)$]$=\indica_{t\le c_w}$ 
      \end{itemize}
      \\            
		\end{tabular}
		\caption{Correct Classification Relative Proportion for $n = 150$  for scenarios \textbf{C0}, \textbf{CA1} and \textbf{CA2}.}
		\label{fig:PVP_n_150}
	\end{figure}
	
	\newpage
	\begin{figure}[ht!]
		\centering
\begin{tabular}{ccc p{2.8cm}} 
 LASSO &  Sign & MCP & \\
			  \raisebox{-\totalheight}{\includegraphics[scale=0.58]{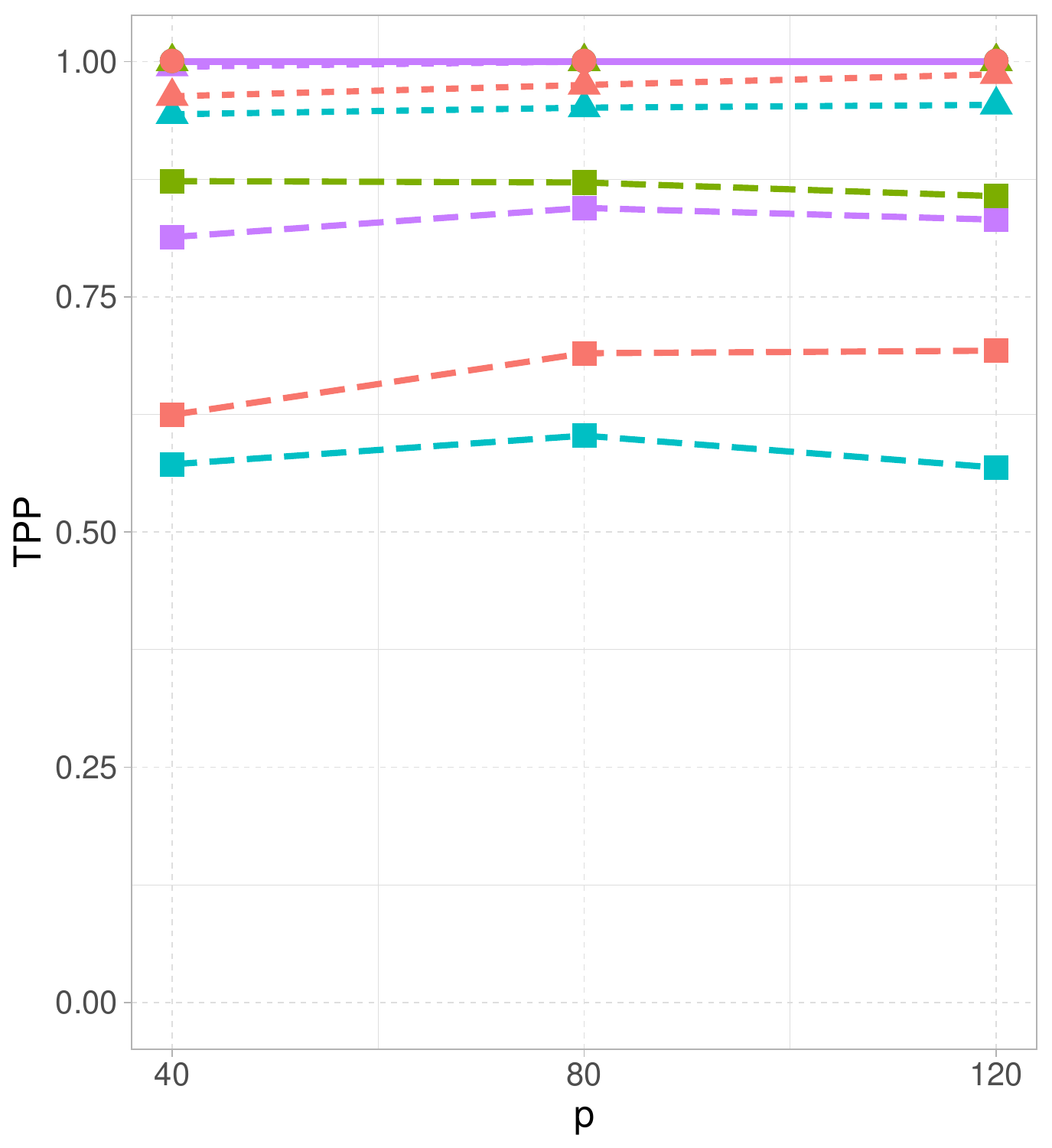}} & 
			  \raisebox{-\totalheight}{\includegraphics[scale=0.58]{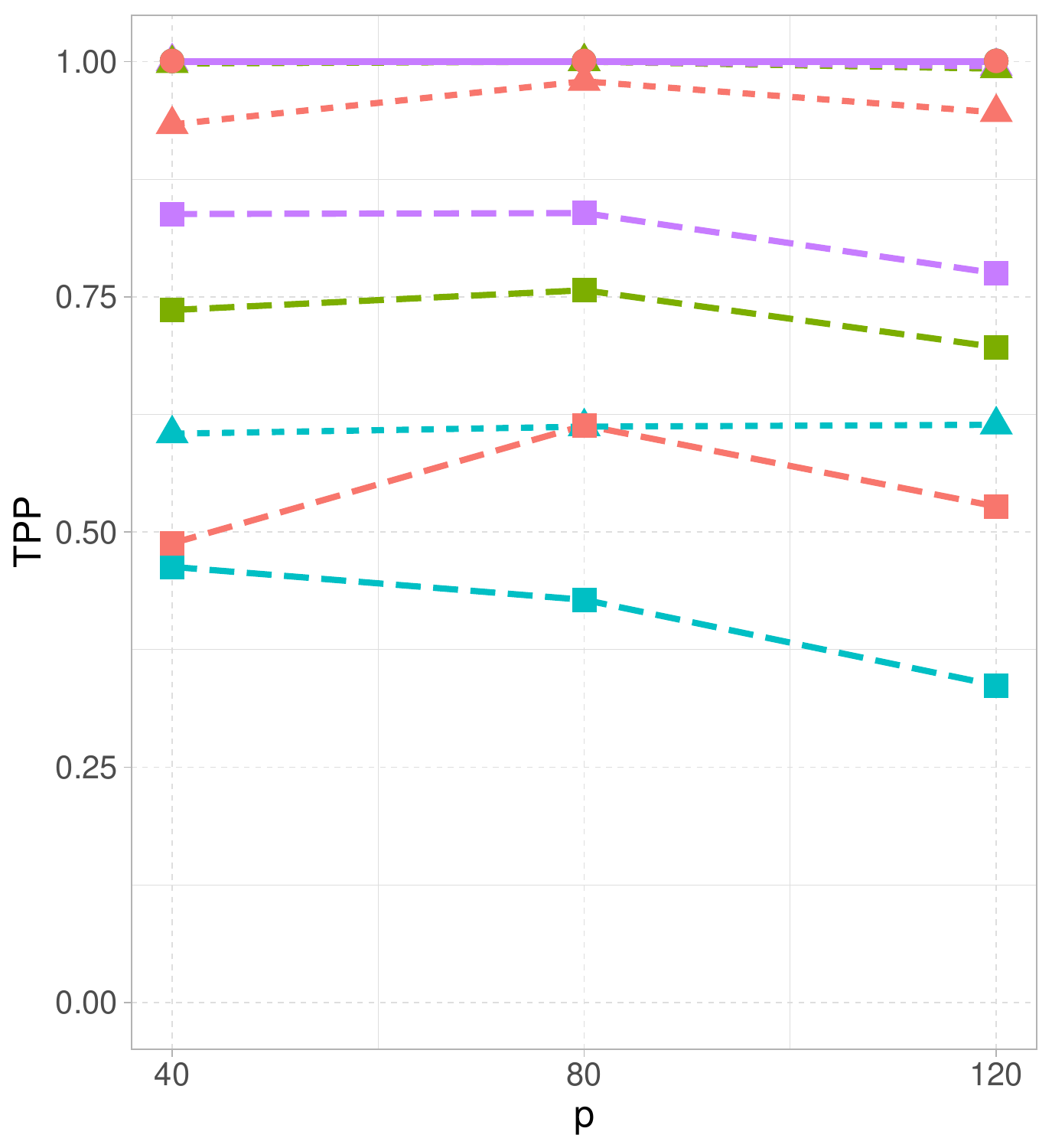}} & 
			  \raisebox{-\totalheight}{\includegraphics[scale=0.58]{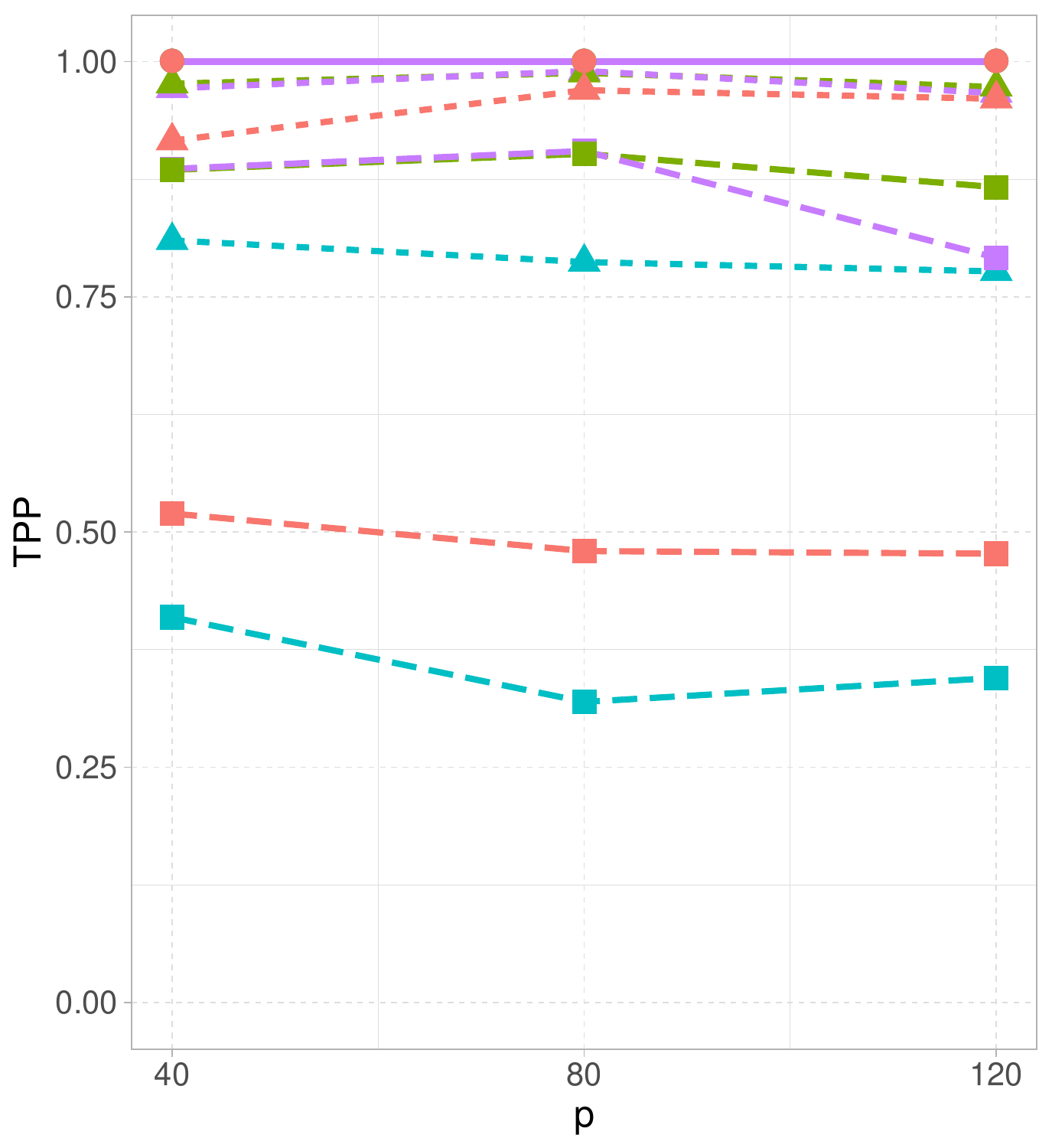}}&  
{\begin{itemize}\item[]\item[]\item[]
      \item[$w\equiv 1$]
      \end{itemize}
      }
      \\
			  \raisebox{-\totalheight}{\includegraphics[scale=0.58]{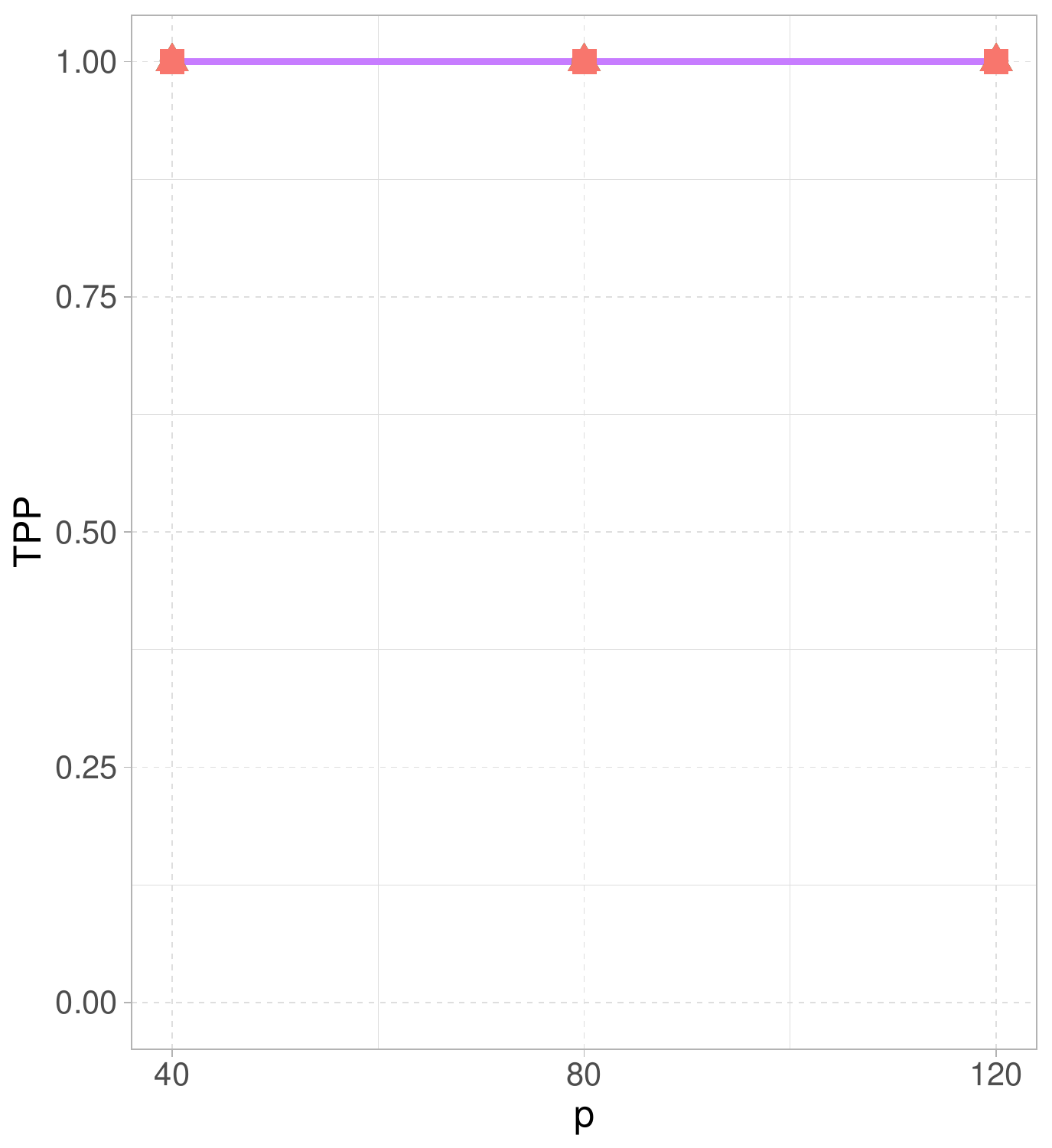}} & 
			  \raisebox{-\totalheight}{\includegraphics[scale=0.58]{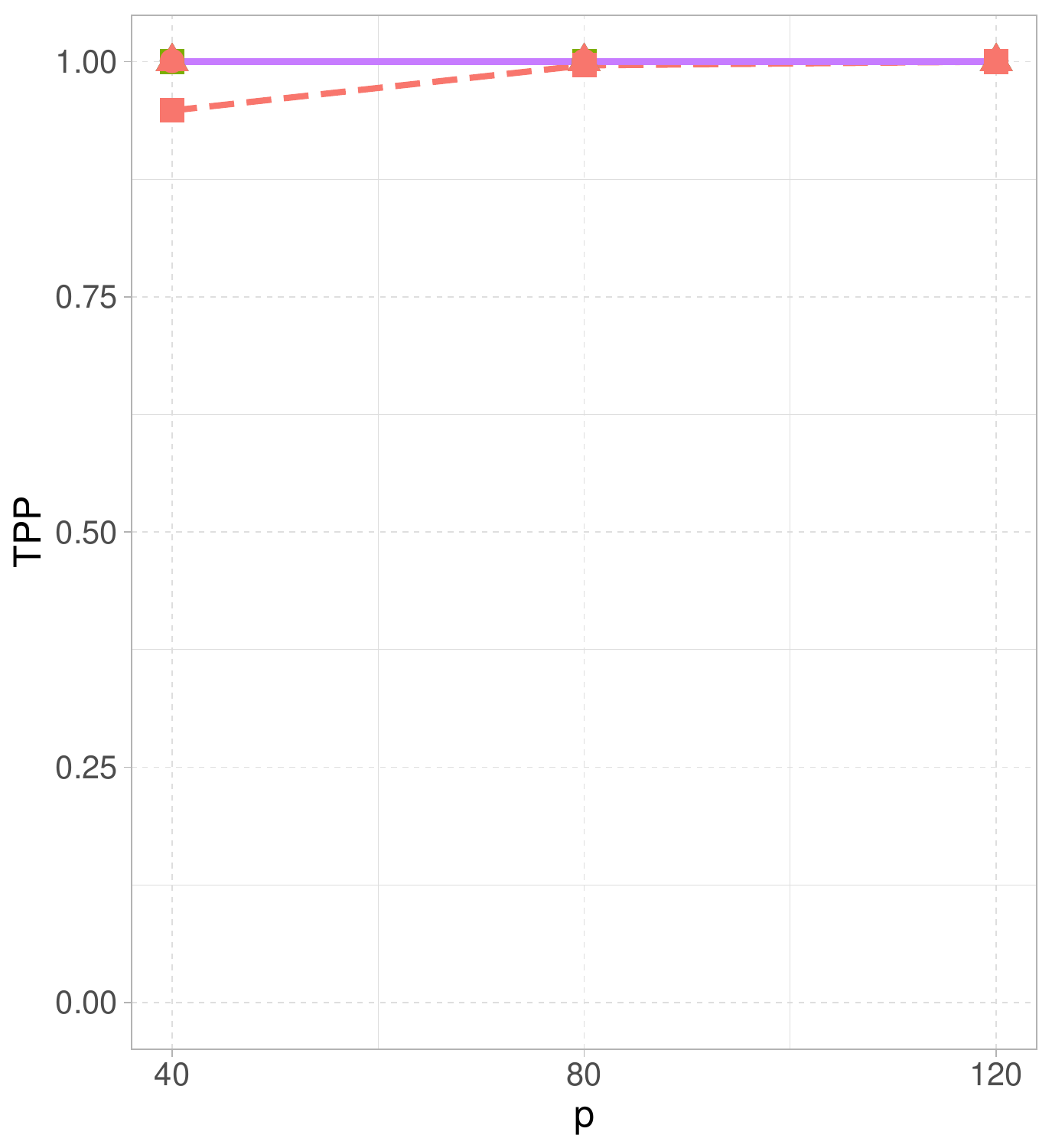}} & 
			  \raisebox{-\totalheight}{\includegraphics[scale=0.58]{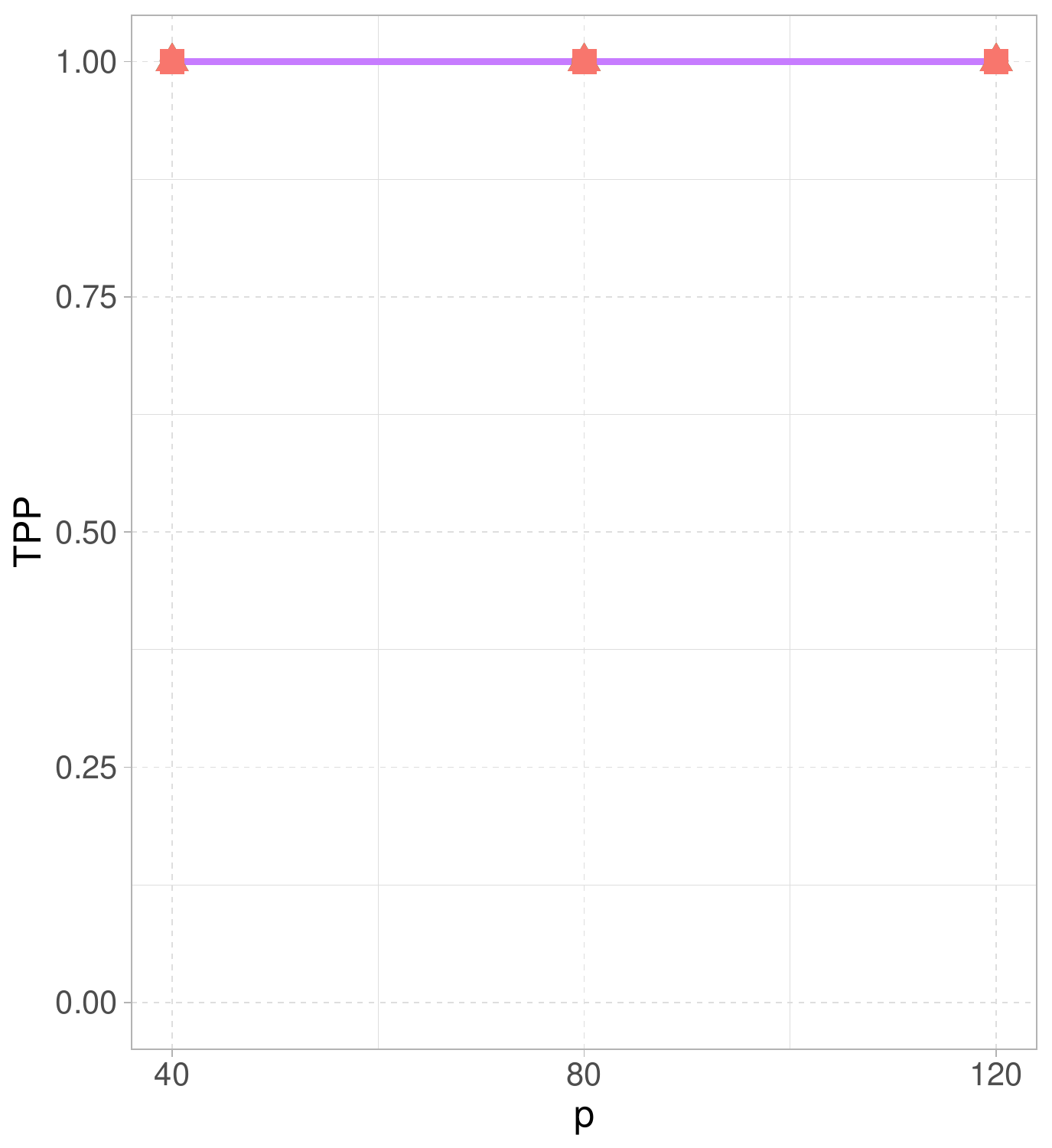}} & 
\begin{itemize}\item[]\item[]\item[]
      \item[$w(t)$]$=\indica_{t\le c_w}$ 
      \end{itemize}
      \\            
		\end{tabular}
		\caption{Correct Classification Relative Proportion for $n = 300$ for scenarios \textbf{C0}, \textbf{CA1} and \textbf{CA2}.}
		\label{fig:PVP_n_300}
	\end{figure}
	
	\newpage

\begin{figure}[ht!]
        \centering
\begin{tabular}{ccc p{2.8cm}} 
 LASSO &  Sign & MCP & \\
 \raisebox{-\totalheight}{\includegraphics[scale=0.58]{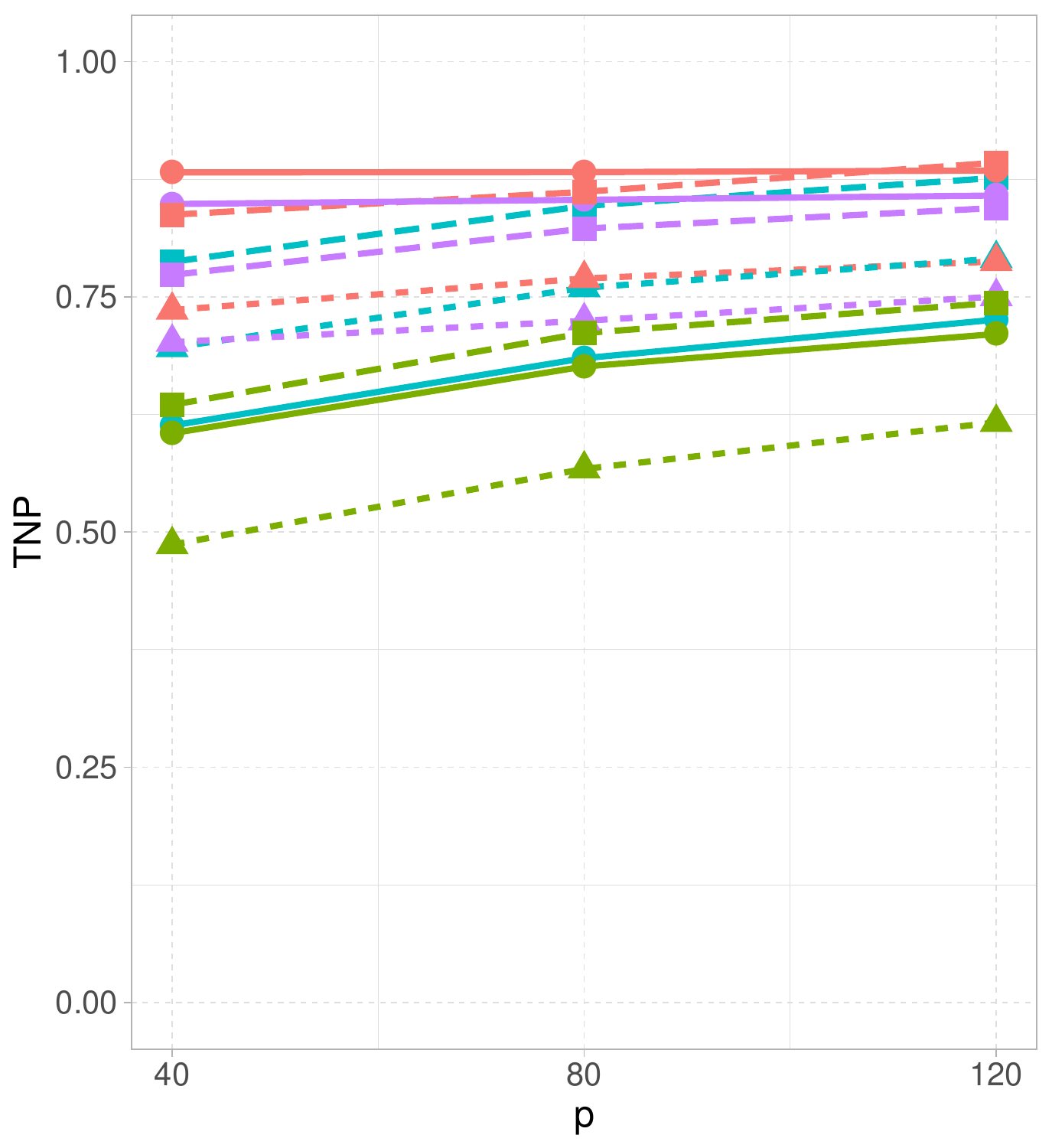}} & 
 \raisebox{-\totalheight}{\includegraphics[scale=0.58]{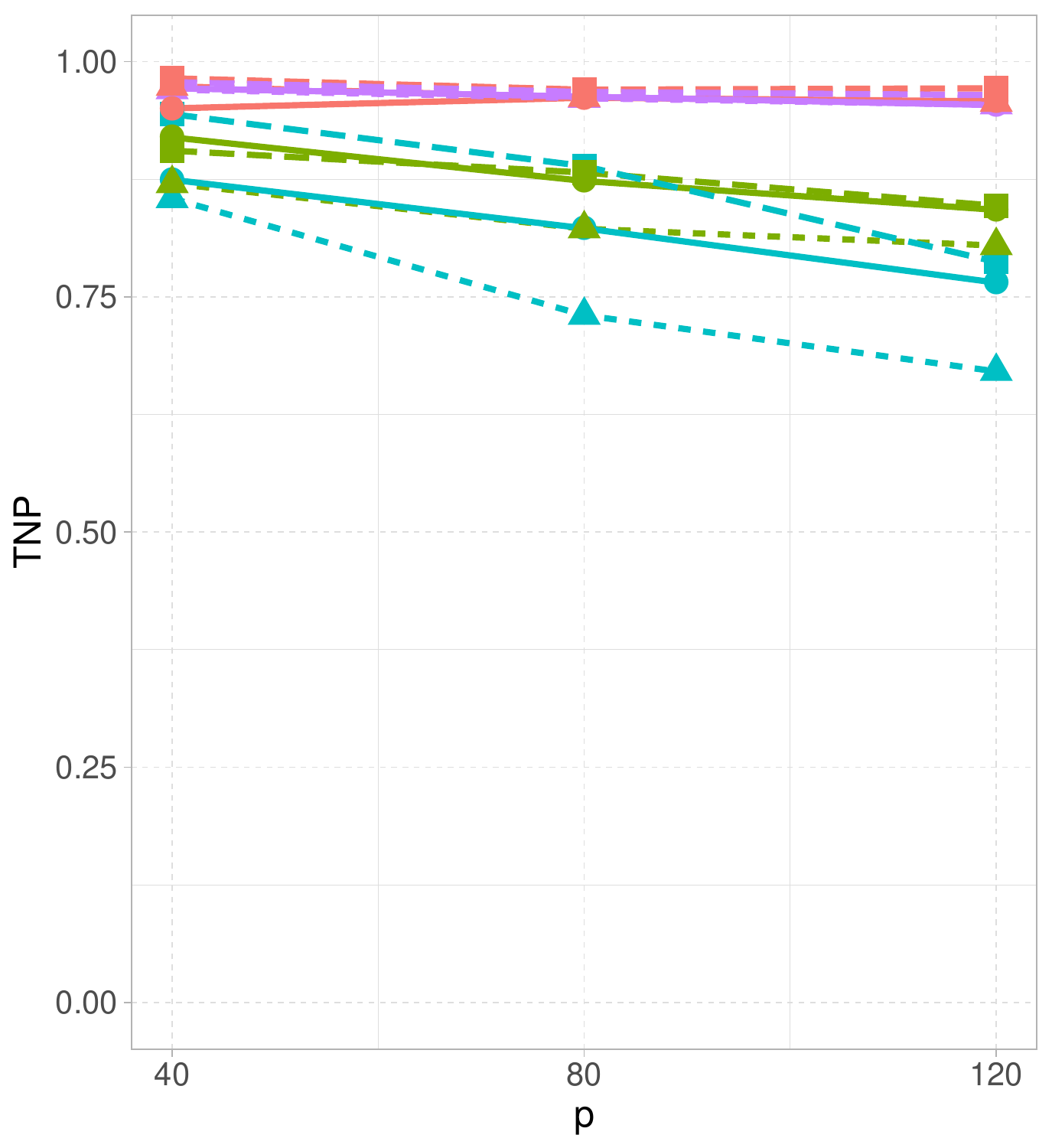}} & 
 \raisebox{-\totalheight}{\includegraphics[scale=0.58]{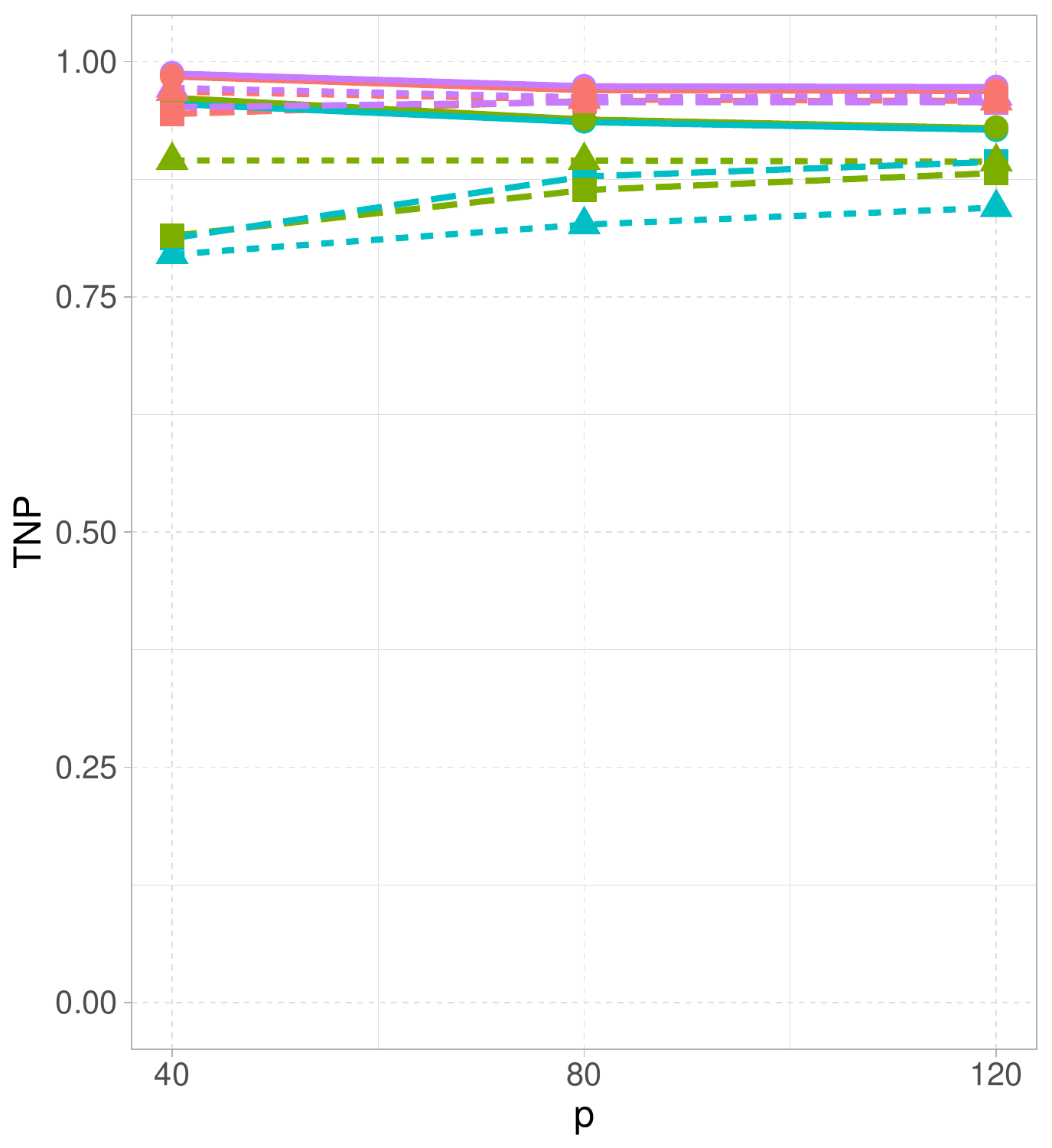}}  &  
{\begin{itemize}\item[]\item[]\item[]
      \item[$w\equiv 1$]
      \end{itemize}
      }
      \\

 \raisebox{-\totalheight}{\includegraphics[scale=0.58]{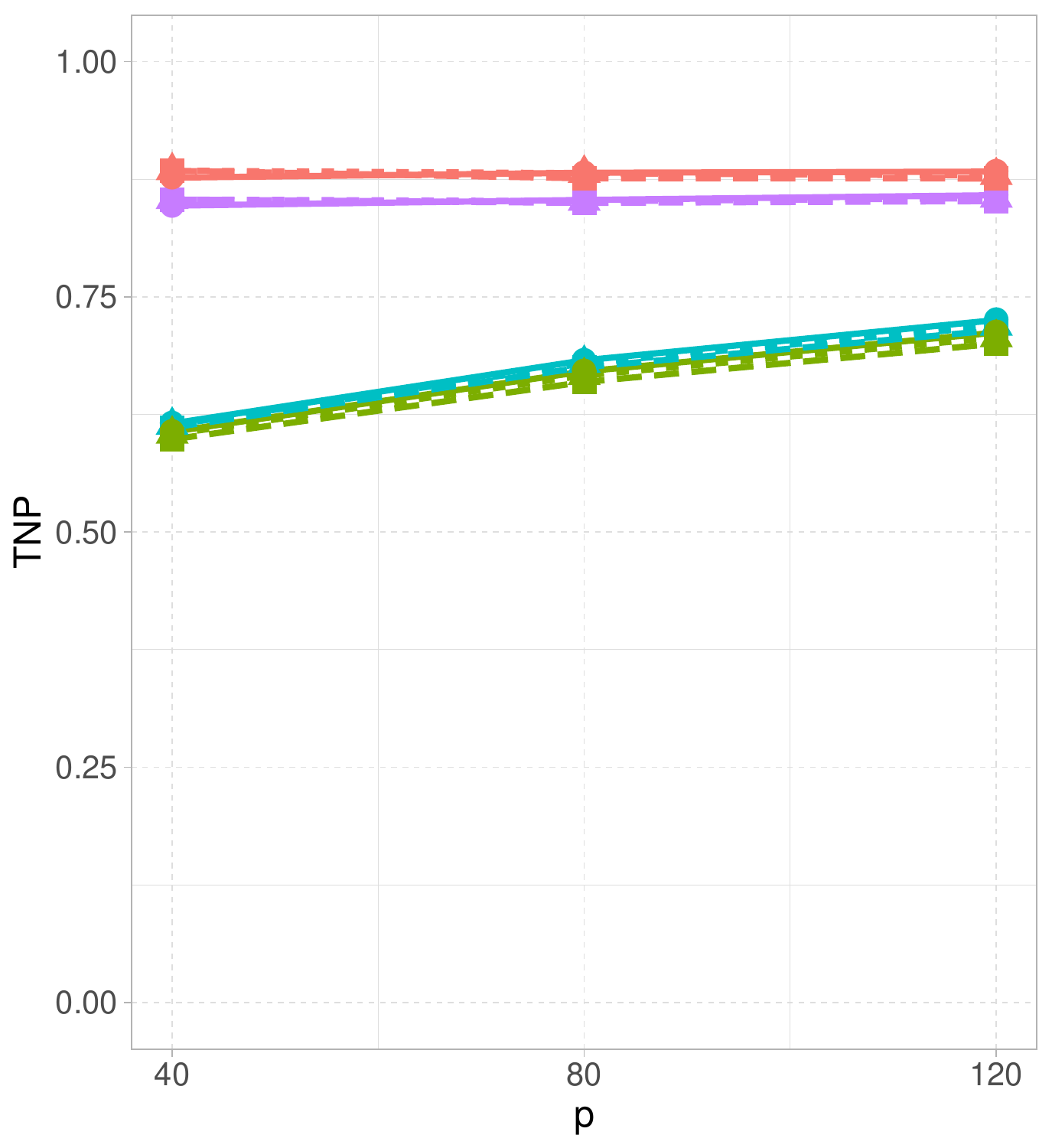}} & 
 \raisebox{-\totalheight}{\includegraphics[scale=0.58]{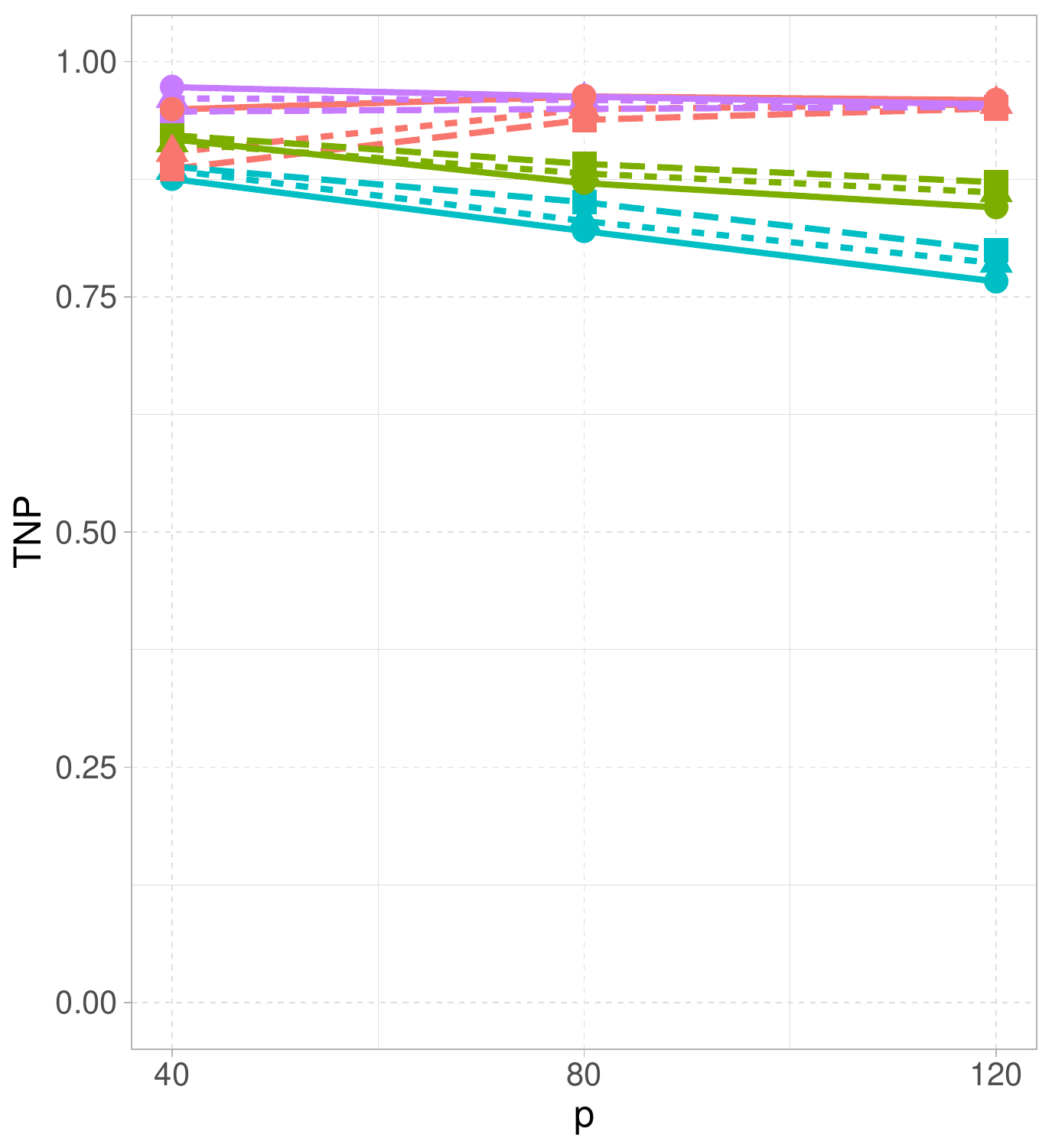}} & 
 \raisebox{-\totalheight}{\includegraphics[scale=0.58]{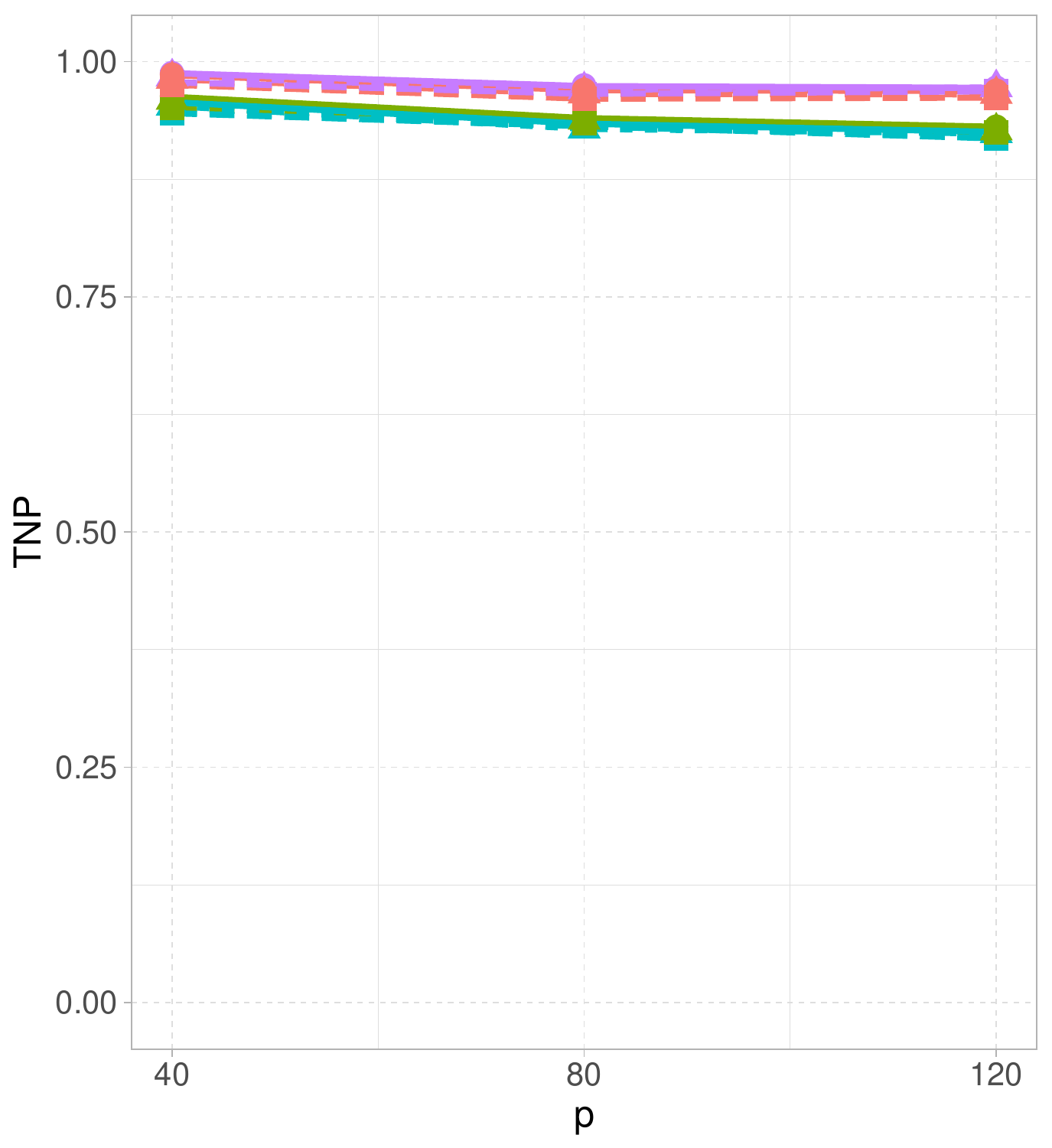}} & 
\begin{itemize}\item[]\item[]\item[]
      \item[$w(t)$]$=\indica_{t\le c_w}$ 
      \end{itemize}
 \\                                    
 		\end{tabular}
		\caption{True Null Proportion for $n = 150$ for scenarios \textbf{C0}, \textbf{CA1} and \textbf{CA2}.}
		\label{fig:PVN_n_150}
	\end{figure}
	
	\newpage
	\begin{figure}[ht!]
		\centering
	\begin{tabular}{ccc p{2.8cm}} 
 LASSO &  Sign & MCP & \\		
			  \raisebox{-\totalheight}{\includegraphics[scale=0.58]{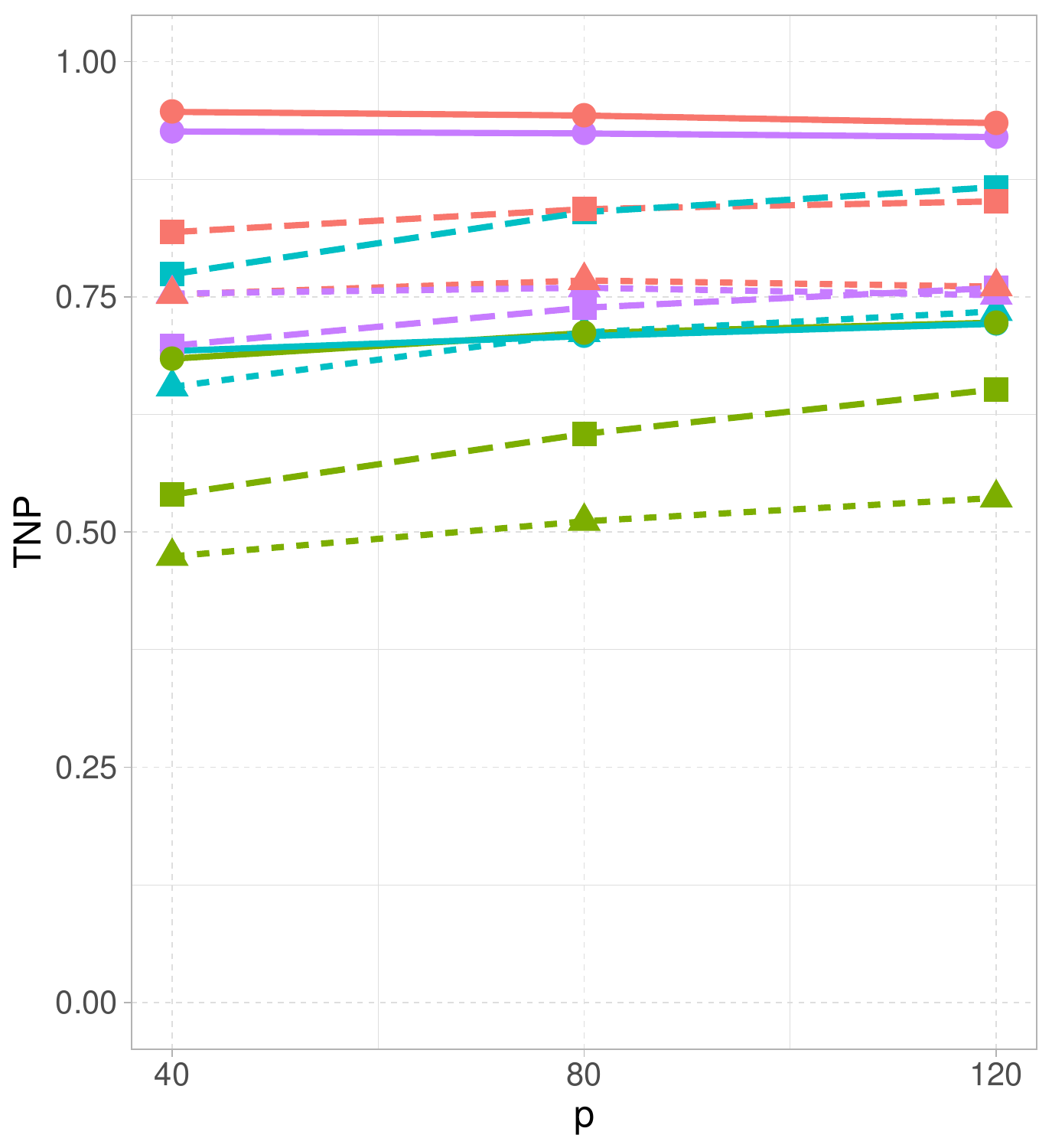}} & 
			  \raisebox{-\totalheight}{\includegraphics[scale=0.58]{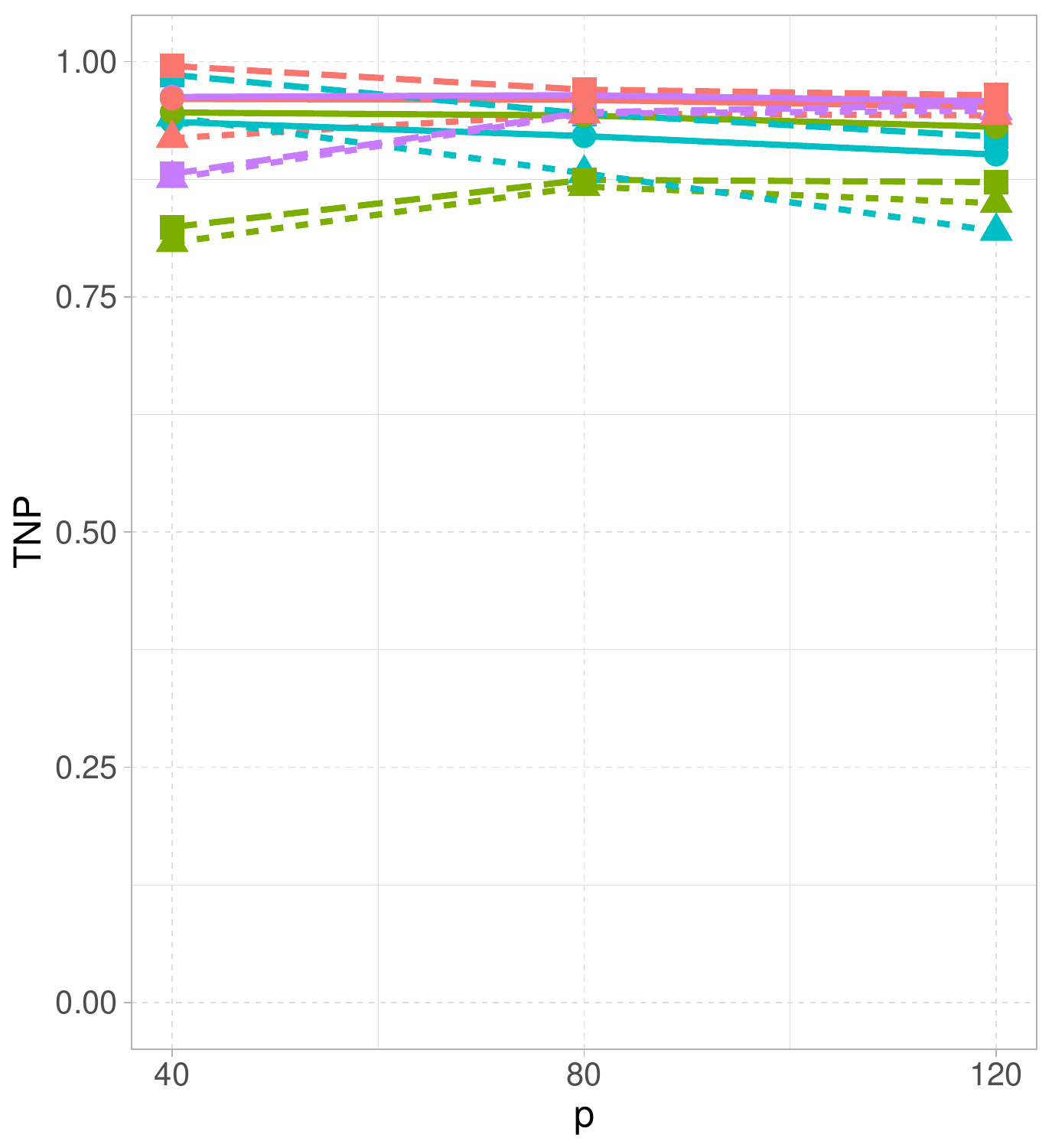}} & 
			  \raisebox{-\totalheight}{\includegraphics[scale=0.58]{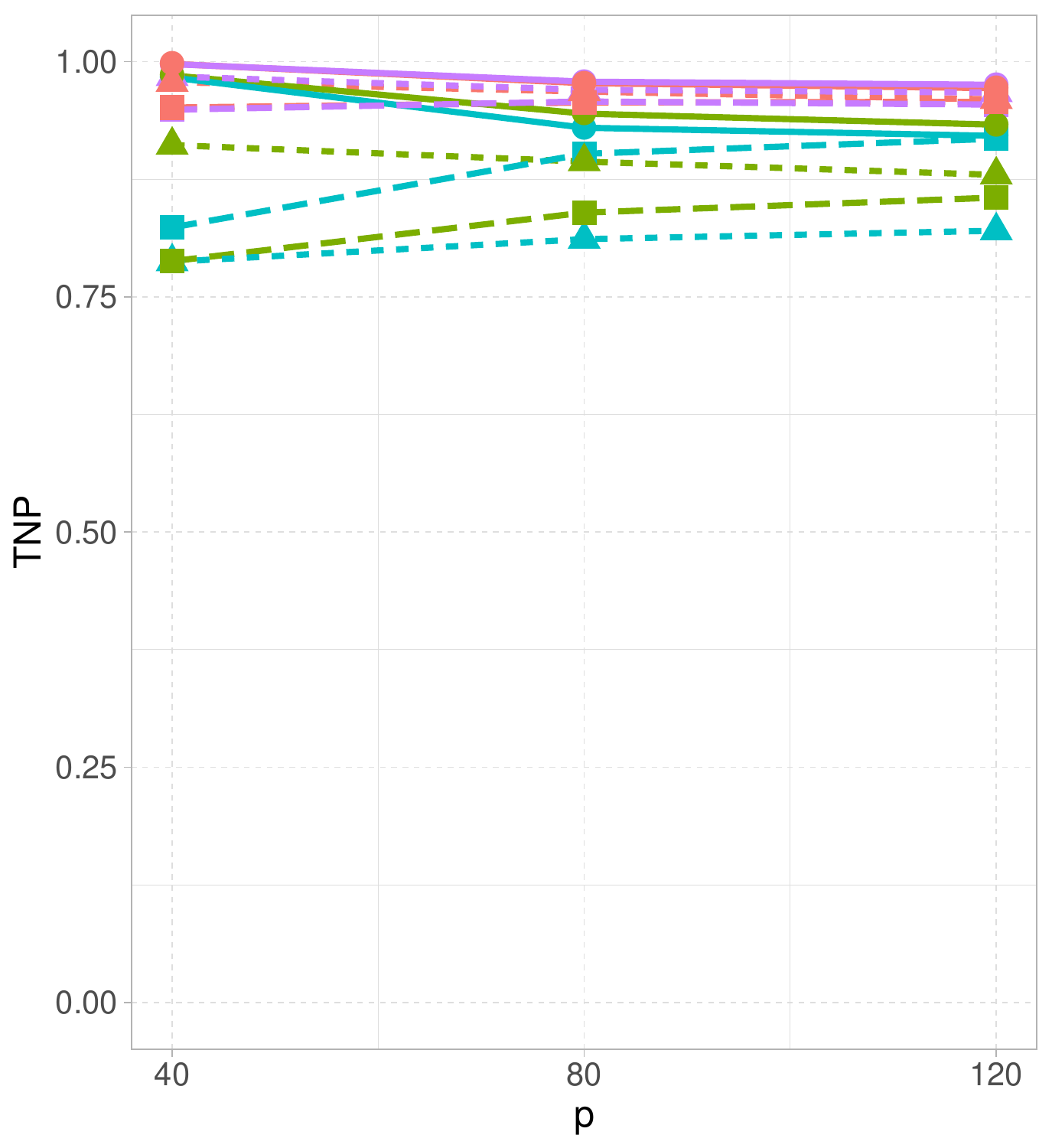}}&  
{\begin{itemize}\item[]\item[]\item[]
      \item[$w\equiv 1$]
      \end{itemize}
      }
      \\
			  \raisebox{-\totalheight}{\includegraphics[scale=0.58]{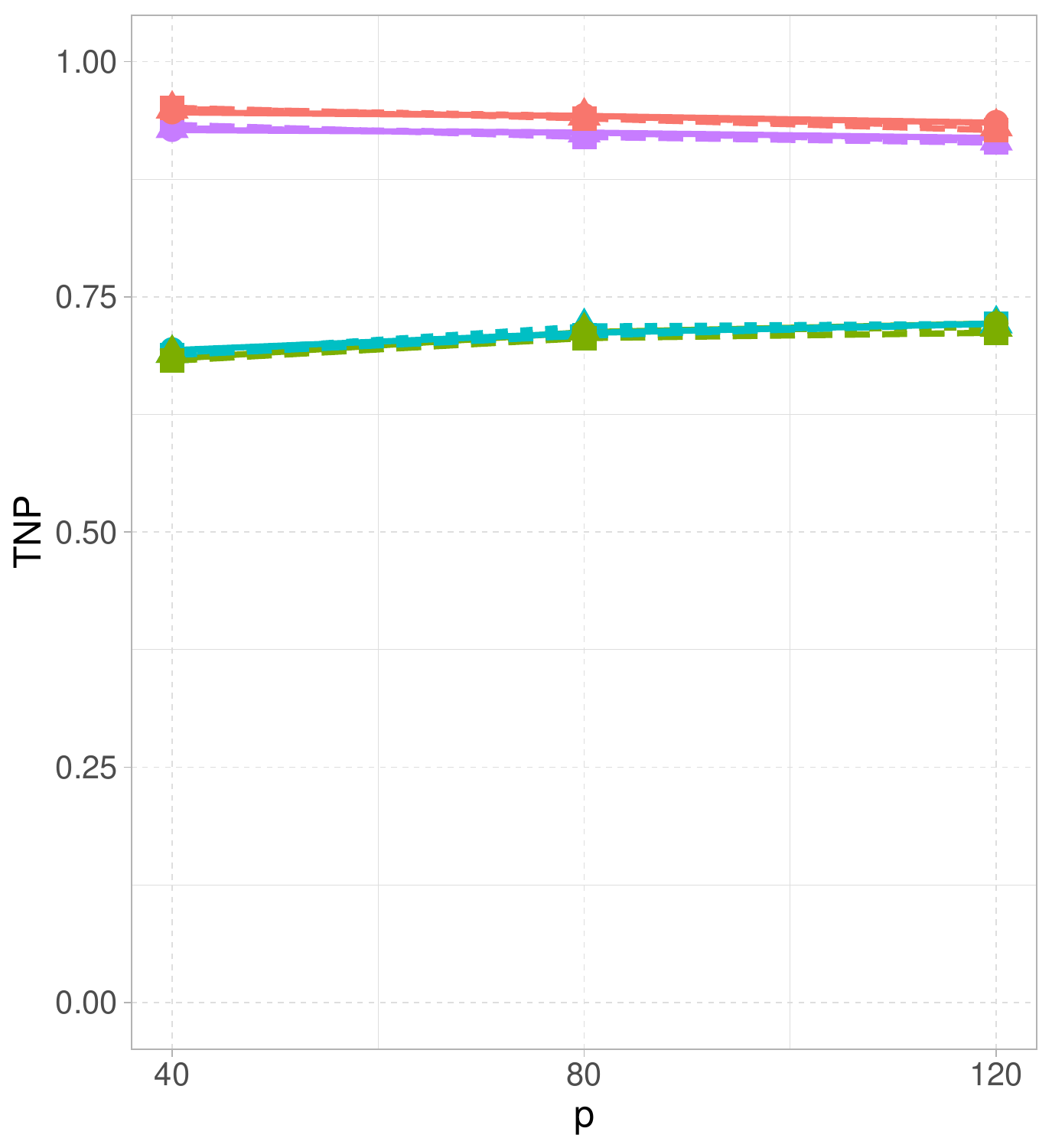}}& 
			  \raisebox{-\totalheight}{\includegraphics[scale=0.58]{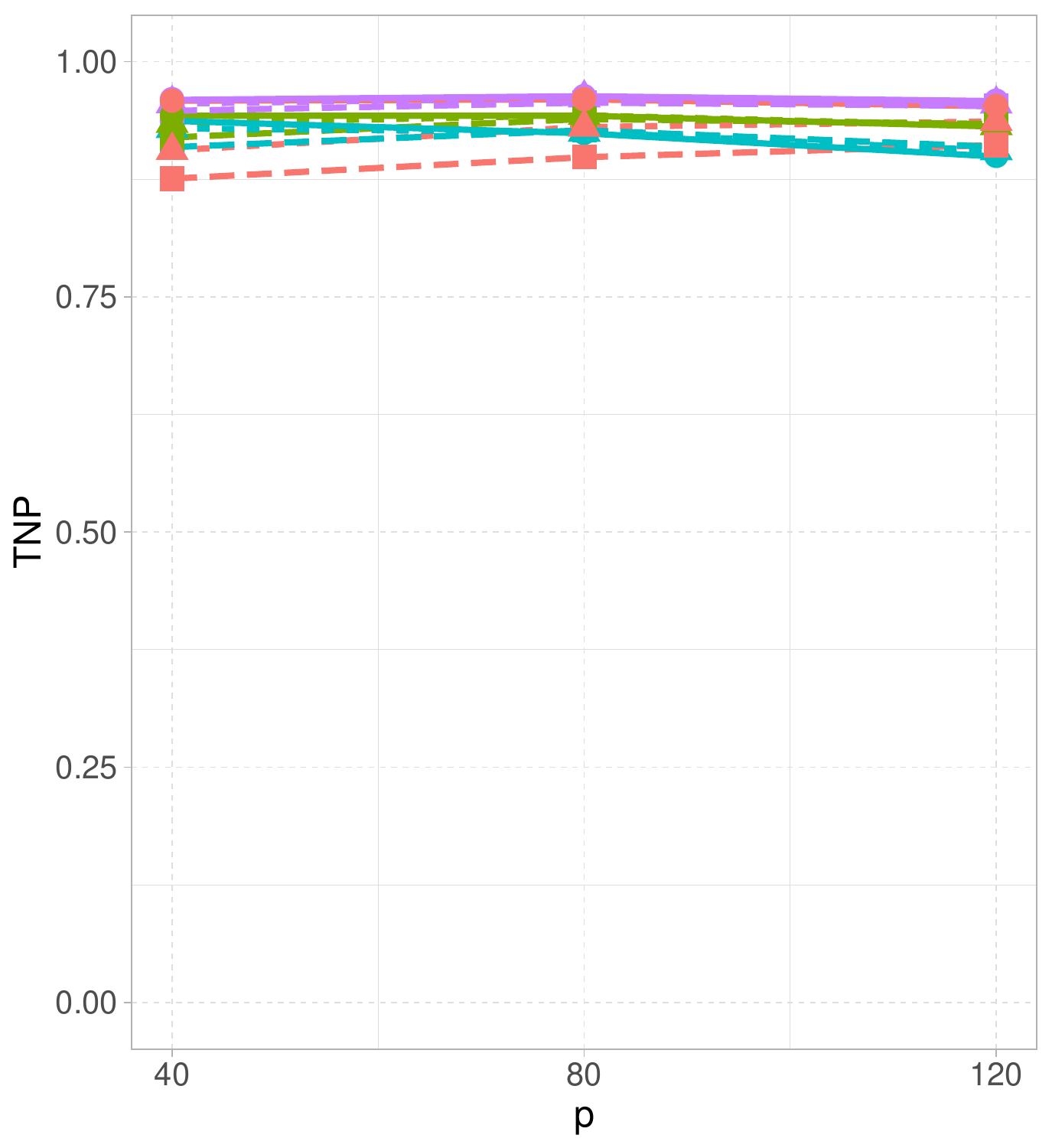}} & 
			  \raisebox{-\totalheight}{\includegraphics[scale=0.58]{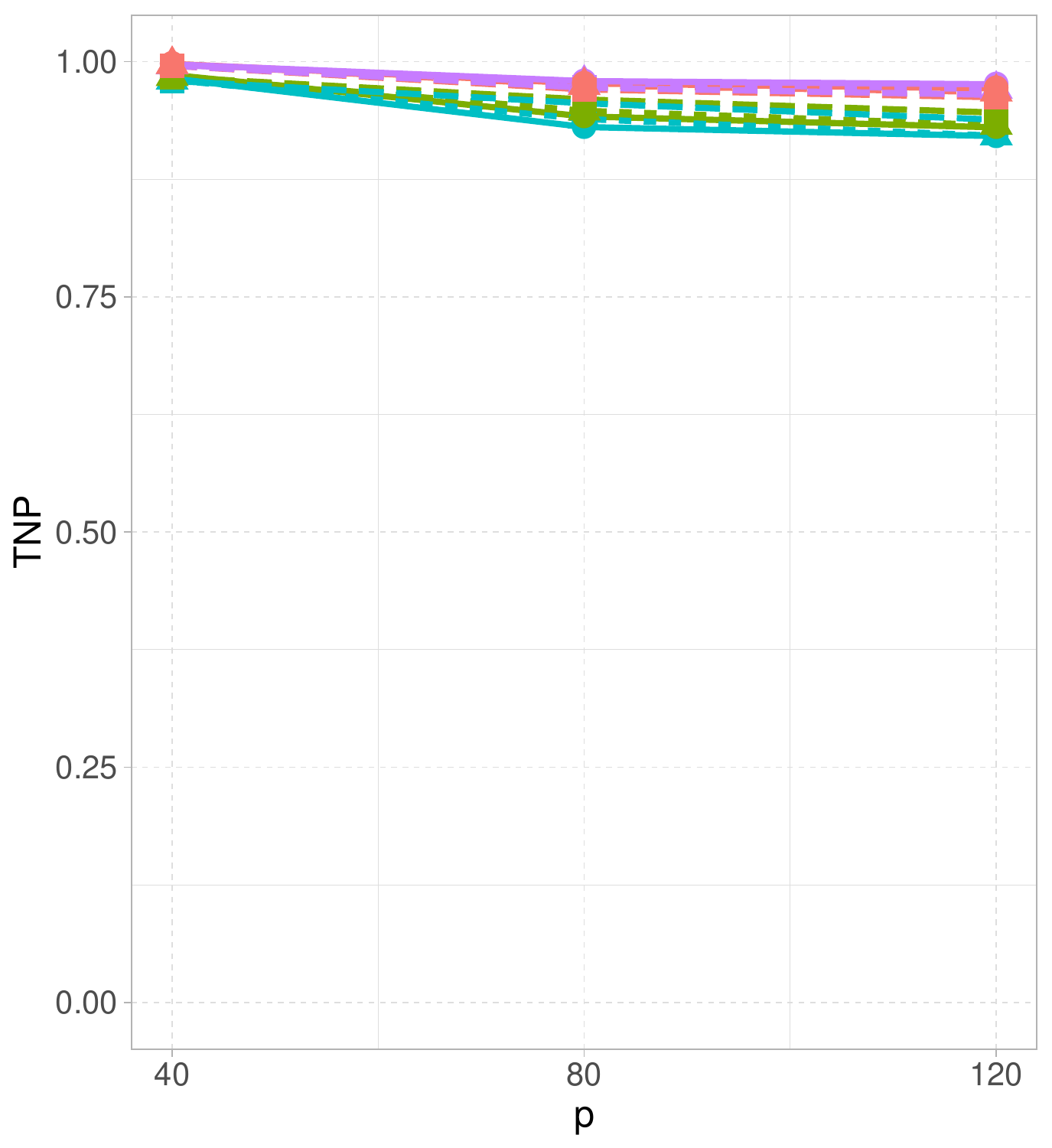}} & 
\begin{itemize}\item[]\item[]\item[]
      \item[$w(t)$]$=\indica_{t\le c_w}$ 
      \end{itemize}
      \\            
		\end{tabular}
		\caption{True Null Proportion for $n = 300$ for scenarios \textbf{C0}, \textbf{CA1} and \textbf{CA2}.}
		\label{fig:PVN_n_300}
	\end{figure}
	
	
	\begin{figure}[ht!]
		\centering
		\begin{tabular}{cc}	
$(n,p)=(150,40)$ & $(n,p)=(150,80)$\\		
			\includegraphics[scale=0.68]{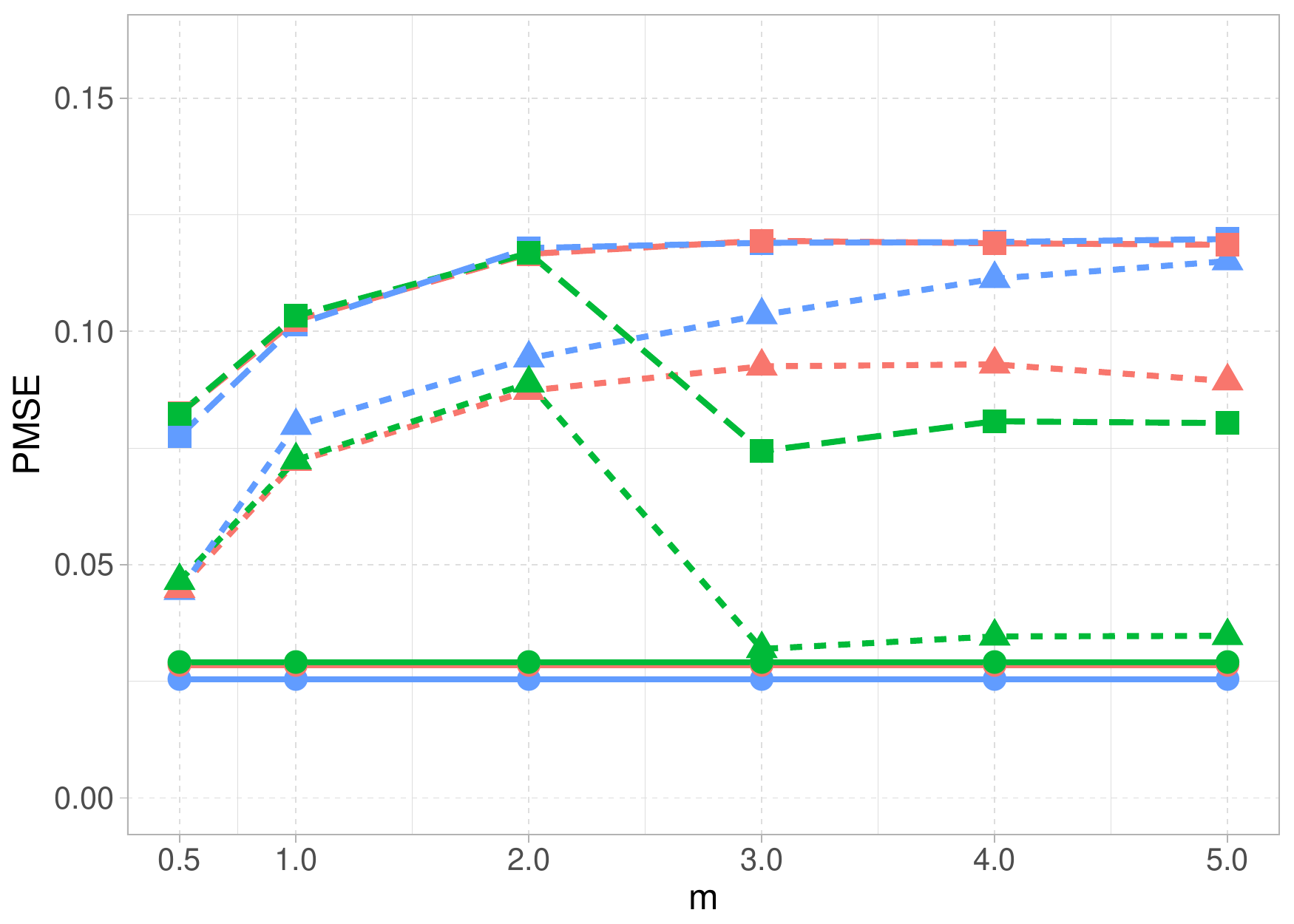} & 
			\includegraphics[scale=0.68]{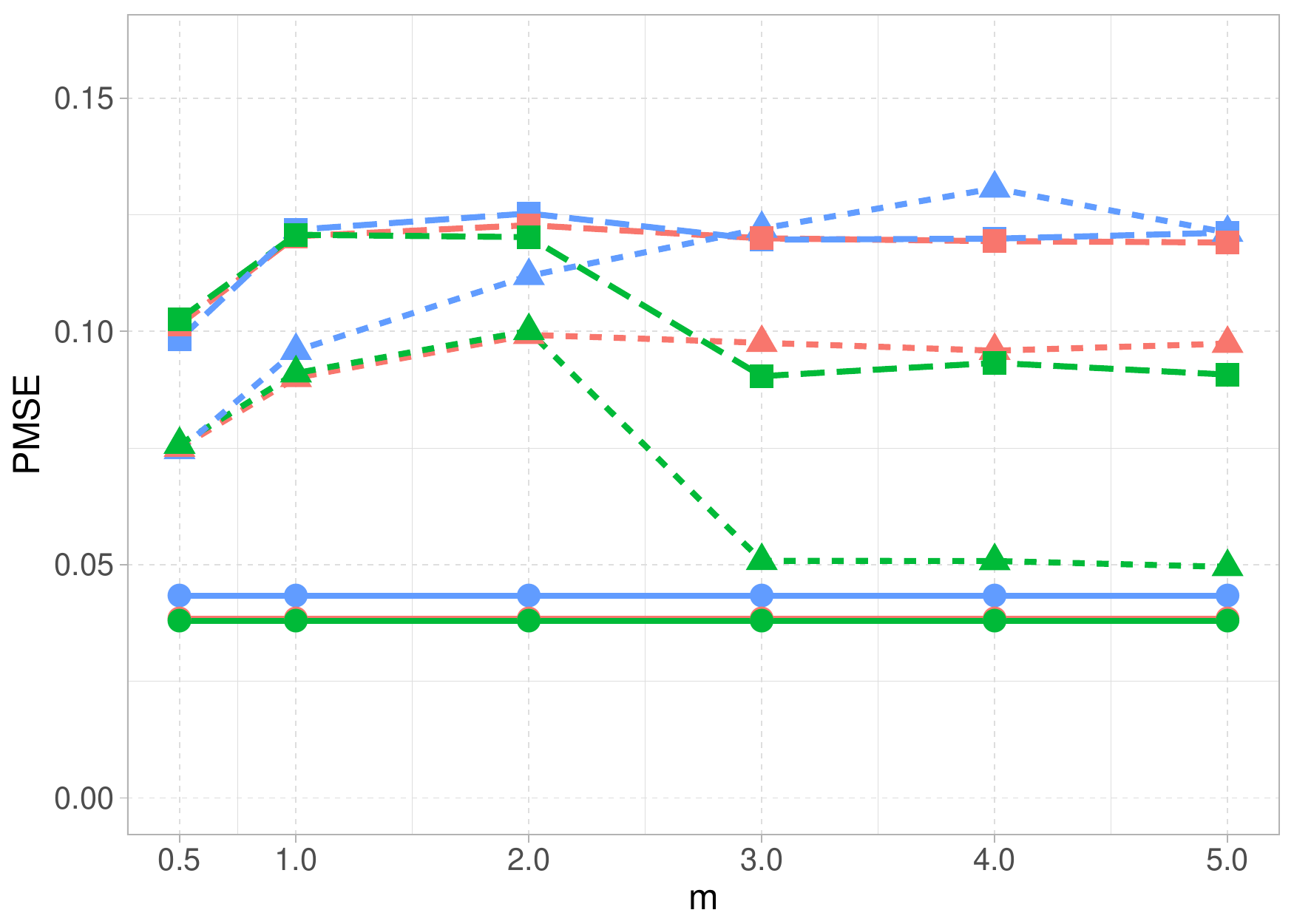}\\
$(n,p)=(300,80)$ & $(n,p)=(300,120)$\\
			\includegraphics[scale=0.68]{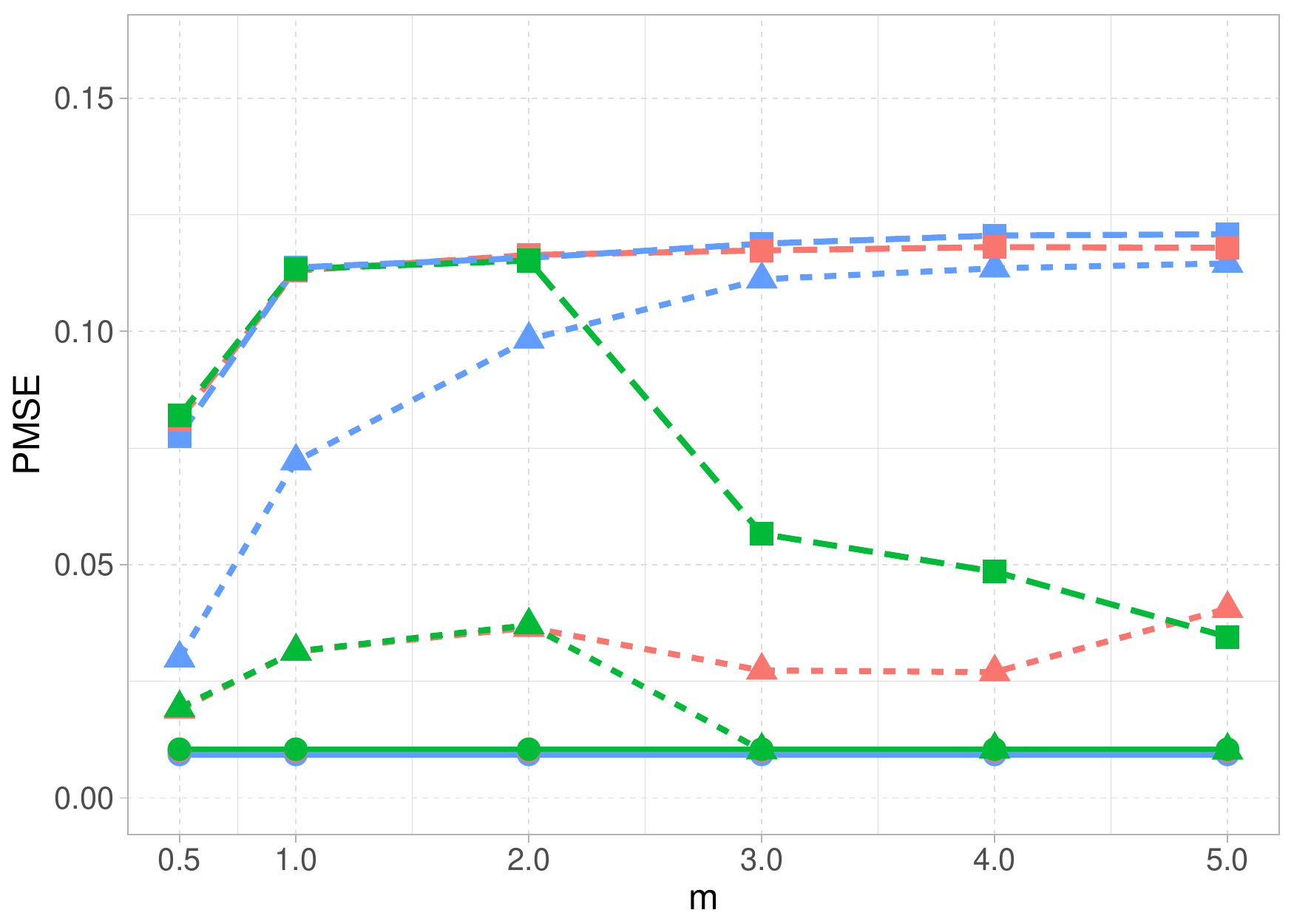} & 
			\includegraphics[scale=0.68]{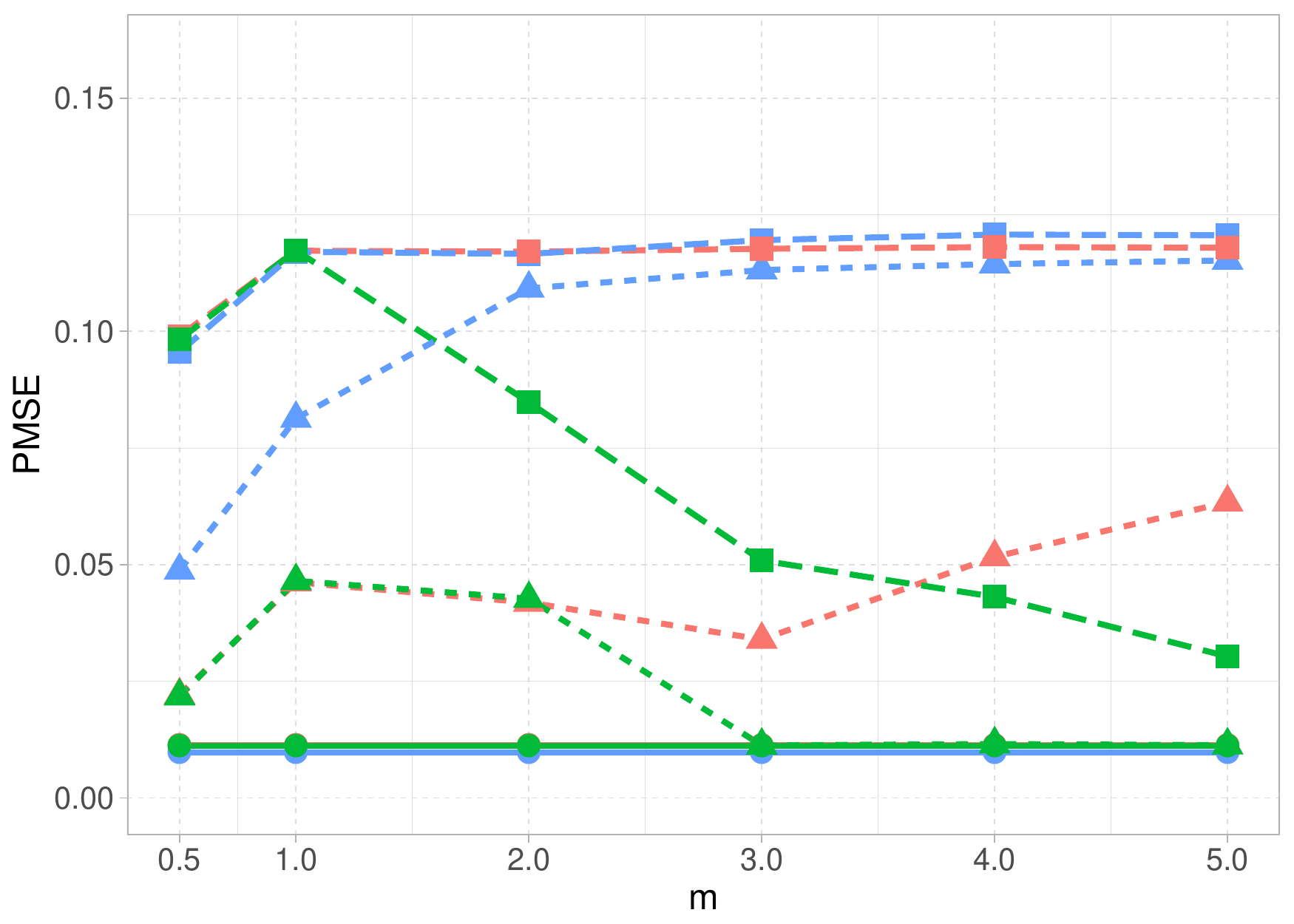}\\            
		\end{tabular}
		\caption{Probabilities Mean Squared Error for Sign penalization for scenarios \textbf{C0}, \textbf{CB1} and \textbf{CB2}.}
		\label{fig:out_B_ECMP_Signo}
	\end{figure}
	
	\newpage
	
	\begin{figure}[ht!]
		\centering
		\begin{tabular}{cc}
$(n,p)=(150,40)$ & $(n,p)=(150,80)$\\				
			\includegraphics[scale=0.68]{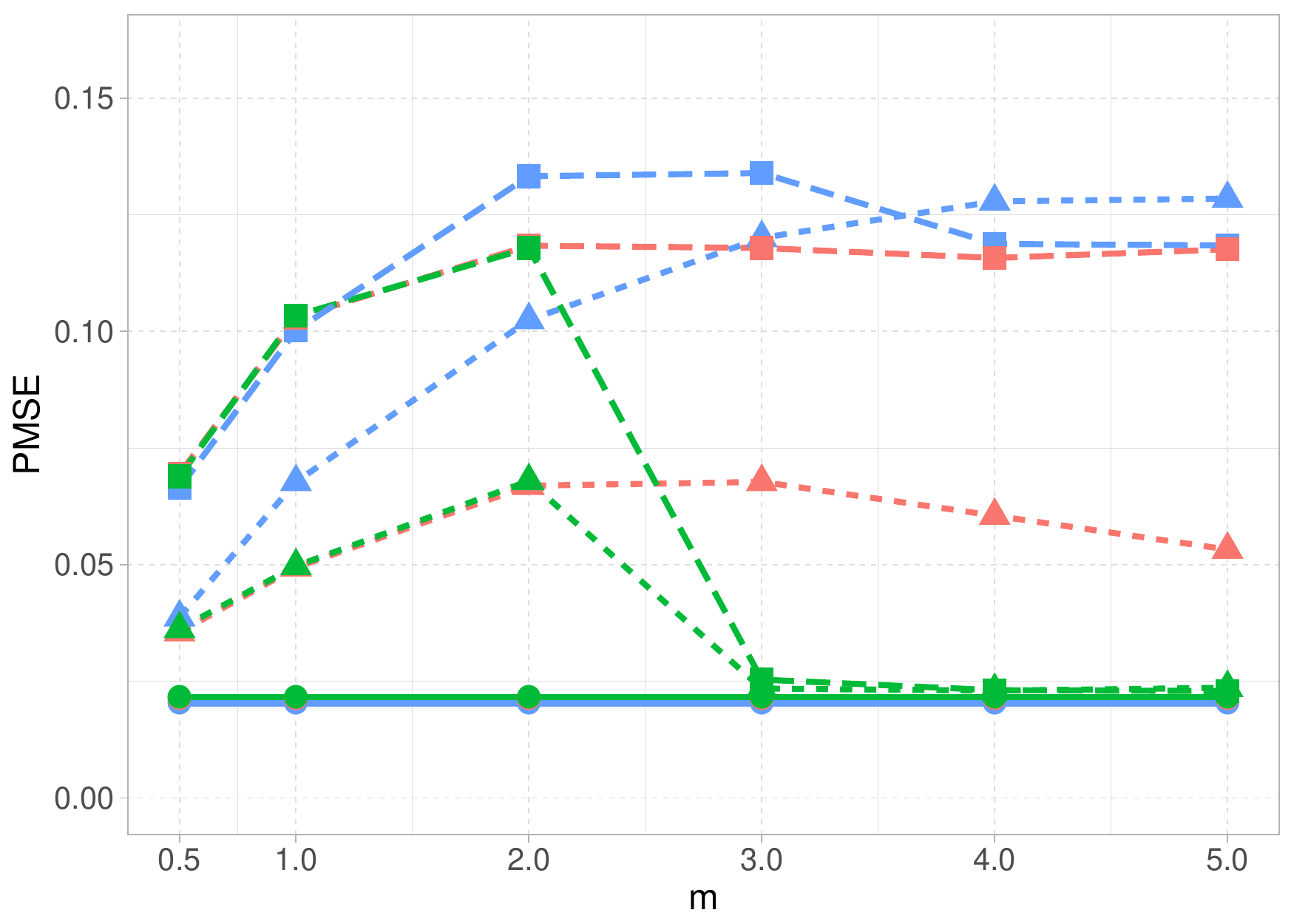} & 
			\includegraphics[scale=0.68]{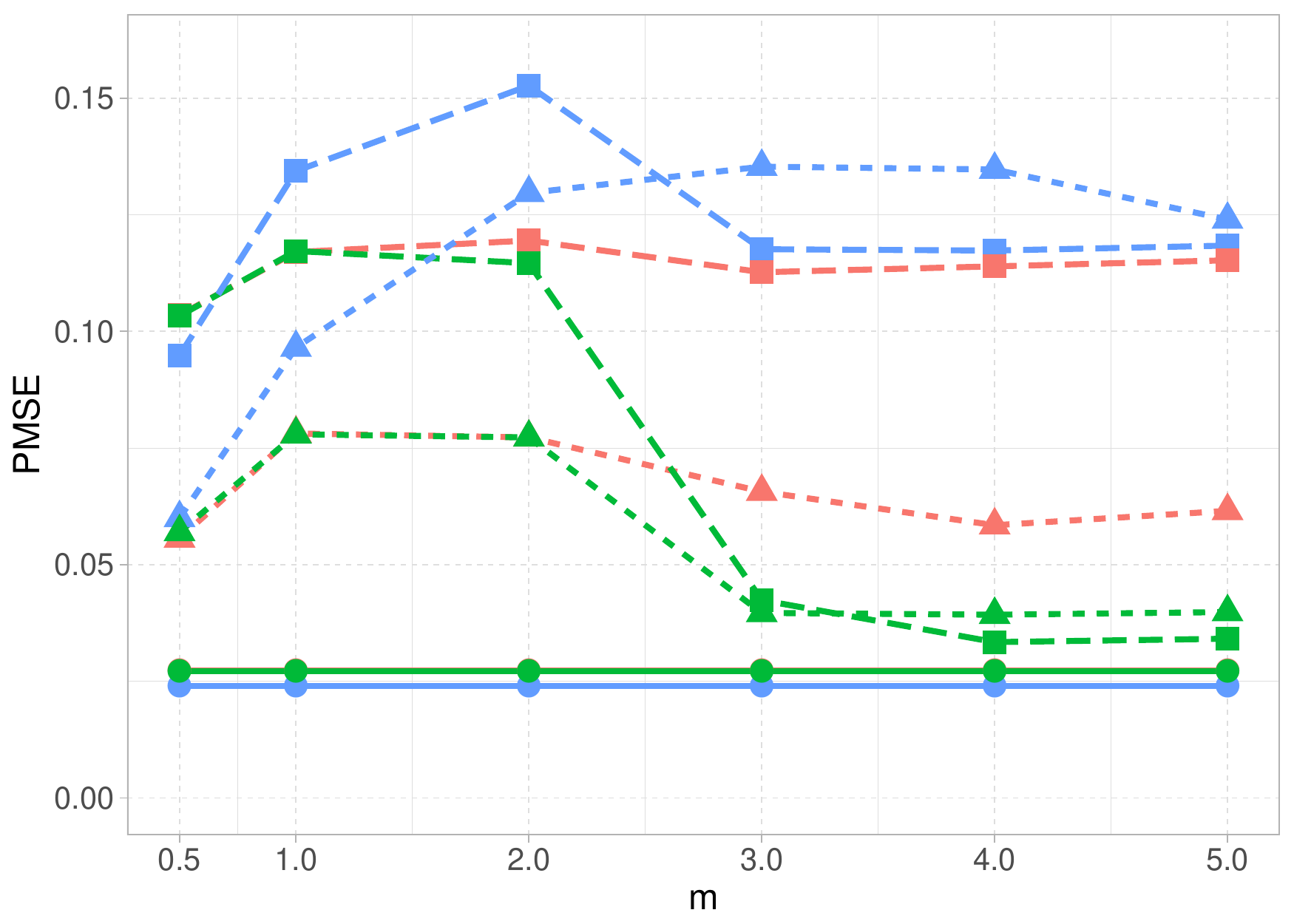}\\
$(n,p)=(300,80)$ & $(n,p)=(300,120)$\\
			\includegraphics[scale=0.68]{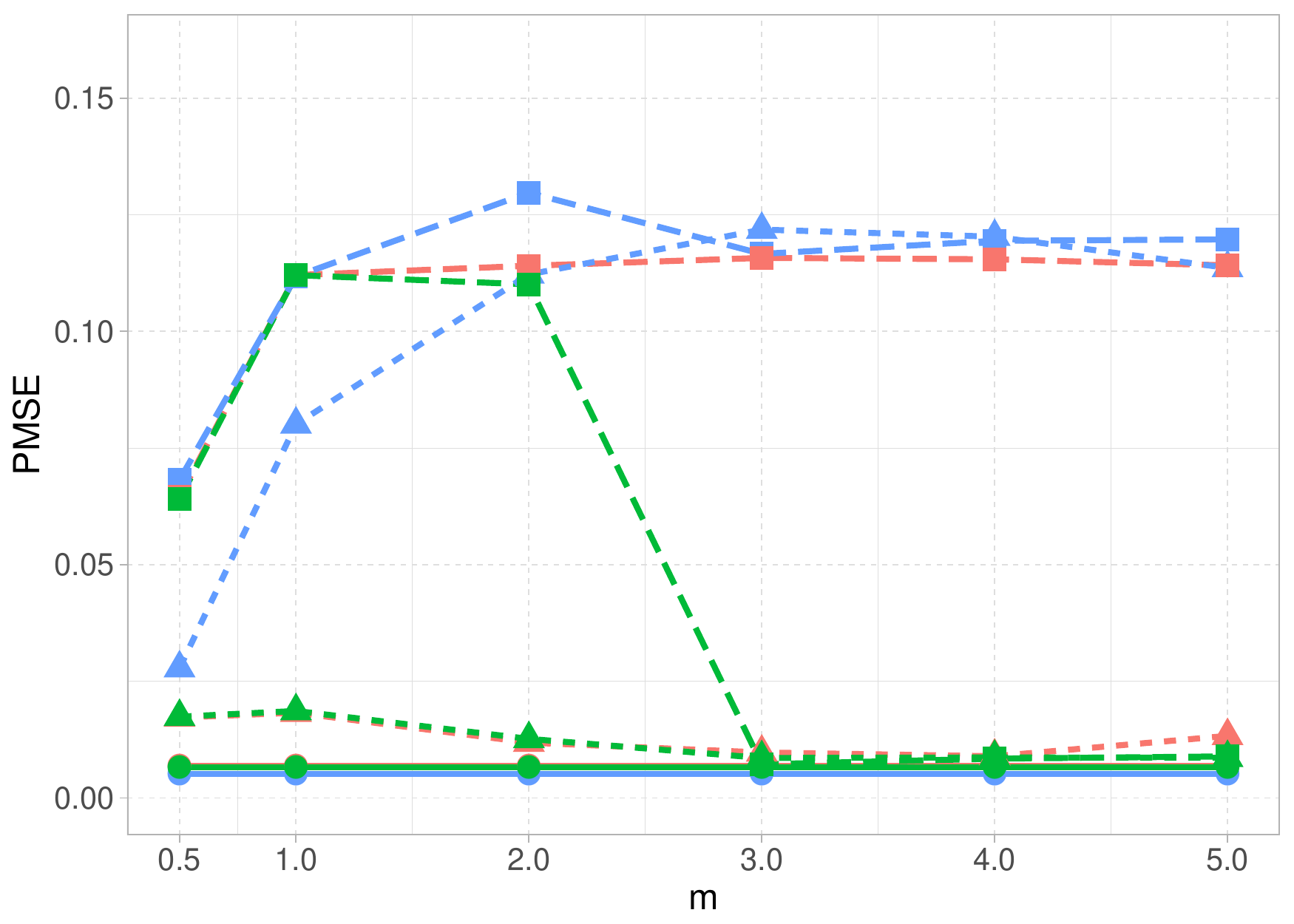} & 
			\includegraphics[scale=0.68]{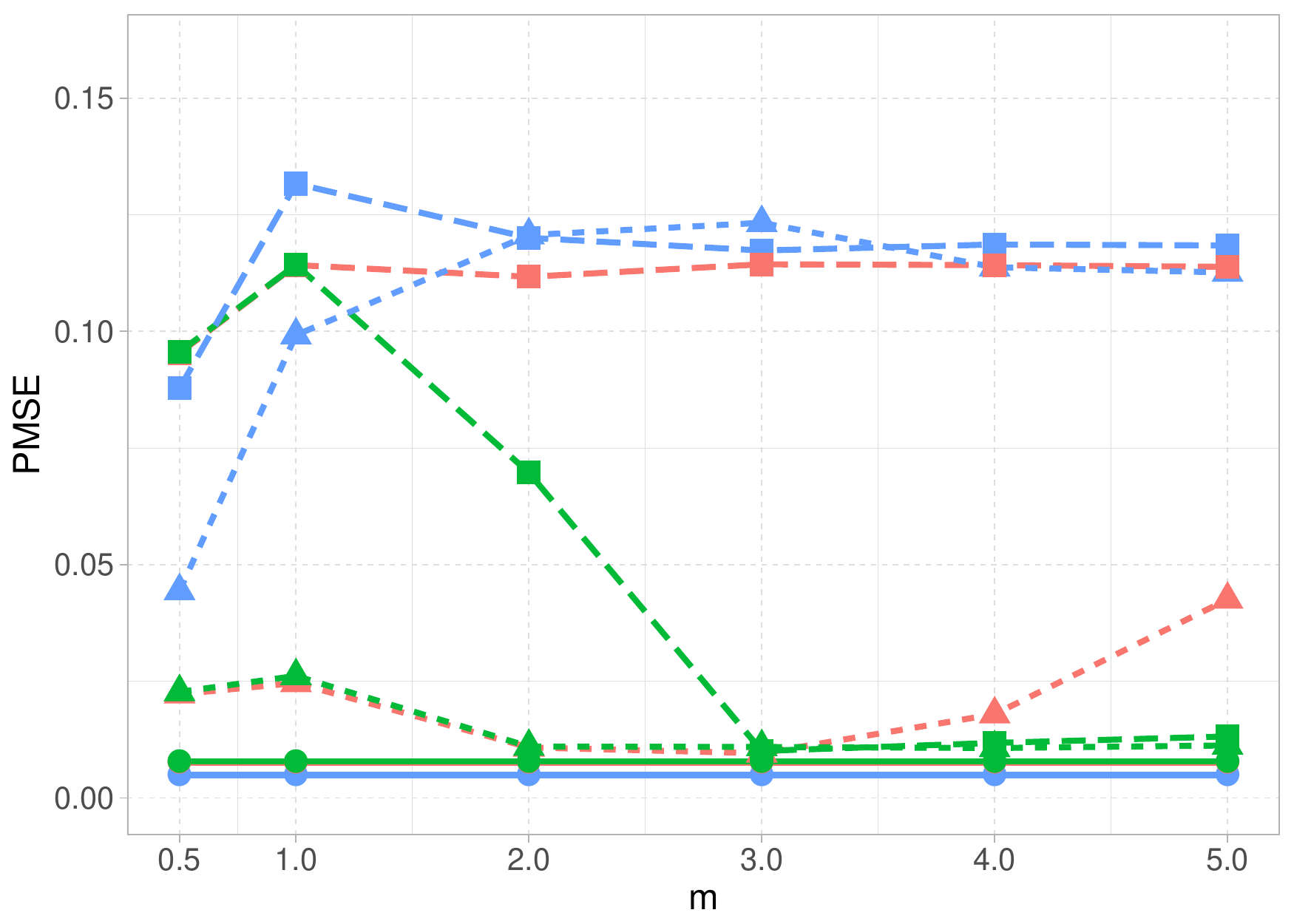}\\            
		\end{tabular}
		\caption{Probabilities Mean Squared Error for MCP penalization for scenarios \textbf{C0}, \textbf{CB1} and \textbf{CB2}.}
		\label{fig:out_B_ECMP_MCP}
	\end{figure}
	
	\newpage
	
	\begin{figure}[ht!]
		\centering
		\begin{tabular}{cc}
$(n,p)=(150,40)$ & $(n,p)=(150,80)$\\			
			\includegraphics[scale=0.68]{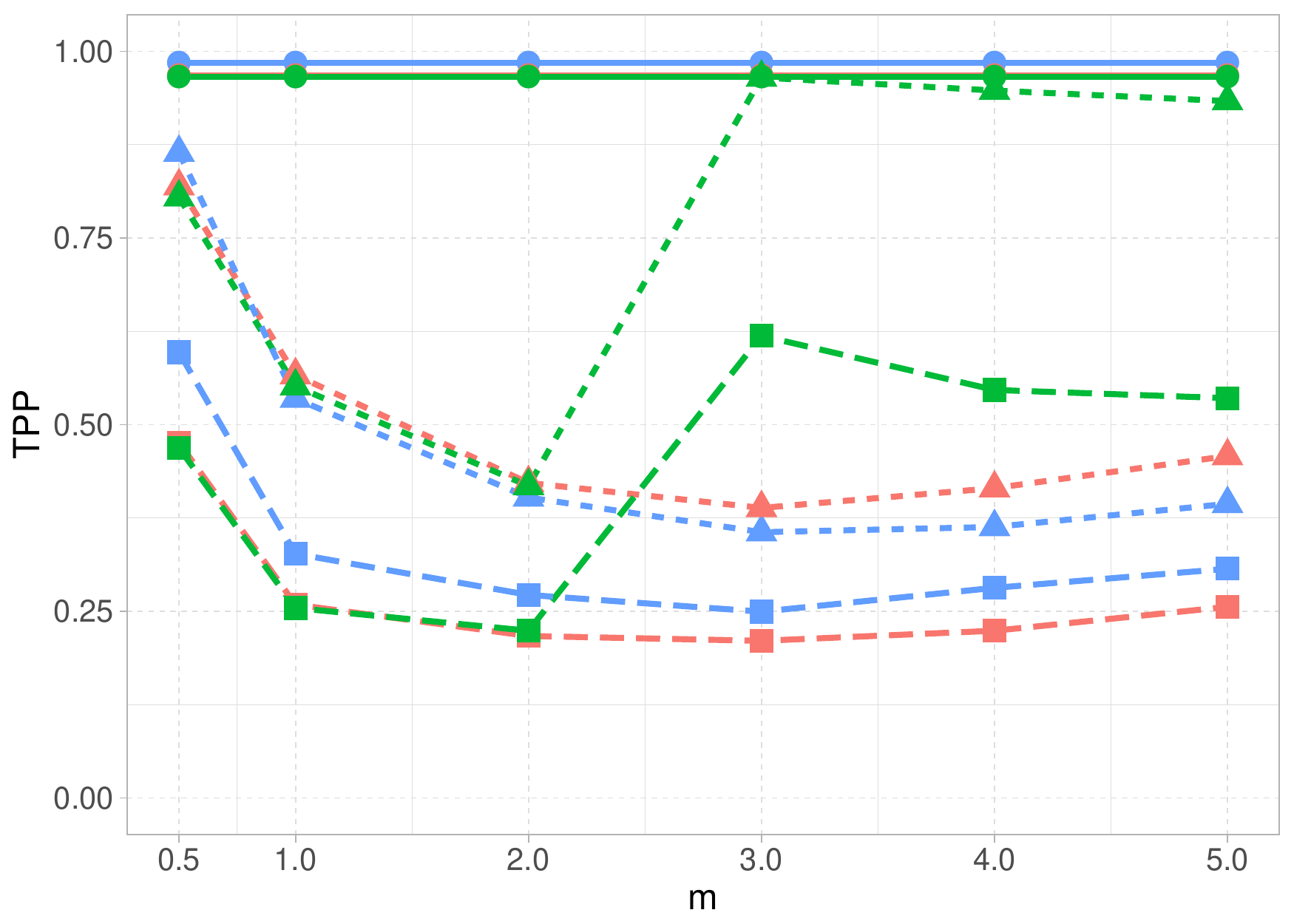} & 
			\includegraphics[scale=0.68]{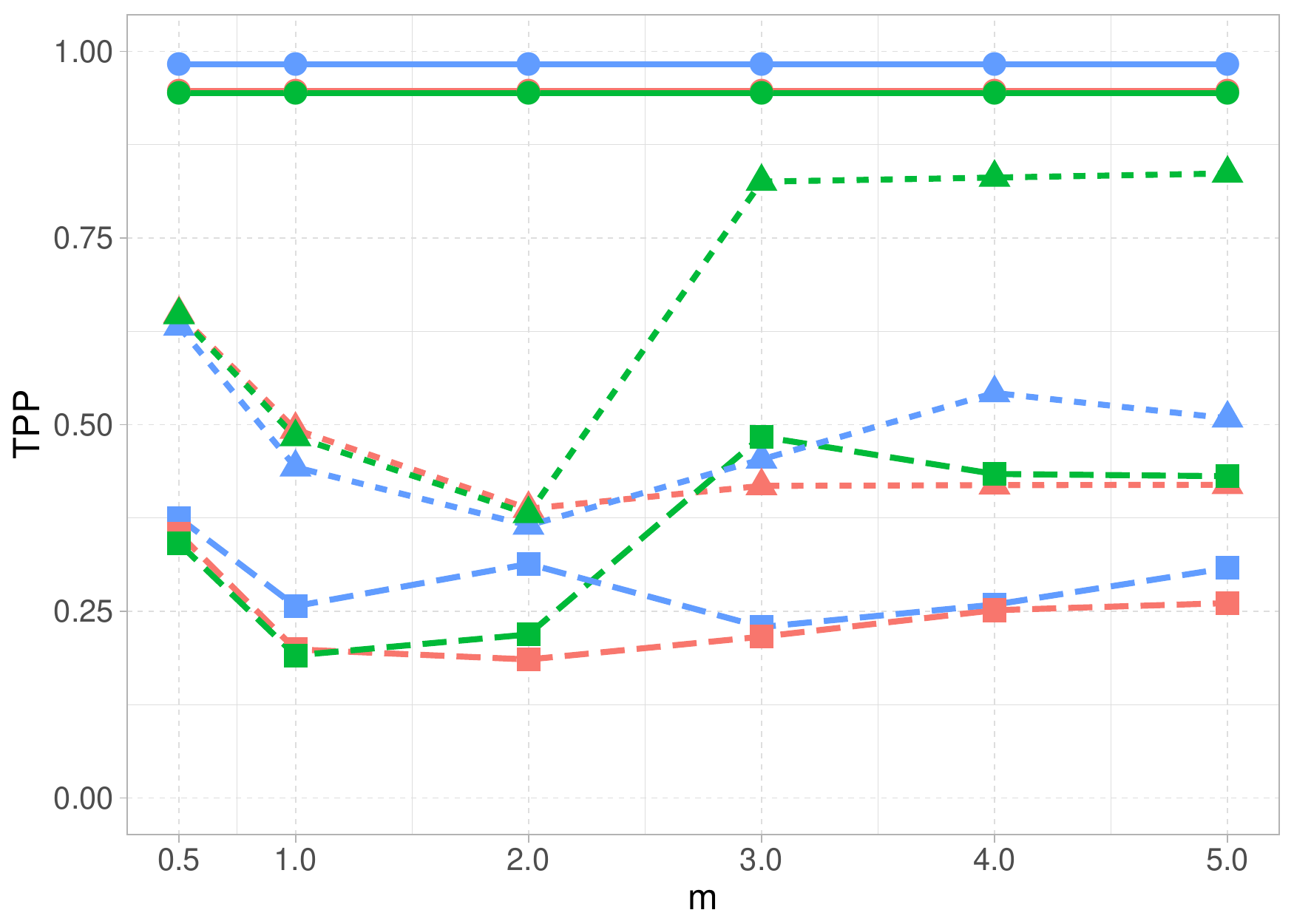}\\
$(n,p)=(300,80)$ & $(n,p)=(300,120)$\\
			\includegraphics[scale=0.68]{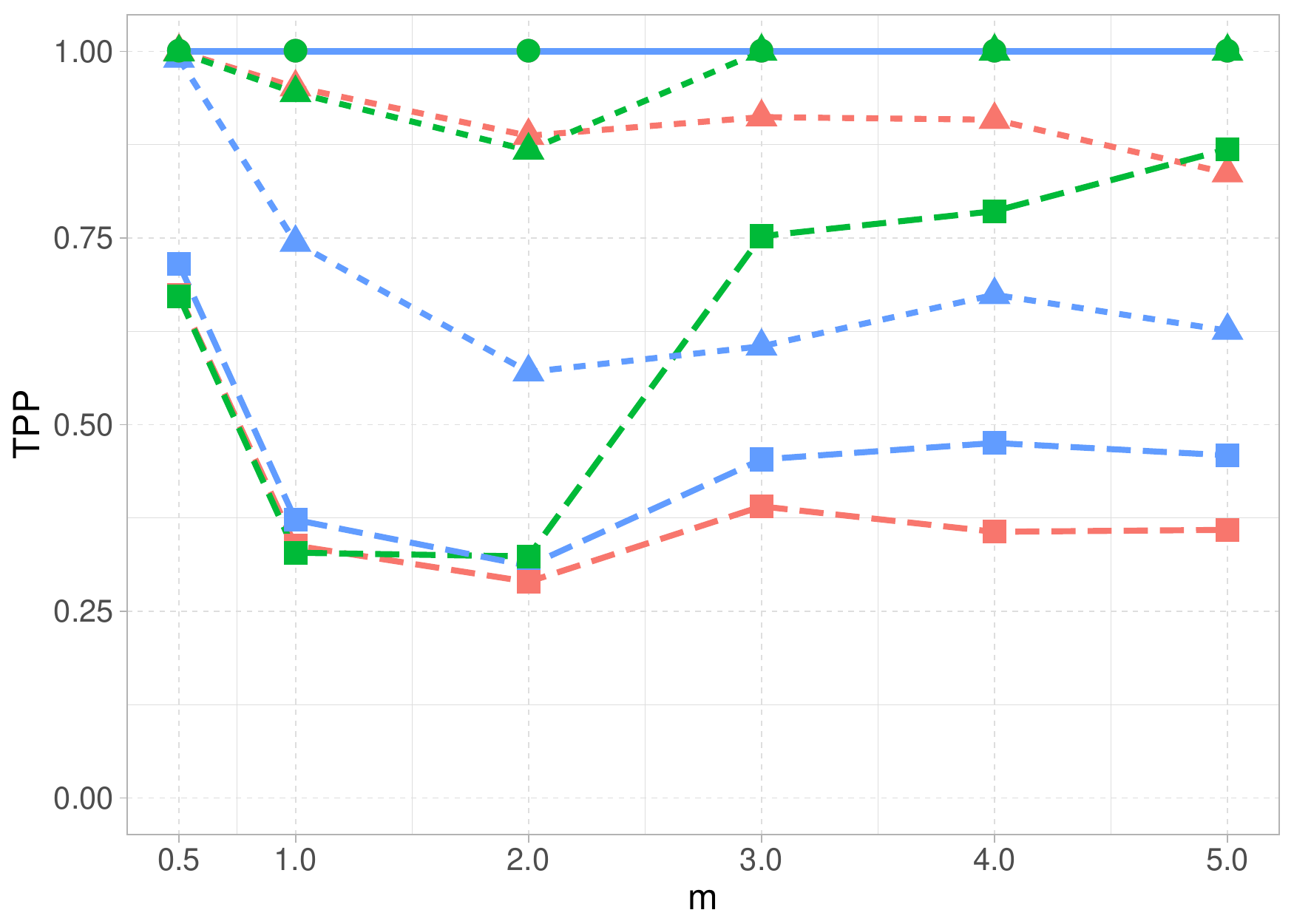} & 
			\includegraphics[scale=0.68]{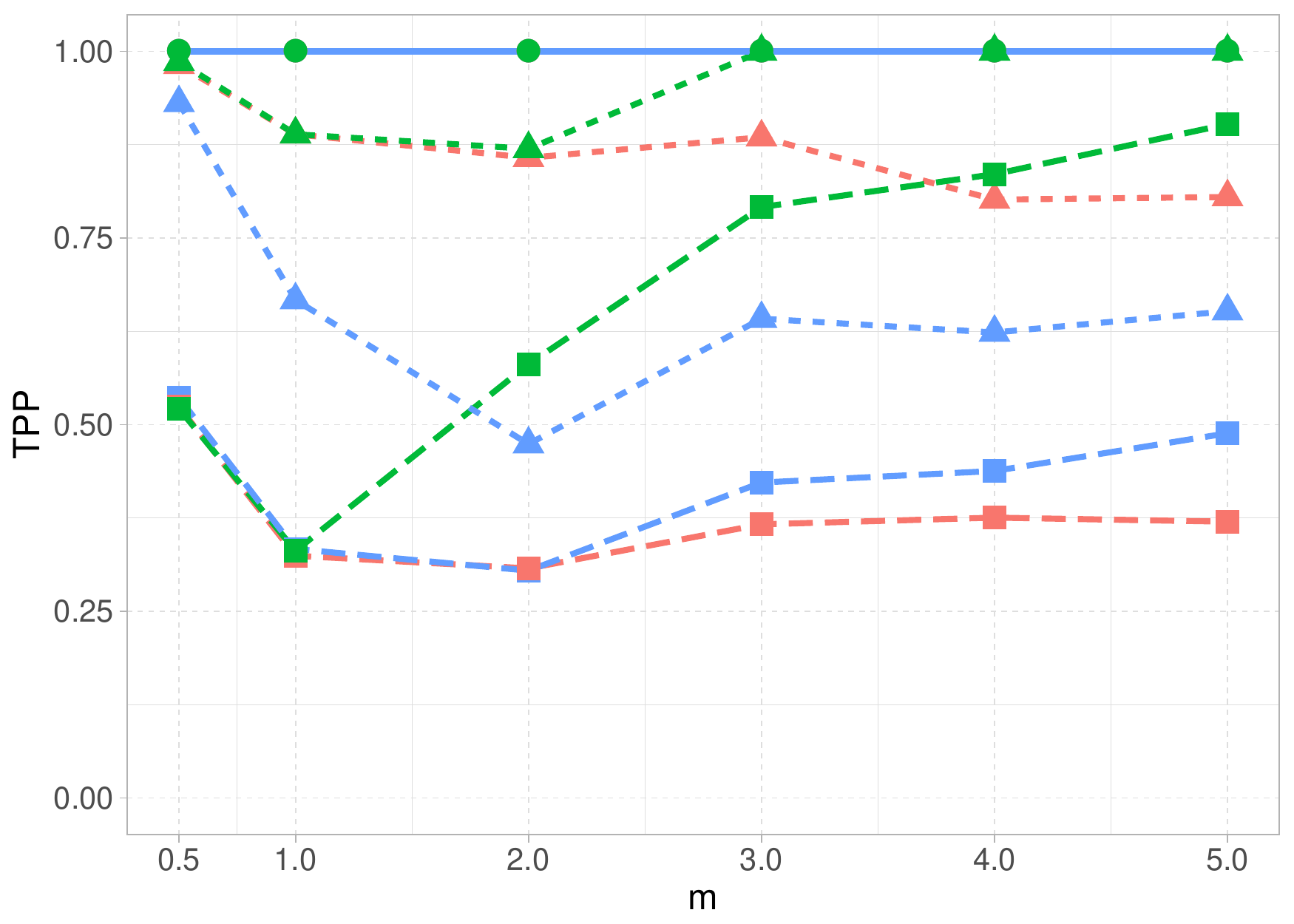}\\            
		\end{tabular}
		\caption{True Positive Proportion for Sign Penalization for scenarios \textbf{C0}, \textbf{CB1} and \textbf{CB2}.}
		\label{fig:out_B_PVP_Signo}
	\end{figure}
	
	\newpage
	
	\begin{figure}[ht!]
		\centering
		\begin{tabular}{cc}
$(n,p)=(150,40)$ & $(n,p)=(150,80)$\\
		\includegraphics[scale=0.68]{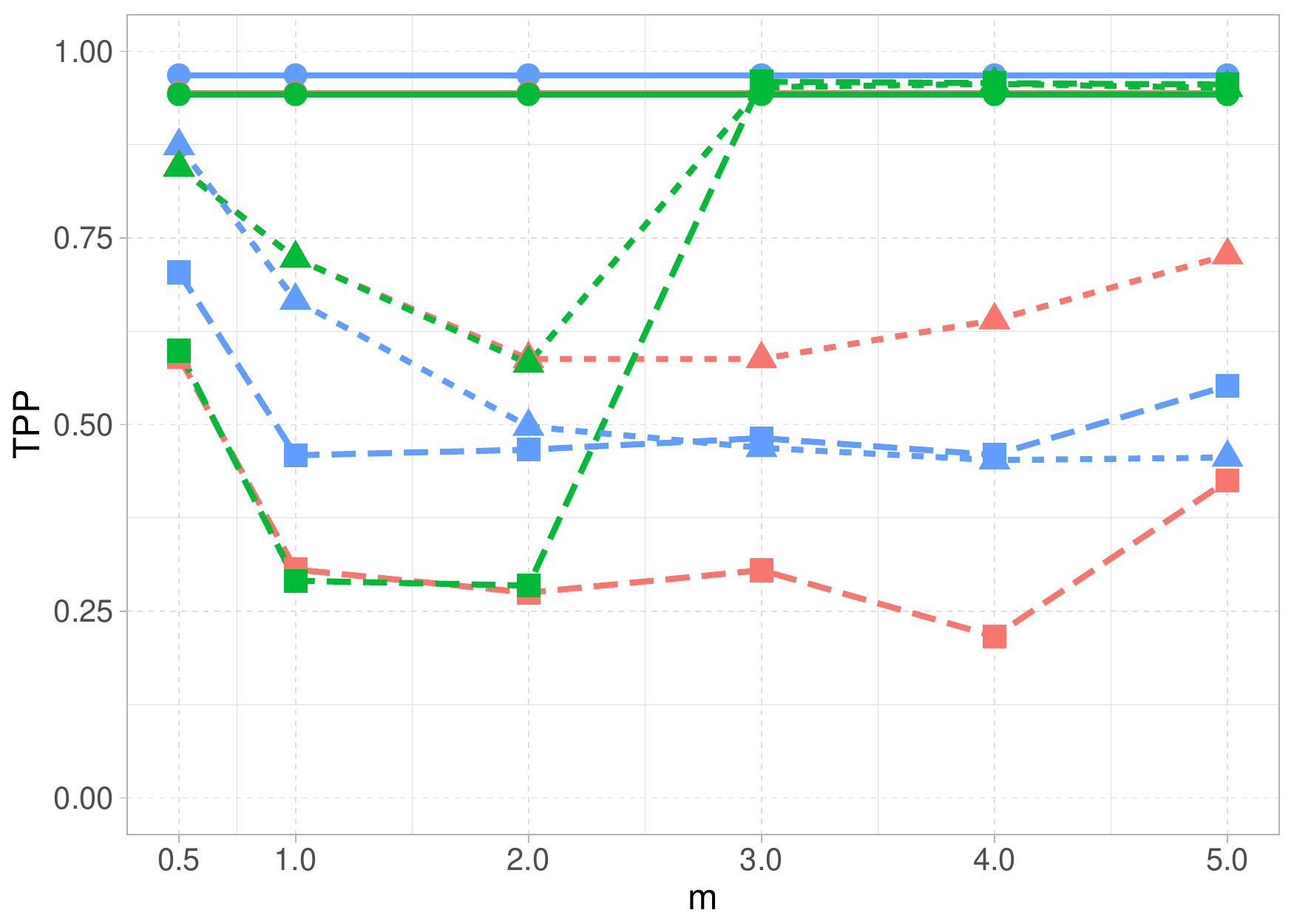} & 
		\includegraphics[scale=0.68]{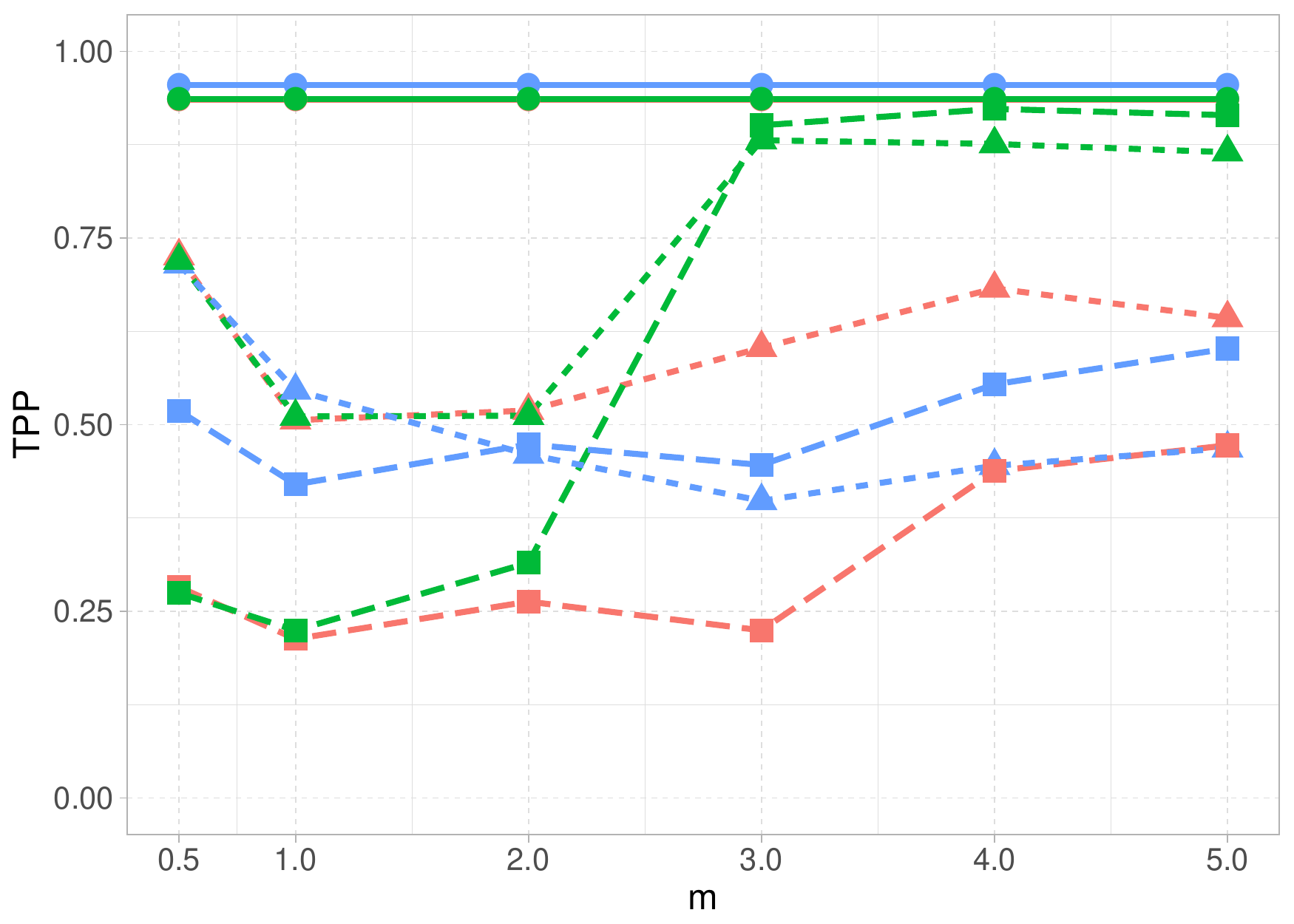}\\
$(n,p)=(300,80)$ & $(n,p)=(300,120)$\\
			\includegraphics[scale=0.68]{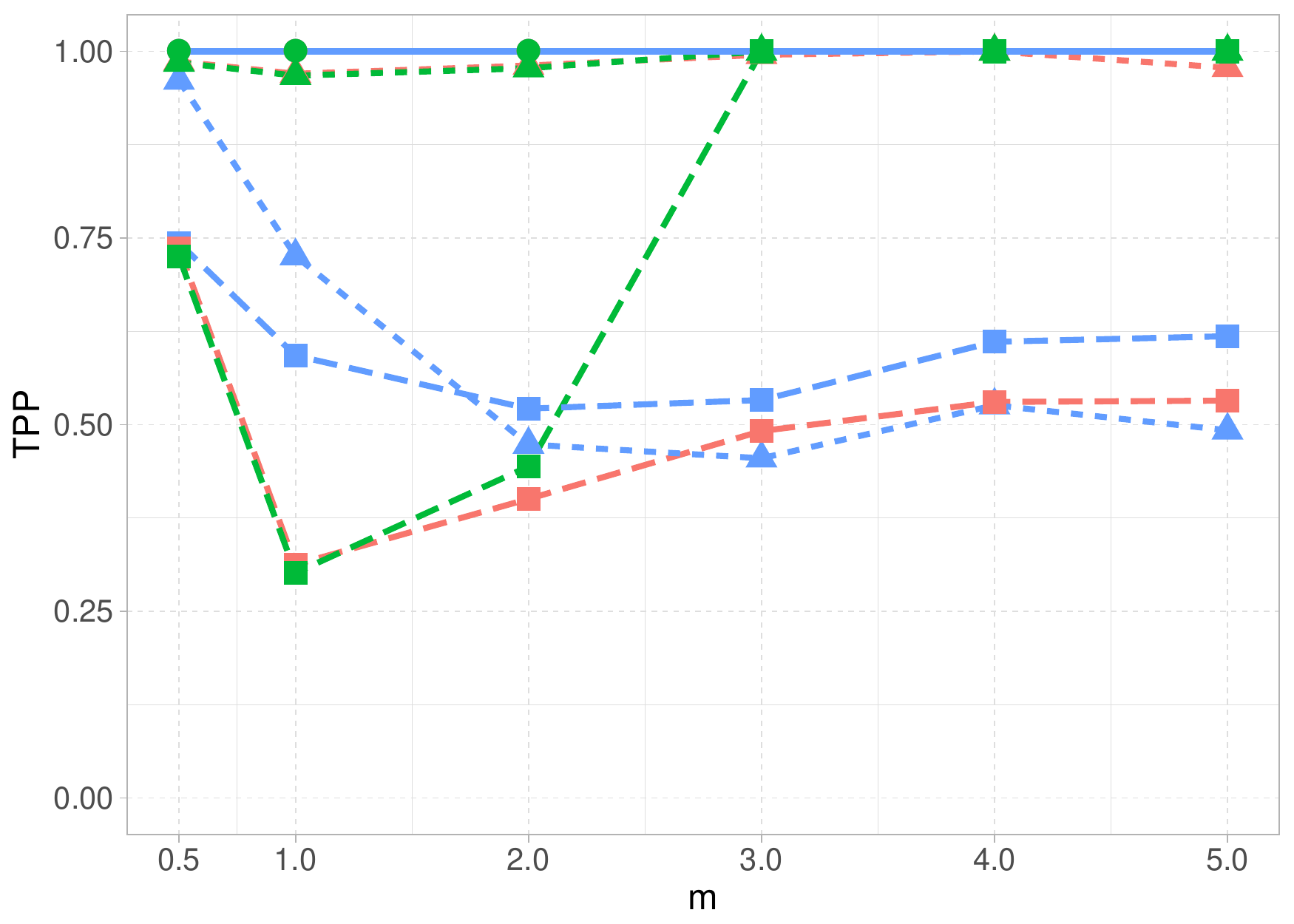} & 
			\includegraphics[scale=0.68]{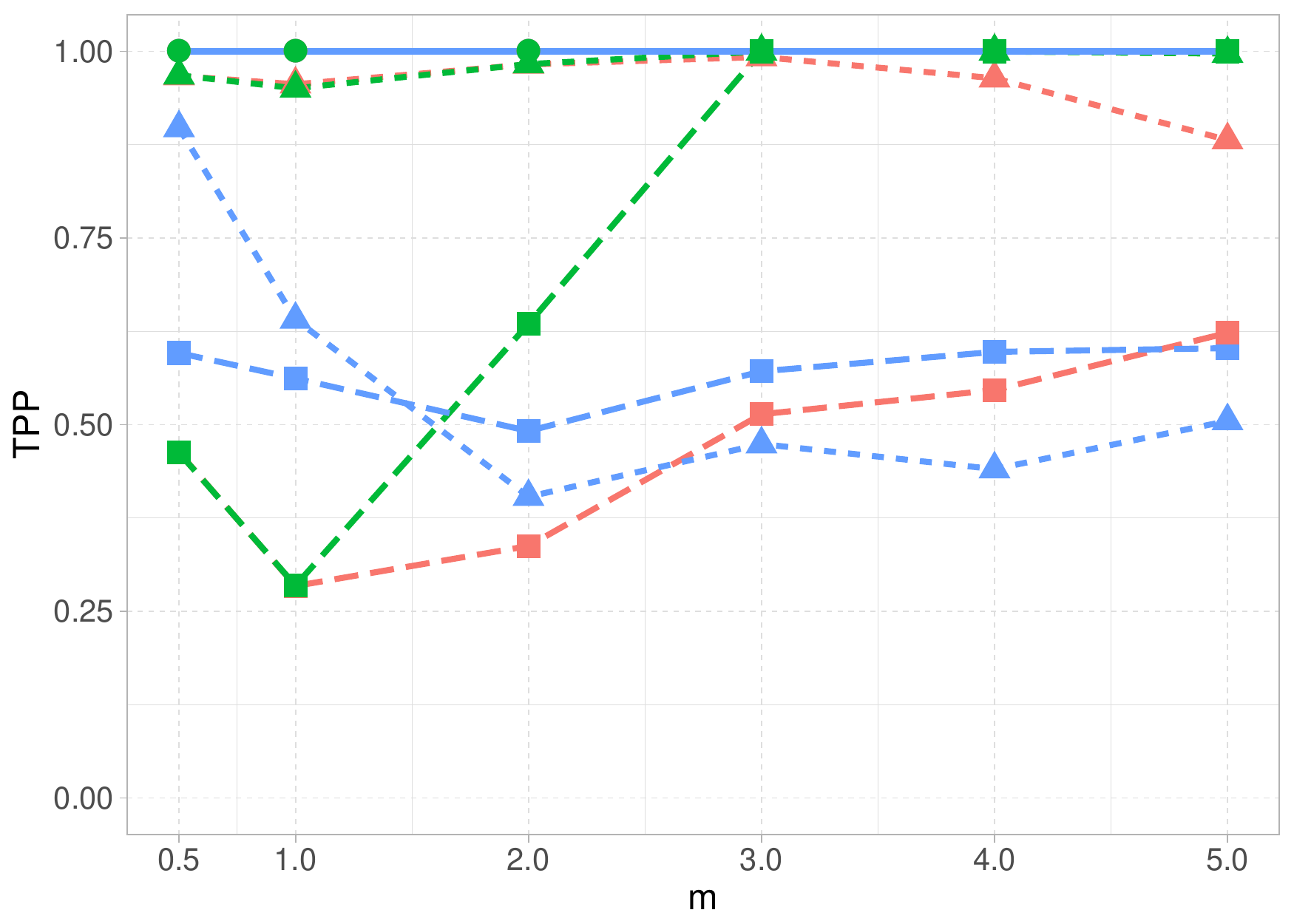}\\            
		\end{tabular}
		\caption{True Positive Proportion for MCP Penalization for scenarios  \textbf{C0}, \textbf{CB1} and \textbf{CB2}.}
		\label{fig:out_B_PVP_MCP}
	\end{figure}
	
	\newpage
	
	\begin{figure}[ht!]
		\centering
		\begin{tabular}{cc}	
$(n,p)=(150,40)$ & $(n,p)=(150,80)$\\		
			\includegraphics[scale=0.68]{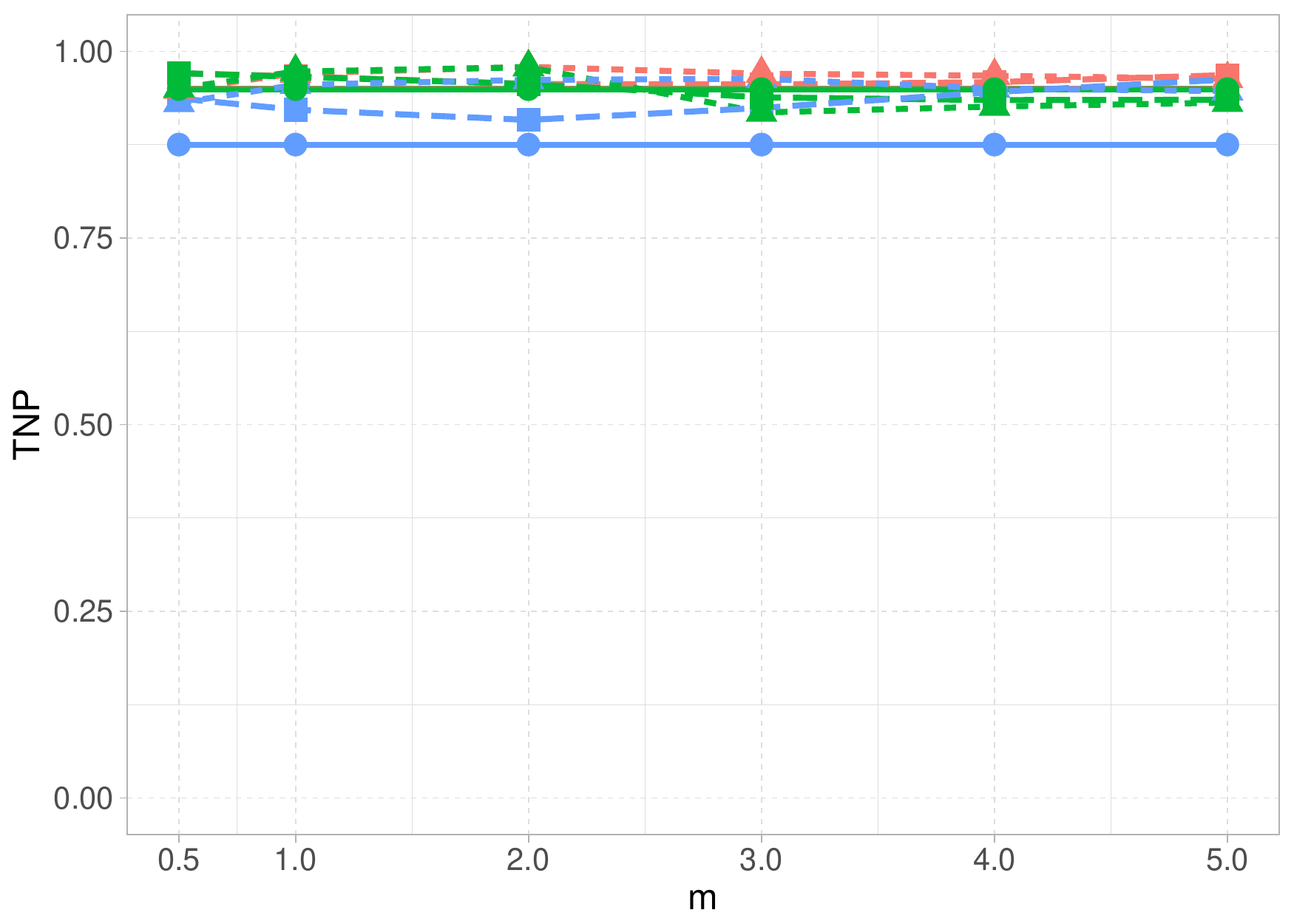} & 
			\includegraphics[scale=0.68]{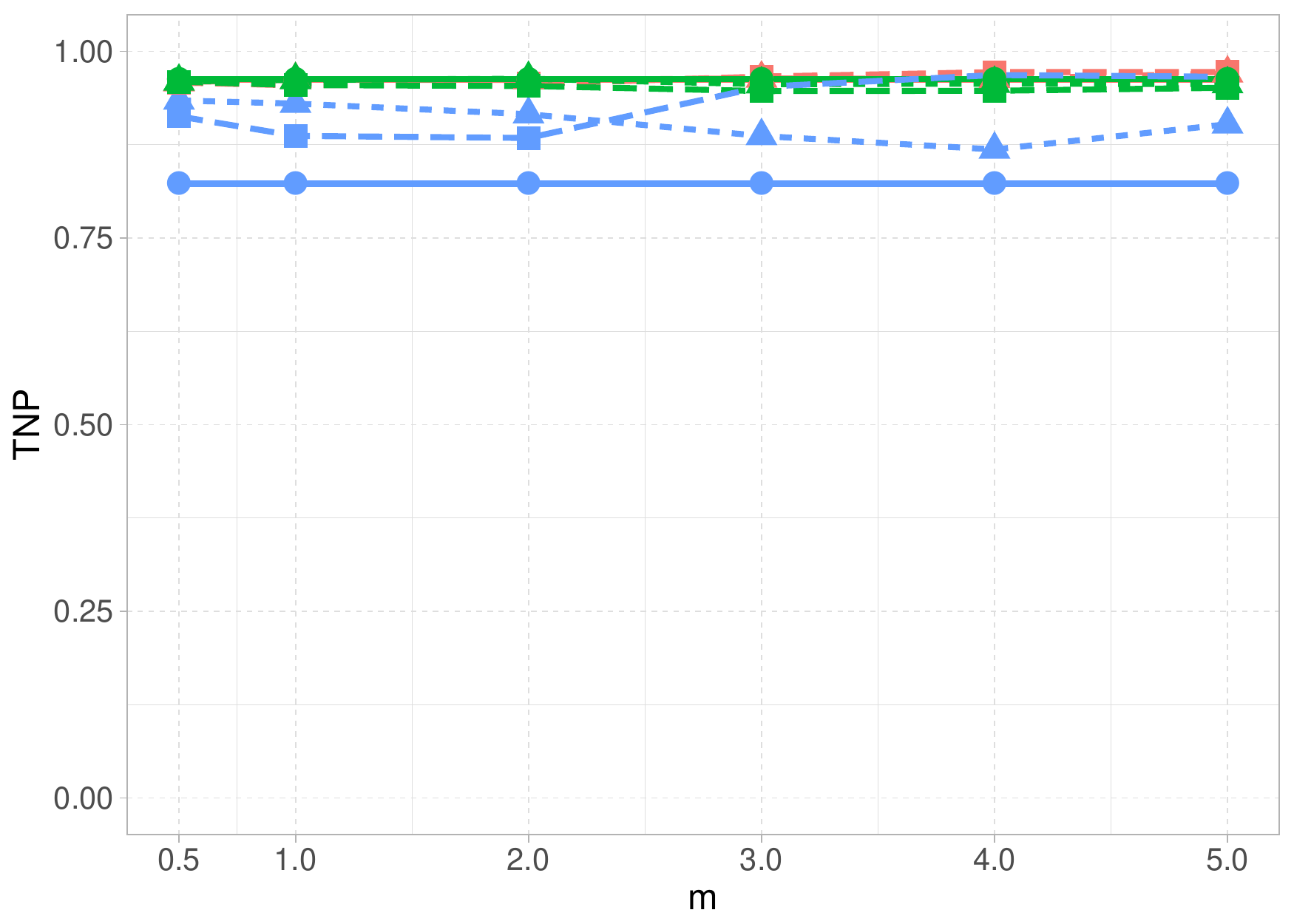}\\
$(n,p)=(300,80)$ & $(n,p)=(300,120)$\\
			\includegraphics[scale=0.68]{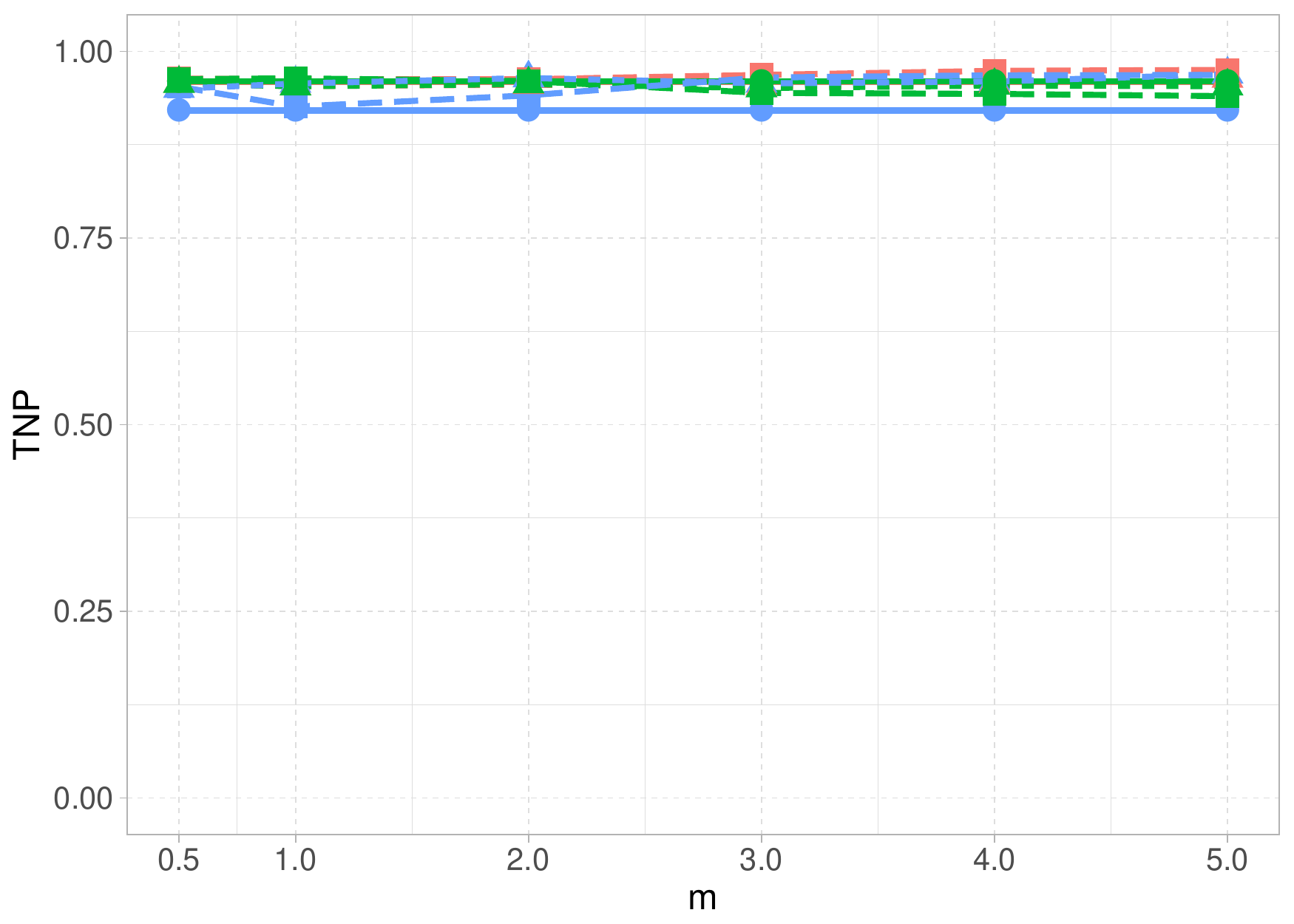} & 
			\includegraphics[scale=0.68]{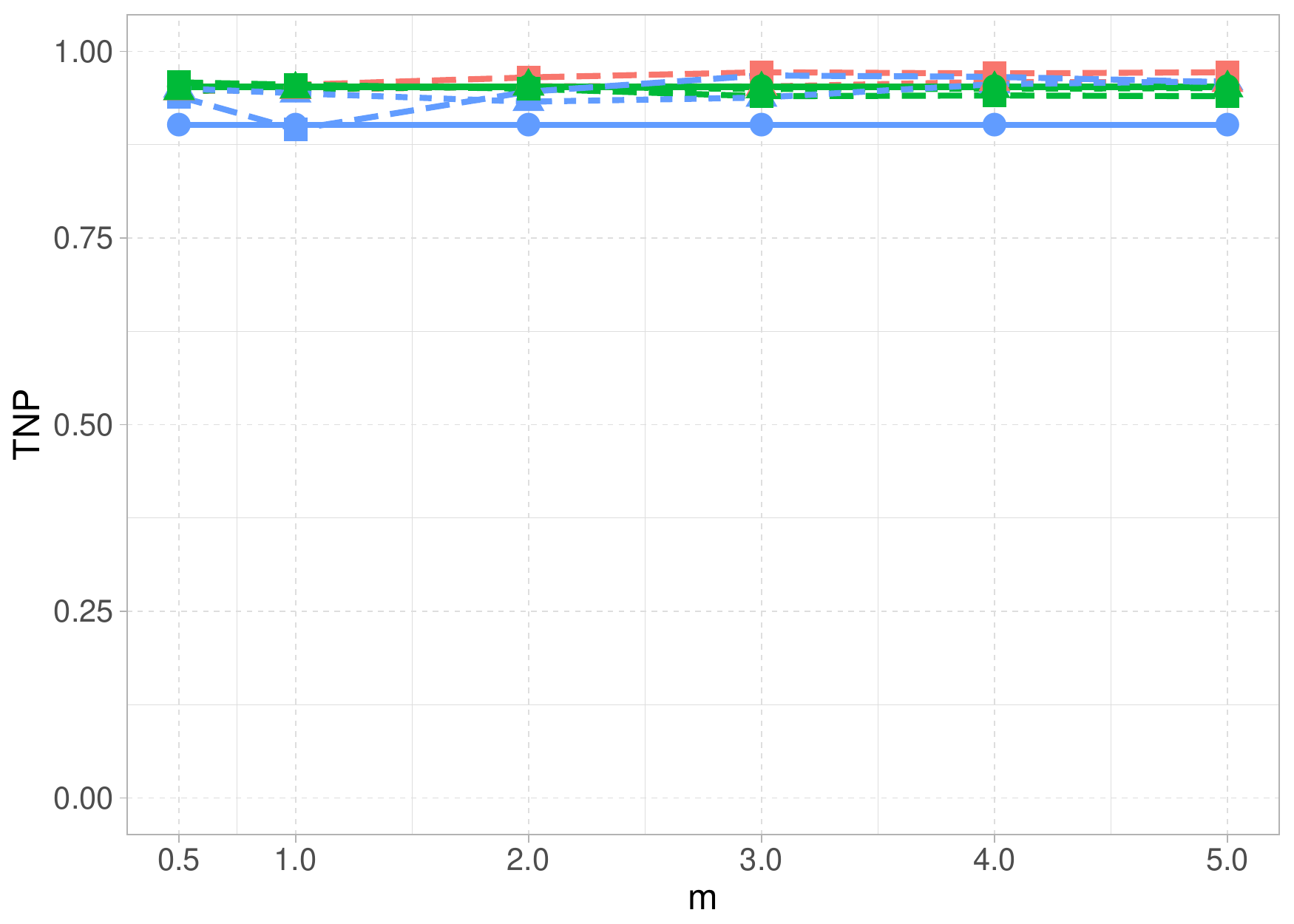}\\            
		\end{tabular}
		\caption{True Null Proportion for Sign Penalization for scenarios \textbf{C0}, \textbf{CB1} and \textbf{CB2}.}
		\label{fig:out_B_PVN_Signo}
	\end{figure}
	
	\newpage
	
	\begin{figure}[ht!]
		\centering
		\begin{tabular}{cc}
$(n,p)=(150,40)$ & $(n,p)=(150,80)$\\
			\includegraphics[scale=0.68]{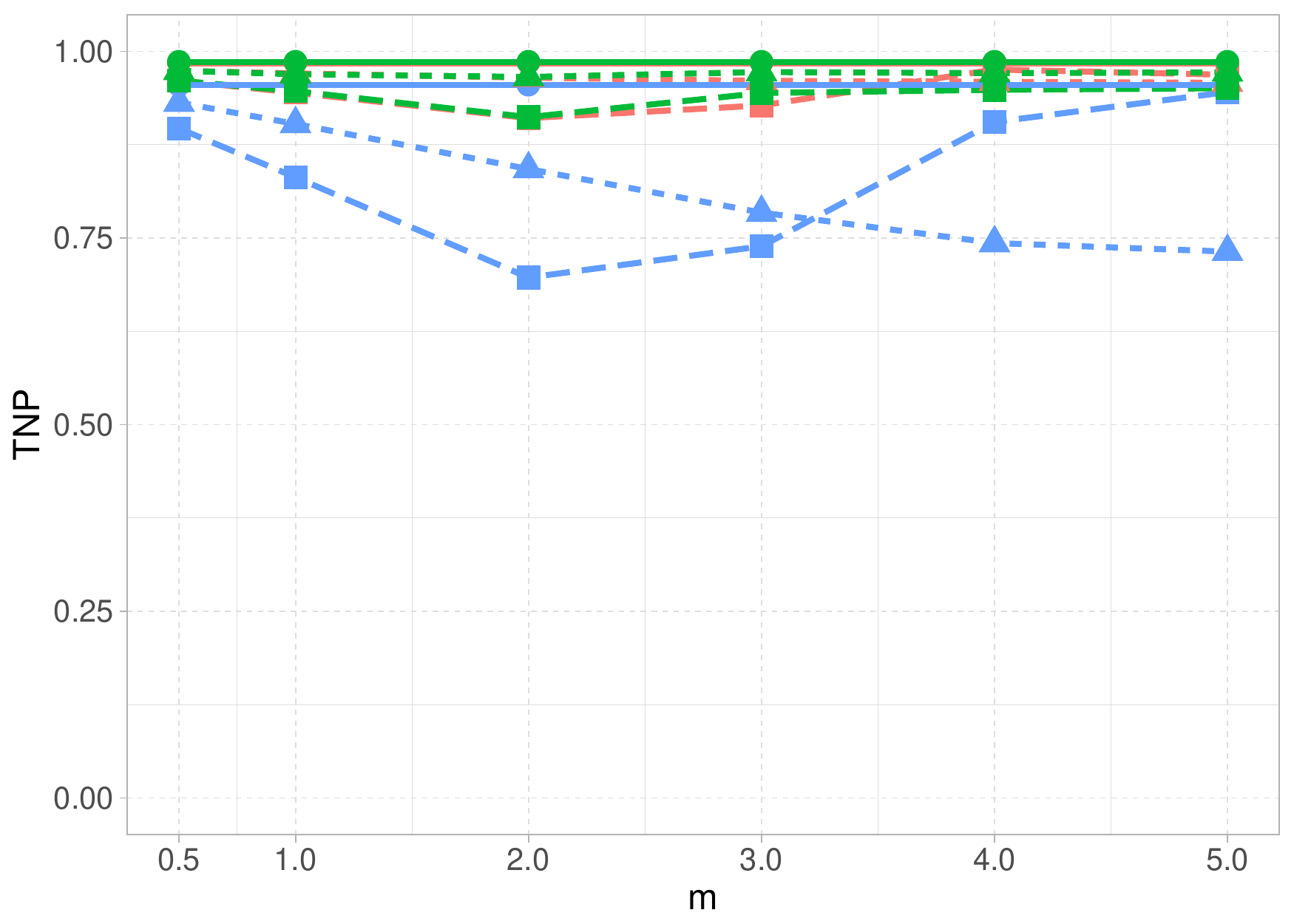} & 
			\includegraphics[scale=0.68]{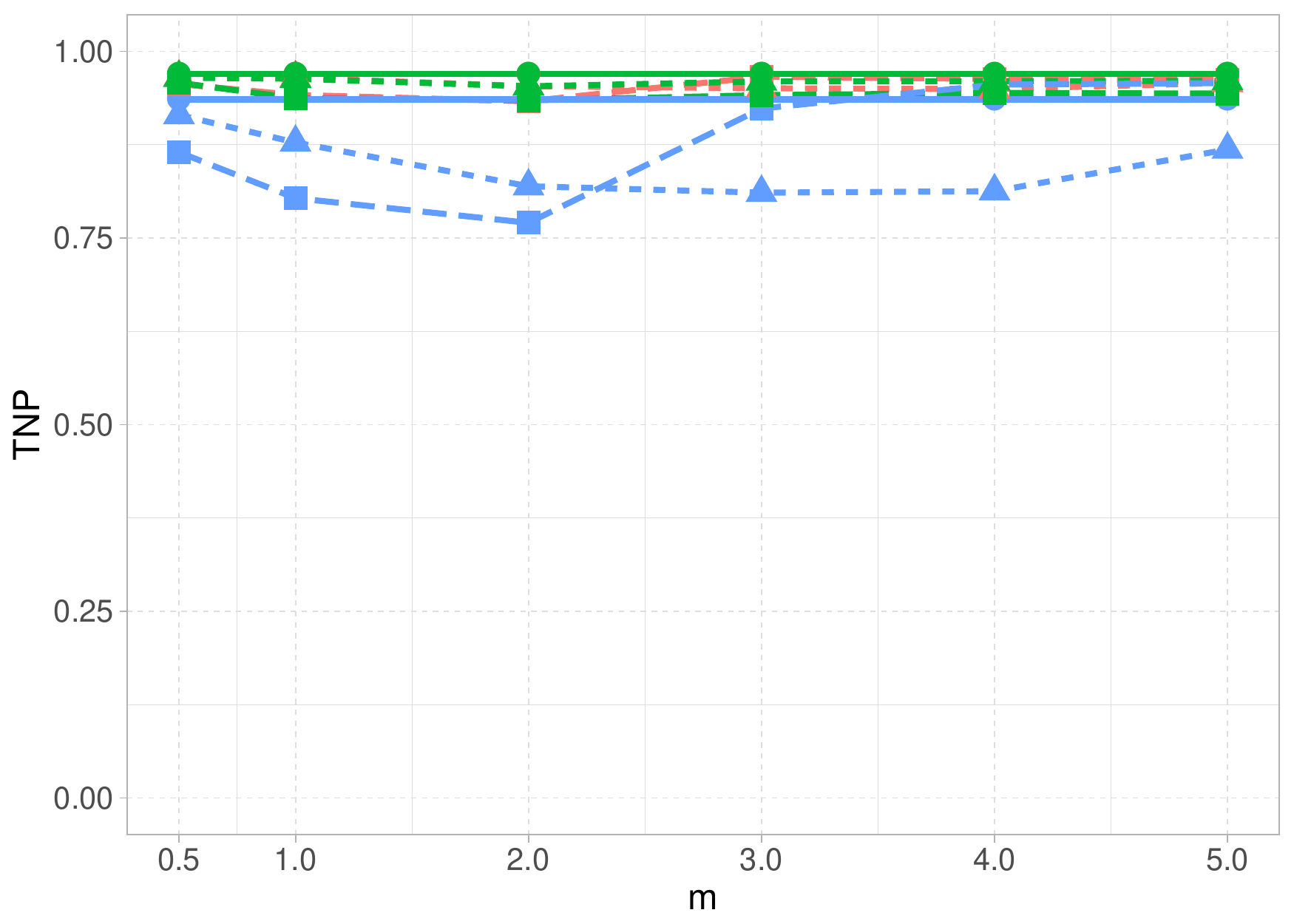}\\
$(n,p)=(300,80)$ & $(n,p)=(300,120)$\\
			\includegraphics[scale=0.68]{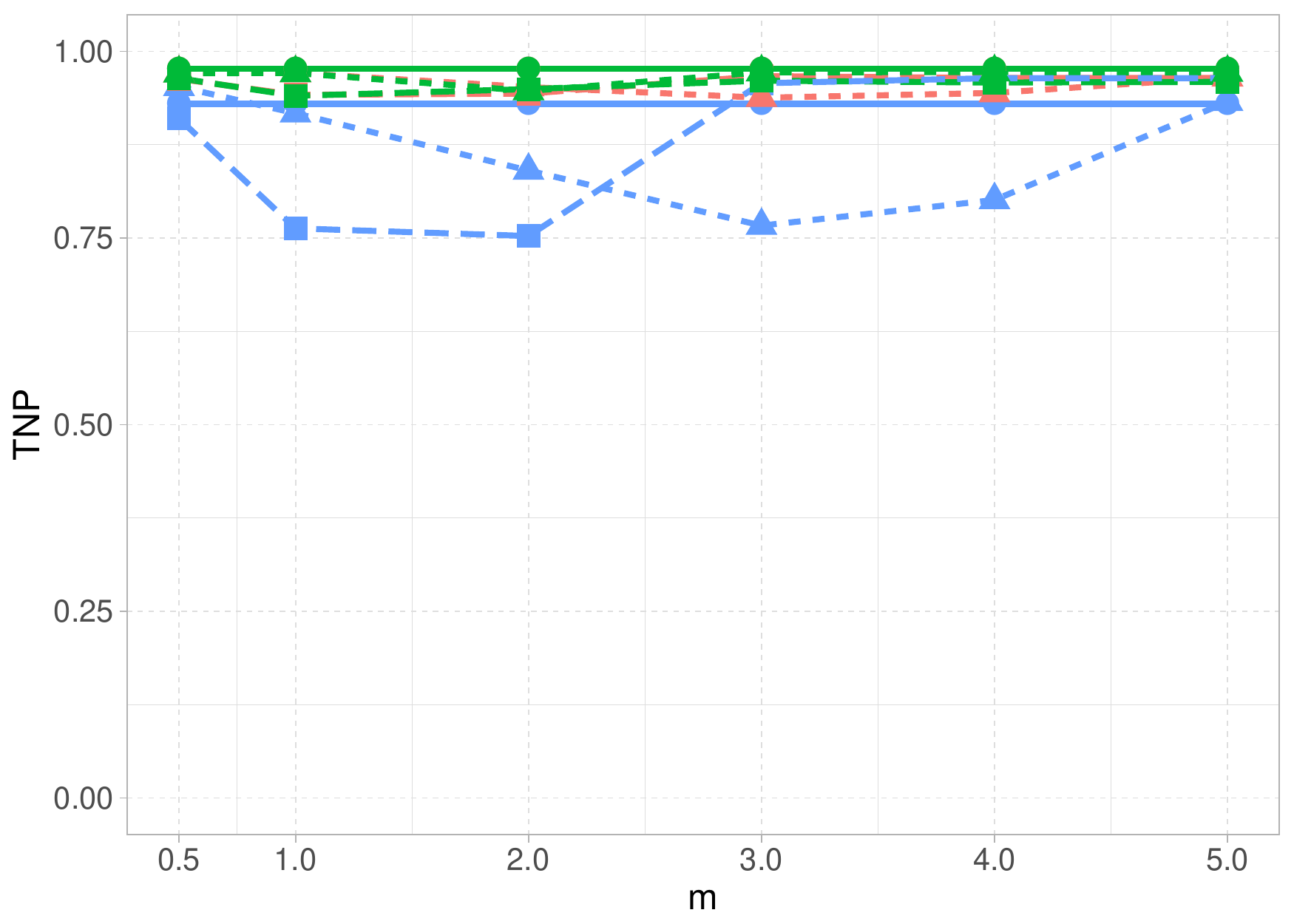} & 
			\includegraphics[scale=0.68]{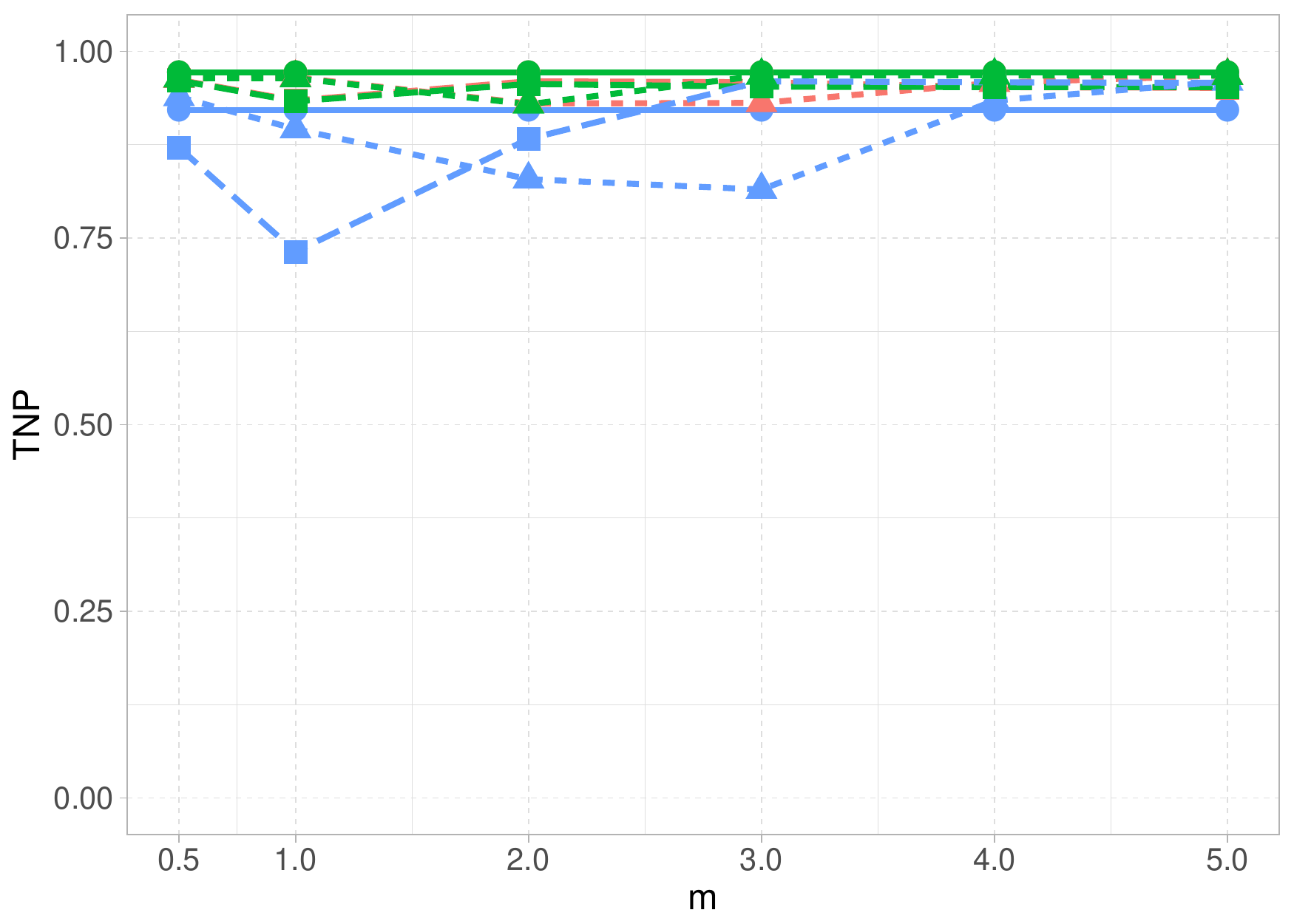}\\            
		\end{tabular}
		\caption{True Null Proportion for MCP Penalization for scenarios \textbf{C0}, \textbf{CB1} and \textbf{CB2}.}
		\label{fig:out_B_PVN_MCP}
	\end{figure}
	
\end{landscape}
\restoregeometry

\small
\section*{References}
\begin{description}

\item  Avella-Medina, M. \&  Ronchetti, E. (2018). Robust and consistent variable selection in high-dimensional generalized linear models. \textsl{Biometrika}, \textbf{105},  31-44.

\item Basu, A., Gosh, A., Mandal, A., Martin, N. \& Pardo, L. (2017). A Wald--type test statistic for testing linear hypothesis in logistic regression
models based on minimum density power divergence estimator. \textsl{Electronic Journal of Statistics}, \textbf{11}, 2741-2772.

\item Bianco, A. \& Boente, G. (2002). On the asymptotic behavior of one-step estimates in heteroscedastic regression
models. \textsl{Statistics and Probability Letters}, \textbf{60}, 33-47.

\item  Bianco, A.  \& Mart\'{\i}nez, E. (2009). Robust testing in the logistic regression model. \textsl{Computational Statistics \&  Data Analysis}, \textbf{53},  4095-4105.

\item Bianco, A. \&  Yohai, V. (1996). Robust estimation in the logistic regression model. \textsl{Lecture Notes in Statistics}, \textbf{109}, 17-34. Springer-Verlag, New York.

 \item Bondell, H. D. (2008). A characteristic function approach to the biased sampling model, with application to robust logistic regression. \textsl{Journal of Statistical Planning and Inference}, \textbf{138}, 742-755.

\item B\"uhlmann, P., \& Van De Geer, S. (2011). \textsl{Statistics for high-dimensional data: methods, theory and applications}.  Springer Science \& Business Media. 

\item Cantoni, E. and Ronchetti, E. (2001), Robust inference for generalized linear models. \textsl{Journal of the American Statistical Association}, \textbf{96}, 1022-1030.

\item Chi, E. C., \& Scott, D. W. (2014). Robust parametric classification and variable selection by a minimum distance criterion. \textsl{Journal of Computational and Graphical Statistics}, \textbf{23}, 111-128.

\item Croux, C., \& Haesbroeck, G. (2003). Implementing the Bianco and Yohai estimator for logistic regression. \textsl{Computational statistics \& data analysis}, \textbf{44}, 273-295.

\item Efron, B., \& Hastie, T. (2016). \textsl{Computer age statistical inference, (Vol. 5)}. Cambridge University Press. 

\item Efron, B., Hastie, T., Johnstone, I. \& Tibshirani, R. (2004).  Least Angle Regression. \textsl{Annals
of Statistics}, \textbf{32}, 407-499.

\item Fan, J., \& Li, R. (2001). Variable selection via non--concave penalized likelihood and its oracle properties. \textsl{Journal of the American Statistical Association}, \textbf{96}, 1348-1360.

\item Fernandes, K., Cardoso, J. S., \& Fernandes, J. (2017, June). Transfer learning with partial observability applied to cervical cancer screening. \textsl{In Iberian conference on pattern recognition and image analysis (pp. 243-250). Springer, Cham.}

\item Frank, L. E., \& Friedman, J. H. (1993). A statistical view of some chemometrics regression tools. \textsl{Technometrics}, \textbf{35}, 109-135.

\item Hastie, T., Tibshirani, R. \& Wainwright, M. (2015). \textsl{Statistical Learning with Sparsity: The Lasso and Generalizations}. Chapman and Hall/CRC Monographs on Statistics and Applied Probability.

\item Hoerl, A. E. \& Kennard, R. W. (1970). Ridge regression: biased estimation for nonorthogonal problems. \textsl{Technometrics}, \textbf{12}, 55-67.

\item Kim, J., \& Pollard, D. (1990). Cube root asymptotics. \textsl{Annals of Statistics}, \textbf{18}, 191-219.

\item Knight, K. \& Fu, W. (2000). Asymptotics for Lasso-type estimators. \textsl{Annals of Statistics}, \textbf{28}, 1356-1378.

\item Kurgan, L. A., Cios, K. J., Tadeusiewicz, R., Ogiela, M., \& Goodenday, L. S. (2001). Knowledge discovery approach to automated cardiac SPECT diagnosis. \textsl{Artificial intelligence in medicine}, \textbf{23(2)}, 149-169.

\item Kurnaz, F. S., Hoffmann, I., \& Filzmoser, P. (2018). Robust and sparse estimation methods for high-dimensional linear and logistic regression. \textsl{Chemometrics and Intelligent Laboratory Systems}, \textbf{172}, 211-222.

\item Meinshausen, N. (2007). Relaxed Lasso. \textsl{Computational Statistics and Data Analysis}, \textbf{52}, 374-393.

\item Park, H., \& Konishi, S. (2016). Robust logistic regression modelling via the elastic net-type regularization and tuning parameter selection. \textsl{Journal of Statistical Computation and Simulation,} \textbf{86}, 1450-1461.

\item Pollard, D. (1989). Asymptotics via Empirical Processes. \textsl{Statistical Science}, \textbf{4}, 341-354. 

\item Pregibon, D. (1982). Resistant fits for some commonly used logistic models with medical applications. \textsl{Biometrics}, \textbf{38}, 485-498.

\item Smucler, E. (2016). \textsl{Estimadores robustos para el modelo de regresi\'on lineal con datos de alta dimensi\'on}. PhD. Thesis (in spanish), Universidad de Buenos Aires. Available at \url{http://cms.dm.uba.ar/academico/carreras/doctorado/Tesis\%20Smucler.pdf}

\item Tibshirani, J., \& Manning, C. D. (2013). Robust Logistic Regression using Shift Parameters. \textsl{Proceedings of the 52nd Annual Meeting of the Association for Computational Linguistics}, 124-129.

\item Tibshirani, R. (1996). Regression shrinkage and selection via the lasso. \textsl{Journal of the Royal Statistical Society, Series B (Methodological)}, \textbf{58},  267-288.


\item Tikhonov, A. N. (1963). Solution of incorrectly formulated problems and the regularization method. \textsl{Soviet Mathematics Doklady}, \textbf{4}, 1035-1038. 

\item Van de Geer, S. (2000). \textsl{Empirical processes in $M-$estimation}.  Cambridge Series in Statistical and Probabilistic Mathematics. 

\item Van Der Vaart, A. W., \& Wellner, J. A. (1996). \textsl{Weak convergence and empirical processes}. Springer, New York, NY. 

\item Zhang, C. H. (2010). Nearly unbiased variable selection under minimax concave penalty. \textsl{The Annals of statistics}, \textbf{38}, 894-942.

\item Zou, H. (2006). The adaptive Lasso and its oracle properties. \textsl{Journal of the American Statistical Association}, \textbf{101}, 1418-1429.

\item Zou, H. \& Hastie, T. (2005). Regularization and variable selection via the elastic net.
\textsl{Journal of the Royal Statistical Society: Series B}, \textbf{67},  301-320.
\end{description}

\end{document}

%% file: definiciones.tex
\newcommand\ba {\mathbf a}
\newcommand\bb {\mathbf b}
\newcommand\bc {\mathbf c}

\newcommand\be {\mathbf e}

\newcommand\bq {\mathbf q}
\newcommand\br {\mathbf r}

\newcommand\bu {\mathbf u}
\newcommand\bv {\mathbf v}
\newcommand\bw {\mathbf w}
\newcommand\bx {\mathbf x}
\newcommand\by {\mathbf y}
\newcommand\bz {\mathbf z}

\newcommand\bA {\mathbf A}
\newcommand\bB {\mathbf B}
\newcommand\bC {\mathbf C}

\newcommand\bI {\mathbf I}

\newcommand\indica {\textbf{1}}

\newcommand\wy {\widehat{y}}

\newcommand\wtu {\widetilde{u}}

\newcommand\wtbx {\widetilde{\bx}}
\newcommand\wtbu {\widetilde{\bu}}

\newcommand\wty {\widetilde{y}}
\newcommand\wtz {\widetilde{z}}
\newcommand\wtbw {\widetilde{\bw}}

\newcommand\wtbA {\widetilde{\bA}}

\newcommand\wtbB {\widetilde{\bB}}

\newcommand\itA {{\mathcal{A}}}
\newcommand\itB {{\mathcal{B}}}
\newcommand\itC {{\mathcal{C}}}
\newcommand\itD {{\mathcal{D}}}
\newcommand\itE {{\mathcal{E}}}
\newcommand\itF {{\mathcal{F}}}

\newcommand\itI {{\mathcal{I}}}

\newcommand\itM {{\mathcal{M}}}
\newcommand\itN {{\mathcal{N}}}

\newcommand\itS {{\mathcal{S}}}
\newcommand\itT {{\mathcal{T}}}

\newcommand\itV {{\mathcal{V}}}

\newcommand\balfa {\mbox{\boldmath $\alpha$}}

\newcommand\bbe {\mbox{\boldmath $\beta$}}

\newcommand\bbech {\mbox{\scriptsize${\bbe}$}}
\newcommand\wbbech {\mbox{\scriptsize${\wbbe}$}}

\newcommand\bmu {\mbox{\boldmath $\mu$}}

\newcommand\bthe {\mbox{\boldmath $\theta$}}

\newcommand\bup {\mbox{\boldmath $\upsilon$}}

\newcommand\bSi {\mbox{\boldmath $\Sigma$}}

\newcommand\wbeta {\widehat{\beta}}
\newcommand\wbbe {\widehat{\bbe}}

\newcommand\wgamma {\widehat{\gamma}}

\newcommand\wbmu {\widehat{\bmu}}

\newcommand\wbthe {\widehat{\bthe}}

\newcommand\wbSi {\widehat{\bSi}}

\newcommand\wtbbe {\widetilde{\bbe}}
\newcommand\wtbbech {\mbox{\scriptsize$\wtbbe$}}

\def\real{\mathbb{R}}
\def\natu{\mathbb{N}}

\newcommand{\esp}{\mathbb{E}}
\newcommand{\prob}{\mathbb{P}}

\newcommand{\var}{\mbox{\sc Var}}

\newcommand{\convpp}{ \buildrel{a.s.}\over\longrightarrow}
\newcommand{\convcs}{ \buildrel{c.s.}\over\longrightarrow}
\newcommand{\convprob}{ \buildrel{p}\over\longrightarrow}

\newcommand{\convdist}{ \buildrel{D}\over\longrightarrow}

\newcommand{\traspi}{{\mbox{\footnotesize \sc t}}}
\newcommand{\trasp}{^{\mbox{\footnotesize \sc t}}}

\newcommand\bcero {{\bf{0}}}
\newcommand\buno {{\bf{1}}}

\def\dst{\displaystyle}

\def\argmin{\mathop{\mbox{\rm argmin}}}

\newcommand{\signo}{{\mbox{\rm sign}}}

\newcommand\noi{\noindent}
\def\dst{\displaystyle}

\def\sqr{\vcenter{
         \hrule height.1mm
         \hbox{\vrule width.1mm height2.2mm\kern2.18mm
\vrule width.1mm}
         \hrule height.1mm}}

\newcommand{\ini}{\mbox{\footnotesize \sc ini}}

\newcommand\scadgrande {\mbox{\sc scad}}
\newcommand\mcpgrande {\mbox{\sc mcp}}

\newcommand\mle {\mbox{\scriptsize\sc ml}}
\newcommand\wmle {\mbox{\scriptsize\sc wml}}

\newcommand\lse {\mbox{\scriptsize\sc ls}}
\newcommand\wmlse {\mbox{\scriptsize\sc wls}}

\newcommand\basu {\mbox{\scriptsize \sc div}}
\newcommand\wbasu {\mbox{\scriptsize \sc wdiv}}

\newcommand{\eme}{\mbox{\scriptsize \sc m}}
\newcommand{\weme}{\mbox{\scriptsize \sc wm}}

\newcommand\mch {\mbox{\scriptsize \sc m}}
\newcommand\wmch {\mbox{\scriptsize \sc wm}}

\newcommand\signpen {\mbox{{\scriptsize\sc s}}}

\newcommand\lpen {\mbox{{\scriptsize\sc l}}}

\newcommand\mcppen {\mbox{{\scriptsize\sc mcp}}}